\documentclass[a4paper,11pt]{article}
\pdfoutput=1
\usepackage{jheppub}
\usepackage[utf8]{inputenc}
\usepackage{url}
\usepackage{mathrsfs}
\usepackage{pdflscape}
\usepackage{refcount}
\usepackage{booktabs}
\usepackage{todonotes}
\usepackage{enumitem}
\usepackage{adjustbox}
\usepackage{afterpage}
\usepackage{mathdots}
\usepackage{ytableau, youngtab}
\usepackage{amsfonts}
\usepackage{bbm}
\usepackage{pdfpages}
\usepackage{caption}
\usepackage{subcaption}
\usepackage{tikz}
\usetikzlibrary{positioning}
\usetikzlibrary{arrows}
\usetikzlibrary{decorations.pathreplacing}
\usetikzlibrary{shapes.geometric}
\usetikzlibrary{calc}
\usetikzlibrary{decorations.pathmorphing}
\usetikzlibrary{shapes.misc}
\usetikzlibrary{positioning,trees,decorations.pathmorphing,decorations.markings,decorations.pathreplacing,calc,shapes,patterns,arrows,chains,arrows.meta,fit,fadings,decorations.markings,graphs,graphs.standard,quotes,plotmarks,calligraphy,decorations.pathreplacing}
\usetikzlibrary{intersections, calc}
\usetikzlibrary{shapes.geometric}
\usetikzlibrary{matrix, arrows.meta}
\usetikzlibrary{shapes.geometric}
\usetikzlibrary{decorations.pathmorphing}
\usetikzlibrary{shapes.misc}
\usetikzlibrary{arrows.meta}
\usetikzlibrary{angles, quotes}
\definecolor{goodgreen}{RGB}{55,169,49}
\definecolor{darkyellow}{RGB}{230,170,10}
\definecolor{brightyellow}{RGB}{255,240,190}
\tikzset{flavour/.style={draw=none,minimum size=0.3mm,fill=white, regular polygon,regular polygon sides=4,draw}}
\tikzset{gaugeBig/.style={inner sep=1mm,draw=none,fill=white,minimum size=2mm,circle, draw}}
\tikzset{bd/.style={circle, draw=black, inner sep=0pt, fill=black, minimum size=2mm}}
\tikzset{wd/.style={circle, draw=black, inner sep=0pt, fill=white, minimum size=2mm}}
\tikzset{Dynkin/.style={circle, draw=black, inner sep=0pt, fill=white, minimum size=2mm}}
\tikzstyle{ligne}=[draw, very thick]
\tikzstyle{gridline}=[draw, gray]
\tikzset{gauge/.style={circle, draw,inner sep=2.5pt}}
\tikzset{gaugeo/.style={circle, draw,inner sep=2.5pt,fill=orange}}
\tikzset{gaugec/.style={circle, draw,inner sep=2.5pt,fill=cyan}}
\tikzset{gauger/.style={circle, draw,inner sep=2.5pt,fill=red}}
\tikzset{gaugeb/.style={circle, draw,inner sep=2.5pt,fill=blue}}
\tikzset{gaugeg/.style={circle, draw,inner sep=2.5pt,fill=green}}
\tikzset{gaugem/.style={circle, draw,inner sep=2.5pt,fill=magenta}}
\tikzset{hasse/.style={circle, fill,inner sep=2pt}}
\tikzset{shrinky/.style={circle, fill,inner sep=1pt}}
\tikzset{sized/.style={circle, draw, inner sep=1.5pt}}
\tikzset{seven/.style={circle, draw,inner sep=3pt}}
\tikzset{dotto/.style={circle, orange, draw,inner sep=1.5pt,fill=orange}}
\tikzset{dottp/.style={circle, purple, draw,inner sep=1.5pt,fill=purple}}
\tikzset{dottc/.style={circle, cyan, draw,inner sep=1.5pt,fill=cyan}}
\tikzset{dottr/.style={circle, red, draw,inner sep=1.5pt,fill=red}}
\tikzset{dottb/.style={circle, blue, draw,inner sep=1.5pt,fill=blue}}
\tikzset{dottg/.style={circle, green, draw,inner sep=1.5pt,fill=green}}
\tikzset{dottm/.style={circle, magenta, draw,inner sep=1.5pt,fill=magenta}}

\tikzset{redgauge/.style={draw=none,minimum size=0.4cm,fill=red,circle, draw}}
\tikzset{gauge3/.style={draw=none,minimum size=0.4cm,fill=white,circle, draw}}
\tikzset{bluegauge/.style={draw=none,minimum size=0.4cm,fill=blue,circle, draw}}
\tikzset{redflavor/.style={draw=none,minimum size=0.6cm,fill=red, regular polygon,regular polygon sides=4,draw}}
\tikzset{blueflavor/.style={draw=none,minimum size=0.6cm,fill=blue, regular polygon,regular polygon sides=4,draw}}
\tikzset{flavour2/.style={draw=none,minimum size=0.6cm,fill=white, regular polygon,regular polygon sides=4,draw}}
\tikzset{rede/.style={line width=0.5mm,red}}
\tikzset{bluee/.style={line width=0.5mm,blue}}
\tikzset{pinkline/.style={line width=0.5mm,purple}}
\pgfdeclarelayer{edgelayer}
\pgfdeclarelayer{nodelayer}
\DeclareMathOperator{\U}{U}
\DeclareMathOperator{\SU}{SU}
\DeclareMathOperator{\SO}{SO}

\DeclareMathOperator{\Sp}{Sp}
\DeclareMathOperator{\USp}{USp}
\DeclareMathOperator{\tr}{Tr}

\DeclareMathOperator{\rank}{rank}

\usepackage{epsfig}
\usepackage{amsmath}
\usepackage{amssymb}
\usepackage{amsfonts}
\usepackage{amsxtra}
\usepackage{amsthm}
\usepackage{mathrsfs}
\usepackage{makeidx}
\usepackage{graphics}
\usepackage{dsfont}
\usepackage{mathtools}
\usepackage{graphicx}
\usepackage{placeins}
\usepackage{bm}
\usepackage[capitalise]{cleveref}
\usepackage{empheq}
\usepackage{colortbl}
\usepackage{xcolor}
\usepackage{titlesec}
\usepackage{longtable}
\usepackage{hhline}
\usepackage{float}
\usepackage{color}
\usepackage{tikz}
\usepackage{xfrac}
\usepackage{footnote}
\usepackage{rotating}
\usepackage{lscape}
\usepackage{makecell}
\usepackage{environ}
\usepackage{tabularx}
\usepackage{subfiles}
\usepackage{ytableau}
\usepackage{tikz-3dplot}
\usepackage{slashed}
\usepackage{pifont}
\usepackage{multirow}
\usepackage{mdframed}
\usepackage{bbm}
\usepackage[normalem]{ulem}
\usepackage{arydshln}
\usepackage{enumitem}
\usepackage{fancybox}
\usetikzlibrary{positioning,trees,decorations.pathmorphing,decorations.markings,decorations.pathreplacing,calc,shapes,patterns,arrows,chains,arrows.meta,fit,fadings,decorations.markings,graphs,graphs.standard,quotes,plotmarks,calligraphy,decorations.pathreplacing}

\usetikzlibrary{positioning}
\usetikzlibrary{chains}
\usetikzlibrary{arrows, arrows.meta ,fit,decorations.pathreplacing}
\tikzstyle{every picture}+=[remember picture]
\tikzstyle{na} = [baseline]
\tikzstyle{ligne}=[draw, thick]

\usetikzlibrary{arrows, decorations.markings, calc, fadings, decorations.pathreplacing, patterns, decorations.pathmorphing, positioning}
\tikzset{>={Latex[width=1.5mm,length=1.5mm]}}
\tikzset{bd/.style={circle, draw=black, inner sep=0pt, fill=black, minimum size=1.2mm}}
\tikzset{bld/.style={circle, draw=blue, inner sep=0pt, fill=blue, minimum size=1.2mm}}
\tikzset{wd/.style={circle, draw=black, inner sep=0pt, fill=white, minimum size=1.2mm}}
\tikzset{rd/.style={circle, draw=red, inner sep=0pt, fill=red, minimum size=.9mm}}
\tikzset{wrd/.style={circle, draw=red, inner sep=0pt, fill=white, minimum size=.9mm}}
\usetikzlibrary{graphs,graphs.standard,quotes}

\tikzstyle{every picture}+=[remember picture]
\tikzstyle{na} = [baseline=-.5ex]

\numberwithin{equation}{section}

\newcommand{\be}{\begin{equation}} \newcommand{\ee}{\end{equation}}
\newcommand{\bea}{\begin{equation} \begin{aligned}} \newcommand{\eea}{\end{aligned} \end{equation}}

\def\tilde{\widetilde}

\def\hat{\widehat}

\def\bar{\overline}

\def\b{\beta}

\def\rt2{\sqrt{2}}

\def\det{\mathop{\rm det}}

\def\Tr{\mathrm{Tr}}
\def\tr{\mathop{\rm tr}}

\def\fa{\mathfrak{a}}
\def\fb{\mathfrak{b}}

\def\1{{\ds 1}}

\newcommand{\fm}{\mathfrak{m}}
\newcommand{\fn}{\mathfrak{n}}

\newcommand{\ID}{\mathds{1}}

\newcommand{\fq}{\mathfrak{q}}
\newcommand{\fc}{\mathfrak{c}}

\def\SO{\mathrm{SO}}

\def\SU{\mathrm{SU}}

\def\Sp{\mathrm{Sp}}
\def\Spin{\mathrm{Spin}}

\def\su{\mathfrak{su}}

\def\so{\mathfrak{so}}
\def\sp{\mathfrak{sp}}
\def\usp{\mathfrak{usp}}

\def\fr{\mathfrak{r}}

\def\repa{\raise4pt\hbox{$\square$}\mkern-14mu\raise-4pt\hbox{$\square$}}
\def\repab{\overline{\raise4pt\hbox{$\square$}\mkern-14mu\raise-4pt\hbox{$\square$}\mkern-1mu}}
\def\smileface{\ensuremath{\hbox{\large$\bigcirc$}\mkern-15mu\raise-1pt\hbox{\scriptsize$\smallsmile$}%
\mkern-10mu\raise4pt\hbox{..}\mkern4mu}}
\def\frownface{\ensuremath{\hbox{\large$\bigcirc$}\mkern-15mu\raise-1pt\hbox{\scriptsize$\smallfrown$}%
\mkern-10mu\raise4pt\hbox{..}\mkern4mu}}

\newcommand{\ba}{\begin{array}}
\newcommand{\ea}{\end{array}}
\newcommand{\bi}{\begin{itemize}}
\newcommand{\ei}{\end{itemize}}

\def\bea#1\eea{\allowdisplaybreaks \begin{align}#1\end{align}}
 \newcommand{\ben}{\begin{enumerate}}
\newcommand{\een}{\end{enumerate}}
\newcommand{\bean}{\begin{eqnarray*}}
\newcommand{\eean}{\end{eqnarray*}}

\newcommand{\sref}[1]{\S\ref{#1}}
\newcommand{\tref}[1]{Table~\ref{#1}}

\newcommand{\tw}{\widetilde{w}}

\newcommand{\BC}{\mathbb{C}}
\newcommand{\BR}{\mathbb{R}}
\newcommand{\BP}{\mathbb{P}}

\newcommand{\BZ}{\mathbb{Z}}
\newcommand{\BF}{\mathbb{F}}

\newcommand{\diag}{\mathrm{diag}}

\newcommand{\adj}{\mathbf{Adj}}

\definecolor{light-gray}{gray}{0.5}
\definecolor{new-green}{rgb}{0,0.7,0.3}
\definecolor{cerulean}{rgb}{0.0, 0.48, 0.65}
\definecolor{claret}{rgb}{0.50, 0.09, 0.20}
\definecolor{darkred}{rgb}{0.7, 0.11, 0.11}
\definecolor{scarlet}{rgb}{1.0, 0.13, 0.0}
\definecolor{orange-red}{rgb}{1.0, 0.27, 0.0}
\definecolor{blue-green}{rgb}{0.0, 0.5, 0.65}
\definecolor{green-red}{rgb}{0.5, 0.65, 0.0}

\def\aup#1 {\overset{#1}{\uparrow} \, \overset{\tilde{#1}}{\downarrow}}

\tikzset{snake it/.style={decorate, decoration={snake, amplitude=.4mm, segment length=2mm,
                       post length=0mm,pre length=0mm}}}

\def\u{\mathfrak{u}}
\DeclareMathAlphabet{\mymathds}{U}{BOONDOX-ds}{m}{n}
\tikzstyle{double_border} = [draw, double, double distance=1pt]

\colorlet{verdescuro}{green!50!black}

\theoremstyle{plain}

\newcommand{\cw}{\mathrm{cw}}
\newcommand{\crt}{\mathrm{cr}}   % coroot lattice subscript (NB: \cr is a TeX primitive)
\def\PSU{\mathrm{PSU}}
\newcommand{\kCS}{k_{\mathrm{CS}}}

\usepackage{array}
\usepackage{makecell}

\newcommand{\ngen}{\fn_{\rm gen}}
\newcommand{\Bmz}{B_{m_0}}
\newcommand{\Fund}{\mathbf{F}}
\newcommand{\Kmat}[4]{\left(\begin{smallmatrix} #1 & #2 \\ #3 & #4 \end{smallmatrix}\right)}
\newcolumntype{L}[1]{>{\raggedright\arraybackslash}p{#1}}
\newcolumntype{C}[1]{>{\centering\arraybackslash}p{#1}}

\title{Generalised global symmetries of 6d SCFTs on a circle from the blow-up equations}

\author[a,b]{William Harding}
\author[b,c]{and Noppadol Mekareeya}
\affiliation[a]{Dipartimento di Fisica, Universit\`a di Milano-Bicocca, Piazza della Scienza 3, I-20126 Milano, Italy}
\affiliation[b]{INFN, sezione di Milano-Bicocca, Piazza della Scienza 3, I-20126 Milano, Italy}
\affiliation[c]{The Institute for Fundamental Study ``The Tah Poe Academia Institute'', Naresuan University, Phitsanulok 65000, Thailand}
\emailAdd{w.harding@campus.unimib.it}
\emailAdd{n.mekareeya@gmail.com}

\abstract{We study 6d $(1,0)$ SCFTs on a circle, possibly twisted, through the blow-up equations of their 5d descriptions. For every theory we consider, we construct the explicit map between the 6d and 5d frames, including the ratio $r$ of the Kaluza--Klein (KK) mass to the instanton mass, which depends on the frame and not only on the twist. Our main result is that the higher-form symmetries of these theories, their mixed anomalies and 2-groups can all be read off from the fluxes through the exceptional curve of the blown-up $\BC^2$. The flux sectors are the 't~Hooft sectors of the 5d 1-form symmetries. On the tensor fluxes, the 6d Green--Schwarz couplings fix the weights, whose fractional exponents we read as the mixed anomalies of the 6d 2-form symmetry with the KK $\U(1)$, gravity, flavour and $\SU(2)_R$. The fluxes forced in each sector decide between a 2-group and a mixed anomaly. For instance, the twisted frames $\su(4)_{0,4}+4\Lambda^2$ carry a non-split $\BZ_4$ 2-group, while, in $\su(4)_2+4\Lambda^2$ and $\su(4)_6+2\Lambda^2$, the instanton symmetry is locked to the flavour centre and only a mixed anomaly remains. The same method applies to the $\SU(2)$ gauging of two, three and four copies of the $E_1$ theory, which describes, respectively, $\SU(4)_0$, pure $\Spin(8)$ and the untwisted $\so(8)$ KK theory. There, the fluxes exhibit the extensions of 1-form symmetries inherited from the $E_1$ 2-groups, identify the states that break each accidental 1-form symmetry and the background to which it is locked, and give the 5d 2-group to which the $S_3$-invariant 6d 3-group reduces.}

\begin{document}
\maketitle

% ===========================================================================
% Section 1: Introduction
% ===========================================================================
\section{Introduction}
\label{sec:intro}

Six-dimensional $(1,0)$ superconformal field theories (SCFTs) have a tensor branch, on which the scalars of $n_T$ tensor multiplets acquire vacuum expectation values and the light degrees of freedom include strings \cite{Witten:1995zh,Ganor:1996mu,Seiberg:1996vs,Seiberg:1996qx,Heckman:2013pva,Heckman:2015bfa}. The string charges form the lattice $\BZ^{n_T}$. Its pairing $\mathcal K$ is positive definite, minus the intersection form of the compact curves of the F-theory base. Surface defects whose charges cannot be screened by the dynamical strings are classified by the defect group $D=\BZ^{n_T}/\mathcal K\BZ^{n_T}$ \cite{DelZotto:2015isa}. It is the group of a discrete 2-form symmetry \cite{Gaiotto:2014kfa}, and this symmetry is anomalous. The 6d theory is relative: it lives on the boundary of a 7d 3-form Chern--Simons theory of level $\mathcal K$ \cite{Gukov:2020btk,Apruzzi:2022dlm}. The 7d theory assigns to each class $\lambda\in D$ a spin $q(\lambda)$, \eqref{eq:qdef}. When a diagonal entry of $\mathcal K$ is odd, the 7d theory needs a Wu structure, and the spin is refined to $q_w(\lambda)$, with $w$ a characteristic vector of $\mathcal K$, \eqref{eq:qw}.

On a circle of radius $1/\tau$, possibly with a twist by a discrete symmetry, a 6d SCFT becomes a 5d Kaluza--Klein (KK) theory \cite{Jefferson:2018irk,Bhardwaj:2018yhy,Bhardwaj:2018vuu,Bhardwaj:2019fzv,Bhardwaj:2020kim}. Twists are of two kinds \cite[\S2.3]{Kim:2020hhh}, \cite[\S3.3.2]{Bhardwaj:2020phs}. A twist of the \emph{first kind} acts as an outer automorphism of the gauge algebra on a curve, e.g.\ charge conjugation of $\su(3)$ on a $-3$ curve. A twist of the \emph{second kind} permutes tensor nodes, e.g.\ it exchanges the two $-2$ curves of the $(2,0)$ $A_2$ theory. Every example in this paper has no twist or a twist of the first kind. Twists of the second kind enter only in remarks (\sref{sec:rrule}, \sref{sec:exMstring} and \sref{sec:conclusions} below), and we leave them open. Many KK theories have one or several 5d gauge-theory descriptions \cite{Rozali:1997cb,Berkooz:1997cq,Seiberg:1997ax,Douglas:2010iu,Lambert:2010iw,Tachikawa:2011ch,Hayashi:2015fsa,Yonekura:2015ksa,Zafrir:2015rga,Hayashi:2015zka,Hayashi:2015vhy,Jefferson:2017ahm,Jefferson:2018irk,Hayashi:2018lyv,Razamat:2018gro,Bhardwaj:2020gyu}, valid in chambers of the extended Coulomb branch. We call each a 5d frame of the KK theory, and the tensor branch on the circle its 6d frame.

A surface defect wrapped on the circle is a line in 5d, so $D$ gives a 1-form symmetry of the KK theory \cite{Bhardwaj:2020phs,Morrison:2020ool,Albertini:2020mdx}. The 7d Chern--Simons theory reduces to a mixed anomaly between this 1-form symmetry and the KK $\U(1)$ \cite[\S4.2]{Apruzzi:2022dlm}. The sector with 't~Hooft flux of class $\lambda$ carries fractional KK momentum $q(\lambda)$. Suppose that, in a 5d frame, the same 1-form symmetry acts as a centre symmetry of the gauge group. Then, the sector is an 't~Hooft flux of the gauge group, and it carries a fractional instanton number $k$ \cite{BenettiGenolini:2020doj,Apruzzi:2021nmk,Harding:2026gzj}. In terms of the KK mass $\tau$, the instanton mass is $m_0=\tau/r$, \eqref{eq:rdef}. So the instanton weight $e^{-km_0}$ of the sector carries KK momentum $k/r$, and one expects $q(\lambda)=k/r$ mod 1. For the untwisted $(2,0)$ theories, the instanton particles are the KK modes, $r=1$, and the fractional instanton numbers reproduce the spins \cite{Gukov:2020btk,BenettiGenolini:2020doj,Apruzzi:2021nmk}. For twisted reductions, and for $(1,0)$ theories in general, it is not known what $r$ is, nor in which precise form the relation holds.

We study this relation, and the global structure of KK theories around it, with the blow-up equations. These express the Nekrasov partition function on $\BC^2$ through that on its blow-up at the origin, $\widehat{\BC}{}^2$, as a sum over magnetic fluxes through the exceptional curve $E$, which is isomorphic to $\BP^1$ \cite{Nakajima:2003pg,Nakajima:2005fg,Keller:2012da}. They determine the BPS spectra of 5d SCFTs, of KK theories and of 6d strings \cite{Huang:2017mis,Gu:2018gmy,Gu:2019dan,Gu:2019pqj,Gu:2020fem,Kim:2019uqw,Kim:2020hhh,Kim:2021gyj,Kim:2023glm}. Each flux $n$ enters with a factor $e^{-V(n)}$ fixed by the effective prepotential, \eqref{eq:Vdef}. The admissible fluxes form a coset, set by the background fluxes on $E$, and a fractional coset is an 't~Hooft flux of a 1-form symmetry. In the companion paper \cite{Harding:2026gzj}, we showed that the exponents of $e^{-V(n)}$ on such cosets encode the mixed anomalies of 5d gauge theories and SCFTs, and decide whether a 1-form symmetry forms a 2-group \cite{Cordova:2018cvg,Benini:2018reh} with the flavour symmetry. For KK theories, two features are new. First, the effective prepotential of the 6d frame is fixed by the 6d Green--Schwarz couplings~\cite{Ohmori:2014kda} and by the KK towers \cite{Kim:2020hhh,Bhardwaj:2019fzv}. Second, the fluxes along the tensor directions alone form a family, the {\it tensor-flux family}, on which $V$ does not depend on the Coulomb moduli. Its sum is a prefactor $\Lambda_B$, one for each sector, labelled by a class $B\in\mathcal K^{-1}\BZ^{n_T}/\BZ^{n_T}\cong D$.

Several ingredients are known. The elliptic blow-up equations of 6d SCFTs \cite{Gu:2018gmy,Gu:2019dan,Gu:2019pqj,Gu:2020fem}, which build on \cite{Huang:2017mis}, have theta functions of the string lattice as prefactors $\Lambda_B$. Those of 5d frames of KK theories are in \cite{Kim:2020hhh,Kim:2021gyj}. 2-groups were studied for 6d SCFTs \cite{Cordova:2020tij,Apruzzi:2021mlh}, little string theories \cite{DelZotto:2020sop}, 5d SCFTs \cite{Apruzzi:2021vcu,DelZotto:2022joo,Cvetic:2022imb} and untwisted circle reductions \cite[\S4.2]{Cvetic:2022imb}. For 5d SCFTs from M-theory on orbifolds and Calabi--Yau cones, the higher-form symmetries and their anomalies also follow from quiver quantum mechanics \cite{DelZotto:2022fnw}, from $\eta$-invariants of the boundary geometry \cite{Cvetic:2025lat} and from the path algebra of the probe quiver \cite{Chakrabhavi:2026iku}.\footnote{See also \cite{Akhond:2026hhl} for a recent analysis of 1-form symmetries from 5-brane webs.} The faithful flavour groups of 6d SCFTs follow from the tensor branch \cite{Heckman:2022suy}. Finally, \cite{Zafrir:2025xca} studied the 6d $\so$-type SCFTs on a circle through an $\SU(2)$ gauging of copies of the $E_1$ SCFT \cite{DelZotto:2015rca,Hayashi:2017jze}.

This paper addresses four questions. In which precise form does the spin of $\lambda\in D$ equal the KK momentum of the matching 5d 't~Hooft sector, and what fixes $r$? What is $\Lambda_B$ in 6d terms, and which anomalies does it encode? What is the faithful 0-form symmetry of a 5d frame, and does the 1-form symmetry form a 2-group with it? When an $\SU(2)$ gauges copies of $E_1$, which 1-form symmetries of this description are genuine, and how do 6d higher groups appear?

\paragraph{Main results.} Let us summarise the main results of this paper as follows.
\begin{itemize}[leftmargin=*,itemsep=2pt]
\item \emph{The frame map and $r$} (\sref{sec:framemap}). The variables of the 6d frame and of a 5d frame are related by a linear map, \eqref{eq:framemapgen}, fixed monomial by monomial (Table~\ref{tab:framemaprecipe}). Without a twist or with a twist of the first kind, $m_0=\delta/k(\delta)$, with $\delta$ the mass of the lightest neutral KK mode and $k(\delta)$ its instanton number, \eqref{eq:m0delta}. So {\it $r$ depends on the frame, not only on the twist.} It is $2$ for the untwisted $\su(4)_2+4\Lambda^2$, and $3$ and $6$ for $\sp(3)_0+2\Lambda^2$ and $\su(4)_6+2\Lambda^2$, two frames of one KK theory.\footnote{We write $\sp(N)=\usp(2N)$ for the algebras.}

\item \emph{The blow-up exponent from 6d} (\sref{sec:prefactor}). On the tensor-flux family, $V$ is quadratic in the tensor fluxes, with the Green--Schwarz couplings as coefficients, \eqref{eq:Vfamily}. The family exists exactly when these couplings obey \eqref{eq:familycond}, which in our examples with a 6d gauge algebra is the cancellation of the mixed gauge--gravitational anomaly, \eqref{eq:anomalygg}. The linear part of $V$ has fractional exponents in the non-trivial sectors, \eqref{eq:sectorexponents}. We read them as mixed anomalies of the 2-form symmetry with gravity and $\SU(2)_R$, and with flavour, $\SU(2)_R$ and Lorentz symmetry together. Conversely, given $r$, the prefactor $\Lambda_B$ of a one-tensor frame determines its Green--Schwarz couplings, \eqref{eq:inversion}.

\item \emph{The prefactor $\Lambda_B$ as a lattice theta function} (\sref{sec:prefactor}). The prefactor $\Lambda_B$ is the theta function of the string lattice of \cite{Gu:2019pqj,Gu:2020fem}, with characteristics $B$ and $\tfrac12w$, \eqref{eq:LambdaTheta}. New is the reading of the characteristics. $B$ is the 't~Hooft class of the 2-form symmetry. $w$ is a Wu vector of $\mathcal K$, supplied by the sign of the blow-up equation, \eqref{eq:signw}, and congruent mod 2 to the gravitational Green--Schwarz coupling, as in 6d supergravity \cite{Monnier:2017oqd,Monnier:2018nfs}. For a 6d gauge algebra whose centre does not survive the twist and the matter, all $|D|$ sectors are expected to be blow-up equations of a single 5d theory, as in the tested cases. This criterion is not necessary (\sref{sec:exso8Z2}). For $|D|\ge3$, the sectors form a linear system at one-string order, solved by Cramer's rule when its determinant is non-zero, \eqref{eq:cramer}, as in \cite[(4.116)]{Kim:2021gyj} for $\SU(3)_9$ and \cite[(4.1)--(4.5)]{Gu:2020fem}. For $\SU(4)_8$, with $|D|=4$, the three sectors $B=0,\pm\tfrac14$ already fix the one-string genus \cite[\S6.1]{Kim:2020hhh}. We check, as a series in $e^{-\tau}$, that it also satisfies the equation of the fourth sector (\sref{sec:exSU48}).

\item \emph{The dictionary} (\sref{sec:pfspin}, \sref{sec:examples}). For no twist or a twist of the first kind, the frame map and the prefactor $\Lambda_B$ give the dictionary \eqref{eq:dictionary}. The KK momentum $k(\nu)/r$ of the 5d 't~Hooft sector matching $\lambda\in D$ equals the spin $q(\lambda)$. Here, $\nu$ is the flux of the sector on the tensor-flux family and $k$ the instanton number \eqref{eq:kQ}. For odd $\mathcal K$, the sign of the blow-up equation refines this to the Wu-refined spin mod 1, \eqref{eq:sequalsqw}. So the 6d spins and the 5d instanton numbers constrain $r$. For $\SU(4)_8$, $k=\tfrac38$ and the generator of $D=\BZ_4$ has spin $\tfrac18$. Since $r$ is a positive integer, this fixes $r=3$, the value the frame map gives. The frame maps of $\su(4)_{0,2,4}+4\Lambda^2$, $\so(7)+5\Fund$, $\sp(3)_0+2\Lambda^2$ and $\su(4)_6+2\Lambda^2$ are new.

\item \emph{Faithful groups and 2-groups} (\sref{sec:faithful}). For seven frames of Table~\ref{tab:master} with a $\BZ_2$ 1-form symmetry of tensor origin, we find the faithful 0-form symmetry and decide between a 2-group and a mixed anomaly. The inputs are the centre twists allowed in 6d, \eqref{eq:centretwist}, the instanton parity read off the frame map, \eqref{eq:parity}, and the forced-flux criterion of \cite{Harding:2026gzj}, \eqref{eq:forcedflux}. The twisted frames $\su(4)_{0,4}+4\Lambda^2$ carry a non-split $\BZ_4$ 2-group, with flavour group $\Sp(4)/\BZ_2$, where $\Sp(N):=\USp(2N)$. It has no 6d 2-group as parent. In $\su(4)_2+4\Lambda^2$ and $\su(4)_6+2\Lambda^2$, the instanton parity locks $-1$ in the flavour group to $e^{i\pi}$ in the instanton symmetry $\U(1)_I$. The 2-group of 5d perturbation theory is then accidental, and a mixed anomaly with $\U(1)_I$ remains. The frames $\sp(3)_0+2\Lambda^2$ and $\su(4)_6+2\Lambda^2$ of one KK theory have non-isomorphic faithful groups (Table~\ref{tab:faithfulresults}).

\item \emph{$\SU(2)$ gauging $N$ copies of $E_1$} (\sref{sec:E1}). An $\SU(2)$ gauging the diagonal $\SO(3)$ instanton symmetry of $N$ copies of $E_1$ describes $\SU(4)_0$ for $N=2$, the untwisted $\so(8)$ KK theory on a $-4$ curve for $N=4$, and pure $\Spin(8)$ for $N=3$ \cite{Hayashi:2017jze,Zafrir:2025xca}. One toolkit, built on the flux through the exceptional curve $E$ seen by the particle on a curve, \eqref{eq:pairingdef}, treats all three. For $N=4$, the frame map gives $r=2$. The $\BZ_4$ 1-form symmetry is the tensor class, the $\BZ_2\times\BZ_2$ is the 6d gauge centre, and a further $\BZ_2$ is accidental, locked to the parity of the KK flux. The $S_3$-invariant form of the 6d 3-group, \eqref{eq:so8threegroup}, reduces to a 2-group with the 0-form $\BZ_2\times\BZ_2$, \eqref{eq:so8twogroup}, which the sector labels reproduce. For $N=2$, a $\BZ_2$ of the $E_1$ description is accidental. For $N=3$, the $\BZ_4$ is. Each is screened by central instantons of odd charge and locked to $\U(1)_I$ (Table~\ref{tab:Nsummary}).
\end{itemize}

\paragraph{Plan.} \sref{sec:setup} fixes the 6d and 5d data and states the dictionary. \sref{sec:framemap} derives the frame map and the rule for $r$. \sref{sec:prefactor} derives the prefactor $\Lambda_B$, the dictionary and the condition under which the sectors give the one-string genus in closed form, and reads the mixed anomalies off its exponents. \sref{sec:examples} works out the frames of Table~\ref{tab:master}, with $\SU(3)_9$ in full. They have twists of order one to three, one or two tensors, and even or odd $\mathcal K$. \sref{sec:faithful} determines the faithful groups and the 2-groups. \sref{sec:E1} sets up a toolkit for $\SU(2)$ gaugings of $E_1$'s and treats two copies, $\SU(4)_0$, as a warm-up, then four copies, the untwisted $\so(8)$ KK theory on a $-4$ curve, and three copies, pure $\Spin(8)$. \sref{sec:conclusions} lists open problems. The appendices explain when a twist kills the gauge centre (Appendix~\ref{app:twists}), give the one-string blow-up equations of $\SU(3)_9$ (Appendix~\ref{app:su39}) and technical details of the $\SU(4)_0$ case of \sref{sec:E1N2} (Appendix~\ref{app:su40}), and collect a finite-difference lemma and the chamber scans (Appendix~\ref{app:technical}).

% ===========================================================================
% Section 2: Setup
% ===========================================================================
\section{Setup}
\label{sec:setup}

Let us summarise the 6d and 5d data which are relevant throughout this paper.

\subsection{6d data: defect group, spin and Wu refinement}
\label{sec:setup6d}

On its tensor branch, a 6d $(1,0)$ SCFT has $n_T$ tensor multiplets, one for each compact curve $C_\alpha$ in the base of its F-theory geometry. String charges $\lambda\in\BZ^{n_T}$ are paired by the negative definite intersection form $\Omega_{\alpha\beta}=C_\alpha\cdot C_\beta$. We use $\mathcal K:=-\Omega$, so a single curve of self-intersection $-3$ has $\mathcal K=(3)$. The 2-form symmetry and its anomaly are described by the 7d Chern--Simons theory \cite[(2.3)]{Gukov:2020btk}
\begin{equation}
\label{eq:CS7d}
S_{\rm 7d}=\frac1{4\pi}\int\mathcal C_3^T\,\mathcal K\,d\mathcal C_3 \,,
\end{equation}
with $\mathcal C_3=(\mathcal C_3^\alpha)$ $\U(1)$-valued 3-form fields, whose boundary values are the backgrounds of the 2-form symmetry. The sign of $\mathcal K$ fixes the orientation. The opposite one would flip every spin below. The operators $e^{i\oint\lambda\cdot\mathcal C_3}$ with $\lambda\in\mathcal K\BZ^{n_T}$ are trivial, so the 3-dimensional operators, and the 't~Hooft sectors of the 2-form symmetry, are labelled by the defect group~\cite{DelZotto:2015isa}, \cite[(2.4)]{Gukov:2020btk}
\begin{equation}
\label{eq:Ddef}
D=\BZ^{n_T}/\mathcal K\BZ^{n_T}\,,\qquad |D|=|\det\mathcal K| \,.
\end{equation}
The braiding of the operators defines the linking pairing $\ell$, which is refined by the spin $q$~\cite[(2.5)--(2.8)]{Gukov:2020btk}:
\begin{align}
\label{eq:qdef}
q(\lambda)&=\tfrac12\,\lambda^T\mathcal K^{-1}\lambda\pmod 1 \,,\\
\label{eq:linking}
\ell(\lambda,\lambda')&=q(\lambda+\lambda')-q(\lambda)-q(\lambda')=\lambda^T\mathcal K^{-1}\lambda'\pmod 1 \,.
\end{align}
The pairing $\ell$ is well defined on $D$, but $q$ need not be. Under a change of representative by $\mathcal K\mu$, with $\mu\in\BZ^{n_T}$,
\begin{equation}
\label{eq:qshift}
q(\lambda+\mathcal K\mu)-q(\lambda)=\lambda^T\mu+\tfrac12\,\mu^T\mathcal K\mu \,,\qquad \mu^T\mathcal K\mu\equiv\sum_\alpha\mathcal K_{\alpha\alpha}\,\mu_\alpha\pmod 2 \,.
\end{equation}
We call $\mathcal K$ \emph{even} if all its diagonal entries are even, and \emph{odd} otherwise.\footnote{The off-diagonal entries $\mathcal K_{\alpha\beta}$ do not matter, since they enter $\mu^T\mathcal K\mu$ in \eqref{eq:qshift} with a factor of $2$, and $\mu_\alpha^2\equiv\mu_\alpha$ mod 2.} For even $\mathcal K$, $q$ is a function on $D$ with values mod $1$. For odd $\mathcal K$, it is only defined mod $\tfrac12$, as for the representatives $1$ and $-2$ of $\mathcal K=(3)$, with $q=\tfrac16$ and $\tfrac23$ (Table~\ref{tab:lattices}). The 7d theory then needs a Wu structure \cite[(2.14), (2.15)]{Gukov:2020btk}. As for spin Chern--Simons theories \cite{Belov:2005ze,Monnier:2018nfs}, the Wu structure enters through a \emph{characteristic} (or \emph{Wu}) vector $w$, which refines the spin to
\begin{equation}
\label{eq:qw}
q_w(\lambda)=\tfrac12\,\lambda^T\mathcal K^{-1}\lambda-\tfrac12\,\lambda^T\mathcal K^{-1}w\pmod 1 \,,\qquad w\in\BZ^{n_T}\,,\quad w_\alpha\equiv\mathcal K_{\alpha\alpha}\pmod 2 \,.
\end{equation}
By \eqref{eq:qshift}, $q_w(\lambda+\mathcal K\mu)-q_w(\lambda)=\lambda^T\mu+\tfrac12\big(\mu^T\mathcal K\mu-w\cdot\mu\big)\in\BZ$.\footnote{\label{fn:wuintegral}By \eqref{eq:qshift} and $w_\alpha\equiv\mathcal K_{\alpha\alpha}$ mod 2, $\mu^T\mathcal K\mu\equiv w\cdot\mu$ mod 2.} So $q_w$ is a function on $D$ with values mod $1$, and it refines $\ell$. Since $q_w(-\lambda)-q_w(\lambda)=\ell(\lambda,w)$, it is invariant under charge conjugation if and only if $w\in\mathcal K\BZ^{n_T}$. Such $w$ always exist.\footnote{For $\xi$ in the kernel of $\mathcal K$ mod 2, $\xi\cdot{\rm diag}\,\mathcal K\equiv\xi^T\mathcal K\xi\equiv0$. Since $\mathcal K$ is symmetric, ${\rm diag}\,\mathcal K$ then lies in the image of $\mathcal K$ mod 2, ${\rm diag}\,\mathcal K\equiv\mathcal K\mu$, and $w=\mathcal K\mu$ is characteristic.} Shifting $w$ by $2\mathcal K\zeta$, with $\zeta\in\BZ^{n_T}$, gives
\begin{equation}
\label{eq:qwshiftw}
q_{w+2\mathcal K\zeta}(\lambda)-q_w(\lambda)=-\tfrac12\,\lambda^T\mathcal K^{-1}(2\mathcal K\zeta)=-\lambda^T\zeta\in\BZ \,.
\end{equation}
So $w$ and $w+2\mathcal K\zeta$ give the same $q_w$. For even $\mathcal K$, $w=0$ gives $q_0=q$. Table~\ref{tab:lattices} lists the lattices of this paper, and \sref{sec:pfspin} derives which refinement the blow-up equations see.

\begin{table}[t]
\centering\small
\setlength{\tabcolsep}{4pt}
\renewcommand{\arraystretch}{1.1}
\begin{tabular}{lcllll}
\toprule
$\mathcal K$ & $D$ & $\lambda$ & $q(\lambda)$ & $q_w(\lambda)$, $w\in\mathcal K\BZ^{n_T}$ & used in \\
\midrule
$(2)$ & $\BZ_2$ & $1$ & $\tfrac14$ & $\tfrac14$ $(w=0)$,\ $\tfrac34$ $(w=2)$ & \S\S\ref{sec:exZ2twist}, \ref{sec:exsu40}, \ref{sec:exsu42}, \ref{sec:exso8tri} \\
$(3)$ & $\BZ_3$ & $1$;\ $-2$ & $\tfrac16$;\ $\tfrac23$ & $\tfrac23$ $(w=3)$ & \sref{sec:exSU39} \\
$(4)$ & $\BZ_4$ & $1$ & $\tfrac18$ & $\tfrac18$ $(w=0)$,\ $\tfrac58$ $(w=4)$ & \S\S\ref{sec:exSU48}, \ref{sec:exso8Z2}, \ref{sec:E1N4} \\
$\Kmat{1}{-1}{-1}{3}$ & $\BZ_2$ & $(0,1)$;\ $(1,0)$ & $\tfrac14$;\ $\tfrac34$ & $\tfrac34$ $(w=(1,1))$,\ $\tfrac14$ $(w=(1,-1))$ & \sref{sec:exsu44} \\
Cartan of $\mathfrak g$ & $Z(\tilde G)$ & weight $\omega$ & $\tfrac12(\omega,\omega)$ & $q$ $(w=0)$ & $(2,0)$ theories \\
\bottomrule
\end{tabular}
\caption{String charge lattices, with $D$ of \eqref{eq:Ddef}, $q$ of \eqref{eq:qdef} and $q_w$ of \eqref{eq:qw} modulo $w\to w+2\mathcal K\BZ^{n_T}$. Odd lattices list two representatives of the generator. $\Kmat{1}{-1}{-1}{3}$ is a $-1$ curve meeting a $-3$ curve with $\su(3)$ \cite[(2.37)]{Bhardwaj:2020gyu}. In the last row, $\tilde G$ is simply connected and $\omega$ is a fundamental weight in the class, e.g.\ $q(\omega_k)=\tfrac{k(N-k)}{2N}$ for $A_{N-1}$ \cite[\S2.1]{Gukov:2020btk}.}
\label{tab:lattices}
\end{table}

\subsection{5d data: the blow-up function}
\label{sec:setup5d}

Consider a 5d $\mathcal N=1$ gauge theory with simple, simply connected gauge group $G$, Coulomb moduli $\phi_i$ in the basis of simple coroots, and hypermultiplets $h$ in representations $\mathbf R_h$ with masses $m_h$ (labelled $f$ in \sref{sec:setupdict}, \sref{sec:framemap} and Appendix~\ref{app:technical}), on the $\Omega$-background $\epsilon_{1,2}$, with $\epsilon_+=\tfrac12(\epsilon_1+\epsilon_2)$ and $p_i=e^{-\epsilon_i}$. Throughout, $m_0$ is the mass of the unit instanton, and $q=e^{-m_0}$. The mass $m_0$ is the inverse gauge coupling, the coefficient of $\tfrac12\tilde K_{ij}\phi_i\phi_j$ in the prepotential \eqref{eq:E5d} below. Here, $\tilde K_{ij}=(\alpha_i^\vee,\alpha_j^\vee)$ is the matrix of inner products of the simple coroots, with the long roots of length squared $2$, \eqref{eq:Ktilde} below. For simply laced $G$, it is the Cartan matrix. The blow-up equations, written explicitly in \eqref{eq:blowupeq} below, involve a sum over gauge fluxes $n$ through the exceptional curve $E$ of the blown-up $\BC^2$. The background fluxes $\Bmz$ and $B_h$ of $m_0$ and $m_h$ are fixed, and each $n$ comes with a weight $e^{-V(n)}$, the blow-up function. Here, $V(n)$, defined in \eqref{eq:Vdef} of \sref{sec:pfblowup}, is a finite difference of the prepotential. A flux $n$ has the canonical instanton number \cite[(4.1), (4.2)]{Harding:2026gzj}
\begin{equation}
\label{eq:kQ}
k(n)=\tfrac12\langle n,n\rangle=\frac{Q(n)}{I_2(F)} \,,\qquad Q(n)=\tfrac12\Tr_F\,n^2 \,,
\end{equation}
with $\langle n,n\rangle=n^T\tilde Kn$ for $n$ in the basis of simple coroots, and $I_2(F)$ the Dynkin index of the defining representation $F$, normalised by $I_2(\adj)=2h^\vee$, with $h^\vee$ the dual Coxeter number. The index $I_2(F)$ is $1$ for $\SU(N)$ and $\USp(2N)$, and $2$ for $\Spin(N)$. The instanton mass, fugacity and flux of \cite{Harding:2026gzj} are $m_0/I_2(F)$, $q^{1/I_2(F)}$ and $\Bmz/I_2(F)$. In our normalisation, the closed form of the blow-up function \cite[(3.3), (3.17)]{Harding:2026gzj} reads
\begin{equation}
\label{eq:expV}
e^{-V(n)}=q^{k(n)}\,(p_1p_2)^{P(n)}\prod_iz_i^{Z_i(n)}\prod_hy_h^{Y_h(n)}  \,,\qquad z_i=e^{\phi_i}\,,\quad y_h=e^{-m_h}\,,
\end{equation}
\begin{equation}
\label{eq:QP}
P(n)=\Bmz\,k(n)+\tfrac16\sum_{\alpha\in\Delta^+}\big(\langle n,\alpha\rangle^3-\langle n,\alpha\rangle\big)+\tfrac{\kCS}6\sum_{j=1}^N\tilde n_j^{\,3}-\tfrac1{12}\sum_h\sum_{\mathsf w\in\mathbf R_h}\big(x_{\mathsf w}^3-c_{\mathsf w}\,x_{\mathsf w}\big)\,.
\end{equation}
Here, $\Delta^+$ are the positive roots, and $Z_i$, $Y_h$ are the exponents $Z_i$, $W_h$ of \cite[(3.17)]{Harding:2026gzj}. The Chern--Simons term, at level $\kCS$, exists only for $\SU(N\ge3)$, with $\tilde n_j$ the components of $n$ in the orthonormal basis $e_j$ of the Cartan of $\mathfrak u(N)$. The last sum runs over the gauge-charged weights $\mathsf w$ of $\mathbf R_h$, with $x_{\mathsf w}=\sigma_{\mathsf w}\big(\langle\mathsf w,n\rangle+B_h\big)$, $\sigma_{\mathsf w}={\rm sign}\big(\langle\mathsf w,\phi\rangle+m_h\big)$ in the chamber, and $c_{\mathsf w}=\tfrac14$ if $\sigma_{\mathsf w}=-\sigma_{-\mathsf w}$, $c_{\mathsf w}=0$ otherwise.

The gauge fluxes $n$ lie in the coweight lattice $\Lambda_\cw$, and $\Lambda_\cw/\Lambda_\crt\cong Z(G)$, with $\Lambda_\crt$ the coroot lattice. A coset is an 't~Hooft sector of the 5d 1-form symmetry if its centre element acts trivially on the matter and is not broken by the Chern--Simons level or the $\theta$ angle~\cite[\S3]{Harding:2026gzj}. Independence of $e^{-V(0)}$ from the gauge fugacities fixes $B_h=\pm\tfrac12$, and integrality of the $Z_i(n)$ on the 't~Hooft sectors constrains $\Bmz$ \cite[\S3.2.2]{Harding:2026gzj}. We write $\ngen$ for the representative of smallest $k$ in a generating coset \cite[\S6.1]{Harding:2026gzj}. The fractional part of $k(\ngen)$ is the mixed anomaly of the 1-form symmetry with the instanton symmetry $\U(1)_I$ \cite[(4.3)]{Harding:2026gzj}. The Chern--Simons part of $P(\ngen)$ is cubic in the flux and has no counterpart in the quadratic theory \eqref{eq:CS7d}. In \sref{sec:exso8Z2} below, the cubic self-anomaly of the 1-form symmetry is read from the cubic part of $P$ and from the gravitational part of its linear term.

\subsection{The factor \texorpdfstring{$r$}{r} and the dictionary}
\label{sec:setupdict}

On a circle of radius $R$, the KK mass is $\tau=1/R$ and the KK fugacity $\fq=e^{-\tau}$. The frame map \eqref{eq:framemapgen}, which we construct in \sref{sec:framemap}, is the linear change of variables between the 6d frame and a 5d gauge-theory frame. In it, $m_0$ depends only on $\tau$ and on the 6d masses $m$. We define $r$ by its $\tau$-coefficient
\begin{equation}
\label{eq:rdef}
m_0=\frac\tau r+u\cdot m \,.
\end{equation}
In all frames of this paper, $u=0$, so $q=\fq^{1/r}$. \sref{sec:rrule} shows that $r$ is a positive integer, fixed by the affine algebra of the 6d gauge algebra, by its simple root that is instantonic in 5d and by the instanton number of that root. For the untwisted $(2,0)$ theories, $r=1$ \cite[\S4]{Tachikawa:2011ch}.

\paragraph{Two readings of the KK momentum of a sector.} On the 5d side, \eqref{eq:expV} and \eqref{eq:rdef} give $q^{k(n)}=\fq^{k(n)/r}$ at fixed 5d masses, so the flux $n$ carries KK momentum $k(n)/r$. On the 6d side, the circle reduction of \eqref{eq:CS7d} contains \cite[(4.10), (4.13)]{Apruzzi:2022dlm}
\begin{equation}
\label{eq:KKterm}
S\supset2\pi\int\tfrac12\,d\mathcal A\wedge\mathcal B^T\mathcal K\,\mathcal B \,,
\end{equation}
with $\mathcal A$ the background of $\U(1)_{\rm KK}$ and $\mathcal B$, the reduction of $\mathcal C_3$, the background of the 1-form symmetry of tensor origin. The periods of $\mathcal B$ lie in $\mathcal K^{-1}\BZ^{n_T}$, and the $\Omega_{ij}$ of \cite{Apruzzi:2022dlm} is our $\mathcal K$.\footnote{\label{fn:Cnorm}Here and in \eqref{eq:GS7d} below, we normalise $\mathcal C_3$ and $\mathcal B$ as in \cite{Apruzzi:2022dlm}, with periods in $\mathcal K^{-1}\BZ^{n_T}$. The $\U(1)$-valued $\mathcal C_3$ of \eqref{eq:CS7d} is $2\pi$ times this.} So a flux $a=\mathcal K^{-1}\lambda$ of $\mathcal B$ carries KK momentum $\tfrac12\lambda^T\mathcal K^{-1}\lambda=q(\lambda)$ mod $1$.

\paragraph{The dictionary.} In the blow-up equations, the flux $a=\mathcal K^{-1}\lambda$ of $\mathcal B$ through $E$ is a flux of the tensor scalars, the \hyperref[par:tensorflux]{shifted tensor flux} that we define in \sref{sec:recipe}. It induces a 5d gauge flux $\nu(a)$. With no gauge flux in the 6d frame, these fluxes form the tensor-flux family of \sref{sec:pfblowup}. The dictionary equates the two readings for this flux. It states that the coset of $\nu(a)$ is the 't~Hooft sector of the class $[\lambda]\in D$, and that, at fixed 5d masses,
\begin{equation}
\label{eq:dictionary}
\frac{k(\nu(a))}{r}=\tfrac12\,\lambda^T\mathcal K^{-1}\lambda \,,\qquad
\frac{k(\nu(a))}{r}+\tfrac12\,w\cdot a\equiv q_w(\lambda)\pmod 1\quad\big(w\in\mathcal K\BZ^{n_T}\big) \,.
\end{equation}
\sref{sec:pfspin} derives it from the blow-up equations, \eqref{eq:dictderived}. \phantomsection\label{par:dictchoices}Equating the two readings involves three choices, of the flux, of the masses held fixed and of the sign.
\begin{itemize}
\item \emph{Flux.} Since $r>1$ in general, $k/r$ mod $1$ depends on $k$ itself, not only on $k$ mod $1$. The tensor-flux family contains $\ngen$ in every example frame of \sref{sec:examples}, listed in Table~\ref{tab:master}, except $\su(4)_6+2\Lambda^2$. There, it meets the 't~Hooft coset at $\nu=\tfrac12(1,2,3)$, with $k/r=\tfrac32/6=q(1)$, and not at $\ngen=(\tfrac12,1,\tfrac12)$ (\sref{sec:exso8tri}). In the geometric frame of \sref{sec:exso8Z2}, the Freed--Witten condition adds a spin$^c$ shift $\tfrac18$ to the tensor flux. There, \eqref{eq:dictionary} holds for the bare flux $a-\tfrac18$, with $\lambda=\mathcal K(a-\tfrac18)$. On $a$ itself, $k/r$ jumps by $\tfrac12$ between neighbouring members of a sector.
\item \emph{Masses.} At fixed 6d masses $m_f$, the shifts of the 5d masses $M_f=m_f+v_f\tau$ in the frame map \eqref{eq:framemapgen} can move KK momentum into the masses (\sref{sec:exsu40}). The background fluxes of the masses also shift $a$, \eqref{eq:adef} below, so two frames of one KK theory can assign the same background to different classes of $D$ (\sref{sec:exso8tri}).
\item \emph{Sign.} For odd $\mathcal K$, the left-hand side of the first relation jumps by $\tfrac12$ between neighbouring members of a sector, by \eqref{eq:qshift}. The blow-up equation weights each member with a sign, the Wu sign $e^{i\pi\,w\cdot a}$ that we derive in \sref{sec:pftheta}, \eqref{eq:signw}. Counted as KK momentum, it adds the $\tfrac12\,w\cdot a$ that makes the left-hand side of the second relation constant mod $1$.
\end{itemize}

\paragraph{Scope and content.}\phantomsection\label{par:scope} We consider 6d SCFTs on a circle with no twist or a twist of the first kind (\sref{sec:intro}). Twists of the second kind are left open (\sref{sec:rrule}). The dictionary concerns the classes of $D$ realised as 't~Hooft sectors of the frame. The frame may have further 't~Hooft sectors, descending from the centre of the 6d gauge group (Appendix~\ref{app:twists}). Frames in which a class of $D$ has no 't~Hooft sector are outside its scope. An example, with a twist of the second kind, is in \sref{sec:conclusions}.

The content of \eqref{eq:dictionary} is twofold. First, the 6d 't~Hooft class labels an 't~Hooft sector of the 5d frame, hence a flux sector of the blow-up equations \eqref{eq:blowupeq} below. Whether its blow-up equation belongs to the same 5d theory is discussed in \sref{sec:pfcramer}. Second, since $r$ is a positive integer, the 6d spins and the 5d instanton numbers on the tensor-flux family constrain $r$, e.g.\ to $r=3$ for $\SU(4)_8$ (\sref{sec:exSU48}). The frame map gives $r$ directly, from the $\tau\phi_0^2$ term of the tensor scalar $\phi_0$ in step 2 of \tref{tab:framemaprecipe}, and \eqref{eq:dictionary} then holds in every frame of Table~\ref{tab:master}, as \sref{sec:pfspin} shows. For the untwisted $(2,0)$ theories, \eqref{eq:dictionary} equates the fractional instanton number of an 't~Hooft flux \cite{BenettiGenolini:2020doj,Apruzzi:2021nmk} with the spin \cite{Gukov:2020btk}.

%%%%%%%%%%%%%%%%%%%%%%%%%%%%%%%%%%%%%%%%%%%%%%%%%%%%%%%%%%%%%%%%%%%%%%
%% Section 3: The two frames and the frame map
%%%%%%%%%%%%%%%%%%%%%%%%%%%%%%%%%%%%%%%%%%%%%%%%%%%%%%%%%%%%%%%%%%%%%%
\section{The two frames and the frame map}
\label{sec:framemap}

Besides the flux set, the blow-up equations of \sref{sec:prefactor} need the effective prepotential $\mathcal E$. The partition function on the $\Omega$-background is the product of $e^{\mathcal E}$ and an index \cite[(2.10), (3.4)]{Kim:2020hhh}. The index enters \eqref{eq:blowupeq} below as $\hat Z$. In each chamber, $\epsilon_1\epsilon_2\,\mathcal E$ is a polynomial in the Coulomb moduli, the masses and $\epsilon_{1,2}$, \eqref{eq:E5d} below. A KK theory typically has two frames in which $\mathcal E$ is known in closed form, a 5d gauge theory (\sref{sec:prepot5d}) and the 6d tensor branch on a circle (\sref{sec:prepot6d}). Since $\mathcal E$ is one function, the two agree after a linear change of variables, the \emph{frame map} (\sref{sec:recipe}). It fixes $r=\tau/m_0$ of \eqref{eq:rdef} (\sref{sec:rrule}).

%======================================================================
\subsection{The 5d frame}
\label{sec:prepot5d}

Consider a 5d gauge theory with simple gauge algebra $\mathfrak g$, inverse coupling $m_0$, Chern--Simons level $\kCS$, and hypermultiplets $f$ in representations $\mathbf R_f$ with masses $m_f$. The Coulomb moduli $\phi_i$ are in the basis of simple coroots, so that a weight $\mathsf w$ has $\mathsf w\cdot\phi=\sum_i\mathsf w_i\phi_i$, with $\mathsf w_i$ its Dynkin labels. The effective prepotential, in the conventions of \cite{Harding:2026gzj}, is \cite[(2.12)]{Kim:2020hhh}
\begin{equation}
\label{eq:E5d}
\begin{aligned}
\mathcal E_{\rm 5d}&=\frac{1}{\epsilon_1\epsilon_2}\Big(\mathcal F+\frac{\epsilon_1^2+\epsilon_2^2}{48}\,C^G\cdot\phi+\frac{\epsilon_+^2}{2}\,C^R\cdot\phi\Big)\,,\\
\mathcal F&=\frac{m_0}{2}\tilde K_{ij}\phi_i\phi_j+\frac{\kCS}{6}\,d_{ijk}\phi_i\phi_j\phi_k+\frac1{12}\Big(\sum_{\alpha\in\Delta}|\alpha\cdot\phi|^3-\sum_f\sum_{\mathsf w\in\mathbf R_f}|\mathsf w\cdot\phi+m_f|^3\Big)\,,\\
C^G_i&=-\partial_i\Big(\sum_{\alpha\in\Delta}|\alpha\cdot\phi|-\sum_f\sum_{\mathsf w\in\mathbf R_f}|\mathsf w\cdot\phi+m_f|\Big)\,,\qquad C^R_i=\tfrac12\,\partial_i\sum_{\alpha\in\Delta}|\alpha\cdot\phi| \,,
\end{aligned}
\end{equation}
with $\Delta$ the roots, $\tilde K$ the matrix of inner products of the simple coroots, \eqref{eq:Ktilde} below, and $d_{ijk}\phi_i\phi_j\phi_k=\tr_{\rm fund}\phi^3$ for $\SU(N)$. $\mathcal F$ is the cubic prepotential. It is a polynomial in each \emph{chamber}, a choice of the signs of all $\alpha\cdot\phi$ and $\mathsf w\cdot\phi+m_f$. $C^G$ and $C^R$ are the levels of the mixed Chern--Simons terms $-\frac1{48}C^G_i\int A_i\wedge p_1(T)$ and $\frac12C^R_i\int A_i\wedge c_2(R)$. Every $C^R_i$ equals $2$. The $\epsilon$-terms of \eqref{eq:E5d} are these Chern--Simons terms with the equivariant values \cite[(2.13)]{Kim:2020hhh}
\begin{equation}
\label{eq:equivrule}
p_1(T)\to-(\epsilon_1^2+\epsilon_2^2)\,,\qquad c_2(R)\to\epsilon_+^2\,,\qquad c_2(L)\to m^2 \,.
\end{equation}
The last entry is the $\SU(2)_R$ rule applied to a flavour $\SU(2)_L$ with Cartan parameter $m$. It makes the two frames of the M-string, $\SU(2)_0+\adj$ and the $(2,0)$ theory of type $A_1$ on a circle, agree under the frame map \eqref{eq:Mstring} below.

\paragraph{Normalisation of $m_0$.} $m_0$ enters $\mathcal E_{\rm 5d}$ only through $\tfrac{m_0}2\tilde K_{ij}\phi_i\phi_j$. We normalise $\Tr$ so that one instanton has $\tfrac14\int\Tr F^2=1$, i.e.\ the long roots have length squared $2$. Since $\tr_{\rm adj}=2h^\vee\,\Tr$ and $\tfrac12\tr_{\rm adj}\phi^2=\sum_{\alpha\in\Delta^+}(\alpha\cdot\phi)^2$,
\begin{equation}
\label{eq:Ktilde}
\tfrac12\tilde K_{ij}\phi_i\phi_j=\frac1{2h^\vee}\sum_{\alpha\in\Delta^+}(\alpha\cdot\phi)^2\,,\qquad\text{i.e.}\qquad \tilde K_{ij}=(\alpha_i^\vee,\alpha_j^\vee)\,.
\end{equation}
So $m_0$ is the mass of one instanton. For simply laced $\mathfrak g$, $\tilde K$ is the Cartan matrix. For $\usp(4)$ and $\mathfrak g_2$, with $\alpha_1$ the short simple root, $\tfrac12\tilde K\phi\phi$ is $2\phi_1^2-2\phi_1\phi_2+\phi_2^2$ and $3\phi_1^2-3\phi_1\phi_2+\phi_2^2$. For $\so(2N+1)$, it is $\tfrac12\sum_ie_i^2$ in orthonormal components $e_i$.

%======================================================================
\subsection{The 6d frame}
\label{sec:prepot6d}

On the tensor branch, the 6d theory has $n_T$ tensor multiplets, with scalars $\phi_0=(\phi_{0,\alpha})$ and string charge pairing $\mathcal K$ (\sref{sec:setup6d}), and a gauge algebra $\mathfrak g_\alpha$, possibly empty, on the $\alpha$-th curve. On a circle of radius $1/\tau$, the holonomies $\phi_{\alpha,i}$ of $\mathfrak g_\alpha$, in the basis of simple coroots, become 5d Coulomb moduli, and so do the $\phi_{0,\alpha}$. $\mathcal E_{\rm 6d}$ has a tree-level and a one-loop part \cite[\S2.3]{Kim:2020hhh}.

\paragraph{Tree level.} The dynamical 2-forms $\mathsf B_\alpha$ couple to 4-forms $X_{4\alpha}$ \cite{Green:1984bx,Sagnotti:1992qw,Sadov:1996zm}:
\begin{equation}
\label{eq:GSdata}
\begin{aligned}
&S\supset\sum_{\alpha,\beta}\mathcal K_{\alpha\beta}\int\mathsf B_\alpha\wedge X_{4\beta}\,,\qquad
X_{4\alpha}=-\tfrac14\fa_\alpha\,p_1(T)+\tfrac14\sum_{\mathfrak h}\fb_{\mathfrak h,\alpha}\Tr F_{\mathfrak h}^2+\fc_\alpha\,c_2(R)\,,\\
&\sum_\beta\mathcal K_{\alpha\beta}X_{4\beta}=I^\alpha=\tfrac14\Big(\sum_\gamma\mathcal K_{\alpha\gamma}\Tr F_\gamma^2-(2-\mathcal K_{\alpha\alpha})\,p_1(T)\Big)+h^\vee_\alpha\,c_2(R)+(\text{flavour})\,,
\end{aligned}
\end{equation}
where $\mathfrak h$ runs over the simple gauge and flavour algebras and $\Tr$ is normalised as in \eqref{eq:Ktilde}. The second line is fixed by anomaly cancellation \cite{Ohmori:2014kda}, with $h^\vee_\alpha=1$ for a curve without gauge algebra. Flavour symmetries enter $I^\alpha$ with the sign opposite to that of gauge symmetries. The overall sign of $X_4$ is the one for which $\fc>0$, as in \cite{Kim:2020hhh}. For a single curve, $\fa=(2-\mathcal K)/\mathcal K$, $\fc=h^\vee/\mathcal K$ and $\fb=1$ for its gauge algebra, e.g.\ $(\fa,\fc)=(1,1)$ for the E-string and $(0,\tfrac12)$ for the M-string. For the $(2,0)$ theory of type $\mathfrak g$, with $\SU(2)_L\times\SU(2)_R\subset\SO(5)_R$, \cite[(2.12)]{Ohmori:2014kda} gives $X_{4\alpha}=\rho_\alpha\big(c_2(R)-c_2(L)\big)$ in this sign convention, with $\rho=\sum_\alpha\rho_\alpha\alpha_\alpha$ the Weyl vector. Since $\mathcal K\rho=(1,\dots,1)^T$, $I^\alpha=c_2(R)-c_2(L)$, which is the rule $h^\vee_\alpha=1$ with $\SU(2)_L$ as a flavour symmetry.

On the circle, each $\mathsf B_\alpha$ gives a 5d $\U(1)$ gauge field with scalar partner $\phi_{0,\alpha}$ and inverse coupling $\tau\mathcal K$, and \eqref{eq:GSdata} becomes the mixed Chern--Simons terms $\sum_\alpha\phi_{0,\alpha}I^\alpha$. With \eqref{eq:equivrule} and $\tfrac14\Tr F_{\mathfrak h}^2\to\tfrac12\tilde K_{\mathfrak h}(\phi_{\mathfrak h},\phi_{\mathfrak h})$, this gives \cite[(2.37)]{Kim:2020hhh}
\begin{equation}
\label{eq:Etree}
\begin{aligned}
\mathcal E_{\rm tree}&=\frac\tau2\,\phi_0^T\mathcal K\phi_0+\sum_{\alpha,\beta}\mathcal K_{\alpha\beta}\,\phi_{0,\alpha}\Big(\frac{\fa_\beta}4(\epsilon_1^2+\epsilon_2^2)+\fc_\beta\,\epsilon_+^2+\sum_{\mathfrak h}\frac{\fb_{\mathfrak h,\beta}}2\tilde K_{\mathfrak h}(\phi_{\mathfrak h},\phi_{\mathfrak h})\Big)\\
&=\frac\tau2\,\phi_0^T\mathcal K\phi_0+\sum_\alpha\phi_{0,\alpha}\Big(\frac{2-\mathcal K_{\alpha\alpha}}4(\epsilon_1^2+\epsilon_2^2)+h^\vee_\alpha\epsilon_+^2\\
&\qquad\qquad\quad\,\,+\sum_\gamma\mathcal K_{\alpha\gamma}\tfrac12\tilde K_\gamma(\phi_\gamma,\phi_\gamma)+\sum_m\kappa_m\,m^2\Big)\,,
\end{aligned}
\end{equation}
where $\phi_{\mathfrak h}$ are the holonomies or masses of $\mathfrak h$, and $\phi_\gamma$ the holonomies of $\mathfrak g_\gamma$. The last sum runs over the masses $m$ of the flavour symmetries on the curve $\alpha$. The coefficient $\kappa_m$ of $m^2\phi_{0,\alpha}$ is the \emph{flavour Green--Schwarz level} of $m$. The first term is the coupling term of \eqref{eq:E5d} for the tensor $\U(1)$ factors, with $m_0\tilde K\to\tau\mathcal K$. The gauge term says that the 6d gauge coupling on the curve $\gamma$ is the string tension $(\mathcal K\phi_0)_\gamma$. The flavour levels are negative, e.g.\ $\kappa_m=-1$ for each pair of fundamentals identified by the twist of $\su(3)+6\Fund$ on a $-2$ curve. If the flavour normalisation is not known in advance, the frame map of \sref{sec:recipe} fixes $\kappa_m$ from the terms $\phi\,mm$ (step 3 of \tref{tab:framemaprecipe}), as in \sref{sec:exsu44}.

\paragraph{Twists.}\phantomsection\label{par:twists} The circle may be twisted by a discrete symmetry $\sigma$ \cite[\S2.3]{Kim:2020hhh}, \cite{Bhardwaj:2019fzv}. A twist of the \emph{first kind} is an outer automorphism of $\mathfrak g_\alpha$, with invariant subalgebra $\mathfrak g^\sigma_\alpha$. Its action on the centre is in Appendix~\ref{app:twists}. A twist of the \emph{second kind} permutes tensor nodes, whose scalars are identified, and folds $\mathcal K$ \cite[(2.43)]{Kim:2020hhh}. A field component on which the lift of $\sigma$ to the gauge group acts by $e^{2\pi i\fr}$ has KK momenta in $\BZ+\fr$, with $\fr$ fixed by the \hyperref[par:liftnorm]{choice of lift} in Appendix~\ref{app:twists}. So a 6d multiplet decomposes into representations $\mathbf R_{\fr}$ of $\mathfrak g^\sigma_\alpha$ with $\fr\in[0,1)$, e.g.\ the decomposition \eqref{eq:su39decomp} of the adjoint of $\su(3)$ under $\su(2)$ in \sref{sec:exSU39}. $\Tr$ in \eqref{eq:GSdata} is then the trace of $\mathfrak g_\alpha$ restricted to $\mathfrak g^\sigma_\alpha$. It equals \eqref{eq:Ktilde} for $\mathfrak g^\sigma_\alpha$ when the embedding index is $1$, as for $\su(2)\subset\su(3)$, $\usp(4)\subset\su(4),\su(5)$ and $\mathfrak g_2,\so(7)\subset\so(8)$ in our examples.

\paragraph{One loop.} A weight $\mathsf w$ of a representation $\mathbf R_{\fr}$ in the decomposition of the adjoint gives a KK tower of masses $p\tau+\mathsf w\cdot\phi$, and a weight of a representation $\mathbf R_{\fr,f}$ of the hypermultiplet $f$ a tower of masses $p\tau+\mathsf w\cdot\phi+m_f$, with $p\in\BZ+\fr$. Each state contributes $\pm\frac1{12}|\text{mass}|^3$, as in \eqref{eq:E5d}. For $p\neq0$ and $|x|\ll\tau$, the part of $|p\tau+x|^3$ quadratic in $x$ is $3|p|\tau x^2$. The Hurwitz zeta function gives $\sum_{p\in\BZ+\fr,\,p\neq0}|p|=-B_2(\fr)$, with $B_2(\fr)=\fr^2-\fr+\tfrac16$. The cubic terms and the levels $C^G$, $C^R$ come from the zero modes $p=0$ only \cite[(2.45)]{Kim:2020hhh}. So, for $\tau\gg|\mathsf w\cdot\phi|,|\mathsf w\cdot\phi+m_f|$ and up to $\phi$-independent terms \cite[(2.44)]{Kim:2020hhh},
\begin{equation}
\label{eq:F1loop6d}
\mathcal F_{\text{1-loop}}=\mathcal F^{(0)}-\frac\tau4\Big[\sum_{\fr}B_2(\fr)\sum_{\mathsf w\in\mathbf R_{\fr}}(\mathsf w\cdot\phi)^2-\sum_f\sum_{\fr}B_2(\fr)\sum_{\mathsf w\in\mathbf R_{\fr,f}}(\mathsf w\cdot\phi+m_f)^2\Big]\,,
\end{equation}
where $\mathcal F^{(0)}$ is the one-loop part of $\mathcal F$ in \eqref{eq:E5d} for the zero modes. The part of the tower term quadratic in the holonomies is proportional to $\tfrac12\tilde K\phi\phi$, the unique invariant form of $\mathfrak g^\sigma_\alpha$. So it shifts the effective gauge coupling $(\mathcal K\phi_0)_\alpha$ by a multiple of $\tau$, e.g.\ to $3\phi_0-\tfrac5{16}\tau$ for $\SU(3)_9$ (\sref{sec:exSU39}).

\paragraph{Total.} The 6d-frame prepotential is \cite[(2.37), (2.46)]{Kim:2020hhh}
\begin{equation}
\label{eq:E6d}
\mathcal E_{\rm 6d}=\frac1{\epsilon_1\epsilon_2}\Big(\mathcal E_{\rm tree}+\mathcal F_{\text{1-loop}}+\frac{\epsilon_1^2+\epsilon_2^2}{48}\,C^G\cdot\phi+\frac{\epsilon_+^2}{2}\,C^R\cdot\phi\Big)\,,
\end{equation}
with $C^G$, $C^R$ given by \eqref{eq:E5d} for the zero modes.

\paragraph{Regime.}\phantomsection\label{par:6dregime} By the \emph{6d frame}, we mean the regime in which \eqref{eq:E6d} holds. It rests on three assumptions. First, the theory is on the tensor branch: the tensor scalars are large, so the string tensions $(\mathcal K\phi_0)_\alpha$ exceed all other masses. Then, every string state keeps the sign of its charge, and no string state contributes to $\mathcal E_{\rm 6d}$, whose dependence on $\phi_0$ is the tree term \eqref{eq:Etree}. Second, the KK modes are heavier than the zero modes, $|\mathsf w\cdot\phi|,|\mathsf w\cdot\phi+m_f|\ll\tau$, so every KK mode keeps the sign of its KK momentum, as \eqref{eq:F1loop6d} assumes. Third, a chamber of the zero modes is chosen: $\phi$ in a Weyl chamber of $\mathfrak g^\sigma_\alpha$, and each mass above or below each pair of weights $\pm\mathsf w$. The cubic terms and $C^G$, $C^R$ depend on this choice.

%======================================================================
\subsection{The frame map}
\label{sec:recipe}

In a given chamber, $\mathcal E_{\rm 6d}$ and $\mathcal E_{\rm 5d}$ coincide after a change of variables, up to terms independent of the Coulomb moduli. In M-theory on the local Calabi--Yau threefold that engineers the theory, both sets of variables are K\"ahler parameters \cite{Intriligator:1997pq,Jefferson:2018irk,Bhardwaj:2019fzv}, normalisable for the Coulomb moduli and non-normalisable for $m_0$, $\tau$ and the masses. So the change of variables is linear, and it maps parameters to parameters and moduli to moduli up to shifts by parameters:
\begin{equation}
\label{eq:framemapgen}
\phi^{\rm 5d}=M\phi^{\rm 6d}+s\,\tau+t\cdot m\,,\qquad m_0=\frac\tau r+u\cdot m\,,\qquad M_f=m_f+v_f\,\tau \,,
\end{equation}
with $\phi^{\rm 6d}=(\phi_{0,\alpha},\phi_{\alpha,i})$, $M$ a unimodular integer matrix, $s,t,u,v$ rational, and $M_f$ the 5d masses, written $m_f$ in \eqref{eq:E5d} (not to be confused with the matrix $M$). The masses must be kept in both frames, since setting them to zero in both is inconsistent when $v\neq0$.

\paragraph{Recipe.} Substitute \eqref{eq:framemapgen} into $\mathcal E_{\rm 5d}$ and compare with $\mathcal E_{\rm 6d}$ monomial by monomial. The degree in the moduli separates the unknowns (\tref{tab:framemaprecipe}).
\begin{table}[t]
\centering\small
\begin{tabular}{c l l L{7.4cm}}
\toprule
step & monomials & unknowns fixed & reason\\
\midrule
1 & $\phi\phi\phi$ & $M$ & independent of $\tau$ and $m$: the shifts drop out\\
2 & $\phi\phi\tau$, $\phi\phi m$ & $1/r$, $s$, $t$, $u$, $v$ & linear in the unknowns once $M$ is known\\
3 & $\phi\tau\tau$, $\phi\tau m$, $\phi mm$ & $v$, $\kappa_m$ if still free & otherwise identities\\
4 & $\phi\,\epsilon^2$ & none & check: the tensor terms of \eqref{eq:Etree} supply the string part of $C^G$, $C^R$\\
\bottomrule
\end{tabular}
\caption{The frame-map recipe: the monomials of $\epsilon_1\epsilon_2\mathcal E$ that fix each set of unknowns of \eqref{eq:framemapgen}. In step 2, $\tfrac{m_0}2\tilde K\phi\phi$ brings $1/r$ and $u$, the shifts inside the 5d cubic bring $s$ and $t$, and mass-dependent cubic terms bring $v$. On the 6d side, the $\tau\times$quadratic terms are the tensor kinetic term of \eqref{eq:Etree} and the tower term of \eqref{eq:F1loop6d}.}
\label{tab:framemaprecipe}
\end{table}
Step 1 is the only non-linear step. We solve it by a scan over unimodular integer matrices and the chambers of both frames, with a bound on $|M_{ij}|$ between $3$ and $6$. \tref{tab:scans} in Appendix~\ref{app:scans} gives the bound for the frames of \sref{sec:examples}. For the map of \sref{sec:E1N2}, between pure $\SU(4)_0$ and an $\SU(2)$ gauging of two $E_1$ theories, it is $3$ (Appendix~\ref{app:su40}). Within these bounds, the solution is unique up to the Weyl group, charge conjugation and redefinitions of the masses.\footnote{\label{fn:so8noscan}For the untwisted $\so(8)$ KK theory as an $\SU(2)$ gauging of four $E_1$ theories (\sref{sec:E1N4}), the bound is $5$, and the solution \eqref{eq:so8M} is unique up to permutations of the four $E_1$'s.} The cubic terms alone can leave an ambiguity, which step 2 removes (\sref{sec:exsu44}). The method applies equally to two 5d frames (\sref{sec:exso8tri}, \sref{sec:E1N2}). \phantomsection\label{it:volmatch}An equivalent method is \emph{volume matching}. A BPS particle is an M2-brane on a holomorphic curve, of mass equal to its volume, so equating the masses of the same states in the two frames gives the same map, provided moduli map to moduli. For $\SU(3)_9$, this forces the exchange of the fibre and the base of a ruled surface (\sref{sec:exSU39}).

The simplest example is the M-string \cite{Haghighat:2013gba}, $\SU(2)_0+\adj$ with adjoint mass $m$, whose 6d frame is the $(2,0)$ $A_1$ theory on a $-2$ curve. In the chamber $2\phi>|m|$, $\epsilon_1\epsilon_2\mathcal E_{\rm 5d}=m_0\phi^2-m^2\phi+\epsilon_+^2\phi$ \cite[(4.7), (4.8)]{Kim:2020hhh}, and \eqref{eq:Etree} gives $\epsilon_1\epsilon_2\mathcal E_{\rm 6d}=\tau\phi_0^2+\phi_0(\epsilon_+^2-m^2)$. There is no cubic term and no shift:
\begin{equation}
\label{eq:Mstring}
\phi=\phi_0\,,\qquad m_0=\tau\,,\qquad r=1 \,.
\end{equation}
The W-boson, of mass $2\phi=\mathcal K\phi_0$, is the wrapped string, and the KK mode is the instanton particle.

\paragraph{Which 5d simple roots are strings.} Let $c_\alpha$ be the column of $M$ along $\phi_{0,\alpha}$. In the mass $\mathsf w\cdot\phi^{\rm 5d}$ of a 5d W-boson with Dynkin labels $\mathsf w$, the coefficient of $\phi_{0,\alpha}$ is $\mathsf w\cdot c_\alpha$. In the 6d frame, it is $(\mathcal K\lambda)_\alpha$, with $\lambda\in\BZ^{n_T}$ the string charge of the state, of tension $\lambda^T\mathcal K\phi_0$. For the $j$-th 5d simple root, $\mathsf w$ is the $j$-th row of the 5d Cartan matrix $C$. Hence
\begin{equation}
\label{eq:stringcolumn}
\sum_iC_{ji}\,(c_\alpha)_i=(\mathcal K\lambda^{(j)})_\alpha \,,
\end{equation}
with $\lambda^{(j)}$ the string charge of the $j$-th 5d simple root. \phantomsection\label{par:rootkinds}In the frames of this paper, a 5d simple root is, in 6d variables, one of three kinds, distinguished by its mass $\mathsf w\cdot\phi^{\rm 5d}$ written through \eqref{eq:framemapgen}:
\begin{itemize}[leftmargin=*,itemsep=2pt]
\item a \emph{wrapped string}, if the mass contains $\phi_0$, i.e.\ $\lambda^{(j)}\neq0$ in \eqref{eq:stringcolumn};
\item the \emph{affine root} $\alpha_0$ of the (possibly twisted) affine algebra $\hat{\mathfrak g}_\alpha$, a KK mode of a 6d gauge boson, if it contains $\tau$ but not $\phi_0$;
\item a \emph{zero mode}, a simple root of $\mathfrak g^\sigma_\alpha$, if it contains neither.
\end{itemize}
The same criterion applies to any 5d BPS state. For a state that is not a string, the coefficient $p$ of $\tau$ is its KK momentum. The remaining simple roots of $\hat{\mathfrak g}_\alpha$ carry instanton charge, as we show in \sref{sec:rrule}. For $\SU(3)_9$, for instance, the $\su(2)$ root $2\phi_1$ is the 5d state $-4\phi_1^{\rm 5d}+2\phi_2^{\rm 5d}+m_0$, of instanton number $1$ (\sref{sec:exSU39}).

\phantomsection\label{par:affinenode}In every single-tensor gauge-theory frame of this paper with a 6d gauge algebra, exactly one 5d simple root is a string. Its tensor column, given by \eqref{eq:stringcolumn}, then turns out to be the comarks $(1,\mathsf a^\vee_1,\dots,\mathsf a^\vee_{\rank\mathfrak g^\sigma})$ of $\hat{\mathfrak g}$ \cite{Kac:1990gs}, with the affine node first.\footnote{\label{fn:so8affinenode}For the untwisted $\so(8)$ KK theory of \sref{sec:E1N4}, the scan returns this column as well, in the order $(2;1,1,1,1)$ of \eqref{eq:so8M}, with the central node first.} This is the \emph{affine-node rule} of \cite[(4.10)]{Bhardwaj:2019fzv} and \cite[(2.38)]{Kim:2020hhh}:
\begin{equation}
\label{eq:affinenode}
\phi^{\rm 5d}_0=\phi_0+\dots\,,\qquad \phi^{\rm 5d}_i=\phi_i+\mathsf a^\vee_i\,\phi_0+\dots\,,\qquad i=1,\dots,\rank\mathfrak g^\sigma \,,
\end{equation}
with the shifts of \eqref{eq:framemapgen} in the dots. For $\SU(3)_9$, $\mathsf a^\vee_1=2$ and $C(1,2)^T=(0,3)^T$, so the first 5d simple root is $\alpha_0$ and the second a single string. \eqref{eq:affinenode} is a shortcut. The cubic terms decide the order of the finite nodes (\sref{sec:exsu40}) and the position of the affine node, which is not always first (\sref{sec:exZ2twist}, \sref{sec:exsu42}).

\paragraph{Flux rule.}\phantomsection\label{it:fluxmap} Fluxes transform like the variables. On the blow-up, each Coulomb modulus is shifted on the two patches by its flux times $\epsilon_1$ or $\epsilon_2$, and each parameter by its background flux, as in the blow-up equation \eqref{eq:blowupeq} of \sref{sec:pfblowup}. So \eqref{eq:framemapgen} also maps the fluxes:
\begin{equation}
\label{eq:fluxmap}
n^{\rm 5d}=M\,n^{\rm 6d}+s\,B_\tau+t\cdot B_m\,,\qquad \Bmz=\frac{B_\tau}r+u\cdot B_m\,,\qquad B_{M_f}=B_{m_f}+v_f\,B_\tau \,,
\end{equation}
with $B_\tau$ the flux of $\tau$, set to zero in \sref{sec:pfblowup}.

\paragraph{The shifted tensor flux $a$.}\phantomsection\label{par:tensorflux} Take a flux with no 6d gauge flux. It consists of the fluxes $n_{0,\alpha}$ of the tensor scalars $\phi_{0,\alpha}$ and the flavour fluxes $B_m$. For each tensor, the frame map contains a 5d Coulomb modulus $\phi_{0,\alpha}+s_\alpha\tau+t_\alpha\cdot m$ with no gauge holonomy, such as $\phi^{\rm 5d}_0$ in \eqref{eq:affinenode}. We define the \emph{shifted tensor flux} $a_\alpha$ as the flux of this modulus. By \eqref{eq:fluxmap}, with $B_\tau=0$,
\begin{equation}
\label{eq:adef}
a_\alpha=n_{0,\alpha}+t_\alpha\cdot B_m \,.
\end{equation}
The shift can be fractional. For example, in \sref{sec:exZ2twist}, $t_\alpha\cdot m=\tfrac14\sum_{i=1}^Nm_i$ for $N$ masses and $B_{m_i}=\tfrac12$, so $a=n_0+\tfrac N8$, \eqref{eq:nushift}. In such a frame, the strings carry the flavour charges $\mathcal Kt$, e.g.\ $\tfrac12$ under each flavour $\U(1)$ in this example. So the flux through $E$ seen by a string of unit charge is $\mathcal Ka$, not $\mathcal Kn_0$, and we call $\lambda=\mathcal Ka$ the string charge of the flux in this frame. This is why \cite[(5.10), (5.12)]{Kim:2020hhh} state their 6d flux sets in shifted variables, which are the 5d ones. The charge $\lambda$ depends on the frame through $t$ (\sref{sec:exso8tri}). The 5d gauge flux $\nu(a)$ of \sref{sec:setupdict} is
\begin{equation}
\label{eq:nuofa}
\nu(a)=\sum_\alpha a_\alpha\,c_\alpha \,,
\end{equation}
with $c_\alpha$ the tensor columns of \eqref{eq:stringcolumn}. Indeed, \eqref{eq:fluxmap} gives $\sum_\alpha n_{0,\alpha}c_\alpha+t\cdot B_m$. For one tensor, $Ct$ lists the flavour charges of the 5d simple roots, as in \eqref{eq:stringcolumn}, and, among them, only the string carries one. So $t$ is proportional to $c_1$. The entry of $c_1$ at the modulus above is $1$, so $t\cdot B_m=(t_1\cdot B_m)\,c_1$. The frame with two tensors, \eqref{eq:su44map} of \sref{sec:exsu44}, has no mass shifts. By \eqref{eq:affinenode}, $\nu(a)$ is $a$ times the comarks of $\hat{\mathfrak g}$, in the order of the 5d nodes, e.g.\ $\nu(a)=a(1,2)$ for $\SU(3)_9$ (\sref{sec:exSU39}) and $\nu(a)=a(2,2,1)$ for $\so(7)+5\Fund$ (\sref{sec:exZ2twist}). These fluxes form the tensor-flux family, which we define in \sref{sec:pfblowup}.

\paragraph{Range of validity.}\phantomsection\label{it:validity} \eqref{eq:F1loop6d} assumes that every KK mode keeps the sign of its KK momentum. A 5d frame reached from the 6d frame only through flops of KK modes, such as $\su(4)_6+2\Lambda^2$, needs a 5d-to-5d map (\sref{sec:exso8tri}). The origin of $\phi_{0,\alpha}$ is a convention, which the regularisation does not fix. Moving it changes the shifts $s$, e.g.\ by the $\tau/16$ of \cite[(2.51)]{Kim:2020hhh} (\sref{sec:exSU39}). So only differences of $\tau$-coefficients between strings of equal charge are meaningful.

%======================================================================
\subsection{What fixes \texorpdfstring{$r$}{r}}
\label{sec:rrule}

We now show that $r$ is fixed by the affine algebra $\hat{\mathfrak g}$, by which of its simple roots is instantonic in 5d, and by the instanton number of that root. We take $u=0$ in \eqref{eq:framemapgen}, as in all frames of this paper (\sref{sec:setupdict}), and no twist or a twist of the first kind. \phantomsection\label{par:secondkind}A twist of the second kind permutes tensor nodes. The combinations of the permuted tensor multiplets that are odd under the permutation are then neutral, with KK momentum in $\BZ+\tfrac12$. Below, we define $\delta$ as the mass of the lightest neutral KK mode, and derive $m_0=\delta/k(\delta)$, \eqref{eq:m0delta}, with $k(\delta)$ the instanton number of this mode. For a twist of the second kind, this would give $\delta=\tfrac\tau2$ and $m_0=\tfrac{\tau}{2k(\delta)}$. However, for $\SU(2)_\pi+\adj$ and $\sp(2)_{0,\pi}+\adj$, frames of the $(2,0)$ theories of type $A_2$, $A_3$ and $A_4$ with the $\BZ_2$ twist \cite{Tachikawa:2011ch}, the folding rule \cite[(2.43)]{Kim:2020hhh} gives instead $m_0=\tau$. We leave this conflict open (\sref{sec:conclusions}).

\paragraph{A neutral state.} Let $\delta$ be the mass of the lightest KK mode of a 6d field with no gauge, flavour or string charge. Without a twist, it is the mode of KK momentum $1$ of a tensor-multiplet scalar or a Cartan gauge field, and $\delta=\tau$. Let the outer automorphism $\sigma$ of a twist of the first kind have order $r_{\rm tw}$. Then, $\delta$ is the mode of KK momentum $1/r_{\rm tw}$ of a zero weight in a twisted part of the adjoint, such as the singlet $\mathbf 1_{1/2}$ of $\su(3)\supset\su(2)$ in \eqref{eq:su39decomp} of \sref{sec:exSU39}, and $\delta=\tau/r_{\rm tw}$. For $A_{2L}$, $r_{\rm tw}=2$ although the lift of $\sigma$ has order four (Appendix~\ref{app:twists}), since the modes of momentum $\tfrac14$ are gauge-charged. On the circle, the 6d gauge fields give the (twisted) loop algebra of $\mathfrak g$, and, together with the KK momentum, $\hat{\mathfrak g}$ \cite{Kac:1990gs}. A mode of $\mathfrak g^\sigma$-weight $\mathsf w$ and KK momentum $p$ has mass $\mathsf w\cdot\phi+p\tau$. The charged modes ($\mathsf w\neq0$) are the real roots of $\hat{\mathfrak g}$, and the neutral ones ($\mathsf w=0$) its imaginary roots, the multiples of $\delta$. So $\delta$ is the smallest imaginary root:
\begin{equation}
\label{eq:deltadef}
\delta=\sum_i\mathsf a_i\,\alpha_i=\frac{\tau}{r_{\rm tw}}\,,
\end{equation}
with $\alpha_i$, $i=0,\dots,\rank\mathfrak g^\sigma$, the masses of the simple roots of $\hat{\mathfrak g}$ and $\mathsf a_i$ their marks. The $\phi$-dependence cancels in the sum.

\paragraph{Its 5d mass.} A 5d BPS state has mass $k\,m_0+\mathsf w\cdot\phi^{\rm 5d}+\sum_f\mathsf f_fM_f$, with $k\in\BZ$ its instanton number, $\mathsf w$ its Dynkin labels and $\mathsf f$ its flavour charges. By definition, $m_0$ is the coefficient of $k$. By \eqref{eq:framemapgen} with $u=0$, the mass in 6d variables is
\begin{equation}
\label{eq:massin6d}
(\mathsf wM)\cdot\phi^{\rm 6d}+\big(\tfrac kr+\mathsf w\cdot s+\mathsf f\cdot v\big)\tau+(\mathsf w\cdot t+\mathsf f)\cdot m \,.
\end{equation}
\phantomsection\label{it:taucoeff}The coefficient of $\tau$ is the KK momentum. So $\U(1)_I$ mixes KK momentum with gauge and flavour charges, and the instanton particle need not carry KK momentum $1/r$. For $\SU(3)_9$, the $k=1$ state $2\phi_1$ above is a zero mode. A flux sector, which sees only $m_0$, still has KK momentum $k(n)/r$ (\sref{sec:setupdict}).

The state of mass $\delta$ has mass $\tau/r_{\rm tw}$ in 6d variables, \eqref{eq:deltadef}, with no $\phi^{\rm 6d}$ and no $m$. We compare this with \eqref{eq:massin6d}. The coefficient of $\phi^{\rm 6d}$ gives $\mathsf wM=0$, so $\mathsf w=0$ since $M$ is invertible. The coefficient of $m$ then gives $\mathsf f=0$, and the shifts $s$, $v$ drop out. So its 5d mass is $k\,m_0=\delta$, with $k\geq1$ since $\delta,m_0>0$. Hence
\begin{equation}
\label{eq:m0delta}
m_0=\frac{\delta}{k(\delta)}\,,\qquad r=\frac\tau{m_0}=r_{\rm tw}\,k(\delta)\,.
\end{equation}
This uses only the linearity of the map and the neutrality of $\delta$, not the regime $\tau\gg\phi,m$. So it also holds for frames reached through flops.

\paragraph{The instanton number of $\delta$.} The instanton number is the coefficient of $m_0$ in the 5d mass, a linear function of the 6d mass. So it is additive, and \eqref{eq:deltadef} gives $k(\delta)=\sum_i\mathsf a_i\,k(\alpha_i)$. Wrapped strings do not enter this sum, since their masses contain $\phi_0$ and they are not roots of $\hat{\mathfrak g}$. Two kinds of 5d simple root in the \hyperref[par:rootkinds]{list} of \sref{sec:recipe}, the affine root and the zero modes, are also simple roots of $\hat{\mathfrak g}$. In a gauge-theory frame, they are W-bosons, like every 5d simple root, so $k=0$. In every such frame of this paper with a 6d gauge algebra, the 5d simple roots that are not strings are $\rank\mathfrak g^\sigma$ of the $\rank\mathfrak g^\sigma+1$ simple roots of $\hat{\mathfrak g}$. So exactly one simple root $\alpha_j$ of $\hat{\mathfrak g}$ is not a 5d simple root. In $\SU(3)_9$, for instance, $\hat{\mathfrak g}=A_2^{(2)}$ has the simple roots $\alpha_0$ and $\alpha_1=2\phi_1$, while the 5d simple roots of $\su(3)$ are $\alpha_0$ and a string, so $\alpha_j=\alpha_1$; see \eqref{eq:SU39roots} of \sref{sec:exSU39} below. One KK theory of this paper, $\so(8)$ on a $-4$ curve with the $\BZ_2$ twist, has no 5d gauge-theory description. \sref{sec:exso8Z2} treats it in a \emph{geometric frame}: four compact ruled surfaces in a local Calabi--Yau threefold, with one mass parameter $m_0$. Their fibres are the four simple roots of $\hat{\mathfrak g}=D_4^{(2)}$. With the non-compact divisor chosen there, only the 5d mass of the affine root $\alpha_0$ contains $m_0$, so $\alpha_j=\alpha_0$, with $k(\alpha_0)=1$. In both kinds of frame, every simple root of $\hat{\mathfrak g}$ other than $\alpha_j$ has $k=0$, so $k(\delta)=\mathsf a_j\,k(\alpha_j)$. Since $k(\delta)\geq1$, as shown above \eqref{eq:m0delta}, $\alpha_j$ is instantonic, and
\begin{equation}
\label{eq:rrule}
k(\delta)=\mathsf a_j\,k(\alpha_j)\,,\qquad r=r_{\rm tw}\,\mathsf a_j\,k(\alpha_j)\,.
\end{equation}
\tref{tab:rvalues} lists the data, and each $r$ agrees with the frame map of the example. In all frames but one, $k(\alpha_j)=1$, so $r$ is the twist order times the mark of the instantonic root. For the untwisted $(2,0)$ theories, which have no 6d gauge algebra, $\delta=\tau$ is itself the instanton particle, so $r=1$, as for the M-string \eqref{eq:Mstring}.
\begin{table}[t]
\centering\footnotesize\setlength{\tabcolsep}{4pt}\renewcommand{\arraystretch}{1.2}
\begin{tabular}{c l c c l}
\toprule
$\hat{\mathfrak g}$ & $\delta$ of \eqref{eq:deltadef} & $r_{\rm tw}$ & $\mathsf a_j$ & 5d frames: $k(\alpha_j)$, $r$\\
\midrule
$A_2^{(2)}$ & $2\alpha_0+\alpha_1$ & $2$ & $1$ & $\SU(3)_9$, $\USp(4)_0+3\Lambda^2$, $\su(4)_4+4\Lambda^2$: $1$, $2$\\
$A_3^{(2)}$ & $\alpha_0+\alpha_1+\alpha_2$ & $2$ & $1$ & $\su(4)_0+4\Lambda^2$: $1$, $2$\\
$A_4^{(2)}$ & $2\alpha_0+2\alpha_1+\alpha_2$ & $2$ & $1$ & $\so(7)+5\Fund$: $1$, $2$\\
$D_4^{(2)}$ & $\alpha_0+\alpha_1+\alpha_2+\alpha_3$ & $2$ & $1$ & geometric frame (\sref{sec:exso8Z2}): $1$, $2$\\
$D_4^{(3)}$ & $\alpha_0+2\alpha_1+\alpha_2$ & $3$ & $1$ & $\SU(4)_8$, $\sp(3)_0+2\Lambda^2$: $1$, $3$;\quad $\su(4)_6+2\Lambda^2$: $2$, $6$\\
$G_2^{(1)}$ & $\alpha_0+3\alpha_1+2\alpha_2$ & $1$ & $2$ & $\su(4)_2+4\Lambda^2$: $1$, $2$\\
$D_4^{(1)}$ & $\alpha_0+\alpha_1+2\alpha_2+\alpha_3+\alpha_4$ & $1$ & $2$ & $\SU(2)$ gauging $4E_1$ (\sref{sec:E1N4}): $1$, $2$\\
\bottomrule
\end{tabular}
\caption{The data of \eqref{eq:rrule} for the frames with a 6d gauge algebra (6d theories in \tref{tab:master}, and in \sref{sec:E1N4} for the last row). $\alpha_0$ is the affine root and $\alpha_{i\geq1}$ are the simple roots of $\mathfrak g^\sigma$, with $\alpha_2$ long for $\usp(4)$ and $\mathfrak g_2$, and central for $\so(8)$. The instantonic root $\alpha_j$ is $\alpha_1$ for $A_2^{(2)}$, $\alpha_0$ for $D_4^{(2)}$, and $\alpha_2$ otherwise. In the last row, $k$ counts instantons of the central $\SU(2)$.}
\label{tab:rvalues}
\end{table}

So {\it $r$ is not the order of the twist.} It is $2$ for the untwisted $\su(4)_2+4\Lambda^2$, whose instantonic root is the long root $\alpha_2$ of $\mathfrak g_2$, of mark $2$ in $G_2^{(1)}$. It also depends on the frame: $\sp(3)_0+2\Lambda^2$ and $\su(4)_6+2\Lambda^2$, frames of one KK theory, share the instantonic root, which has instanton number $1$ in the first and $2$ in the second (\sref{sec:exso8tri}).

% ===========================================================================
% Section 4: The blow-up prefactor from the 6d frame
% ===========================================================================
\section{\texorpdfstring{The blow-up prefactor $\Lambda_B$ from the 6d frame}{The blow-up prefactor from the 6d frame}}
\label{sec:prefactor}

This section derives the prefactor $\Lambda_B$ of the blow-up equations of a KK theory from its 6d frame, for untwisted theories and twists of the first kind. On the tensor-flux family, the fluxes labelled by the flux of the tensor scalars alone (\sref{sec:pfblowup}), the exponent of the blow-up function is fixed by the Green--Schwarz couplings (\sref{sec:pffamily}). The tensor-flux family exists only if a relation among these couplings holds, which we identify with 6d anomaly cancellation (\sref{sec:pffamilycond}). Summed over the tensor-flux family, the prefactor $\Lambda_B$ is a theta function of the string charge lattice (\sref{sec:pftheta}). Its sign refines the spin and gives the dictionary \eqref{eq:dictionary} (\sref{sec:pfspin}). Three sectors of one theory determine the one-string genus (\sref{sec:pfcramer}). We read the linear part of the exponent as mixed anomalies of the 2-form symmetry (\sref{sec:pfanomalies}).

The theta functions below are, up to a shift of the first characteristic for odd $\mathcal K$ (footnote~\ref{fn:Guthetaconv} below), those of the elliptic blow-up equations of 6d theories \cite{Gu:2018gmy,Gu:2019dan,Gu:2019pqj,Gu:2020fem}, which build on the blow-up equations of refined topological strings \cite{Huang:2017mis}. There, their argument was fixed by modularity and explained heuristically through the Green--Schwarz four-form \cite[\S3.4]{Gu:2020fem}. We derive them from the effective prepotential of a KK theory, twisted or not. What is new is the reading of the characteristic as an 't~Hooft class of the 2-form symmetry of the 6d theory, the refined spin and the dictionary that follow from it, and the mixed anomalies.

% ---------------------------------------------------------------------------
\subsection{The blow-up equation}
\label{sec:pfblowup}

Let $\widehat{\mathbb C}{}^2$ be $\mathbb C^2$ blown up at the origin, and let $E\cong\BP^1$ be its exceptional curve, with $E\cdot E=-1$. A background on $\widehat{\mathbb C}{}^2$ is specified by the fluxes through $E$ of all variables of the effective prepotential $\mathcal E$. We write $n$ for the fluxes of the Coulomb moduli $\phi$ and $B_m$ for the flux of a parameter $m$, which may be a mass, the inverse coupling $m_0$ or the KK mass $\tau$. Localising on the two fixed points $N$ and $S$ of $E$ gives the blow-up equation \cite{Nakajima:2003pg,Nakajima:2005fg,Keller:2012da,Huang:2017mis,Kim:2019uqw}. In the form of \cite[(3.2), (3.7)]{Kim:2020hhh}, it reads
\begin{equation}
\label{eq:blowupeq}
\begin{aligned}
\Lambda_B\,\hat Z(\phi,m;\epsilon_1,\epsilon_2)=\sum_{n\in\mathbb F_B}(-1)^{|n|}\,e^{-V(n)}\,
&\hat Z(\phi+n\epsilon_1,m+B_m\epsilon_1;\epsilon_1,\epsilon_2-\epsilon_1)\\
\times\,&\hat Z(\phi+n\epsilon_2,m+B_m\epsilon_2;\epsilon_1-\epsilon_2,\epsilon_2) \,.
\end{aligned}
\end{equation}
Here, $\hat Z$ is the index part of the partition function, defined with $(-1)^{2J_R}$ \cite[(3.3)--(3.5)]{Kim:2020hhh}, and $|n|$ is the sum of the components of $n$ in the basis of simple coroots of the 5d gauge algebra. The sum runs over a set $\mathbb F_B$ of fluxes, called a \emph{sector}, and $\Lambda_B$ is its \emph{prefactor}, independent of the Coulomb moduli. The exponent is the finite difference of $\mathcal E$ of \eqref{eq:E5d} or \eqref{eq:E6d}:
\begin{equation}
\label{eq:Vdef}
\begin{aligned}
V(n)=\mathcal E(\phi,m;\epsilon_1,\epsilon_2)&-\mathcal E(\phi+n\epsilon_1,m+B_m\epsilon_1;\epsilon_1,\epsilon_2-\epsilon_1)\\
&-\mathcal E(\phi+n\epsilon_2,m+B_m\epsilon_2;\epsilon_1-\epsilon_2,\epsilon_2) \,.
\end{aligned}
\end{equation}
We set the flux of $\tau$ to zero, $B_\tau=0$, as in \cite{Kim:2020hhh,Kim:2021gyj}. The flux of $m_0$ then follows from the \hyperref[it:fluxmap]{flux rule} of \sref{sec:framemap}.

Expand both sides of \eqref{eq:blowupeq} in the string fugacities of the 6d frame. At leading order, the three factors $\hat Z$ contribute $1$,\footnote{At vanishing gauge flux, and with flavour fluxes $B_m=\pm\frac12$, the perturbative factors of the two patches combine into that of $\mathbb C^2$ \cite[(3.7)--(3.12)]{Gu:2020fem}. For $\SU(3)_9$, this is $R_0=1$ in the one-string equation \eqref{eq:BU394} of Appendix~\ref{app:su39}. There, $R_b$ is the ratio of the combined perturbative factors of the two patches to that of $\mathbb C^2$, at 6d gauge flux $b$.} and only the fluxes whose $V$ does not depend on the Coulomb moduli survive:
\begin{equation}
\label{eq:Lambdalead}
\Lambda_B=\sum_{n\in\mathbb F_B,\ \partial_\phi V(n)=0}(-1)^{|n|}\,e^{-V(n)} \,.
\end{equation}
In all the KK theories of \sref{sec:examples}, these fluxes form one family, the \emph{tensor-flux family}. On it, the gauge flux of the 6d frame vanishes, and the \hyperref[par:tensorflux]{shifted tensor fluxes} $a=(a_\alpha)$, $\alpha=1,\dots,n_T$, of \eqref{eq:adef} are free. We hold the 5d masses fixed. In a gauge-theory frame, the 5d gauge flux $\nu(a)$ of the member $a$, \eqref{eq:nuofa}, lies in a coset of the coweight lattice of the 5d gauge algebra whose centre element acts trivially on the matter (\sref{sec:setup5d}). In every such frame of \sref{sec:examples} below, this makes the string charge $\lambda=\mathcal Ka$ of the flux integral. So the members of one sector form a coset $a\in\BZ^{n_T}+B$, with $B=\mathcal K^{-1}\lambda\bmod\BZ^{n_T}$. The labels $B$ form $\mathcal K^{-1}\BZ^{n_T}/\BZ^{n_T}\cong D$, the defect group \eqref{eq:Ddef}. Frames without a gauge-theory description can carry a shift. In the geometric frame of $\so(8)$ with the $\BZ_2$ twist (\sref{sec:exso8Z2}), the Freed--Witten condition adds a spin$^c$ shift $\tfrac18$ to the whole family, so $a\in\BZ+B+\tfrac18$. There, $\mathcal Ka$ is half-integral, $B$ is the class of the bare flux $a-\tfrac18$, and the first characteristic of \eqref{eq:LambdaTheta} below is $B+\tfrac18$. In the elliptic frame of the untwisted rank-one theories with odd $\mathcal K$, $\mathcal Ka\in\BZ+\tfrac12$ \cite[(3.4)]{Gu:2020fem}, \cite[(4.93)]{Kim:2021gyj} (footnote~\ref{fn:Guthetaconv} below).

% ---------------------------------------------------------------------------
\subsection{The exponent on the tensor-flux family}
\label{sec:pffamily}

The tensor scalars enter $\mathcal E_{\rm 6d}$ of \eqref{eq:E6d} only through its tree-level part, the second form of \eqref{eq:Etree}. Their coefficients are the Green--Schwarz couplings: the tensor kinetic term $\tfrac\tau2\phi_0^T\mathcal K\phi_0$, and, on the node $\alpha$, the couplings $h^\vee_\alpha$ to $\SU(2)_R$, $2-\mathcal K_{\alpha\alpha}$ to gravity, $\kappa_m$ to the flavour masses $m$ on that node, and the 6d gauge coupling. Appendix~\ref{app:lemma} evaluates \eqref{eq:Vdef} monomial by monomial. At vanishing gauge flux, the gauge moduli drop out, provided every charged hypermultiplet sees a flavour flux $\pm\frac12$. The tensor terms survive through the shifts $\phi_{0,\alpha}\to\phi_{0,\alpha}+a_\alpha\epsilon$ and $m\to m+B_m\epsilon$, with $\epsilon=\epsilon_{1}$ on $N$ and $\epsilon_2$ on $S$. We evaluate them in the 6d variables and masses, in which the tensor flux differs from $a$ by a constant. In terms of $a$ and the 5d masses, the result keeps its form, up to $a$-independent terms, as \sref{sec:exZ2twist} below shows explicitly. The lemma \eqref{eq:lemma} of Appendix~\ref{app:lemma} lists the contribution of each monomial of $\mathcal E$ to $V$. Summing them gives
\begin{equation}
\label{eq:Vfamily}
\begin{aligned}
V\big|_a&=\frac\tau2\,a^T\mathcal Ka+\sum_\alpha a_\alpha x_\alpha+\sum_\alpha\Delta_\alpha\,\phi_{0,\alpha}\,,\\
x_\alpha&=-\tfrac12\big(h^\vee_\alpha-2+\mathcal K_{\alpha\alpha}\big)\,\epsilon_++\sum_{m\text{ on }\alpha}2\kappa_mB_m\,(m+B_m\epsilon_+)\,,\\
\Delta_\alpha&=\tfrac14\,h^\vee_\alpha+\tfrac34\,\big(2-\mathcal K_{\alpha\alpha}\big)+\sum_{m\text{ on }\alpha}\kappa_mB_m^2 \,.
\end{aligned}
\end{equation}
The first term comes from the tensor kinetic term, and $x_\alpha$ from the Green--Schwarz couplings to $\SU(2)_R$, gravity and flavour. The last term is the subject of \sref{sec:pffamilycond}. The twist does not enter \eqref{eq:Vfamily}, since only the Green--Schwarz couplings do. For the example frames of \sref{sec:examples}, \eqref{eq:Vfamily} with $\Delta_\alpha=0$ and $B_m=\frac12$, both imposed below, gives the argument $x$ listed in Table~\ref{tab:master} below. For untwisted rank-one theories, $-x$ is the argument $k_F\lambda_F\cdot m_F+y(\epsilon_1+\epsilon_2)$ of the theta function in \cite[(3.1)--(3.2)]{Gu:2020fem}, up to the normalisation of the flavour fluxes. Without masses, on a curve of self-intersection $-n$, both $-x$ and $y(\epsilon_1+\epsilon_2)$ equal $\frac14(n-2+h^\vee)(\epsilon_1+\epsilon_2)$.

% ---------------------------------------------------------------------------
\subsection{The family condition and anomaly cancellation}
\label{sec:pffamilycond}

The prefactor $\Lambda_B$ does not depend on the Coulomb moduli, so the last term of \eqref{eq:Vfamily} must vanish:
\begin{equation}
\label{eq:familycond}
\Delta_\alpha=\tfrac14\,h^\vee_\alpha+\tfrac34\,\big(2-\mathcal K_{\alpha\alpha}\big)+\sum_{m\text{ on }\alpha}\kappa_mB_m^2=0\qquad\text{for every }\alpha \,.
\end{equation}
Without it, there is no tensor-flux family and no $\Lambda_B$. For one curve of self-intersection $-n$, it reads
\begin{equation}
\label{eq:familycondsingle}
h^\vee+6-3n=-4\sum_m\kappa_mB_m^2 \,.
\end{equation}
Without masses, it is $h^\vee=3n-6$, the relation obeyed by the single-curve non-Higgsable clusters without matter, $n=3,4,5,6,8,12$, from $\su(3)$ on $-3$ to $\mathfrak e_8$ on $-12$ \cite{Morrison:2012np}. With masses, the right-hand side is positive, since $\kappa_m<0$. The hypermultiplets require $B_m=\pm\frac12$ (Appendix~\ref{app:lemma}) and, in every example of \sref{sec:examples}, this value solves \eqref{eq:familycond}. This is the value used in \cite[(3.15)]{Gu:2020fem} and \cite{Harding:2026gzj}. We take $B_m=\frac12$ from now on.

The right-hand side of \eqref{eq:familycondsingle} has a 6d meaning. We observe that, in every example of \sref{sec:examples} with matter, it is the total Dynkin index of the 6d matter:
\begin{equation}
\label{eq:kappaindex}
-4\sum_m\kappa_mB_m^2 = -\sum_m\kappa_m =\sum_{\mathbf R}\mathrm{ind}_{\mathbf R}\,,\qquad \mathrm{ind}_{\adj}=h^\vee \,,
\end{equation}
with the values $3,4,4,5,6$ for $\su(3)+6\Fund$, $\su(4)+8\Fund$, $\mathfrak g_2+4\cdot\mathbf 7$, $\su(5)+10\Fund$ and $\so(8)+2(\mathbf 8_v+\mathbf 8_s+\mathbf 8_c)$. This is consistent with gauge anomaly cancellation, since the flavour Green--Schwarz levels $\kappa_m$ cancel the mixed gauge--flavour anomaly of the hypermultiplets, which is proportional to their Dynkin indices.\footnote{\label{fn:kappaper}Per mass, $\kappa_m=-\sum_h\mathrm{ind}_{\mathbf R_h}\,q_{h,m}^2$, summed over the hypermultiplets $h$ of charge $q_{h,m}$ under $m$. For example, $\kappa_m=-1$ for a pair of fundamentals identified by the twist, and $\kappa_m=-3$ for the three $\mathbf 8$'s permuted by triality.} With \eqref{eq:kappaindex}, \eqref{eq:familycondsingle} becomes
\begin{equation}
\label{eq:anomalygg}
h^\vee-\sum_{\mathbf R}\mathrm{ind}_{\mathbf R}=3n-6 \,,
\end{equation}
the cancellation of the mixed gauge--gravitational anomaly on the tensor branch \cite{Ohmori:2014kda,Heckman:2015bfa}. So, in our examples, the existence of the tensor-flux family is a 6d anomaly-cancellation condition. The E- and M-strings have no gauge anomaly. For them, \eqref{eq:familycondsingle}, with $h^\vee=1$, is equivalent to the modularity constraint $(r_m/2,r_m/2)=\frac12(7-3n)$ of \cite[(3.22)]{Gu:2019pqj}.

% ---------------------------------------------------------------------------
\subsection{\texorpdfstring{The prefactor $\Lambda_B$ as a lattice theta function}{The prefactor as a lattice theta function}}
\label{sec:pftheta}

With \eqref{eq:familycond}, each member of the tensor-flux family contributes $e^{-V}=\fq^{\frac12a^T\mathcal Ka}\,e^{-a\cdot x}$, with $\fq=e^{-\tau}$. The remaining input is the sign $(-1)^{|\nu(a)|}$ of \eqref{eq:blowupeq}. We derive it for one curve in three steps.
\begin{itemize}
\item By \eqref{eq:nuofa}, $\nu(a)=a\,c_1$, and $c_1$ lists the comarks $\mathsf a^\vee_j$ of the affine algebra. Their sum is the dual Coxeter number $h^\vee$ of the 6d gauge algebra, also for twisted affine algebras. So $(-1)^{|\nu(a)|}=(-1)^{h^\vee a}$.
\item By \eqref{eq:anomalygg}, $h^\vee\equiv n+\sum_{\mathbf R}\mathrm{ind}_{\mathbf R}\bmod 2$.
\item By \eqref{eq:kappaindex}, the factor $(-1)^{a\sum_{\mathbf R}\mathrm{ind}_{\mathbf R}}$ is what the half-period shift $m\to m-i\pi$ of every mass does to $e^{-a\cdot x}$. This shift exchanges the conventions $(-1)^F$ and $(-1)^{2J_R}$ for the hypermultiplets, as in the replacement \cite[(3.3)]{Kim:2020hhh}, which they implement by $\epsilon_1\to\epsilon_1+2\pi i$. \sref{sec:exMstring} below shows its effect for the M-string.
\end{itemize}
The argument applies node by node to several tensors, provided each tensor column $c_\alpha$ of \eqref{eq:nuofa} sums to $h^\vee_\alpha$ mod $2$, as in the two-tensor frame of \sref{sec:exsu44} below, where $|c_1|=1$ and $|c_2|=3$. Hence, up to a constant on each sector and the half-period shift of the masses,
\begin{equation}
\label{eq:signw}
(-1)^{|\nu(a)|}=e^{i\pi\,w\cdot a}\,,\qquad w_\alpha\equiv\mathcal K_{\alpha\alpha}\bmod 2 \,,
\end{equation}
with $w$ a Wu vector of the lattice, as in \eqref{eq:qw}. The same sign appears in \cite[(3.38)]{Gu:2019pqj}. Summing over the tensor-flux family gives the prefactor $\Lambda_B$:
\begin{equation}
\label{eq:LambdaTheta}
\Lambda_B=\Theta\big[{\textstyle\genfrac{}{}{0pt}{}{B}{w/2}}\big](\mathcal K\tau,x):=\sum_{a\in\BZ^{n_T}+B}e^{i\pi\,w\cdot a}\,\fq^{\frac12a^T\mathcal Ka}\,e^{-a\cdot x} \,.
\end{equation}
For one tensor, $\Lambda_B=\vartheta[\genfrac{}{}{0pt}{}{B}{\mathcal K/2}](\mathcal K\tau,x)$.\footnote{\label{fn:thetaconv}Conventions for theta functions differ in the literature. We keep the KK mass $\tau$ as the argument, with $\fq=e^{-\tau}$, so the modular parameter is $i\tau/2\pi$, and define $\vartheta[\genfrac{}{}{0pt}{}{\alpha}{\beta}](\tau,x):=\sum_{n\in\BZ+\alpha}e^{2\pi i\beta n}\,\fq^{\frac12n^2}\,e^{-nx}$, which is \cite[(A.2)]{Kim:2021gyj} in these variables. The Jacobi theta functions are $\theta_3=\vartheta[\genfrac{}{}{0pt}{}{0}{0}]$, $\theta_2=\vartheta[\genfrac{}{}{0pt}{}{1/2}{0}]$, $\theta_4=\vartheta[\genfrac{}{}{0pt}{}{0}{1/2}]$ and, up to a constant phase, $\theta_1=\vartheta[\genfrac{}{}{0pt}{}{1/2}{1/2}]$. \eqref{eq:LambdaTheta} is the version for several tensors.} The first characteristic $B$ is the 't~Hooft flux of the 2-form symmetry through $E$, so each sector of the blow-up equation is a class of $D$. The second characteristic is half the Wu vector.\footnote{\label{fn:Guthetaconv}With $w=\mathrm{diag}\,\mathcal K$, \eqref{eq:LambdaTheta} has the form of the generalised theta function of \cite[(3.38)]{Gu:2019pqj}, see also \cite[(3.69)--(3.70)]{Gu:2020fem}. There, the first characteristic is taken in $\mathcal K^{-1}\big(\BZ^{n_T}+\frac12\mathrm{diag}\,\mathcal K\big)$ \cite[(3.39)]{Gu:2019pqj}, i.e.\ it is our $B$ shifted by $\frac12\mathcal K^{-1}\mathrm{diag}\,\mathcal K$. For even $\mathcal K$, this only relabels the sectors. For odd $\mathcal K$, it amounts to a shift of the elliptic argument by $\frac12\mathcal K^{-1}\mathrm{diag}\,\mathcal K\,\tau$. The two families are then inequivalent: for $\mathcal K=3$ and $x=-2\epsilon_+$, the determinant of \eqref{eq:cramer} below vanishes identically for the shifted characteristics, which is the exception of \cite[fn.~27]{Gu:2020fem}, but not for ours. With our $B$, the 't~Hooft class, \eqref{eq:LambdaTheta} gives the theta functions of \cite[(4.116)]{Kim:2021gyj} for $\SU(3)_9$.} For even $\mathcal K$, we may take $w\in2\BZ^{n_T}$, e.g.\ $w=0$. Then, $e^{i\pi w\cdot a}=e^{i\pi w\cdot B}$ is constant on the sector $a\in\BZ^{n_T}+B$, and the second characteristic only contributes this phase. For odd $\mathcal K$, some $w_\alpha$ is odd, and moving $a_\alpha$ by $1$ within the sector flips the sign. So the sign alternates within a sector.

% ---------------------------------------------------------------------------
\subsection{The Wu sign, the refined spin and the dictionary}
\label{sec:pfspin}

For odd $\mathcal K$, the exponent $\frac12a^T\mathcal Ka$ is not constant mod $1$ on a sector. By \eqref{eq:qshift}, it jumps by $\frac12$ between neighbouring points. For $\mathcal K=(3)$ and $B=\frac13$, the points $a=\frac13$ and $a=-\frac23$ give $\frac12a^T\mathcal Ka=\frac16$ and $\frac23$. The sign \eqref{eq:signw} alternates in exactly the same way. Count it as half a unit of KK momentum, $e^{i\pi w\cdot a}=e^{2\pi i\cdot\frac12w\cdot a}$, and define the refined KK momentum
\begin{equation}
\label{eq:sdef}
p(a)=\tfrac12\,a^T\mathcal Ka+\tfrac12\,w\cdot a\pmod 1 \,.
\end{equation}
In terms of $\lambda=\mathcal Ka$, $p(a)=q_{-w}(\lambda)$, with $q_w$ the Wu-refined spin \eqref{eq:qw}. Since $-w_\alpha\equiv w_\alpha$ mod $2$, $-w$ is also a Wu vector. Moving $a$ by $\mu\in\BZ^{n_T}$ moves $\lambda$ by $\mathcal K\mu$. So, by the integrality shown below \eqref{eq:qw}, $p$ is constant on each sector. Since $q_{-w}(\lambda)-q_w(\lambda)=w^T\mathcal K^{-1}\lambda=(\mathcal K^{-1}w)\cdot\lambda$, $p(a)=q_{-w}(\lambda)$ and $q_w(\lambda)$ agree mod $1$ on all of $D$ if and only if $\mathcal K^{-1}w$ is integral:
\begin{equation}
\label{eq:sequalsqw}
p(a)\equiv q_w(\mathcal Ka)\pmod 1\qquad\text{for } w\in\mathcal K\BZ^{n_T} \,.
\end{equation}
Such $w$ exist for every $\mathcal K$ (\sref{sec:setup6d}).

The prefactor $\Lambda_B$ fixes $w$ only mod $2$. Let $\mu\in\BZ^{n_T}$, so that $w+2\mu$ is again a Wu vector. Under $w\to w+2\mu$, the sign $e^{i\pi w\cdot a}$ changes by $e^{2\pi i\mu\cdot a}=e^{2\pi i\mu\cdot B}$, since $a\in\BZ^{n_T}+B$. This is a constant on each sector, which the blow-up equations do not test, since it can be absorbed into the prefactor $\Lambda_B$ of that sector. The refined spin changes by $q_{w+2\mu}(\lambda)-q_w(\lambda)=-\ell(\lambda,\mu)$, with $\ell$ of \eqref{eq:linking}. This vanishes for all $\lambda$ if and only if $\mu\in\mathcal K\BZ^{n_T}$, as in \eqref{eq:qwshiftw}. So the refinements compatible with $\Lambda_B$ are labelled by $\mu\in D$. For $\mathcal K=(3)$, $w=1,3,5$ give $(q_w(1),q_w(2))=(0,\frac13)$, $(\frac23,\frac23)$, $(\frac13,0)$. Only $w=\mathcal K$, the charge-conjugation-invariant choice, gives the conjugate classes equal spin (Table~\ref{tab:lattices}).

\paragraph{The dictionary.} Compare the two frames on the tensor-flux family. In the 5d frame, at fixed 5d masses, $\tau$ enters $V$ only through $m_0=\tau/r$ of \eqref{eq:rdef}. By \eqref{eq:expV}, $e^{-V}\supset q^{k(\nu)}=\fq^{k(\nu)/r}$, with $k$ the instanton number \eqref{eq:kQ}. In the 6d frame, the same exponent is $\frac12a^T\mathcal Ka$, by \eqref{eq:Vfamily}. Adding the sign \eqref{eq:signw} and using \eqref{eq:sequalsqw}, with $\lambda=\mathcal Ka$,
\begin{equation}
\label{eq:dictderived}
\frac{k(\nu(a))}{r}=\tfrac12\,\lambda^T\mathcal K^{-1}\lambda\,,\qquad
\frac{k(\nu(a))}{r}+\tfrac12\,w\cdot a\equiv q_w(\lambda)\pmod 1\quad\big(w\in\mathcal K\BZ^{n_T}\big) \,.
\end{equation}
This proves the dictionary \eqref{eq:dictionary}, with the \hyperref[par:dictchoices]{three choices} and the \hyperref[par:scope]{scope} stated in \sref{sec:setupdict}. For $\SU(3)_9$, $\nu(a)=(a,2a)$, $k=3a^2$ and $r=2$ give $\frac16$ at $a=\frac13$. The sign adds $\frac12$, so $p=\frac23=q_3(1)$.

% ---------------------------------------------------------------------------
\subsection{Three sectors determine the one-string genus}
\label{sec:pfcramer}

All sectors share the same $\hat Z$, which makes them a linear system at one-string order. Take one tensor and write $\hat Z=\hat Z_{\rm pert}\,(1+e^{-\mathcal K\phi_0}\hat Z_1+\dots)$, with $\hat Z_{\rm pert}$ the perturbative part in the 6d frame. Let $\hat Z_1^N$ and $\hat Z_1^S$ be the values of $\hat Z_1$ on the two patches of \eqref{eq:blowupeq}. On the patch $N$, the flux shifts the one-string fugacity to $e^{-\mathcal K\phi_0}p_1^{\mathcal Ka}$, with $p_i=e^{-\epsilon_i}$, and, on $S$, to $e^{-\mathcal K\phi_0}p_2^{\mathcal Ka}$. These factors shift the argument of $\Lambda_B$ by $\mathcal K\epsilon_{1,2}$. The fluxes off the tensor-flux family with one unit of string charge give a source $S_B$, built from theta functions and perturbative data. So, at one-string order,
\begin{equation}
\label{eq:sectoreq}
\Theta_B(x)\,\hat Z_1=\Theta_B(x+\mathcal K\epsilon_1)\,\hat Z_1^N+\Theta_B(x+\mathcal K\epsilon_2)\,\hat Z_1^S+S_B\,,\qquad
\Theta_B(x):=\Theta\big[{\textstyle\genfrac{}{}{0pt}{}{B}{w/2}}\big](\mathcal K\tau,x) \,.
\end{equation}
Three sectors $B_1,B_2,B_3$ of the same theory determine the three unknowns by Cramer's rule,
\begin{equation}
\label{eq:cramer}
\hat Z_1=\frac{\det\big(S,\ \Theta(x+\mathcal K\epsilon_1),\ \Theta(x+\mathcal K\epsilon_2)\big)}{\det\big(\Theta(x),\ \Theta(x+\mathcal K\epsilon_1),\ \Theta(x+\mathcal K\epsilon_2)\big)}\,,
\end{equation}
where each column is a vector over the three sectors, $\Theta(x)=(\Theta_{B_i}(x))_{i=1,2,3}$ and $S=(S_{B_i})_{i=1,2,3}$. For $\SU(3)_9$, with $x=-\epsilon_1-\epsilon_2$, this is the closed form \cite[(4.116)]{Kim:2021gyj} (Appendix~\ref{app:su39}). For untwisted rank-one theories with $n\ge3$ that have no unpaired half-hypermultiplets, it is, with the shifted characteristics of footnote~\ref{fn:Guthetaconv}, the recursion \cite[(4.1)--(4.5)]{Gu:2020fem} of the elliptic frame.

We say that the sectors \emph{close} when, as here, they determine the one-string genus $\hat Z_1$ in closed form. So the sectors close only with $|D|\ge3$ sectors of one theory and a non-vanishing determinant. Consider a frame with a 6d gauge algebra whose centre does not survive the twist and the matter (Appendix~\ref{app:twists}). Then, the 1-form symmetry of the frame is $D$, all $|D|$ classes are 't~Hooft sectors of one 5d theory, and each should give a blow-up equation of it. This is shown for $\SU(3)_9$, $\USp(4)_0+3\Lambda^2$ and three sectors of $\SU(4)_8$, and checked at one-string order for its fourth sector (\sref{sec:exSU48} below). Elsewhere, it is a prediction (\sref{sec:examples} below). This criterion is not necessary. In the geometric frame of $\so(8)$ with the $\BZ_2$ twist, treated in \sref{sec:exso8Z2} below, the subgroup $\{1,v\}$ of the centre of $\Spin(8)$ survives, and all four sectors of $D=\BZ_4$ are still admissible flux sets of that frame. Three of them fix the one-string genus \cite[\S6.2]{Kim:2020hhh}. What decides closure when a gauge centre survives is open (\sref{sec:conclusions}). The criterion says nothing about frames without a 6d gauge algebra. For the M-string, both sectors are nevertheless blow-up equations of $\SU(2)_0+\adj$ (\sref{sec:exMstring} below). With $|D|=2$, the two sector equations cannot fix three unknowns and are consistency conditions, as for class B of \cite{Gu:2020fem}. Table \eqref{eq:closuretable} collects the cases.
\begin{equation}
\label{eq:closuretable}
\mbox{\footnotesize\begin{tabular}{L{2.9cm}cL{4.7cm}L{3.9cm}}
\toprule
theory & $D$ & sectors & one-string genus\\
\midrule
$\SU(3)_9$ & $\BZ_3$ & all three \cite[(4.115)]{Kim:2021gyj} & fixed by \eqref{eq:cramer}, \cite[(4.116)]{Kim:2021gyj}\\
$\SU(4)_8$ & $\BZ_4$ & $B=0,\pm\frac14$ \cite[(6.9)]{Kim:2020hhh}; $B=\frac12$ checked at one-string order (\sref{sec:exSU48}) & fixed by the first three \cite[\S6.1]{Kim:2020hhh}\\
$\so(8)$ on $-4$, $\BZ_2$ twist & $\BZ_4$ & $B=0,\pm\frac14$, i.e.\ $a\in\BZ+B+\frac18$ in \cite[(6.13)]{Kim:2020hhh}; $B=\frac12$ untested & fixed by these three \cite[\S6.2]{Kim:2020hhh}, although $\{1,v\}$ survives\\
untwisted rank one, $n\ge3$, no unpaired half-hypermultiplets & $\BZ_n$ & all $n$, elliptic frame & fixed \cite[(4.5)]{Gu:2020fem}, except for $\su(3)$ on $-3$ \cite[fn.~27]{Gu:2020fem}\\
M-string & $\BZ_2$ & both, \cite[(3.26)]{Gu:2019pqj} and \sref{sec:exMstring} & not fixed; input from \cite[(4.71)]{Kim:2021gyj}\\
$\USp(4)_0+3\Lambda^2$ & $\BZ_2$ & both \cite[\S5.1.1]{Kim:2020hhh} & consistency only\\
$\su(4)_{0,2,4}+4\Lambda^2$, $\so(7)+5\Fund$, $\sp(3)_0+2\Lambda^2$, $\su(4)_6+2\Lambda^2$ & $\BZ_2$ & both, as a prediction (\sref{sec:examples} below) & consistency only\\
\bottomrule
\end{tabular}}
\end{equation}

% ---------------------------------------------------------------------------
\subsection{Mixed anomalies of the 2-form symmetry}
\label{sec:pfanomalies}

The 7d symmetry theory of \cite[(2.8)]{Apruzzi:2022dlm} contains the gauge part of the Green--Schwarz coupling. With the full four-form $X_4=(X_{4\alpha})$ of \eqref{eq:GSdata}, this coupling is
\begin{equation}
\label{eq:GS7d}
2\pi\int\mathcal C_3^T\,\mathcal K\,X_4 \,,
\end{equation}
with $\mathcal C_3$ the background of the 2-form symmetry. Its R-symmetry, gravitational and flavour terms are absent from \cite{Apruzzi:2022dlm}. On the circle, $\mathcal C_3$ reduces to the background $\mathcal B$ of the 1-form symmetry of tensor origin \eqref{eq:KKterm}. Then, \eqref{eq:GS7d} becomes a mixed anomaly of this 1-form symmetry with $\SU(2)_R$, gravity and flavour, with the same coefficients. If a 1-form symmetry of gauge origin survives the twist, the gauge term gives its mixed anomaly with the 1-form symmetry of tensor origin (\sref{sec:exso8Z2}). The linear part of \eqref{eq:Vfamily} is where the blow-up equations see it.

\paragraph{The universal form.} With $B_m=\frac12$, \eqref{eq:familycond} gives $\frac12(h^\vee_\alpha-2+\mathcal K_{\alpha\alpha})=2(\mathcal K_{\alpha\alpha}-2)-\frac12\sum_{m\text{ on }\alpha}\kappa_m$. So $x_\alpha=\sum_{m\text{ on }\alpha}\kappa_m(m+\epsilon_+)-2(\mathcal K_{\alpha\alpha}-2)\,\epsilon_+$, and each member of the tensor-flux family contributes
\begin{equation}
\label{eq:Vuniversal}
e^{-V}\big|_a=\fq^{\frac12a^T\mathcal Ka}\,\prod_m \tilde y_m^{-\kappa_m\,a_{\alpha(m)}}\,(p_1p_2)^{\tw\cdot a}\,,\qquad \tilde y_m=e^{m+\epsilon_+}\,,\quad \tw_\alpha=2-\mathcal K_{\alpha\alpha}\,,
\end{equation}
where $\alpha(m)$ is the node that carries $m$. Two consequences are immediate. First, once \eqref{eq:familycond} holds, the $\SU(2)_R$ coupling $h^\vee_\alpha$ is fixed by $\mathcal K$ and the $\kappa_m$. Besides $\mathcal K$, which also fixes $\tw$, the prefactor $\Lambda_B$ carries only the flavour levels $\kappa_m$. If the anomaly-cancellation form of the couplings is not assumed, as in the inversion below, $\tw$ is a separate output. Second, every mass enters together with $\epsilon_+$.

\paragraph{Gravity.} The exponent $\tw_\alpha=2-\mathcal K_{\alpha\alpha}=(\mathcal K\mathfrak a)_\alpha$ is the gravitational coupling of \eqref{eq:Etree}, as in \cite[(3.65)]{Heckman:2022suy}. It vanishes on $-2$ curves. By adjunction, $\tw_\alpha=-K_{\rm base}\cdot C_\alpha$, which is characteristic by the Wu formula $C\cdot C\equiv K_{\rm base}\cdot C\bmod 2$ \cite{Monnier:2017oqd,Monnier:2018nfs}. So $\tw\equiv w\bmod 2$, and, up to a phase per sector, the sign and the $\epsilon_+$-dependence combine into $(-p_1p_2)^{\tw\cdot a}$. For $\mathcal K=(3)$, $\tw=-1$ gives the same refinement as $w=5$, by \eqref{eq:qwshiftw}, which is not the charge-conjugation-invariant one (\sref{sec:pfspin}).

\paragraph{Flavour and $\epsilon_+$.} Locally, the $\Omega$-background twists $\SU(2)_R$ with the rotations $\SU(2)_r$ of $\mathbb C^2$, whose Cartan parameter is $\epsilon_+$. This is the equivariant rule $c_2(R)\to\epsilon_+^2$ of \eqref{eq:equivrule}, and the shift $m_F\to m_F+\lambda_F\epsilon_+$ of \cite[(3.78)]{Gu:2020fem}. Globally, $\widehat{\mathbb C}{}^2$ is not spin, since $c_1\cdot E=2+E\cdot E=1$ by adjunction. The hypermultiplet fermions are $\SU(2)_R$ singlets, but spacetime spinors. So their gauge and flavour bundle must have $w_2$ equal to $w_2(\widehat{\mathbb C}{}^2)$. On the tensor-flux family, the gauge flux vanishes. So the flavour bundle carries $w_2(\widehat{\mathbb C}{}^2)$, which is $B_m=\frac12$. Then, $m+\epsilon_+$ arises in two halves. The shift $m\to m+B_m\epsilon$ gives $\kappa_ma(m+\frac12\epsilon_+)$ by the last row, $m^2\phi_{0,\alpha}$, of the lemma \eqref{eq:lemma} in Appendix~\ref{app:lemma}, and the family condition \eqref{eq:familycond} supplies the other $\frac12\epsilon_+$ through $h^\vee$. The faithful groups behind these twists are the subject of \sref{sec:faithful}.

\paragraph{The sectors.} Table \eqref{eq:sectorexponents} lists the fractional exponents of \eqref{eq:Vuniversal} in the non-trivial sectors, mod $1$.
\begin{equation}
\label{eq:sectorexponents}
\mbox{\footnotesize\begin{tabular}{L{6.3cm}lcc}
\toprule
theory & sector $B$ & $\tilde y_m$-exponent & $p_1p_2$-exponent\\
\midrule
M-string; $\USp(4)_0+3\Lambda^2$; $\su(4)_{0,2}+4\Lambda^2$; $\so(7)+5\Fund$; $\sp(3)_0+2\Lambda^2$, $\su(4)_6+2\Lambda^2$ & $\frac12$ & $\frac12$ & $0$\\
$\su(4)_4+4\Lambda^2$ & $(\frac12,\frac12)$ & $\frac12$ & $0$\\
$\SU(3)_9$ & $\frac13,\ \frac23$ & -- & $\frac23,\ \frac13$\\
$\SU(4)_8$ & $\frac14,\ \frac12,\ \frac34$ & -- & $\frac12,\ 0,\ \frac12$\\
\bottomrule
\end{tabular}}
\end{equation}
Local BPS states carry integer exponents. A fractional exponent in the sector $B$ is a fractional charge of its states, which no redefinition of local operators removes. We read it as a {\it mixed 't~Hooft anomaly of the 2-form symmetry of the 6d theory.} A fractional $\tilde y_m$-exponent comes from $\kappa_m$. It pairs the 2-form symmetry with flavour, $\SU(2)_R$ and Lorentz together. In $\su(4)_0+4\Lambda^2$ and $\su(4)_4+4\Lambda^2$, it signals a {\it 2-group} instead (\sref{sec:faithful}). A fractional $p_1p_2$-exponent comes from $\tw$ and pairs the 2-form symmetry with gravity and $\SU(2)_R$. For $\SU(4)_8$, it is of order $2$ in $\BZ_4$. As \sref{sec:exso8Z2} below shows for the same 6d theory, it splits into a gravitational part $\tfrac14\tw\cdot a$ and an $\SU(2)_R$ part $\tfrac34\tw\cdot a$, which give the cubic self-anomaly and a mixed anomaly with $\SU(2)_R$, $\tfrac12$ each on the generator. There, the reading is checked on the bare flux against the anomalies of the geometry \cite[(4.21)]{Apruzzi:2021nmk} and against the circle reduction of \eqref{eq:CS7d} and \eqref{eq:GS7d}. This check does not test the correlation with $w_2(\widehat{\BC}{}^2)$. The trivial sector has integer exponents, so the table does not see anomalies without the 2-form symmetry.

This reading is an interpretation. It becomes a derivation once \eqref{eq:GS7d} is evaluated on $\widehat{\mathbb C}{}^2$, with flavour and $\SU(2)_R$ backgrounds correlated with $w_2(\widehat{\mathbb C}{}^2)$. For the flavour part, the blow-up cannot separate $\mathcal B\cup\mathcal B$ from $\mathcal B\cup w_2(\widehat{\mathbb C}{}^2)$, since $\mathcal B$ and $w_2(\widehat{\mathbb C}{}^2)$ have the same value on $E$.

\paragraph{Inversion.} Conversely, the prefactor $\Lambda_B$ determines the Green--Schwarz data of one tensor. $\mathcal K$ is read from the $\fq$-exponent, and $\kappa_m$ from the coefficient $2\kappa_mB_m$ of $a\,m$. Let $\xi$ be the coefficient of $-a\,\epsilon_+$ in $V$. Take the tensor couplings in the first form of \eqref{eq:Etree}, with $\mathfrak a$ and $\mathfrak c$ not yet fixed by anomaly cancellation. Then, the $\epsilon_+$-part of \eqref{eq:Vfamily} and the family condition \eqref{eq:familycond} give
\begin{equation}
\label{eq:inversion}
\tfrac12\,\mathcal K(\mathfrak c-\mathfrak a)-2\sum_m\kappa_mB_m^2=\xi\,,\qquad
\tfrac14\,\mathcal K\mathfrak c+\tfrac34\,\mathcal K\mathfrak a+\sum_m\kappa_mB_m^2=0 \,.
\end{equation}
Three inputs are assumed: $r$ from \eqref{eq:m0delta}, since the 5d data alone give only $\mathcal Kr$, the identification of the primitive direction of the tensor-flux family with the tensor flux, and the Green--Schwarz form of the linear term. For $\SU(3)_9$, the 5d frame gives $V=3m_0a^2-2a\epsilon_+$. There are no masses, so the $\kappa_m$-terms are absent and $\xi=2$. With $m_0=\tau/r$ and $r=2$, the $\fq$-exponent is $\frac32a^2\tau=\frac12\mathcal Ka^2\tau$, so $\mathcal K=3$. Then, \eqref{eq:inversion} reads $\mathfrak c-\mathfrak a=\frac43$ and $\mathfrak c+3\mathfrak a=0$, so $(\mathfrak c,\mathfrak a)=(1,-\frac13)$, in agreement with the second form of \eqref{eq:Etree}. So the tensor sectors of the blow-up equations, together with the family condition \eqref{eq:familycond}, determine $(\mathcal K,\mathfrak a,\kappa_m,\mathfrak c)$. These are all the mixed anomalies of the 2-form symmetry with the KK $\U(1)$, gravity, flavour and $\SU(2)_R$.

% ===========================================================================
% Section 5, first half: master table, SU(3)_9, M-string,
% su(N)+2NF with the Z_2 twist (N=3,5), SU(4)_8.
% ===========================================================================
\section{Examples}
\label{sec:examples}

This section applies \sref{sec:framemap} and \sref{sec:prefactor} to eleven 5d frames of ten KK theories. Table~\ref{tab:master} collects the outcome. Ten frames are gauge theories. The last one, in \sref{sec:exso8Z2} below, is the geometric frame of \sref{sec:rrule}, for $\so(8)$ on a $-4$ curve with the $\BZ_2$ twist. Each example with a gauge-theory frame goes through the same steps: (i) the 6d data and $\mathcal E_{\rm 6d}$ \eqref{eq:E6d} in a stated chamber, (ii) $\mathcal E_{\rm 5d}$ \eqref{eq:E5d} in a stated chamber, (iii) the frame map \eqref{eq:framemapgen} by Table~\ref{tab:framemaprecipe}, which pairs the two chambers (\tref{tab:scans}), (iv) the 5d states in 6d variables, with $\delta$ and $k(\delta)$, which fix $r$ by \eqref{eq:m0delta}, (v) the prefactor $\Lambda_B$ on the tensor-flux family, \eqref{eq:Vfamily} and \eqref{eq:LambdaTheta}, with the 5d flux coset of each sector, and (vi) the dictionary \eqref{eq:dictionary}. \sref{sec:exSU39} carries out all six steps. The later examples show only what is new.

\paragraph{States in 6d variables.} We sort 5d BPS states into strings, KK modes and zero modes by the \hyperref[par:rootkinds]{criterion} of \sref{sec:recipe}. String masses are written through the effective coupling $\Phi=\mathcal K\phi_0-c\,\tau$ of \eqref{eq:F1loop6d}, the coefficient of $\tfrac12\tilde K\phi\phi$ in $\mathcal E_{\rm 6d}$. We write $M_i=m_i+v_i\tau$ for the 5d masses of \eqref{eq:framemapgen}, and $e_i$ for the orthonormal components of the 5d Coulomb moduli.

\begin{table}[tbp]
\centering\footnotesize
\setlength{\tabcolsep}{2.3pt}\renewcommand{\arraystretch}{1.25}
\begin{tabular}{l c l c c c c c l c}
\toprule
5d frame & KK & 6d theory & $\mathcal K$;\ $D$ & $q$ & $k(\nu)$ & $r$ & $w/2$ & $x$ & \S\\
\midrule
$\SU(2)_0+\adj$ & (2.1) & $(2,0)$ $A_1$ & $2$;\ $\BZ_2$ & $\tfrac14$ & $\tfrac14$ & $1$ & $0$ & $-m_1-\epsilon_+$ & \ref{sec:exMstring}\\
\midrule
$\SU(3)_9$ & (2.23) & $\su(3)$ on $-3$ & $3$;\ $\BZ_3$ & $\tfrac16^{\,*}$ & $\tfrac13$ & $2$ & $\tfrac12$ & $-2\epsilon_+$ & \ref{sec:exSU39}\\
$\USp(4)_0+3\Lambda^2$ & (2.25) & $\su(3)+6\Fund$ on $-2$ & $2$;\ $\BZ_2$ & $\tfrac14$ & $\tfrac12$ & $2$ & $0$ & $-{\textstyle\sum}M-3\epsilon_+$ & \ref{sec:exZ2twist}\\
$\so(7)+5\Fund$ & (2.20) & $\su(5)+10\Fund$ on $-2$ & $2$;\ $\BZ_2$ & $\tfrac14$ & $\tfrac12$ & $2$ & $0$ & $-{\textstyle\sum}M-5\epsilon_+$ & \ref{sec:exZ2twist}\\
$\SU(4)_8$ & (2.29) & $\so(8)$ on $-4$ & $4$;\ $\BZ_4$ & $\tfrac18$ & $\tfrac38$ & $3$ & $0$ & $-4\epsilon_+$ & \ref{sec:exSU48}\\
$\su(4)_0+4\Lambda^2$ & (2.35) & $\su(4)+8\Fund$ on $-2$ & $2$;\ $\BZ_2$ & $\tfrac14$ & $\tfrac12$ & $2$ & $0$ & $-{\textstyle\sum}M-4\epsilon_+$ & \ref{sec:exsu40}\\
$\su(4)_4+4\Lambda^2$ & (2.37) & E-string $+$ $\su(3)$ on $-3$ & $\Kmat{1}{-1}{-1}{3}$;\ $\BZ_2$ & $\tfrac14^{\,*}$ & $\tfrac12$ & $2$ & $(\tfrac12,\tfrac12)$ & $(-2\epsilon_+-{\textstyle\sum}M,\,-2\epsilon_+)$ & \ref{sec:exsu44}\\
$\su(4)_2+4\Lambda^2$ & (2.36) & $\mathfrak g_2+4\cdot\mathbf 7$ on $-2$ & $2$;\ $\BZ_2$ & $\tfrac14$ & $\tfrac12$ & $2$ & $0$ & $-{\textstyle\sum}M-4\epsilon_+$ & \ref{sec:exsu42}\\
$\sp(3)_0+2\Lambda^2$ & (2.38) & $\so(8)+6\cdot\mathbf 8$ on $-2$ & $2$;\ $\BZ_2$ & $\tfrac14$ & $\tfrac34$ & $3$ & $0$ & $-3{\textstyle\sum}M-6\epsilon_+$ & \ref{sec:exso8tri}\\
$\su(4)_6+2\Lambda^2$ & (2.30) & $\so(8)+6\cdot\mathbf 8$ on $-2$ & $2$;\ $\BZ_2$ & $\tfrac14$ & $\tfrac32^{\,\dagger}$ & $6$ & $0$ & $3{\textstyle\sum}M'-6\epsilon_+$ & \ref{sec:exso8tri}\\
\midrule
geometric & --- & $\so(8)$ on $-4$, $\BZ_2$ twist & $4$;\ $\BZ_4$ & $\tfrac18$ & $\tfrac14^{\,\ddagger}$ & $2$ & $0$ & $-4\epsilon_+$ & \ref{sec:exso8Z2}\\
\bottomrule
\end{tabular}
\caption{The examples, one row per frame. The last row is the geometric frame of \sref{sec:exso8Z2}, which has no gauge-theory description. Column KK labels the KK theory as in \cite[\S2.1]{Bhardwaj:2020gyu}, and the rows (2.1) and (2.20) are the members $m=2$ and $m=7$ of these families. Here, $6\cdot\mathbf 8$ is $2(\mathbf 8_v\oplus\mathbf 8_s\oplus\mathbf 8_c)$. The entry $q$ is the spin \eqref{eq:qdef} of the generator of $D$, $\nu$ the 5d gauge flux of the generating 't~Hooft sector on the tensor-flux family, $k(\nu)$ its instanton number \eqref{eq:kQ}, and $r=\tau/m_0$. The affine algebra, $\delta$ and $k(\delta)$ are in \tref{tab:rvalues}. The prefactor is $\Lambda_B=\Theta\big[\genfrac{}{}{0pt}{1}{B}{w/2}\big](\mathcal K\tau,x)$ of \eqref{eq:LambdaTheta}, with $B$ the class of the shifted tensor flux and $x$ in the 5d masses, $\sum M:=\sum_iM_i$ ($M'_i$ for $\su(4)_6+2\Lambda^2$). The characteristic $w/2$ is listed modulo $1$. $^*$Odd lattice, $q$ defined modulo $\tfrac12$. $^\dagger$Above the minimum $\tfrac12$ of the coset. $^\ddagger$On the bare flux $\nu(\tfrac14)$. The fluxes of this frame carry the spin$^c$ shift $\tfrac18$, and the first characteristic of $\Lambda_B$ is $B+\tfrac18$.}
\label{tab:master}
\end{table}

\paragraph{Sectors.} Every row of Table~\ref{tab:master} satisfies \eqref{eq:dictionary} with the $r$ of \eqref{eq:m0delta}. For the M-string, both sectors are blow-up equations of $\SU(2)_0+\adj$ (\sref{sec:exMstring} below). In every other gauge-theory row, all sectors should be flux sets of one 5d theory, since no 1-form symmetry of gauge origin survives (\sref{sec:pfcramer}). This is shown for $\SU(3)_9$ by the closed form \eqref{eq:cramer} of its one-string genus, for $\USp(4)_0+3\Lambda^2$ by the spectrum \cite{Kim:2020hhh}, and for the sectors $B=0,\pm\tfrac14$ of $\SU(4)_8$ in \cite[\S6.1]{Kim:2020hhh}. The equation of the sector $B=\tfrac12$ of $\SU(4)_8$ is checked at one-string order (\sref{sec:exSU48} below). Elsewhere, this is a prediction. In the last row, the subgroup $\{1,v\}$ of the gauge centre survives, and all four sectors are still admissible flux sets of the frame. There, \eqref{eq:dictionary} holds for the bare flux (\sref{sec:exso8Z2} below).

% ---------------------------------------------------------------------------
\subsection{\texorpdfstring{$\SU(3)_9$}{SU(3)9}: the frame map step by step}
\label{sec:exSU39}

The 6d theory is minimal $\su(3)$ on a $-3$ curve, with the outer automorphism of $\su(3)$ around the circle \cite[\S4.2]{Jefferson:2018irk}, \cite[(2.23)]{Bhardwaj:2020gyu}. Its 5d frame is $\SU(3)_9$ \cite[\S3.2.3]{Kim:2020hhh}, with a 5-brane web in \cite[\S5]{Hayashi:2018lyv}. The frame map appears in \cite[(2.51)]{Kim:2020hhh} as the substitutions $\phi_0\to\phi_0-\tfrac\tau{16}$ and $\phi_1\to\phi_1-2\phi_0+\tfrac\tau4$. Here, we derive it.

\paragraph{Data.} $\mathcal K=3$, $D=\BZ_3$, and $(\mathfrak a,\mathfrak b,\mathfrak c)=(-\tfrac13,1,1)$ in \eqref{eq:GSdata}. The twisted affine algebra is $A_2^{(2)}$. The invariant subalgebra is $\su(2)$, embedded with index one, with modulus $\phi_1$ and $\tfrac12\tilde K\phi\phi=\phi_1^2$. With the order-four lift of the twist (Appendix~\ref{app:twists}) \cite[(2.47)]{Kim:2020hhh}
\begin{equation}
\label{eq:su39decomp}
\mathbf 8\ \to\ \mathbf 3_0\oplus\mathbf 2_{1/4}\oplus\mathbf 2_{3/4}\oplus\mathbf 1_{1/2}\,,
\end{equation}
with the KK momentum $\mathfrak r$ as subscript. The twist inverts the centre of $\SU(3)$ and the doublets screen that of $\SU(2)$. So the 5d $\BZ_3^{(1)}$ is of tensor origin (Appendix~\ref{app:twists}).

\paragraph{6d frame.} The tree-level part \eqref{eq:Etree} is $\tfrac32\tau\phi_0^2+3\phi_0\big(\phi_1^2-\tfrac1{12}(\epsilon_1^2+\epsilon_2^2)+\epsilon_+^2\big)$. The towers of \eqref{eq:su39decomp}, with weights $\mathsf w$, enter \eqref{eq:F1loop6d} with $B_2(\mathfrak r)=\mathfrak r^2-\mathfrak r+\tfrac16$:
\begin{equation}
\label{eq:su39towers}
\begin{array}{c|c|c|c|c}
\text{tower} & \mathfrak r & \mathsf w\cdot\phi & B_2(\mathfrak r) & \sum_{\mathsf w} B_2(\mathfrak r)\,(\mathsf w\cdot\phi)^2\\ \hline
\mathbf 3_0 & 0 & \pm2\phi_1,\ 0 & \tfrac16 & \tfrac43\phi_1^2\\
\mathbf 2_{1/4},\ \mathbf 2_{3/4} & \tfrac14,\ \tfrac34 & \pm\phi_1 & -\tfrac1{48} & -\tfrac1{24}\phi_1^2\ \text{ each}\\
\mathbf 1_{1/2} & \tfrac12 & 0 & -\tfrac1{12} & 0
\end{array}
\end{equation}
The sum is $\tfrac54\phi_1^2$, so the towers give $-\tfrac\tau4\cdot\tfrac54\phi_1^2=-\tfrac5{16}\tau\phi_1^2$. The zero modes $\mathbf 3_0$ give the cubic $\tfrac43\phi_1^3$ and the levels $C^G_1=-4$, $C^R_1=2$. Altogether \cite[(2.50)]{Kim:2020hhh}
\begin{equation}
\label{eq:E250}
\epsilon_1\epsilon_2\,\mathcal E_{\rm 6d}=\tfrac32\tau\phi_0^2-\tfrac5{16}\tau\phi_1^2+3\phi_0\phi_1^2+\tfrac43\phi_1^3
-\big(\tfrac14\phi_0+\tfrac1{12}\phi_1\big)(\epsilon_1^2+\epsilon_2^2)+(3\phi_0+\phi_1)\,\epsilon_+^2\,.
\end{equation}

\paragraph{5d frame.} In the basis of simple coroots and the chamber of \cite[(3.87)]{Kim:2020hhh}, \eqref{eq:E5d} gives
\begin{equation}
\label{eq:E5dSU39}
\begin{aligned}
\epsilon_1\epsilon_2\,\mathcal E_{\rm 5d}&=m_0\big[(\phi^{\rm 5d}_1)^2-\phi^{\rm 5d}_1\phi^{\rm 5d}_2+(\phi^{\rm 5d}_2)^2\big]+\mathcal C(\phi^{\rm 5d}_1,\phi^{\rm 5d}_2)
-(\phi^{\rm 5d}_1+\phi^{\rm 5d}_2)\big(\tfrac{\epsilon_1^2+\epsilon_2^2}{12}-\epsilon_+^2\big)\,,\\
\mathcal C(\varphi_1,\varphi_2)&:=\tfrac16\big(8\varphi_1^3+24\varphi_1^2\varphi_2-30\varphi_1\varphi_2^2+8\varphi_2^3\big)\,.
\end{aligned}
\end{equation}

\paragraph{The map.} The ansatz \eqref{eq:framemapgen} is $(\phi^{\rm 5d}_1,\phi^{\rm 5d}_2)^T=M\,(\phi_0,\phi_1)^T+(s_1,s_2)^T\tau$ and $m_0=\tau/r$. We substitute it into \eqref{eq:E5dSU39} and compare with \eqref{eq:E250} in the order of Table~\ref{tab:framemaprecipe}.
\begin{itemize}[leftmargin=*,topsep=2pt,itemsep=1pt]
\item \phantomsection\label{it:su39step1}\emph{Step 1: the cubic terms fix $M$.} The condition $\mathcal C(M\phi)=3\phi_0\phi_1^2+\tfrac43\phi_1^3$ has, among unimodular matrices with $|M_{ij}|\le6$, the single solution $M=\Kmat{1}{0}{2}{1}$. This is the affine-node rule \eqref{eq:affinenode} with the comark $2$ of the $\su(2)$ node.
\item \phantomsection\label{it:su39step2}\emph{Step 2: the $\tau\times$quadratic terms fix $r$ and $s$.} The coupling gives $\tfrac\tau r(3\phi_0^2+3\phi_0\phi_1+\phi_1^2)$. The shifts inside the cubic give $\tau\big[s_1(-12\phi_0\phi_1-5\phi_1^2)+s_2(6\phi_0\phi_1+4\phi_1^2)\big]$. Matching $\tfrac32\tau\phi_0^2-\tfrac5{16}\tau\phi_1^2$ of \eqref{eq:E250},
\begin{equation}
\label{eq:SU39mapeqs}
\phi_0^2:\ \ \frac3r=\frac32\,,\qquad
\phi_0\phi_1:\ \ \frac3r-12s_1+6s_2=0\,,\qquad
\phi_1^2:\ \ \frac1r-5s_1+4s_2=-\frac5{16}\,,
\end{equation}
with the unique solution $r=2$, $s_1=\tfrac1{16}$, $s_2=-\tfrac18$.
\end{itemize}
The linear terms are a check. $\phi^{\rm 5d}_1+\phi^{\rm 5d}_2=3\phi_0+\phi_1-\tfrac\tau{16}$ reproduces those of \eqref{eq:E250} up to constants. The result is
\begin{equation}
\label{eq:SU39map}
\phi_1^{\rm 5d}=\phi_0+\tfrac{\tau}{16}\,,\qquad \phi_2^{\rm 5d}=2\phi_0+\phi_1-\tfrac{\tau}{8}\,,\qquad m_0=\tfrac\tau2\,,\qquad r=2\,.
\end{equation}
Solved for the 6d variables, it is the composition of the substitutions of \cite[(2.51)]{Kim:2020hhh}.

\paragraph{Reading of the map.} Let $e=(\phi_1^{\rm 5d},\,\phi_2^{\rm 5d}-\phi_1^{\rm 5d},\,-\phi_2^{\rm 5d})$ and $\Phi:=3\phi_0-\tfrac5{16}\tau$, the coefficient of $\phi_1^2$ in \eqref{eq:E250}, i.e.\ the effective $\su(2)$ coupling of \sref{sec:examples}. By \eqref{eq:SU39map}, the positive roots of $\su(3)$ and the instantonic W-boson are
\begin{equation}
\label{eq:SU39roots}
\begin{array}{l|l|l|l}
\text{5d state} & \text{5d variables} & \text{6d variables} & \text{6d state}\\ \hline
e_1-e_2 & 2\phi_1^{\rm 5d}-\phi_2^{\rm 5d} & \tfrac\tau4-\phi_1=\alpha_0 & \text{lowest state of }\mathbf 2_{1/4}\\
e_2-e_3 & -\phi_1^{\rm 5d}+2\phi_2^{\rm 5d} & \Phi+2\phi_1 & \text{string, }\su(2)\text{ weight }2\\
e_1-e_3 & \phi_1^{\rm 5d}+\phi_2^{\rm 5d} & \Phi+\phi_1+\tfrac\tau4 & \text{string, }\su(2)\text{ weight }1\\ \hline
m_0-2(e_1-e_2) & -4\phi_1^{\rm 5d}+2\phi_2^{\rm 5d}+m_0 & 2\phi_1=\alpha_1 & \su(2)\text{ W-boson, }k=1
\end{array}
\end{equation}
The simple roots of $\su(3)$ are the affine root $\alpha_0$ of $A_2^{(2)}$ and a wrapped string. The $\su(2)$ root $\alpha_1$ carries one unit of instanton charge, but no KK momentum. With the marks $(2,1)$ of $A_2^{(2)}$, $\delta=2\alpha_0+\alpha_1=\tfrac\tau2$. In 5d variables, the same combination is $2(e_1-e_2)+\big[m_0-2(e_1-e_2)\big]=m_0$. So $k(\delta)=1$, and \eqref{eq:m0delta} gives $m_0=\tfrac\tau2$, as Step~2 found.

\paragraph{Volumes and the shift \texorpdfstring{$\tau/16$}{tau/16}.} \hyperref[it:volmatch]{Volume matching} (\sref{sec:recipe}) of the curves of \cite[(3.88), (3.90)]{Kim:2020hhh} gives \eqref{eq:SU39map}, written in the tensor scalar $\phi_0^{\rm geom}=\phi_0+\tfrac\tau{16}$ of \cite[(3.90)]{Kim:2020hhh}. Since moduli must map to moduli, it forces the exchange of the fibre and the base of the ruled surface $\mathbb F_0$ noted below \cite[(3.86)]{Kim:2020hhh}. The shift $\tfrac\tau{16}$ has a simple origin. Take the string $e_2-e_3$ and subtract the instantonic W-boson $m_0-2(e_1-e_2)$, both in \eqref{eq:SU39roots}. In 5d variables, the difference is $3\phi_1^{\rm 5d}-m_0$, the string fugacity of \cite[(3.91)]{Kim:2020hhh}. In 6d variables, it is $(\Phi+2\phi_1)-2\phi_1=\Phi$. Equating the two expressions, with $\mathcal K=3$ and $m_0=\frac\tau2$,
\begin{equation}
\label{eq:tau16}
\phi_1^{\rm 5d}=\phi_0+\frac{m_0-\tfrac5{16}\tau}{\mathcal K}=\phi_0+\frac\tau{16}\,.
\end{equation}
The maps of \cite[(3.90)]{Kim:2020hhh} and \cite[(4.114)]{Kim:2021gyj} are \eqref{eq:SU39map} with other origins of $\phi_0$ (\hyperref[it:validity]{range of validity} of \sref{sec:framemap}). The same reading gives the shift $\tfrac{5\tau}{32}$ of the $\USp(4)_0+3\Lambda^2$ map \eqref{eq:usp4map} below, derived in \sref{sec:exZ2twist}. It does not give the shifts of the $\SU(4)_8$ map of \sref{sec:exSU48} below. There, the string $e_3-e_4$ minus the instantonic root $\alpha_2$ of $D_4^{(3)}$ in \eqref{eq:su48roots} below is $\Phi+2\phi_1$ rather than $\Phi$.

\paragraph{Prefactor $\Lambda_B$ and sectors.} By the \hyperref[it:fluxmap]{flux rule} of \sref{sec:framemap}, the fluxes $(a,b)$ of $(\phi_0,\phi_1)$ map to $n^{\rm 5d}=(a,2a+b)$, with $B_\tau=\Bmz=0$. The exponent is independent of the Coulomb moduli only for $b=0$ (Appendix~\ref{app:su39}). On this tensor-flux family, $\nu(a)=a(1,2)$ with $a\in\BZ+B$, and $B\in\{0,\tfrac13,\tfrac23\}$ runs over the 't~Hooft classes of $\SU(3)$, i.e.\ over $D$. Since $k(\nu(a))=3a^2$ by \eqref{eq:kQ}, the instanton factor is $q^{k}=\fq^{3a^2/2}$, and \eqref{eq:Vfamily}, \eqref{eq:LambdaTheta} give
\begin{equation}
\label{eq:SU39Lambda}
V(a,0)=\tfrac32\tau a^2-a(\epsilon_1+\epsilon_2)\,,\qquad
\Lambda_B=\vartheta\big[\genfrac{}{}{0pt}{1}{B}{1/2}\big](3\tau,\,-\epsilon_1-\epsilon_2)\,.
\end{equation}
The characteristic $\tfrac12$ is the Wu sign \eqref{eq:signw} of the odd lattice. The sector $a=\tfrac13$ is $\ngen=(\tfrac13,\tfrac23)$, with $k=\tfrac13$ and leading power $\fq^{1/6}$. With $r=2$ from \eqref{eq:SU39map}, $k/r=\tfrac16=q(1)$ of \eqref{eq:qdef}, modulo $\tfrac12$ since $\mathcal K$ is odd \eqref{eq:qshift}. This is the first relation of \eqref{eq:dictionary}. With the Wu sign, the refined KK momentum $p(\tfrac13)$ of \eqref{eq:sdef} is $\tfrac16+\tfrac12=\tfrac23=q_w(1)$, with $w=3$, as \eqref{eq:sequalsqw} states. All three sectors are blow-up equations of $\SU(3)_9$. At one-string order, they close, i.e.\ they determine the one-string genus, and Cramer's rule \eqref{eq:cramer} gives the closed form \cite[(4.116)]{Kim:2021gyj} (Appendix~\ref{app:su39}).

% ---------------------------------------------------------------------------
\subsection{A known case: the M-string}
\label{sec:exMstring}

Here $r$, the dictionary and the prefactor $\Lambda_B$ are known in advance \cite{Gukov:2020btk,BenettiGenolini:2020doj,Kim:2021gyj}, so this example tests the method. The $(2,0)$ $A_1$ theory on a circle is $\SU(2)_0+\adj$ \cite{Rozali:1997cb,Berkooz:1997cq,Seiberg:1997ax,Douglas:2010iu,Lambert:2010iw}, \cite[\S4.1.2]{Kim:2020hhh}, the KK theory of the M-string \cite{Haghighat:2013gba}. Its frame map is \eqref{eq:Mstring}, with $r=1$, and $(\mathfrak a,\mathfrak c)=(0,\tfrac12)$. Every flux lies on the tensor-flux family, and \eqref{eq:Vfamily} with the adjoint mass $m_1$ (the $m$ of \sref{sec:recipe}), $\kappa_{m_1}=-1$ and $B_{m_1}=\tfrac12$ gives $V(a)=\tau a^2-a(m_1+\epsilon_+)$, so
\begin{equation}
\label{eq:LambdaMstring}
\Lambda_0=\theta_3(2\tau,x)\,,\qquad \Lambda_{1/2}=\theta_2(2\tau,x)\,,\qquad x=-m_1-\epsilon_+\,.
\end{equation}
Up to a constant phase, these are the $\theta_4$, $\theta_1$ of \cite[(4.69)]{Kim:2021gyj} without the shift $x\to x-\epsilon_1$ that the Wilson loop adds in \cite[(4.68)]{Kim:2021gyj}, after the half-period shift $m_1\to m_1-i\pi$ \cite[(A.8)]{Kim:2021gyj}, which accounts for their alternating sign. $\Lambda_{1/2}$ starts at $\fq^{1/4}=\fq^{q(1)}$, and $k(\nu(\tfrac12))/r=\tfrac14$.

The sectors do not close, i.e.\ they do not determine the one-string genus (\sref{sec:pfcramer}). Both flux sets, $a\in\BZ$ and $a\in\BZ+\tfrac12$, give blow-up equations of $\SU(2)_0+\adj$ \cite[(3.26)]{Gu:2019pqj}. The second is the 't~Hooft sector of its $\BZ_2$ 1-form symmetry. With $|D|=2$, these are two equations for three unknowns, and \eqref{eq:cramer} does not apply. The set $a\in\BZ+\tfrac12$ is also the flux set of $\SU(2)_\pi+\adj$ \cite[(4.12)]{Kim:2020hhh}, a different KK theory with the same $\mathcal E_{\rm 5d}$. So the equation of the sector $B=\tfrac12$ alone does not distinguish the two theories. The theory $\SU(2)_\pi+\adj$ descends from the $(2,0)$ $A_2$ theory with a twist of the second kind \cite[\S\S5.2, 6]{Tachikawa:2011ch}, \cite[\S4.1.3]{Kim:2020hhh}, which we leave open (\sref{sec:conclusions} below).

% ---------------------------------------------------------------------------
\subsection{\texorpdfstring{$\su(N)+2N\Fund$ on a $-2$ curve with the $\BZ_2$ twist: $\USp(4)_0+3\Lambda^2$ and $\so(7)+5\Fund$}{su(N)+2NF on a -2 curve with the Z2 twist: USp(4)0+3AS and so(7)+5F}}
\label{sec:exZ2twist}

For $N=3$, the KK theory is $\USp(4)_0+3\Lambda^2$ \cite[\S6]{Hayashi:2015vhy}, \cite[\S4.2]{Jefferson:2018irk}, \cite[(2.25)]{Bhardwaj:2020gyu}, with both frames in \cite[\S5.1.1]{Kim:2020hhh}. For $N=5$, it is $\so(7)+5\Fund$ \cite[\S6]{Hayashi:2015vhy}, the member $m=7$ of \cite[(2.20)]{Bhardwaj:2020gyu}. New here are hypermultiplets, so the map shifts the masses and the flavour fluxes shift the tensor flux. Since $\mathcal K=2$ is even, there is no Wu sign. The twist pairs $\mathbf N$ with $\bar{\mathbf N}$ and leaves $N$ masses $m_i$. The decompositions are \cite[(2.42), (5.7), (A.46)]{Kim:2020hhh}:\footnote{In \cite[(5.7)]{Kim:2020hhh}, the singlet $\mathbf 1_{3/2}$ should read $\mathbf 1_{3/4}$, the conjugate of $\mathbf 1_{1/4}$.}
\begin{equation}
\label{eq:Z2data}
\mbox{\small\setlength{\arraycolsep}{3pt}$\begin{array}{l|l|l}
 & N=3 & N=5\\ \hline
\hat{\mathfrak g};\ \text{invariant subalgebra};\ (\mathfrak a,\mathfrak c) & A_2^{(2)};\ \su(2);\ (0,\tfrac32) & A_4^{(2)};\ \usp(4);\ (0,\tfrac52)\\
\text{adjoint} & \eqref{eq:su39decomp} & \mathbf{10}_0\oplus\mathbf 4_{1/4}\oplus\mathbf 4_{3/4}\oplus\mathbf 5_{1/2}\oplus\mathbf 1_{1/2}\\
\mathbf N\oplus\bar{\mathbf N},\ \text{each }i & \mathbf 2_0\oplus\mathbf 1_{1/4}\oplus\mathbf 2_{1/2}\oplus\mathbf 1_{3/4} & \mathbf 4_0\oplus\mathbf 1_{1/4}\oplus\mathbf 4_{1/2}\oplus\mathbf 1_{3/4}\\
\text{tower sum: vectors}-\text{hypers} & \tfrac54-\tfrac12=\tfrac34 & \tfrac{19}{12}-\tfrac56=\tfrac34
\end{array}$}
\end{equation}
The last row is the bracket of the tower term in \eqref{eq:F1loop6d}, $\sum B_2(\fr)(\mathsf w\cdot\phi)^2$ over the vector multiplet minus the same sum over the hypermultiplets, in units of $\mathsf q_N:=\tfrac12\tilde K\phi\phi$ of the invariant subalgebra. The vector part is computed as in \eqref{eq:su39towers}, and, for $N=3$, it is the $\tfrac54$ found there. Each of the $N$ pairs of hypermultiplets contributes $\big[B_2(0)+B_2(\tfrac12)\big] \times 2=\tfrac16$, from its $\mathbf 2_0$ and $\mathbf 2_{1/2}$ (resp.\ $\mathbf 4_0$ and $\mathbf 4_{1/2}$). The difference is $\tfrac34$ for both $N=3$ and $N=5$, so the towers give $-\tfrac\tau4\cdot\tfrac34\,\mathsf q_N=-\tfrac3{16}\tau\,\mathsf q_N$ and $\Phi=2\phi_0-\tfrac3{16}\tau$.

\paragraph{6d frame.} Take $\phi$ in the Weyl chamber of the invariant subalgebra and $|m_i|<\mathsf w\cdot\phi$ for the positive weights $\mathsf w$ of $\mathbf 2_0$ (resp.\ $\mathbf 4_0$). In this chamber, the zero modes, which alone fix the cubic terms and $C^G$, $C^R$, have definite signs, and \eqref{eq:E6d} gives
\begin{equation}
\label{eq:Z2E6d}
\epsilon_1\epsilon_2\,\mathcal E_{\rm 6d}=\tau\phi_0^2+2\phi_0\Big(\mathsf q_N-\tfrac12\sum_im_i^2+\tfrac N2\epsilon_+^2\Big)-\tfrac3{16}\tau\,\mathsf q_N+\mathcal C_N+\tfrac{\epsilon_1^2+\epsilon_2^2}{48}\,C^G\!\cdot\phi+\epsilon_+^2\sum_j\phi_j\,,
\end{equation}
with $\mathsf q_N$ as above and
\begin{equation}
\label{eq:Z2cubics}
\begin{aligned}
N=3:&\ \ \mathsf q_3=\phi_1^2\,,\ \ C^G=2\,,\ \ \mathcal C_3=\tfrac56\phi_1^3-\tfrac12\textstyle\sum_im_i^2\,\phi_1\,,\\
N=5:&\ \ \mathsf q_5=2\phi_1^2-2\phi_1\phi_2+\phi_2^2\,,\ \ C^G=(-4,6)\,,\\
&\ \ \mathcal C_5=\tfrac43\phi_1^3-\tfrac12\phi_1^2\phi_2-\tfrac12\phi_1\phi_2^2+\tfrac12\phi_2^3-\tfrac12\textstyle\sum_im_i^2\,\phi_2\,.
\end{aligned}
\end{equation}
For $\usp(4)$, $\alpha_1=(2,-1)$ is short and $\alpha_2=(-2,2)$ long. The term $-\phi_0\sum_im_i^2$ is the Green--Schwarz flavour coupling, $\kappa_{m_i}=-1$. For $N=3$, \eqref{eq:Z2E6d} is \cite[(5.8)]{Kim:2020hhh}.

\paragraph{5d frames and maps.} For $N=3$, the 5d frame is \cite[(5.3)--(5.4)]{Kim:2020hhh}, each $\Lambda^2$ being the $\mathbf 5$ of $\usp(4)$. Step~1 has the single solution $M=\Kmat{1}{0}{2}{1}$ with $|M_{ij}|\le6$. Steps~2 and~3 now also fix the mass mixings and the mass shift (Table~\ref{tab:framemaprecipe}). The result is \cite[(5.9)]{Kim:2020hhh}, solved for the 5d variables:
\begin{equation}
\label{eq:usp4map}
\begin{aligned}
\phi_1^{\rm 5d}&=\phi_0+\tfrac{5\tau}{32}+\tfrac14\sum_im_i\,,\qquad &\phi_2^{\rm 5d}&=2\phi_0+\phi_1+\tfrac{\tau}{16}+\tfrac12\sum_im_i\,,\\
m_0&=\tfrac\tau2\,,\qquad &M_i&=m_i+\tfrac\tau4\,.
\end{aligned}
\end{equation}
For $N=5$, the 5d frame is $\so(7)+5\Fund$, with $e=(\phi^{\rm 5d}_1,\,\phi^{\rm 5d}_2-\phi^{\rm 5d}_1,\,2\phi^{\rm 5d}_3-\phi^{\rm 5d}_2)$. By \eqref{eq:Ktilde} and \eqref{eq:kQ}, the vector-class flux has $k=\tfrac12$. We scan the 6d and 5d mass chambers together, with $|M_{ij}|\le6$ (\tref{tab:scans} of Appendix~\ref{app:scans}). Step~1 always returns the affine-node matrix with the comarks $(1,2,2)$ of $A_4^{(2)}$, the affine node being the third 5d node. The 6d chamber above maps to the 5d chamber $e_3<e_2<M_i<e_1$. There, steps~2 and~3 have two solutions, related by $m_i\to m_i-\tfrac25\sum_jm_j$. The one without mass mixing is
\begin{equation}
\label{eq:so7map}
\begin{aligned}
\phi_1^{\rm 5d}&=2\phi_0+\phi_2+\tfrac{\tau}{16}+\tfrac12\sum_im_i\,,\qquad
&\phi_2^{\rm 5d}&=2\phi_0+\phi_1+\tfrac{5\tau}{16}+\tfrac12\sum_im_i\,,\\
\phi_3^{\rm 5d}&=\phi_0+\tfrac{9\tau}{32}+\tfrac14\sum_im_i\,,\qquad
&m_0&=\tfrac\tau2\,,\qquad M_i=m_i+\tfrac\tau4\,.
\end{aligned}
\end{equation}
With these maps, the 5d prepotentials equal \eqref{eq:Z2E6d} up to $\phi$-independent terms.

\paragraph{Reading of the maps.} For $N=3$, let $e=(\phi^{\rm 5d}_1,\,\phi^{\rm 5d}_2-\phi^{\rm 5d}_1)$, and let $\Sigma:=\tfrac12\sum_jm_j$. By \eqref{eq:usp4map} and \eqref{eq:so7map}, the simple roots of $\hat{\mathfrak g}$ are
\begin{equation}
\label{eq:Z2roots}
\begin{array}{l|l|l|l}
 & \alpha_0 & \alpha_1 & \alpha_2\\ \hline
N=3 & e_1-e_2=\tfrac\tau4-\phi_1 & m_0-2e_1+2e_2=2\phi_1 & \text{---}\\
N=5 & e_3=\tfrac\tau4-\phi_1 & e_2-e_3=2\phi_1-\phi_2 & m_0-2e_2=2\phi_2-2\phi_1
\end{array}
\end{equation}
The instantonic root is $\alpha_1$ for $N=3$ and the long root $\alpha_2$ of $\usp(4)$ for $N=5$. The remaining 5d simple root is a string, $2e_2=\Phi+2\phi_1+\Sigma$ for $N=3$ and $e_1-e_2=\Phi-\phi_1+2\phi_2+\Sigma$ for $N=5$. So strings carry charge $\tfrac12$ under each flavour $\U(1)$. The zero weight of each $\Lambda^2$ (resp.\ vector) is the singlet $\mathbf 1_{1/4}$, of mass $M_i=m_i+\tfrac\tau4$, the origin of the mass shift. The marks $(2,1)$ of $A_2^{(2)}$ and $(2,2,1)$ of $A_4^{(2)}$ give $\delta=\tfrac\tau2=m_0$. So $k(\delta)=1$ and $r=2$ by \eqref{eq:m0delta}.

\paragraph{Fluxes.} Here, the family condition \eqref{eq:familycond} reads $\tfrac N4-NB_m^2=0$, so $B_{m_i}=\pm\tfrac12$, and we take $\tfrac12$. The \hyperref[it:fluxmap]{flux rule} of \sref{sec:framemap} carries these fluxes into the gauge fluxes. The affine-node variable, $\phi^{\rm 5d}_1$ of \eqref{eq:usp4map} for $N=3$ and $\phi^{\rm 5d}_3$ of \eqref{eq:so7map} for $N=5$, is $\phi_0+\tfrac{5\tau}{32}+\tfrac14\sum_im_i$ (resp.\ $\tfrac{9\tau}{32}$). So its flux is $n_0+\tfrac{5}{32}B_\tau+\tfrac14\sum_iB_{m_i}$ (resp.\ $\tfrac9{32}B_\tau$), with $n_0$ the flux of the unshifted $\phi_0$. This is the 5d flux along the tensor direction, i.e.\ the shifted tensor flux $a$ of \eqref{eq:adef}. With $B_\tau=0$, it is
\begin{equation}
\label{eq:nushift}
a=n_0+\tfrac14\sum_{i=1}^NB_{m_i}=n_0+\tfrac N8\,.
\end{equation}
The string charge $\lambda=\mathcal Ka=2a$ of \sref{sec:pfblowup} must be an integer, so $a\in\BZ$ or $a\in\BZ+\tfrac12$. In the same way as for $a$, with the gauge fluxes $n_{1,2}\in\BZ$ of the invariant subalgebra, the other variables of \eqref{eq:usp4map} and \eqref{eq:so7map} give the 5d fluxes below. The arrow gives the 5d flux coset that each set of $a$ produces:
\begin{equation}
\label{eq:Z2sectors}
\begin{aligned}
N=3:&\ \ n^{\rm 5d}=(a,\,2a+n_1)\,, &&a\in\BZ\leftrightarrow\BZ^2\,, &&a\in\BZ+\tfrac12\leftrightarrow(\tfrac12,0)+\BZ^2\,,\\
N=5:&\ \ n^{\rm 5d}=(2a+n_2,\,2a+n_1,\,a)\,, &&a\in\BZ\leftrightarrow\BZ^3\,, &&a\in\BZ+\tfrac12\leftrightarrow(0,0,\tfrac12)+\BZ^3\,.
\end{aligned}
\end{equation}
Both cosets are allowed 5d fluxes: $\Lambda_\crt$ and its 't~Hooft coset, the centre class of $\USp(4)$ for $N=3$ and the vector class of $\Spin(7)$ for $N=5$. So the sector label $B\in\{0,\tfrac12\}$ is the fractional part of $a$, and the unshifted tensor flux $n_0$ lies in $\BZ+B-\tfrac N8$.

\paragraph{Prefactor $\Lambda_B$ and dictionary.} On the tensor-flux family, \eqref{eq:Vfamily} gives $V=\tau n_0^2-n_0\big(\sum_im_i+N\epsilon_+\big)$. With $n_0=a-\tfrac N8$ and $m_i=M_i-\tfrac\tau4$, the terms linear in $\tau a$ \mbox{cancel}, and
\begin{equation}
\label{eq:Z2Lambda}
\begin{gathered}
V=\tau a^2-a\Big(\sum_iM_i+N\epsilon_+\Big)+\text{const}\,,\\
\Lambda_B=\vartheta\big[\genfrac{}{}{0pt}{1}{B}{0}\big](2\tau,\,x_M)\,,\qquad x_M=-\sum_iM_i-N\epsilon_+\,.
\end{gathered}
\end{equation}
For $N=3$, the same $V$ follows from the 5d frame alone, with the flux set \cite[(5.6)]{Kim:2020hhh} on the tensor-flux family $\nu(a)=(a,2a)$. The 't~Hooft sector $B=\tfrac12$ has $\Lambda_{1/2}=\theta_2(2\tau,x_M)$, with leading power $\fq^{1/4}=\fq^{q(1)}$, and $k(\nu(\tfrac12))/r=\tfrac14$ for both $N=3$ and $N=5$. For $N=3$, the blow-up equations of both sectors, $a\in\BZ$ ($B=0$) and $a\in\BZ+\tfrac12$ ($B=\tfrac12$), reproduce the spectrum of the single theory $\USp(4)_0+3\Lambda^2$ \cite{Kim:2020hhh}. For $N=5$, both sectors lie in the allowed 5d cosets \eqref{eq:Z2sectors} and obey the pairing rule of \cite[(3.9)]{Kim:2020hhh} (footnote~\ref{fn:faithfulfluxes} below), but the blow-up equations beyond $\Lambda_B$ and the members $m\ge8$ of \cite[(2.20)]{Bhardwaj:2020gyu} have not been tested.

% ---------------------------------------------------------------------------
\subsection{\texorpdfstring{$\SU(4)_8$: why $r=3$}{SU(4)8: why r=3}}
\label{sec:exSU48}

The 6d theory is $\so(8)$ on a $-4$ curve with the triality twist \cite{Razamat:2018gro}, \cite[(2.29)]{Bhardwaj:2020gyu}, and its 5d frame is $\SU(4)_8$ \cite[\S6.1]{Kim:2020hhh}. It is the first example with a twist of order three, and with $\mathcal K=4$, $D=\BZ_4$ and $q(\lambda)=\tfrac{\lambda^2}8$.

\paragraph{Data, frames and map.} Here $(\mathfrak a,\mathfrak c)=(-\tfrac12,\tfrac32)$, without matter, and $\hat{\mathfrak g}=D_4^{(3)}$. The invariant subalgebra is $\mathfrak g_2$, with the short root $\alpha_1=(2,-1)$, the long root $\alpha_2=(-3,2)$, and $\tfrac12\tilde K\phi\phi=\mathsf q:=3\phi_1^2-3\phi_1\phi_2+\phi_2^2$. The adjoint decomposes as $\mathbf{28}\to\mathbf{14}_0\oplus\mathbf 7_{1/3}\oplus\mathbf 7_{2/3}$ \cite[(2.42)]{Kim:2020hhh}, the $\mathbf 7$ being the six short roots and a zero weight. The towers give $\tfrac83\mathsf q$ from $\mathbf{14}_0$ and $-\tfrac29\mathsf q$ from each $\mathbf 7$.\footnote{\label{fn:g2towers}By \eqref{eq:F1loop6d}, each $\mathbf R_{\fr}$ contributes $B_2(\fr)\sum_{\mathsf w\in\mathbf R}(\mathsf w\cdot\phi)^2$ to the tower bracket. In Dynkin labels, the six short roots of $\mathfrak g_2$ are $\pm(2,-1)$, $\pm(-1,1)$, $\pm(1,0)$, and the six long roots are $\pm(-3,2)$, $\pm(3,-1)$, $\pm(0,1)$. Their squares $(\mathsf w\cdot\phi)^2$ sum to $4\mathsf q$ and $12\mathsf q$. So $\sum_{\mathbf{14}}(\mathsf w\cdot\phi)^2=16\mathsf q$, and, since the non-zero weights of the $\mathbf 7$ are the short roots, $\sum_{\mathbf 7}(\mathsf w\cdot\phi)^2=4\mathsf q$. With $B_2(0)=\tfrac16$ and $B_2(\tfrac13)=B_2(\tfrac23)=-\tfrac1{18}$, this gives $\tfrac83\mathsf q$ and $-\tfrac29\mathsf q$. The total $\tfrac83-\tfrac49=\tfrac{20}9$, times the factor $-\tfrac\tau4$ of \eqref{eq:F1loop6d}, is $-\tfrac59\tau\,\mathsf q$.} So $\Phi=4\phi_0-\tfrac59\tau$, and in the fundamental chamber of $\mathfrak g_2$ \cite[(6.7)]{Kim:2020hhh}
\begin{equation}
\label{eq:su48E6d}
\begin{split}
\epsilon_1\epsilon_2\,\mathcal E_{\rm 6d}=2\tau\phi_0^2 &+\Phi\,\mathsf q+\tfrac43\phi_1^3+3\phi_1^2\phi_2-4\phi_1\phi_2^2+\tfrac43\phi_2^3
\\ &+(6\phi_0+\phi_1+\phi_2)\,\epsilon_+^2-\big(\tfrac12\phi_0+\tfrac{\phi_1+\phi_2}{12}\big)(\epsilon_1^2+\epsilon_2^2)\,.
\end{split}
\end{equation}
The 5d frame is \cite[(6.4)]{Kim:2020hhh}.\footnote{The coupling term in $6\mathcal F$ of \cite[(6.4)]{Kim:2020hhh} lacks a factor $6$. It should read $6m_0\big(\phi_1^2-\phi_1\phi_2+\phi_2^2-\phi_2\phi_3+\phi_3^2\big)$, in their 5d variables, i.e.\ $\frac{m_0}2\tilde K_{ij}\phi_i\phi_j$ of \eqref{eq:E5d}. The frame map confirms this.} Step~1 gives the affine-node matrix with the comarks $(1,2,3)$ of $D_4^{(3)}$ \cite[below (6.7)]{Kim:2020hhh}. Step~2, six linear equations for $(1/r,s_1,s_2,s_3)$, has the unique solution \cite[(6.8)]{Kim:2020hhh}
\begin{equation}
\label{eq:su48map}
\phi_1^{\rm 5d}=\phi_0+\tfrac\tau9\,,\qquad \phi_2^{\rm 5d}=2\phi_0+\phi_1-\tfrac\tau9\,,\qquad
\phi_3^{\rm 5d}=3\phi_0+\phi_2-\tfrac\tau3\,,\qquad m_0=\tfrac\tau3\,.
\end{equation}

\paragraph{Reading of the map.} Let $e=(\phi_1^{\rm 5d},\,\phi_2^{\rm 5d}-\phi_1^{\rm 5d},\,\phi_3^{\rm 5d}-\phi_2^{\rm 5d},\,-\phi_3^{\rm 5d})$, and recall that $\Phi=4\phi_0-\tfrac59\tau$ is the coefficient of $\mathsf q$ in \eqref{eq:su48E6d}. By \eqref{eq:su48map}, the positive roots of $\su(4)$ and the instantonic root are, in 6d variables,
\begin{equation}
\label{eq:su48roots}
\mbox{\footnotesize\setlength{\arraycolsep}{1.4pt}$\begin{array}{l|l|l|l}
\text{root of }\su(4) & \text{in }\phi_i^{\rm 5d} & \text{in 6d variables} & \text{6d state}\\ \hline
e_1-e_2\ \text{(simple)} & 2\phi_1^{\rm 5d}-\phi_2^{\rm 5d} & \tfrac\tau3-\phi_1=\alpha_0 & \text{lowest state of $\mathbf 7_{1/3}$}\\
e_2-e_3\ \text{(simple)} & -\phi_1^{\rm 5d}+2\phi_2^{\rm 5d}-\phi_3^{\rm 5d} & 2\phi_1-\phi_2=\alpha_1 & \text{zero mode, short root}\\
e_3-e_4\ \text{(simple)} & -\phi_2^{\rm 5d}+2\phi_3^{\rm 5d} & \Phi-\phi_1+2\phi_2 & \text{string}\\
e_1-e_3 & \phi_1^{\rm 5d}+\phi_2^{\rm 5d}-\phi_3^{\rm 5d} & \tfrac\tau3+\phi_1-\phi_2 & \text{state of $\mathbf 7_{1/3}$}\\
e_2-e_4 & -\phi_1^{\rm 5d}+\phi_2^{\rm 5d}+\phi_3^{\rm 5d} & \Phi+\phi_1+\phi_2 & \text{string}\\
e_1-e_4 & \phi_1^{\rm 5d}+\phi_3^{\rm 5d} & \Phi+\phi_2+\tfrac\tau3 & \text{string}\\ \hline
m_0-(e_1+e_2-2e_3) & m_0-3\phi_2^{\rm 5d}+2\phi_3^{\rm 5d} & -3\phi_1+2\phi_2=\alpha_2 & \text{long root, $k=1$}
\end{array}$}
\end{equation}
The coefficients of $(\phi_1,\phi_2)$ are $\mathfrak g_2$ Dynkin labels. The roots with $\Phi$ contain $4\phi_0=\mathcal K\phi_0$, so they are strings. The roots $e_1-e_2$ and $e_1-e_3$ have the form $\tfrac\tau3+\mathsf w\cdot\phi$, with $\mathsf w=(-1,0)$ and $(1,-1)$. These are short roots of $\mathfrak g_2$, hence weights of the $\mathbf 7$, and $(-1,0)$ is its lowest weight. So both are states of $\mathbf 7_{1/3}$. Hence, the simple roots of $\su(4)$ are the affine root $\alpha_0$ of $D_4^{(3)}$,\footnote{Recall from the \hyperref[par:rootkinds]{criterion} of \sref{sec:recipe} that the affine root is the simple root whose mass contains $\tau$, but not $\phi_0$. This is $e_1-e_2$. The root $e_1-e_3=\alpha_0+\alpha_1$ has the same property, but is not simple.} the zero mode $\alpha_1$ and a string, as for $N=5$ in \sref{sec:exZ2twist}. The long root $\alpha_2$ of $\mathfrak g_2$ is instantonic, with $k=1$. With the marks $(1,2,1)$ of $D_4^{(3)}$, $\delta=\alpha_0+2\alpha_1+\alpha_2=\tfrac\tau3=m_0$, and $\alpha_2$ has mark $1$, as \eqref{eq:rrule} requires. So $k(\delta)=1$ and $r=3$ by \eqref{eq:m0delta}.

\paragraph{Prefactor $\Lambda_B$ and sectors.} The $\phi_0$-dependent part of $V(a,n_1,n_2)$ is $4\phi_0\,\mathsf q(n)$, positive definite, so the tensor-flux family is $n_1=n_2=0$. On it, $\nu(a)=a(1,2,3)$, and \eqref{eq:Vfamily}, \eqref{eq:LambdaTheta} give
\begin{equation}
\label{eq:su48Lambda}
V(a,0,0)=2\tau a^2-4a\epsilon_+\,,\qquad
\Lambda_B=\vartheta\big[\genfrac{}{}{0pt}{1}{B}{0}\big](4\tau,\,-4\epsilon_+)\,,\qquad B\in\{0,\tfrac14,\tfrac12,\tfrac34\}\,.
\end{equation}
The lattice is even, so the sign $(-1)^{|\nu(a)|}$ is constant on each sector. Since $k(\nu(a))=6a^2$, the leading member $a=\tfrac\lambda4$ of the sector of $\lambda\in\{0,\pm1,2\}$ has $k/r=2a^2=\tfrac{\lambda^2}8=q(\lambda)$. The generating sector contains $\ngen=\tfrac14(1,2,3)$, with $k=\tfrac38$. The flux sets of \cite[(6.9)]{Kim:2020hhh} are the sectors $B\in\{0,\pm\tfrac14\}$. They reproduce the spectrum, and \cite[\S6.1]{Kim:2020hhh} solve them for the elliptic genera. The fourth sector, $B=\tfrac12$, generates $\BZ_2\subset\BZ_4$ and has spin $\tfrac12$. No gauge centre survives the twist, so it should be a blow-up equation of the same theory, as for $\SU(3)_9$. It is, at one-string order, to the orders we checked. The one-string genus fixed by the sectors $B=0,\pm\tfrac14$ satisfies the equation \eqref{eq:sectoreq} of the sector $B=\tfrac12$. We checked this through relative order $\fq$ with the other fugacities kept exact, and through relative order $\fq^{24}$ at generic rational values of them. In the 5d frame, the blow-up equation of the flux set $\tfrac12(1,2,3)+\BZ^3$, with $\Bmz=0$, holds with the spectrum of \cite[Table~30]{Kim:2020hhh} at every order that the table covers completely.

\paragraph{Why \texorpdfstring{$r=3$}{r=3}.} The 6d side gives $q(1)=\tfrac18$, the 5d side $k(\ngen)=\tfrac38$. By \eqref{eq:m0delta}, $r=(\tau/\delta)\,k(\delta)$ is a positive integer for a twist of the first kind. So $k/r=\tfrac3{8r}$ lies in $(0,\tfrac38]$, and it equals $\tfrac18$ modulo $1$ only for $r=3$. The dictionary, with $r$ an integer, thus gives the $r$ of the map. The value $r=1$ would give $\tfrac38$, the spin of the $(2,0)$ $A_3$ theory, and $r=2$ would give $\tfrac3{16}$, which is not a $\BZ_4$ spin.

% ===========================================================================
% Section 5, second half: su(4)_0, su(4)_4, su(4)_2 + 4 Lambda^2 and the
% so(8) triality pair sp(3)_0 / su(4)_6 + 2 Lambda^2.
% ===========================================================================

\subsection{\texorpdfstring{$\su(4)_0+4\Lambda^2$}{su(4)0+4AS}: a new frame map}
\label{sec:exsu40}

The 6d theory is $\su(4)+8\Fund$ on a $-2$ curve with the outer-automorphism twist, so $\hat{\mathfrak g}=A_3^{(2)}$, and its 5d frame is $\su(4)_0+4\Lambda^2$ \cite[\S6]{Hayashi:2015vhy}, \cite[(2.35)]{Bhardwaj:2020gyu}. Neither its 6d-frame prepotential nor its frame map is in \cite{Kim:2020hhh}. The 5d gauge algebra $\su(4)$ is not the invariant subalgebra $\usp(4)$, so neither $M$ nor the 5d chamber is known in advance.

\paragraph{6d frame.} We have $\mathcal K=2$ and $(\mathfrak a,\mathfrak b,\mathfrak c)=(0,1,2)$. The twist pairs $\mathbf 4$ with $\bar{\mathbf 4}$ and leaves four masses $m_i$ with $\kappa_{m_i}=-1$. The fundamentals screen the gauge centre, so $D=\BZ_2$ is of tensor origin. The $\usp(4)$ moduli $(\phi_1,\phi_2)$ are Dynkin components, with $\alpha_1=(2,-1)$ short, $\alpha_2=(-2,2)$ long and $\tfrac12\tilde K\phi\phi=2\phi_1^2-2\phi_1\phi_2+\phi_2^2$ \eqref{eq:Ktilde}. With $\mathbf{15}\to\mathbf{10}_0\oplus\mathbf 5_{1/2}$ and $\mathbf 4\oplus\bar{\mathbf 4}\to\mathbf 4_0\oplus\mathbf 4_{1/2}$ \cite[(2.42), (A.34)]{Kim:2020hhh}, the towers of \eqref{eq:E6d} give $12B_2(0)+4B_2(\tfrac12)=\tfrac53$ from the vectors and $8\big[B_2(0)+B_2(\tfrac12)\big]=\tfrac23$ from the hypermultiplets, in units of $\tfrac12\tilde K\phi\phi$.\footnote{\label{fn:indexsum}In these units, $\sum_{\mathsf w\in\mathbf R}(\mathsf w\cdot\phi)^2=4\,\mathrm{ind}_{\mathbf R}$, with $\mathrm{ind}_{\adj}=h^\vee$, by \eqref{eq:Ktilde}. For $\usp(4)$, $h^\vee=3$, so the $\mathbf{10}$ gives $12$, the $\mathbf 5$ (index $1$) gives $4$, and each $\mathbf 4$ (index $\tfrac12$) gives $2$, for each of the four masses.} They renormalise the $\usp(4)$ coupling to $\Phi$. In the $\usp(4)$ chamber with $0<m_i$ below every $|\mathsf w\cdot\phi|$,
\begin{equation}
\label{eq:su40E6d}
\begin{aligned}
\epsilon_1\epsilon_2\,\mathcal E_{\rm 6d}&=\tau\phi_0^2+\Phi\big(2\phi_1^2-2\phi_1\phi_2+\phi_2^2\big)-\phi_0\textstyle\sum_im_i^2+4\phi_0\epsilon_+^2+\tfrac43\phi_1^3-\phi_1\phi_2^2+\tfrac23\phi_2^3\\
&\quad-\tfrac12\textstyle\sum_im_i^2\,\phi_2-\tfrac1{12}(\phi_1-\phi_2)(\epsilon_1^2+\epsilon_2^2)+(\phi_1+\phi_2)\,\epsilon_+^2\,,\qquad \Phi:=2\phi_0-\tfrac\tau4\,.
\end{aligned}
\end{equation}

\paragraph{5d frame.} We write the prepotential \eqref{eq:E5d} of $\su(4)_{\kCS}+4\Lambda^2$ once, for use also in \sref{sec:exsu44} and \sref{sec:exsu42} below. With the Dynkin variables $P$ and the orthonormal components $e=(P_1,P_2-P_1,P_3-P_2,-P_3)$, in the Weyl chamber $e_1>e_2>e_3>e_4$,
\begin{equation}
\label{eq:su40E5d}
\begin{aligned}
\epsilon_1\epsilon_2\,\mathcal E_{\rm 5d}&=m_0\,k(P)+\tfrac{\kCS}6\textstyle\sum_ae_a^3+\tfrac16\sum_{a<b}(e_a-e_b)^3+\tfrac1{48}(\epsilon_1^2+\epsilon_2^2)\,C^G\cdot P\\
&\quad-\tfrac1{12}\textstyle\sum_{i=1}^4\sum_{a<b}\left[s_{ab}(e_a+e_b+M_i)\right]^3+\epsilon_+^2\,(P_1+P_2+P_3)\,,
\end{aligned}
\end{equation}
with $k(P)=\tfrac12\sum_ae_a^2$ the form of \eqref{eq:kQ}, $M_i$ the 5d masses, $C^G$ of \eqref{eq:E5d}, and chamber signs $s_{ab}=\pm1$, the signs of $e_a+e_b+M_i$. A pair $\pm(e_a+e_b)$ is mass-dominated if $M_i>|e_a+e_b|$, so that both its signs are $+1$. The eight sign patterns, five of them distinct, are described in Appendix~\ref{app:scans}.

\paragraph{The map.} The cubic terms of \tref{tab:framemaprecipe}, with the masses off, match in one of the five distinct patterns only (Appendix~\ref{app:scans}). In it, the pairs $\pm(e_1+e_3)$ and $\pm(e_1+e_4)$ are mass-dominated and $\pm(e_1+e_2)$ is not. There, $M$ is unique up to the charge conjugation $P_1\leftrightarrow P_3$:
\begin{equation}
\label{eq:su40M}
M=\begin{pmatrix}1&0&0\\2&0&1\\1&1&0\end{pmatrix}\,,\qquad P_1=\phi_0+\dots\,,\quad P_2=2\phi_0+\phi_2+\dots\,,\quad P_3=\phi_0+\phi_1+\dots\,.
\end{equation}
This is the affine-node rule \eqref{eq:affinenode} for $A_3^{(2)}$, whose comarks are $1$ on the affine node, $2$ on the long and $1$ on the short simple root. The finite nodes enter in the order (long, short), the reverse of the $\usp(4)$ labelling. The terms with masses have two solutions, related by $m_i\to m_i-\tfrac12\sum_jm_j$. Only the one without mass mixing, of the form \eqref{eq:framemapgen}, maps the hypermultiplet states to 6d hypermultiplet weights, \eqref{eq:su40hypers} below. It is
\begin{equation}
\label{eq:su40map}
\begin{aligned}
P_1&=\phi_0+\tfrac\tau4+\tfrac14\textstyle\sum_im_i\,,\qquad P_2=2\phi_0+\phi_2+\tfrac12\textstyle\sum_im_i\,,\\
P_3&=\phi_0+\phi_1+\tfrac14\textstyle\sum_im_i\,,\qquad m_0=\tfrac\tau2\,,\qquad M_i=m_i+\tfrac\tau4\,.
\end{aligned}
\end{equation}
With it, \eqref{eq:su40E5d} at $\kCS=0$ equals \eqref{eq:su40E6d} up to $\phi$-independent terms, so $r=2$.

\paragraph{States.} By \eqref{eq:su40map}, with $\Sigma:=\tfrac12\sum_im_i$ and $\Phi$ of \eqref{eq:su40E6d}, the positive roots of $\su(4)$ and the instantonic root are, in 6d variables,
\begin{equation}
\label{eq:su40W}
\mbox{\footnotesize\setlength{\arraycolsep}{1.4pt}$\begin{array}{l|l|l|l}
\text{root of }\su(4) & \text{in }P_i & \text{in 6d variables} & \text{6d state}\\ \hline
e_1-e_2\ \text{(simple)} & 2P_1-P_2 & \tfrac\tau2-\phi_2=\alpha_0 & \text{lowest state of $\mathbf 5_{1/2}$}\\
e_2-e_3\ \text{(simple)} & -P_1+2P_2-P_3 & \Phi-\phi_1+2\phi_2+\Sigma & \text{string}\\
e_3-e_4\ \text{(simple)} & -P_2+2P_3 & 2\phi_1-\phi_2=\alpha_1 & \text{zero mode, short root}\\
e_1-e_3 & P_1+P_2-P_3 & \Phi-\phi_1+\phi_2+\Sigma+\tfrac\tau2 & \text{string}\\
e_2-e_4 & -P_1+P_2+P_3 & \Phi+\phi_1+\phi_2+\Sigma & \text{string}\\
e_1-e_4 & P_1+P_3 & \Phi+\phi_1+\Sigma+\tfrac\tau2 & \text{string}\\ \hline
m_0-(e_1-e_2+e_3-e_4) & m_0-2(P_1-P_2+P_3) & -2\phi_1+2\phi_2=\alpha_2 & \text{long root, $k=1$}
\end{array}$}
\end{equation}
The coefficients of $(\phi_1,\phi_2)$ are $\usp(4)$ Dynkin labels. The roots with $\Phi$ contain $2\phi_0=\mathcal K\phi_0$, so they are strings, each with charge $\tfrac12$ under every flavour $\U(1)$. With the marks $(1,1,1)$ of $A_3^{(2)}$, $\delta=\alpha_0+\alpha_1+\alpha_2=m_0=\tfrac\tau2$, and the instantonic root $\alpha_2$ has mark $1$, as \eqref{eq:rrule} requires. For each mass, the six weights $e_a+e_b$ of $\Lambda^2$ give
\begin{equation}
\label{eq:su40hypers}
\mbox{\footnotesize\setlength{\arraycolsep}{1.4pt}$\begin{array}{l|l|l|l}
\text{hypermultiplet state} & \text{in }P_i & \text{in 6d variables} & \text{6d state}\\ \hline
e_2+e_3+M_i & -P_1+P_3+M_i & m_i+\phi_1 & \text{zero mode of $\mathbf 4_0$}\\
e_2+e_4+M_i & -P_1+P_2-P_3+M_i & m_i-\phi_1+\phi_2 & \text{zero mode of $\mathbf 4_0$}\\
e_1+e_3+M_i & P_1-P_2+P_3+M_i & m_i+\phi_1-\phi_2+\tfrac\tau2 & \text{KK mode of $\mathbf 4_{1/2}$}\\
e_1+e_4+M_i & P_1-P_3+M_i & m_i-\phi_1+\tfrac\tau2 & \text{KK mode of $\mathbf 4_{1/2}$}\\
e_1+e_2+M_i & P_2+M_i & \Phi+\phi_2+\Sigma+m_i+\tfrac\tau2 & \text{string}\\
e_3+e_4+M_i & -P_2+M_i & -(\Phi+\phi_2+\Sigma)+m_i & \text{anti-string}
\end{array}$}
\end{equation}
In the 6d chamber, with $0<m_i$ small and $\tau$, $\Phi$ large, only $e_3+e_4+M_i$ is negative. This is the chamber pattern of the map above.

\paragraph{Prefactor $\Lambda_B$ and sectors.} At $B_\tau=0$, the gauge fluxes enter the $\phi_0$-dependent part of $V$ of \eqref{eq:Vdef} through $2\phi_0(2n_1^2-2n_1n_2+n_2^2)$, so the tensor-flux family is $n_1=n_2=0$. There, \eqref{eq:familycond} reads $1-4B_m^2=0$, so $B_{m_i}=\tfrac12$. By the \hyperref[it:fluxmap]{flux rule} of \sref{sec:framemap}, \eqref{eq:su40map} then gives $B_{M_i}=\tfrac12$ and $\nu(a)=(a,2a,a)$ with $a=n_0+\tfrac12$. The four flavour fluxes add $\tfrac12$ to the flux of $P_1$, as in \sref{sec:exZ2twist}, so $a\in\BZ+B$ is the shifted tensor flux, and \eqref{eq:Vfamily} gives
\begin{equation}
\label{eq:su40Lambda}
V\big|_a=\tau n_0^2-n_0\Big(\textstyle\sum_im_i+4\epsilon_+\Big)=\tau a^2+a\,x_M+\text{const}\,,\qquad x_M:=-\textstyle\sum_iM_i-4\epsilon_+\,.
\end{equation}
So $\Lambda_B=\vartheta\big[\genfrac{}{}{0pt}{1}{B}{0}\big](2\tau,x_M)$, and the two sectors are
\begin{equation}
\label{eq:su40sectors}
\begin{array}{c|c|c|c|c}
B & \text{6d flux }n_0 & \text{5d flux} & \Lambda_B & \text{leading power}\\ \hline
0 & \BZ+\tfrac12 & \BZ^3 & \theta_3(2\tau,x_M) & \fq^0\\
\tfrac12 & \BZ & \omega_2+\BZ^3 & \theta_2(2\tau,x_M) & \fq^{1/4}
\end{array}
\qquad \omega_2=(\tfrac12,0,\tfrac12)\,.
\end{equation}
The 't~Hooft sector has KK momentum $\tfrac14=q(1)=k(\omega_2)/r$, with $k(\omega_2)=\tfrac12$ \cite[\S3.3]{Harding:2026gzj}. This holds at fixed 5d masses, as in \eqref{eq:dictionary}. At fixed 6d masses, the $\fq$-exponent of \eqref{eq:su40Lambda} is $n_0^2$, so $n_0\in\BZ+\tfrac12$ ($B=0$) gives $\theta_2$ and $n_0\in\BZ$ ($B=\tfrac12$) gives $\theta_3$. Substituting $m_i=M_i-\tfrac\tau4$ adds $\tau n_0$ to $V$, a quarter from each of the four masses, and the exponent becomes $n_0^2+n_0=a^2-\tfrac14$. Since the two sectors differ by $\tfrac12$ in $n_0$, their relative KK momentum changes by $\tfrac12$, and $\theta_2$ and $\theta_3$ trade places. No 1-form symmetry of gauge origin survives, since the fundamentals screen the centre (see the 6d frame above), so both flux sets should be blow-up equations of one theory, but with $|D|=2$ they cannot determine the one-string genus (\sref{sec:pfcramer}).

\subsection{\texorpdfstring{$\su(4)_4+4\Lambda^2$}{su(4)4+4AS}: two tensors}
\label{sec:exsu44}

The 6d theory \cite[(2.37)]{Bhardwaj:2020gyu} is the E-string on a $-1$ curve, with tensor scalar $\phi_{0,1}$, glued to $\su(3)$ on a $-3$ curve, with tensor scalar $\phi_{0,2}$. The $\BZ_2$ twist acts on $\su(3)$ alone, as in \sref{sec:exSU39}, and leaves the $\su(2)$ modulus $\phi_1$. With $\phi_0:=(\phi_{0,1},\phi_{0,2})$,
\begin{equation}
\label{eq:su44K}
\mathcal K=\begin{pmatrix}1&-1\\-1&3\end{pmatrix}\,,\qquad \mathcal K^{-1}=\frac12\begin{pmatrix}3&1\\1&1\end{pmatrix}\,,\qquad T:=\mathcal K\phi_0=\big(\phi_{0,1}-\phi_{0,2},\ 3\phi_{0,2}-\phi_{0,1}\big)\,,
\end{equation}
so $D=\BZ_2$, and a string of charge $\lambda\in\BZ^2$ has tension $\lambda\cdot T$.

Gauging $\su(3)$ leaves $E_6$ of the E-string $E_8$, the subgroup that commutes with $\SU(3)$ in $E_8\supset\SU(3)\times E_6$, on which the twist acts with fixed subalgebra $F_4$ or $C_4=\usp(8)$. The geometry of \cite[(4.121)--(4.122)]{Bhardwaj:2020gyu} selects $C_4$. There, each pair $(x_i,y_i)$ of $\Lambda^2$ curves is a doublet of the $\su(2)_i$ of the flavour root $x_i-y_i$ alone, so that the root is twice a hypermultiplet weight, which $F_4$ does not allow. The masses $m_{1,\dots,4}$ are Cartan parameters of $\usp(8)$.

\paragraph{6d frame and map.} In \eqref{eq:Etree}, each $\phi_{0,\alpha}$ couples to the Green--Schwarz four-form of its node \cite{Ohmori:2014kda}, with $h^\vee=(1,3)$ and $2-\mathcal K_{\alpha\alpha}=(1,-1)$. So the $\su(2)$ coupling is $T_2$, the tension of the $-3$ string, and the E-string couples to the $\usp(8)$ instanton density through a quadratic form $\mathsf q_F(m)$. Its normalisation depends on the embedding of $C_4$ in $E_8$, so we leave it unknown. It enters only the terms $\phi\,mm$, i.e.\ step~3 of \tref{tab:framemaprecipe}, so it does not affect the map, and the matching below fixes it. The one-loop terms are those of \sref{sec:exSU39}, and the E-string node has none. Altogether
\begin{equation}
\label{eq:su44E6d}
\begin{aligned}
\epsilon_1\epsilon_2\,\mathcal E_{\rm 6d}&=\tfrac\tau2\,\phi_0^T\mathcal K\phi_0+T_2\,\phi_1^2-\phi_{0,1}\,\mathsf q_F(m)+\tfrac43\phi_1^3-\tfrac5{16}\tau\phi_1^2\\
&\quad+\left[\tfrac14(\phi_{0,1}-\phi_{0,2})-\tfrac1{12}\phi_1\right](\epsilon_1^2+\epsilon_2^2)+(\phi_{0,1}+3\phi_{0,2}+\phi_1)\,\epsilon_+^2\,.
\end{aligned}
\end{equation}
The 6d cubic has no terms quadratic or cubic in the tensor scalars. With this constraint, the cubic terms of \eqref{eq:su40E5d} at $\kCS=4$ match only in the small-mass chamber with $e_1+e_4<0$. Within the scan bound $|M_{ij}|\le3$, they give three matrices (Appendix~\ref{app:scans}), acting on $(\phi_{0,1},\phi_{0,2},\phi_1)$:
\begin{equation}
\label{eq:su44Mn}
\setlength{\arraycolsep}{5pt}
M^{(n)}=\begin{pmatrix}0&1&-n\\0&2&1-2n\\1&0&-3n\end{pmatrix}\,,\qquad n=0,\pm1\,.
\end{equation}
They are related by $\phi_{0,1}\to\phi_{0,1}-3n\phi_1$, $\phi_{0,2}\to\phi_{0,2}-n\phi_1$. This substitution preserves $T_2$, hence the 6d cubic, but not $\tfrac\tau2\phi_0^T\mathcal K\phi_0$, and the $\tau\times$quadratic terms exclude $n=\pm1$. Appendix~\ref{app:scans} shows this from the coupling term $m_0\,k(P)$ of each matrix, \eqref{eq:su44Q}. For $n=0$, the member without mass mixing is
\begin{equation}
\label{eq:su44map}
P_1=\phi_{0,2}+\tfrac\tau{16}\,,\qquad P_2=2\phi_{0,2}+\phi_1-\tfrac\tau8\,,\qquad P_3=\phi_{0,1}\,,\qquad m_0=\tfrac\tau2\,,\qquad M_i=m_i\,.
\end{equation}
With it, \eqref{eq:su40E5d} at $\kCS=4$ equals \eqref{eq:su44E6d} up to $\phi$-independent terms, so $r=2$. The $-3$ node reproduces \eqref{eq:SU39map}. The matching fixes $\mathsf q_F=\sum_im_i^2=\tfrac12\tr_{\mathbf 8}m^2$, the Killing form of $C_4$. So the coefficient of $\phi_{0,1}m_i^2$ in \eqref{eq:su44E6d}, the flavour Green--Schwarz level of \eqref{eq:Etree} on the E-string node, is $\kappa_{m_i}=-1$, which confirms $\usp(8)$.

\paragraph{States.} A string of charge $\lambda$ has tension $\lambda\cdot T$, so $\lambda$ is read off from the coefficients of $T$. By \eqref{eq:su44map}, the positive roots of $\su(4)$ and the instantonic root are, in 6d variables,
\begin{equation}
\label{eq:su44roots}
\mbox{\footnotesize\setlength{\arraycolsep}{1.4pt}$\begin{array}{l|l|l|l}
\text{root of }\su(4) & \text{in }P_i & \text{in 6d variables} & \text{6d state}\\ \hline
e_1-e_2\ \text{(simple)} & 2P_1-P_2 & \tfrac\tau4-\phi_1=\alpha_0 & \text{lowest state of $\mathbf 2_{1/4}$}\\
e_2-e_3\ \text{(simple)} & -P_1+2P_2-P_3 & T_2+2\phi_1-\tfrac{5\tau}{16} & \text{string, }\lambda=(0,1)\\
e_3-e_4\ \text{(simple)} & -P_2+2P_3 & 2T_1-\phi_1+\tfrac\tau8 & \text{string, }\lambda=(2,0)\\
e_1-e_3 & P_1+P_2-P_3 & T_2+\phi_1-\tfrac\tau{16} & \text{string, }\lambda=(0,1)\\
e_2-e_4 & -P_1+P_2+P_3 & 2T_1+T_2+\phi_1-\tfrac{3\tau}{16} & \text{string, }\lambda=(2,1)\\
e_1-e_4 & P_1+P_3 & 2T_1+T_2+\tfrac\tau{16} & \text{string, }\lambda=(2,1)\\ \hline
m_0-2(e_1-e_2) & m_0-4P_1+2P_2 & 2\phi_1=\alpha_1 & \text{$\su(2)$ W-boson, $k=1$}
\end{array}$}
\end{equation}
With $\Phi:=T_2-\tfrac5{16}\tau$, the coefficient of $\phi_1^2$ in \eqref{eq:su44E6d}, the rows without $T_1$ are, in 6d variables, those of \eqref{eq:SU39roots}, since the $-3$ node reproduces \eqref{eq:SU39map}. So the simple roots of $\su(4)$ are the affine root $\alpha_0$ of $A_2^{(2)}$ and two strings, of charges $(0,1)$ and $(2,0)$. As for $\SU(3)_9$, the instantonic root $\alpha_1$ is a root of $\su(2)$, but not of $\su(4)$. The marks $(2,1)$ of $A_2^{(2)}$ give $\delta=2\alpha_0+\alpha_1=\tfrac\tau2=m_0$, with $\alpha_1$ of mark $1$, as \eqref{eq:rrule} requires. The $\Lambda^2$ states are $e_a+e_b+M_i$, with $M_i=m_i$:
\begin{equation}
\label{eq:su44states}
\mbox{\footnotesize\setlength{\arraycolsep}{1.4pt}$\begin{array}{l|l|l|l}
\text{state} & \text{in }P_i & \text{in 6d variables} & \text{6d state}\\ \hline
e_1+e_2+M_i & P_2+M_i & T_1+T_2+\phi_1-\tfrac\tau8+m_i & \text{string, }\lambda=(1,1)\\
e_3+e_4+M_i & -P_2+M_i & -T_1-T_2-\phi_1+\tfrac\tau8+m_i & \text{string, }\lambda=(-1,-1)\\
e_1+e_3+M_i & P_1-P_2+P_3+M_i & T_1-\phi_1+\tfrac{3\tau}{16}+m_i & \text{string, }\lambda=(1,0)\\
e_2+e_4+M_i & -P_1+P_2-P_3+M_i & -T_1+\phi_1-\tfrac{3\tau}{16}+m_i & \text{string, }\lambda=(-1,0)\\
e_2+e_3+M_i & -P_1+P_3+M_i & T_1-\tfrac\tau{16}+m_i & \text{string, }\lambda=(1,0)\\
e_1+e_4+M_i & P_1-P_3+M_i & -T_1+\tfrac\tau{16}+m_i & \text{string, }\lambda=(-1,0)
\end{array}$}
\end{equation}
On the tensor branch, $T_1$ is large, so $e_1+e_4=-T_1+\tfrac\tau{16}<0$, the chamber found above. Every state has E-string charge $\lambda_1=\pm1$. A state of negative charge is minus a state of positive charge with $m_i\to-m_i$, i.e.\ its antiparticle. So each of the three states with $\lambda_1=1$ comes with the eight weights $\pm m_i$ of the $\mathbf 8$ of $\usp(8)$. These are E-string states, so they have no 6d one-loop term. Since $m_0$ and $M_i$ in \eqref{eq:su44map} contain no $\phi$, a state of $\su(4)$ weight $\mathsf w=0$ has no string charge and no weight under the invariant subalgebra $\su(2)^\sigma\subset\su(3)$.

\paragraph{Prefactor $\Lambda_B$.} The fluxes are $a=(a_1,a_2)$ for the tensor scalars $(\phi_{0,1},\phi_{0,2})$, $b$ for the $\su(2)$ modulus $\phi_1$ and $B_{m_i}$ for the masses, with $B_\tau=0$. The map \eqref{eq:su44map} has no mass shifts, so $a$ is also the shifted tensor flux \eqref{eq:adef}. Evaluating \eqref{eq:Vdef} on \eqref{eq:su44E6d}, the terms of $V$ that involve $b$ are
\begin{equation}
\label{eq:su44Vb}
V(a,b)-V(a,0)=b^2\,\Phi+2b\,(\lambda_2+2b)\,\phi_1+\tfrac b3\,\big(3\lambda_2b+4b^2-1\big)(\epsilon_1+\epsilon_2)\,,
\end{equation}
with $\Phi=T_2-\tfrac5{16}\tau$ as above and $\lambda_2=3a_2-a_1$ the second component of the string charge $\lambda=\mathcal Ka$. They depend on the moduli unless $b=0$. At $b=0$, $V$ is \eqref{eq:Vfamily}, whose moduli dependence $\sum_\alpha\Delta_\alpha\phi_{0,\alpha}$ vanishes by the family condition \eqref{eq:familycond}. With $h^\vee=(1,3)$, $2-\mathcal K_{\alpha\alpha}=(1,-1)$ and $\kappa_{m_i}=-1$ on node 1, the condition reads $\Delta_1=\tfrac14+\tfrac34-4B_m^2=0$ and $\Delta_2=\tfrac34-\tfrac34=0$. So node 2 satisfies it identically, and node 1 forces $B_{m_i}=\tfrac12$. The tensor-flux family is therefore $b=0$, with $B_{m_i}=\tfrac12$. On it, \eqref{eq:Vfamily} gives $x_1=-\sum_i(m_i+\tfrac12\epsilon_+)$, from the four masses alone since $h^\vee_1-2+\mathcal K_{11}=0$, and $x_2=-\tfrac12(3-2+3)\epsilon_+=-2\epsilon_+$, with $2\epsilon_+=\epsilon_1+\epsilon_2$. The sign \eqref{eq:signw} needs a Wu vector, $w\equiv\mathrm{diag}\,\mathcal K\equiv(1,1)$ mod $2$. We take $w=(1,1)=\mathcal K(2,1)^T$, which lies in $\mathcal K\BZ^2$ as \eqref{eq:sequalsqw} requires. Altogether
\begin{align}
V\big|_a&=\tfrac\tau2\,a^T\mathcal Ka-(a_1+a_2)(\epsilon_1+\epsilon_2)-a_1\textstyle\sum_im_i\,,\label{eq:su44Lambda}\\
\Lambda_B&=\sum_{a\in\BZ^2+B}e^{i\pi(a_1+a_2)}\,\fq^{\frac12a^T\mathcal Ka}\,e^{-a\cdot x}\,,\qquad x=\Big(-\epsilon_1-\epsilon_2-\textstyle\sum_im_i,\ -\epsilon_1-\epsilon_2\Big)\,.\label{eq:Lambda237}
\end{align}
This is \eqref{eq:LambdaTheta} for $n_T=2$, with period matrix $\mathcal K\tau$ and characteristic $(B;\tfrac12,\tfrac12)$. In the basis of the nodes, the cross term $\fq^{-a_1a_2}$ in $\fq^{\frac12a^T\mathcal Ka}$ couples $a_1$ and $a_2$. The lattice is nevertheless a sum of two rank-one lattices. In the basis $u=a_1-a_2$, $v=a_2$ of $\BZ^2$, $\tfrac12a^T\mathcal Ka=\tfrac12u^2+v^2$, i.e.\ $\mathcal K\cong(1)\oplus(2)$. Varying $u$ moves $a$ along the E-string direction $(1,0)$, of norm $1$, and varying $v$ along $(1,1)$, of norm $2$ and orthogonal to it. Both sectors have $u\in\BZ$ and $v\in\BZ+B_2$, and $e^{i\pi(a_1+a_2)}=(-1)^u\,e^{2\pi iv}$. Hence
\begin{equation}
\label{eq:su44factor}
\Lambda_B=e^{2\pi iB_2}\,\theta_4(\tau,x_1)\,\vartheta\big[\genfrac{}{}{0pt}{1}{B_2}{0}\big](2\tau,\,x_1+x_2)\,,
\end{equation}
i.e.\ $\theta_4(\tau,x_1)\,\theta_3(2\tau,x_1+x_2)$ for $B=0$ and $-\theta_4(\tau,x_1)\,\theta_2(2\tau,x_1+x_2)$ for $B=(\tfrac12,\tfrac12)$. The first factor is the same in both sectors. The $\BZ_2$ sits in the second factor, which is $\Lambda_B$ of \eqref{eq:su40sectors} for $\mathcal K=(2)$, with argument $x_1+x_2$: $\theta_3(2\tau,\cdot)$ in the trivial sector and $\theta_2(2\tau,\cdot)$ in the 't~Hooft sector.

\paragraph{Sectors.} By the \hyperref[it:fluxmap]{flux rule}, \eqref{eq:su44map} maps the fluxes to $n^{\rm 5d}=(a_2,\,2a_2+b,\,a_1)$, with no offset since there are no mass shifts. On the tensor-flux family, this is $\nu(a)=(a_2,2a_2,a_1)$, i.e.\ \eqref{eq:nuofa} with the tensor columns $c_1=(0,0,1)$ and $c_2=(1,2,0)$. The string charge $\lambda=\mathcal Ka$ must be integral, so $a\in\BZ^2+B$ with $B\in\mathcal K^{-1}\BZ^2/\BZ^2=\{0,(\tfrac12,\tfrac12)\}\cong D=\BZ_2$. The non-trivial sector is $B=\mathcal K^{-1}\lambda$ with $\lambda=(0,1)$, the generator of $D$. Its member $a=B$ has $\nu(B)=(\tfrac12,1,\tfrac12)\in\omega_2+\BZ^3$, so this is the 5d 't~Hooft sector of \eqref{eq:su40sectors}. Its leading members are $a=\pm(\tfrac12,\tfrac12)$, both with $\tfrac12a^T\mathcal Ka=\tfrac14$ and sign $e^{i\pi(a_1+a_2)}=-1$. So the leading term of $\Lambda_B$ is $-\fq^{1/4}\big(e^{\frac12(x_1+x_2)}+e^{-\frac12(x_1+x_2)}\big)$. Its KK momentum is $\tfrac14=q(0,1)=k(\omega_2)/r$, with $k(\omega_2)=\tfrac12$ and $r=2$, as \eqref{eq:dictionary} states. The sign $-1$ counts as half a unit of KK momentum, \eqref{eq:sdef}, so the refined KK momentum is $\tfrac14+\tfrac12w\cdot B=\tfrac34\equiv-\tfrac14$. This is $q_w(0,1)$ for $w=(1,1)$ (\tref{tab:lattices}), as \eqref{eq:sequalsqw} states. The other choice $w=(1,-1)=\mathcal K(1,0)^T$ removes the sign and gives $\tfrac14$. The two choices differ by a constant sign on this sector, which the blow-up equations do not test (\sref{sec:pfspin}). Since $M_i=m_i$, all this holds at fixed 5d and 6d masses alike, unlike in \sref{sec:exsu40}.

\subsection{\texorpdfstring{$\su(4)_2+4\Lambda^2$}{su(4)2+4AS}: \texorpdfstring{$r=2$}{r=2} without a twist}
\label{sec:exsu42}

The 6d theory is $\mathfrak g_2$ with four $\mathbf 7$'s on a $-2$ curve, reduced without a twist, and its 5d frame is $\su(4)_2+4\Lambda^2$ \cite[(2.36)]{Bhardwaj:2020gyu}.

\paragraph{6d frame and map.} We have $\mathcal K=2$, $(\mathfrak a,\mathfrak b,\mathfrak c)=(0,1,2)$, and the $\mathfrak g_2$ basis of \sref{sec:exSU48}, with $\alpha_1=(2,-1)$ short, $\alpha_2=(-3,2)$ long and $\tfrac12\tilde K\phi\phi=\mathsf q:=3\phi_1^2-3\phi_1\phi_2+\phi_2^2$. The matter is a half-hypermultiplet in the $(\mathbf 7,\mathbf 8)$ of $G_2\times\USp(8)$, with masses in $\usp(8)$, and the weights of the $\mathbf 7$ are $0$ and $\pm\mathsf w_k$, with $\mathsf w_1=(2,-1)$, $\mathsf w_2=(-1,1)$, $\mathsf w_3=(1,0)$. $G_2$ has no centre, so $D=\BZ_2$ is of tensor origin. Without a twist, the vector and hypermultiplet towers cancel, since $h^\vee=4$ is four times the index of the $\mathbf 7$, so $2\phi_0$ is not renormalised. In the $\mathfrak g_2$ chamber with $m_i>\mathsf w_k\cdot\phi$,
\begin{equation}
\label{eq:su42E6d}
\begin{aligned}
\epsilon_1\epsilon_2\,\mathcal E_{\rm 6d}&=\tau\phi_0^2+\Big(2\phi_0-\textstyle\sum_im_i\Big)\mathsf q-\phi_0\,\mathsf q_F(m)+\tfrac43\phi_1^3+3\phi_1^2\phi_2-4\phi_1\phi_2^2+\tfrac43\phi_2^3\\
&\quad-\tfrac1{12}(\phi_1+\phi_2)(\epsilon_1^2+\epsilon_2^2)+(4\phi_0+\phi_1+\phi_2)\,\epsilon_+^2\,,
\end{aligned}
\end{equation}
with the flavour form $\mathsf q_F$ unknown. The 5d frame is \eqref{eq:su40E5d} at $\kCS=2$. Even with the masses off, the 6d cubic depends on the 6d chamber, i.e.\ on which short-root pairs the masses dominate. With the masses below all short-root pairs, no matrix in the scan range matches. Scanning both sides gives three matching pairs of chambers (Appendix~\ref{app:scans}), one of them with the 6d chamber above, always with
\begin{equation}
\label{eq:su42M}
M=\begin{pmatrix}1&1&0\\2&0&1\\1&0&0\end{pmatrix}\,,\qquad P_1=\phi_0+\phi_1+\dots\,,\quad P_2=2\phi_0+\phi_2+\dots\,,\quad P_3=\phi_0+\dots\,.
\end{equation}
This is the affine-node rule \eqref{eq:affinenode} for $G_2^{(1)}$, with comarks $1$ on the affine node, $1$ on the short and $2$ on the long root, and the affine node third. It is the charge conjugate of \eqref{eq:su40M}, and $\kCS=2$ selects this orientation. In the pair with all short-root pairs dominated in 6d and none, with $e_1+e_4<0$, in 5d, the member without mass mixing, which also holds in the other two pairs, is
\begin{equation}
\label{eq:su42map}
\begin{aligned}
P_1&=\phi_0+\phi_1+\tfrac\tau4-\tfrac12\textstyle\sum_im_i\,,\quad P_2=2\phi_0+\phi_2+\tfrac\tau2-\textstyle\sum_im_i\,,\\
P_3&=\phi_0+\tfrac{3\tau}4-\tfrac12\textstyle\sum_im_i\,,\quad m_0=\tfrac\tau2\,,\quad M_i=m_i-\tfrac\tau2\,,
\end{aligned}
\end{equation}
with $\mathsf q_F=\sum_im_i^2$. With it, \eqref{eq:su40E5d} at $\kCS=2$ equals \eqref{eq:su42E6d} up to $\phi$-independent terms, so $r=2$. The large 6d masses are small 5d masses, by $M_i=m_i-\tfrac\tau2$.

\paragraph{Why $r=2$.} Let $\Phi:=2\phi_0-\sum_im_i$, the coefficient of $\mathsf q$ in \eqref{eq:su42E6d}. Without a twist, the towers do not shift it, but, in this chamber, the masses do. By \eqref{eq:su42map}, the positive roots of $\su(4)$ and the instantonic root are, in 6d variables,
\begin{equation}
\label{eq:su42roots}
\mbox{\footnotesize\setlength{\arraycolsep}{1.4pt}$\begin{array}{l|l|l|l}
\text{root of }\su(4) & \text{in }P_i & \text{in 6d variables} & \text{6d state}\\ \hline
e_1-e_2\ \text{(simple)} & 2P_1-P_2 & 2\phi_1-\phi_2=\alpha_1 & \text{zero mode, short root}\\
e_2-e_3\ \text{(simple)} & -P_1+2P_2-P_3 & \Phi-\phi_1+2\phi_2 & \text{string}\\
e_3-e_4\ \text{(simple)} & -P_2+2P_3 & \tau-\phi_2=\alpha_0 & \text{KK mode of the W-boson of $-\theta$}\\
e_1-e_3 & P_1+P_2-P_3 & \Phi+\phi_1+\phi_2 & \text{string}\\
e_2-e_4 & -P_1+P_2+P_3 & \Phi-\phi_1+\phi_2+\tau & \text{string}\\
e_1-e_4 & P_1+P_3 & \Phi+\phi_1+\tau & \text{string}\\ \hline
m_0-(2e_1-e_2+e_3) & m_0-3P_1+2P_2-P_3 & -3\phi_1+2\phi_2=\alpha_2 & \text{long root, $k=1$}
\end{array}$}
\end{equation}
The coefficients of $(\phi_1,\phi_2)$ are $\mathfrak g_2$ Dynkin labels. The roots with $\Phi$ contain $2\phi_0=\mathcal K\phi_0$, so they are strings. The highest root of $\mathfrak g_2$ is $\theta=3\alpha_1+2\alpha_2=(0,1)$, so $\theta\cdot\phi=\phi_2$, and $e_3-e_4=\tau-\theta\cdot\phi$ is the affine root $\alpha_0$ of $G_2^{(1)}$, with KK momentum $1$. So the simple roots of $\su(4)$ are $\alpha_0$, the zero mode $\alpha_1$ and a string, with the affine node now third. The long root $\alpha_2$ of $\mathfrak g_2$ is not a root of $\su(4)$. It is the instantonic root, with $k=1$. With the marks $(1,3,2)$ of $G_2^{(1)}$, $\delta=\alpha_0+3\alpha_1+2\alpha_2=\tau$. In 5d, $\alpha_0$ and $\alpha_1$ are W-bosons, with $k=0$, so $k(\delta)=2$, the mark of $\alpha_2$, as \eqref{eq:rrule} requires. Then \eqref{eq:m0delta} gives $m_0=\delta/2=\tfrac\tau2$, i.e.\ $r=2$ without a twist.

\paragraph{Hypermultiplet states.} For each mass, the six weights $e_a+e_b$ of $\Lambda^2$ give, by \eqref{eq:su42map},
\begin{equation}
\label{eq:su42hypers}
\mbox{\footnotesize\setlength{\arraycolsep}{1.4pt}$\begin{array}{l|l|l|l}
\text{hypermultiplet state} & \text{in }P_i & \text{in 6d variables} & \text{6d state}\\ \hline
e_1+e_3+M_i & P_1-P_2+P_3+M_i & m_i+\phi_1-\phi_2 & \text{zero mode, weight $-\mathsf w_2$}\\
e_2+e_3+M_i & -P_1+P_3+M_i & m_i-\phi_1 & \text{zero mode, weight $-\mathsf w_3$}\\
e_1+e_4+M_i & P_1-P_3+M_i & m_i+\phi_1-\tau & \text{KK mode, weight $\mathsf w_3$}\\
e_2+e_4+M_i & -P_1+P_2-P_3+M_i & m_i-\phi_1+\phi_2-\tau & \text{KK mode, weight $\mathsf w_2$}\\
e_1+e_2+M_i & P_2+M_i & m_i+\Phi+\phi_2 & \text{string}\\
e_3+e_4+M_i & -P_2+M_i & m_i-\Phi-\phi_2-\tau & \text{anti-string}\\ \hline
M_i+m_0 & M_i+m_0 & m_i & \text{zero mode, weight $0$, $k=1$}
\end{array}$}
\end{equation}
with $\mathsf w_k$ the weights of the $\mathbf 7$ above. The two KK modes and the anti-string have KK momentum $-1$. The last row is not a $\Lambda^2$ state. Since $m_i=M_i+m_0$ by \eqref{eq:su42map}, the zero mode of the zero weight of the $\mathbf 7$, of mass $m_i$, is the 5d state $M_i+m_0$, with instanton number $1$ (used in \sref{sec:faithful} below).

\paragraph{Prefactor $\Lambda_B$ and sectors.} As in \sref{sec:exsu40}, the tensor-flux family has zero $\mathfrak g_2$ flux and $B_{m_i}=\tfrac12$, with $V|_a$ in 6d masses as in \eqref{eq:su40Lambda}. On the tensor-flux family, \eqref{eq:su42map} gives $\nu(a)=a(1,2,1)$ with $a=n_0-1$, since the offset $-\tfrac12\sum_iB_{m_i}=-1$ is now an integer. So the 't~Hooft sector $B=\tfrac12$ is $n_0\in\BZ+\tfrac12$, the opposite assignment to \eqref{eq:su40sectors}. At fixed 5d masses, $V=\tau a^2+a\,x_M+\text{const}$, with $x_M$ of \eqref{eq:su40Lambda}, so $\Lambda_B$ is as in \eqref{eq:su40sectors}, and since $a-n_0\in\BZ$ the labels agree in both mass conventions.

\subsection{\texorpdfstring{$\so(8)$}{so(8)} with the triality twist: \texorpdfstring{$\sp(3)_0+2\Lambda^2$}{sp(3)0+2AS} and \texorpdfstring{$\su(4)_6+2\Lambda^2$}{su(4)6+2AS}}
\label{sec:exso8tri}

The 6d theory is $\so(8)$ with $2(\mathbf 8_v\oplus\mathbf 8_s\oplus\mathbf 8_c)$ on a $-2$ curve, reduced with the $\BZ_3$ triality twist. It has two 5d frames, $\sp(3)_0+2\Lambda^2$ and $\su(4)_6+2\Lambda^2$ \cite[(2.38), (2.30), \S2.5.1, (2.211), (4.93)]{Bhardwaj:2020gyu}. Triality leaves the diagonal $\sp(2)$ of the 6d flavour symmetry $\sp(2)^3$, with masses $m_{1,2}$. The pair tests whether $r$ belongs to the KK theory or to the frame.

\paragraph{6d frame.} We have $\mathcal K=2$ and $(\mathfrak a,\mathfrak b,\mathfrak c)=(0,1,3)$. The invariant subalgebra is $\mathfrak g_2$, with the basis and $\mathsf q$ of \sref{sec:exsu42}, and $\mathbf{28}\to\mathbf{14}_0\oplus\mathbf 7_{1/3}\oplus\mathbf 7_{2/3}$. Each $\mathbf 8$ is $\mathbf 7\oplus\mathbf 1$, and the eigencombinations of the three permuted $\mathbf 8$'s carry KK charges $0,\tfrac13,\tfrac23$ \cite[Table~34]{Kim:2020hhh}. In units of $\tau\cdot\tfrac12\tilde K\phi\phi$, the vector towers give $-\tfrac59$ (\sref{sec:exSU48}) and the hypermultiplet towers $+\tfrac19$, from $B_2(0)+B_2(\tfrac13)+B_2(\tfrac23)=\tfrac1{18}$. In the fundamental $\mathfrak g_2$ chamber with $m_i$ below the short roots,\footnote{\label{fn:g2chamber}That is, $\alpha_1\cdot\phi=2\phi_1-\phi_2>0$, $\alpha_2\cdot\phi=-3\phi_1+2\phi_2>0$ and $|m_i|<2\phi_1-\phi_2$. The last bound suffices, since $2\phi_1-\phi_2$ is the smallest of the $\mathsf w_k\cdot\phi$ of \sref{sec:exsu42}: $\mathsf w_2\cdot\phi-\mathsf w_1\cdot\phi=\alpha_2\cdot\phi$ and $\mathsf w_3\cdot\phi-\mathsf w_2\cdot\phi=\alpha_1\cdot\phi$ are positive.}
\begin{equation}
\label{eq:so8triE6d}
\begin{aligned}
\epsilon_1\epsilon_2\,\mathcal E_{\rm 6d}&=\tau\phi_0^2+\big(2\phi_0-\tfrac49\tau\big)\mathsf q+6\phi_0\epsilon_+^2-\phi_0\,\mathsf q_F(m)-\tfrac43\phi_1^3+6\phi_1^2\phi_2-5\phi_1\phi_2^2+\tfrac43\phi_2^3\\
&\quad-\textstyle\sum_im_i^2\,\phi_1+\tfrac\tau9\textstyle\sum_im_i^2+\tfrac1{12}(\phi_1-\phi_2)(\epsilon_1^2+\epsilon_2^2)+(\phi_1+\phi_2)\,\epsilon_+^2\,.
\end{aligned}
\end{equation}

\paragraph{The $\sp(3)_0$ frame.} With orthonormal moduli $e_i$, $\tfrac12\tilde K\,ee=\sum_ie_i^2$, and the $\mathbf{14}$ has the weights $\pm e_i\pm e_j$, $i<j$, and two zero weights. In the chamber $e_1>e_2>e_3>0$ with the masses $M_i$ below every $e_i\pm e_j$,
\begin{equation}
\label{eq:so8triE5d}
\begin{aligned}
\epsilon_1\epsilon_2\,\mathcal E_{\rm 5d}&=m_0\textstyle\sum_ie_i^2+\tfrac23e_1^3+e_2^3+\tfrac43e_3^3-e_1e_2^2-e_1e_3^2-e_2e_3^2-(2e_1+e_2)\textstyle\sum_iM_i^2\\
&\quad+\tfrac1{12}(e_1-e_3)(\epsilon_1^2+\epsilon_2^2)+(3e_1+2e_2+e_3)\,\epsilon_+^2\,.
\end{aligned}
\end{equation}
Together, the cubic and the $\epsilon_+^2$ terms match only in two chamber pairs (Appendix~\ref{app:scans}), and the remaining terms fix the map up to an $\mathrm O(2)$ rotation of $(m_1,m_2)$. Its canonical member is
\begin{equation}
\label{eq:so8trimap}
e_1=\phi_0+\tfrac\tau9\,,\quad e_2=\phi_0+\phi_1-\tfrac{2\tau}9\,,\quad e_3=\phi_0-\phi_1+\phi_2-\tfrac{2\tau}9\,,\quad m_0=\tfrac\tau3\,,\quad M_i=m_i\,,
\end{equation}
with $\mathsf q_F=3\sum_im_i^2$. With it, \eqref{eq:so8triE5d} equals \eqref{eq:so8triE6d} up to $\phi$-independent terms, and $r=3$. In the Dynkin variables $P:=(e_1,\,e_1+e_2,\,e_1+e_2+e_3)$ of $\sp(3)$, the map is the affine-node rule with the comarks $(2,3)$ of $D_4^{(3)}$, and it coincides with the $\SU(4)_8$ map \eqref{eq:su48map}. Each permuted $\mathbf 8$ contributes $-1$ to $\kappa_{m_i}=-3$.

\paragraph{States.} Let $\Phi:=2\phi_0-\tfrac49\tau$, the coefficient of $\mathsf q$ in \eqref{eq:so8triE6d}. By \eqref{eq:so8trimap}, the positive roots of $\sp(3)$ and the instantonic root are, in 6d variables,
\begin{equation}
\label{eq:so8triroots}
\mbox{\footnotesize\setlength{\arraycolsep}{1.4pt}$\begin{array}{l|l|l|l}
\text{root of }\sp(3) & \text{in }P_i & \text{in 6d variables} & \text{6d state}\\ \hline
e_1-e_2\ \text{(simple)} & 2P_1-P_2 & \tfrac\tau3-\phi_1=\alpha_0 & \text{lowest state of $\mathbf 7_{1/3}$}\\
e_2-e_3\ \text{(simple)} & -P_1+2P_2-P_3 & 2\phi_1-\phi_2=\alpha_1 & \text{zero mode, short root}\\
2e_3\ \text{(simple)} & -2P_2+2P_3 & \Phi-2\phi_1+2\phi_2 & \text{string}\\
e_1-e_3 & P_1+P_2-P_3 & \tfrac\tau3+\phi_1-\phi_2 & \text{state of $\mathbf 7_{1/3}$}\\
e_2+e_3 & -P_1+P_3 & \Phi+\phi_2 & \text{string}\\
e_1+e_3 & P_1-P_2+P_3 & \Phi-\phi_1+\phi_2+\tfrac\tau3 & \text{string}\\
2e_2 & -2P_1+2P_2 & \Phi+2\phi_1 & \text{string}\\
e_1+e_2 & P_2 & \Phi+\phi_1+\tfrac\tau3 & \text{string}\\
2e_1 & 2P_1 & \Phi+\tfrac{2\tau}3 & \text{string}\\ \hline
m_0-(e_1+e_2-2e_3) & m_0-3P_2+2P_3 & -3\phi_1+2\phi_2=\alpha_2 & \text{long root, $k=1$}
\end{array}$}
\end{equation}
The roots with $\Phi$ contain $2\phi_0=\mathcal K\phi_0$, so they are strings. As for $\SU(4)_8$ in \eqref{eq:su48roots}, the simple roots are the affine root $\alpha_0$ of $D_4^{(3)}$, the zero mode $\alpha_1$ and a string, and the long root $\alpha_2$ of $\mathfrak g_2$ is instantonic, with $k=1$. With the marks $(1,2,1)$, $\delta=\alpha_0+2\alpha_1+\alpha_2=\tfrac\tau3=m_0$, and $\alpha_2$ has mark $1$, as \eqref{eq:rrule} requires. For each mass, the weights $\mathsf w$ of the $\mathbf{14}$ give the states $\mathsf w+M_i$, with $M_i=m_i$:
\begin{equation}
\label{eq:so8trihypers}
\mbox{\footnotesize\setlength{\arraycolsep}{1.4pt}$\begin{array}{l|l|l|l}
\text{state} & \text{in }P_i & \text{in 6d variables} & \text{6d state}\\ \hline
e_2-e_3+M_i & -P_1+2P_2-P_3+M_i & 2\phi_1-\phi_2+m_i & \text{zero mode of }\mathbf 7_0\\
-(e_2-e_3)+M_i & P_1-2P_2+P_3+M_i & -2\phi_1+\phi_2+m_i & \text{zero mode of }\mathbf 7_0\\
e_1-e_2+M_i & 2P_1-P_2+M_i & \tfrac\tau3-\phi_1+m_i & \text{KK mode of }\mathbf 7_{1/3}\\
e_1-e_3+M_i & P_1+P_2-P_3+M_i & \tfrac\tau3+\phi_1-\phi_2+m_i & \text{KK mode of }\mathbf 7_{1/3}\\
-(e_1-e_2)+M_i & -2P_1+P_2+M_i & -\tfrac\tau3+\phi_1+m_i & \text{KK mode of }\mathbf 7_{2/3}\\
-(e_1-e_3)+M_i & -P_1-P_2+P_3+M_i & -\tfrac\tau3-\phi_1+\phi_2+m_i & \text{KK mode of }\mathbf 7_{2/3}\\
\pm(e_2+e_3)+M_i & \pm(P_3-P_1)+M_i & \pm(\Phi+\phi_2)+m_i & \text{string, }\lambda=\pm1\\
\pm(e_1+e_3)+M_i & \pm(P_1-P_2+P_3)+M_i & \pm(\Phi-\phi_1+\phi_2+\tfrac\tau3)+m_i & \text{string, }\lambda=\pm1\\
\pm(e_1+e_2)+M_i & \pm P_2+M_i & \pm(\Phi+\phi_1+\tfrac\tau3)+m_i & \text{string, }\lambda=\pm1\\
M_i\ (\text{twice}) & M_i & m_i & \text{zero weight of }\mathbf 7_0,\ \text{singlet }\mathbf 1_0
\end{array}$}
\end{equation}
A state that is not a string lies in the tower $\mathbf R_{\fr}$ with $\fr$ its KK momentum mod $1$, read at flavour weight $+m_i$. So the rows of momentum $-\tfrac13$ lie in $\mathbf 7_{2/3}$. Their negatives are states of $\mathbf 7_{1/3}$ with flavour weight $-m_i$, which is how the flop table \eqref{eq:so8triflops} below reads $e_1-e_2-M_i$. By \eqref{eq:so8trimap}, $\tau=3m_0$, $\phi_1=m_0-(e_1-e_2)$ and $\phi_2=2m_0-2(e_1-e_2)-(e_2-e_3)$. So a 6d state $\mathsf w\cdot\phi+p\tau+m_i$, with $\mathsf w=(j_1,j_2)$ in Dynkin labels, has instanton number $j_1+2j_2+3p$ in the $\sp(3)_0$ frame. On the towers of $\mathbf 7\oplus\mathbf 1$, it vanishes for exactly one state per weight, the eight states in the rows above without $\Phi$. All other KK modes carry instanton charge, e.g.\ the zero mode $\phi_1+m_i$ of $\mathbf 7_0$ has $k=1$.

\paragraph{Prefactor $\Lambda_B$ and sectors.} The family condition \eqref{eq:familycond}, $\tfrac32-6B_m^2=0$, gives $B_{m_i}=\tfrac12$, and \eqref{eq:Vfamily} gives $V|_a=\tau a^2-a\big[3(m_1+m_2)+6\epsilon_+\big]$. Since $M_i=m_i$, the tensor-flux family $\nu(a)=(a,a,a)$ in the basis $e_i$, i.e.\ $a(1,2,3)$ in $P_i$, has no offset, and $a=n_0$. So $B=\tfrac12$ is the 't~Hooft coset $(\tfrac12,\tfrac12,\tfrac12)+\BZ^3$ of the centre of $\USp(6)$, with $k=\tfrac34$ and $k/r=\tfrac14=q(1)$.

\paragraph{The $\su(4)_6$ frame.} Primes denote its variables: the Dynkin variables $P'_i$ of $\su(4)$, with $e'=(P'_1,P'_2-P'_1,P'_3-P'_2,-P'_3)$, the instanton mass $m_0'$ and the masses $M'_i$. This frame is reached from the 6d frame only through flops of KK modes (\hyperref[it:validity]{range of validity} in \sref{sec:recipe}). So we first map it to the $\sp(3)_0$ frame, with $P'=Ne+\dots$ and the level $\kCS$ free.

Both 5d prepotentials depend on the mass chamber. The non-zero weights of the $\mathbf{14}$ of $\sp(3)$ come in six pairs $\pm\mathsf w$, which we label
\begin{equation}
\label{eq:sp3pairs}
1,\dots,6:\qquad \mathsf w\cdot e=e_1+e_2,\ e_1+e_3,\ e_2+e_3,\ e_1-e_2,\ e_1-e_3,\ e_2-e_3\,.
\end{equation}
We say that the masses \emph{dominate} the pair $\pm\mathsf w$ if $|M_i|>\mathsf w\cdot e$. Then, the two hypermultiplet masses $\pm\mathsf w\cdot e+M_i$ have the same sign, that of $M_i$. Otherwise, they have opposite signs. Each case gives a different polynomial for the terms $|\pm\mathsf w\cdot e+M_i|^3$ of \eqref{eq:E5d}. We label a chamber by its set of dominated pairs. So $\{\}$ is the chamber of \eqref{eq:so8triE5d}, with the masses below every $\mathsf w\cdot e$, and in $\{4,5,6\}$ the masses lie above the three differences $e_a-e_b$, but below the three sums. Ten such chambers are realised (Appendix~\ref{app:scans}). For $\su(4)$, the weights of $\Lambda^2$ come in three pairs, which we label
\begin{equation}
\label{eq:su4pairs}
1',2',3':\qquad \pm(e'_1+e'_2),\quad \pm(e'_1+e'_3),\quad \pm(e'_1+e'_4)\,,
\end{equation}
and a chamber of $\su(4)$ is labelled by the set of these pairs that the masses $M'_i$ dominate. The primes distinguish these labels from those of \eqref{eq:sp3pairs}. The two sets label the chambers of different frames, and Appendix~\ref{app:scans} relates them. The 6d frame is the \hyperref[par:6dregime]{regime} of \sref{sec:prepot6d}, here with $|m_i|$ below the short roots, as in \eqref{eq:so8triE6d}. By \eqref{eq:so8triroots}, the sums $e_a+e_b$ then exceed $|M_i|=|m_i|$ through $\Phi=2\phi_0-\tfrac49\tau$, the differences $e_1-e_2$ and $e_1-e_3$, KK modes of $\mathbf 7_{1/3}$, through $\tfrac\tau3$, and $e_2-e_3=\alpha_1$ as a short root. So no pair is dominated, and the 6d frame is the $\sp(3)$ chamber $\{\}$.

One relation holds in every chamber. By \eqref{eq:E5d}, $C^R_i=\tfrac12\partial_i\sum_{\alpha\in\Delta}|\alpha\cdot\phi|$ involves only the roots, so it does not depend on the masses, and every $C^R_i$ equals $2$. In Dynkin variables, the term $\tfrac12\epsilon_+^2C^R\cdot\phi$ is therefore $\epsilon_+^2$ times the sum of the variables. That is $\epsilon_+^2(P_1+P_2+P_3)=\epsilon_+^2(3e_1+2e_2+e_3)$ for $\sp(3)$, and $\epsilon_+^2(P'_1+P'_2+P'_3)$ for $\su(4)$. Equating the two under $P'=Ne+\dots$ requires the columns of $N$ to sum to $(3,2,1)$. With this constraint, the scan over $\kCS$ and the ten chambers of $\sp(3)$ finds a solution only for $\kCS=6$ and $\sp(3)$ in the chamber $\{4,5,6\}$ or in a chamber with more dominated pairs (Appendix~\ref{app:scans}). Up to charge conjugation and $\mathrm O(2)$, it is
\begin{equation}
\label{eq:so8trifmap5d}
\begin{aligned}
P'_1&=e_1+\tfrac14m_0-\tfrac12(M_1+M_2)\,,\qquad P'_2=e_1+e_2+\tfrac12m_0-(M_1+M_2)\,,\\
P'_3&=e_1+e_2+e_3+\tfrac34m_0-\tfrac32(M_1+M_2)\,,\qquad m_0'=\tfrac12m_0\,,\qquad M'_i=\tfrac12m_0-M_i\,.
\end{aligned}
\end{equation}
With it, the two prepotentials agree up to $\phi$-independent terms, and the large-mass chamber $\{4,5,6\}$ of $\sp(3)$ is the small-mass chamber $\{\}$ of $\su(4)$. The correspondence of all chambers is \eqref{eq:so8trichambers} in Appendix~\ref{app:scans}.

\paragraph{Flops.} When the mass $x$ of a BPS state goes from negative to positive on a wall, its term $-\tfrac1{12}|x|^3$ in \eqref{eq:E5d} changes the prepotential by
\begin{equation}
\label{eq:flopjump}
\Delta\big(\epsilon_1\epsilon_2\,\mathcal E\big)=-\tfrac16x^3+(\text{terms linear in the moduli})\,.
\end{equation}
A gauge-theory prepotential contains this jump only if the state is a perturbative weight. Between $\{\}$ and $\{4,5,6\}$, the flopping states are, for each mass,
\begin{equation}
\label{eq:so8triflops}
\mbox{\footnotesize$\begin{array}{l|l|l}
\sp(3)_0\ (\text{perturbative}) & \su(4)_6\ (k=-1) & \text{6d}\\ \hline
e_1-e_2-M_i & e'_1-e'_2+M'_i-m_0' & \tfrac\tau3-\phi_1-m_i,\ \text{KK mode of }\mathbf 7_{1/3}\\
e_1-e_3-M_i & e'_1-e'_3+M'_i-m_0' & \tfrac\tau3+\phi_1-\phi_2-m_i,\ \text{KK mode of }\mathbf 7_{1/3}\\
e_2-e_3-M_i & e'_2-e'_3+M'_i-m_0' & 2\phi_1-\phi_2-m_i,\ \text{zero mode of }\mathbf 7_0
\end{array}$}
\end{equation}
In $\su(4)_6$, they have instanton number $-1$, since $M_i=m_0'-M'_i$, so the $\su(4)$ prepotential holds only past the walls. In 6d the third state is a zero mode, whose flop the 6d prepotential follows. The first two are KK modes of momentum $\tfrac13$, whose signs the 6d frame takes as fixed (the second assumption of the \hyperref[par:6dregime]{regime} in \sref{sec:prepot6d}), so past their walls \eqref{eq:E6d} needs \eqref{eq:flopjump}. These four KK modes separate the tensor branch from the $\su(4)_6$ phase.

\paragraph{The $\su(4)_6$ frame in 6d variables.} Composing the map \eqref{eq:so8trimap} from the 6d frame to $\sp(3)_0$ with the map \eqref{eq:so8trifmap5d} from $\sp(3)_0$ to $\su(4)_6$, and writing $\hat\phi_0:=\phi_0-\tfrac12\sum_im_i$, we find
\begin{equation}
\label{eq:so8trisu4map}
\begin{gathered}
P'_1=\hat\phi_0+\tfrac{7\tau}{36}\,,\quad P'_2=2\hat\phi_0+\phi_1+\tfrac{\tau}{18}\,,\quad P'_3=3\hat\phi_0+\phi_2-\tfrac{\tau}{12}\,,\\
m_0'=\tfrac\tau6\,,\quad M'_i=\tfrac\tau6-m_i\,.
\end{gathered}
\end{equation}
With it, the $\su(4)_6$ prepotential equals \eqref{eq:so8triE6d} continued past the flops of \eqref{eq:so8triflops}. This requires $m_i>\tfrac\tau3-\phi_1$ and $m_i>\tfrac\tau3+\phi_1-\phi_2$, i.e.\ masses of the order of the KK scale. In \eqref{eq:so8trisu4map}, the masses enter the $P'_i$ only through $\hat\phi_0$. The positive roots of $\su(4)$ and the instantonic root are, in 6d variables,
\begin{equation}
\label{eq:so8trisu4roots}
\mbox{\footnotesize\setlength{\arraycolsep}{1.4pt}$\begin{array}{l|l|l|l}
\text{root of }\su(4) & \text{in }P'_i & \text{in 6d variables} & \text{6d state}\\ \hline
e'_1-e'_2\ \text{(simple)} & 2P'_1-P'_2 & \tfrac\tau3-\phi_1=\alpha_0 & \text{lowest state of $\mathbf 7_{1/3}$}\\
e'_2-e'_3\ \text{(simple)} & -P'_1+2P'_2-P'_3 & 2\phi_1-\phi_2=\alpha_1 & \text{zero mode, short root}\\
e'_3-e'_4\ \text{(simple)} & -P'_2+2P'_3 & 4\hat\phi_0-\phi_1+2\phi_2-\tfrac{2\tau}9 & \text{strings, }\lambda=2\\
e'_1-e'_3 & P'_1+P'_2-P'_3 & \tfrac\tau3+\phi_1-\phi_2 & \text{state of $\mathbf 7_{1/3}$}\\
e'_2-e'_4 & -P'_1+P'_2+P'_3 & 4\hat\phi_0+\phi_1+\phi_2-\tfrac{2\tau}9 & \text{strings, }\lambda=2\\
e'_1-e'_4 & P'_1+P'_3 & 4\hat\phi_0+\phi_2+\tfrac\tau9 & \text{strings, }\lambda=2\\ \hline
2m_0'-(e'_1+e'_2-2e'_3) & 2m_0'-3P'_2+2P'_3 & -3\phi_1+2\phi_2=\alpha_2 & \text{long root, $k=2$}
\end{array}$}
\end{equation}
The roots with $4\hat\phi_0=2\mathcal K\hat\phi_0$ are states of two strings. So the simple roots are again the affine root $\alpha_0$, the zero mode $\alpha_1$ and a string state, now of two strings. The instantonic root $\alpha_2$ now has instanton number $2$. With the marks $(1,2,1)$, $\delta=\alpha_0+2\alpha_1+\alpha_2=\tfrac\tau3$. Since $\alpha_0$ and $\alpha_1$ have $k=0$, $k(\delta)$ is the mark $1$ of $\alpha_2$ times $k(\alpha_2)=2$, as \eqref{eq:rrule} requires. So \eqref{eq:m0delta} gives $m_0'=\delta/2=\tfrac\tau6$, i.e.\ $r=6$. Every weight $e'_a+e'_b$ of $\Lambda^2$ contains $\pm2\hat\phi_0$, so every hypermultiplet state of this frame is a string or an anti-string. With $M'_i=\tfrac\tau6-m_i$,
\begin{equation}
\label{eq:so8trisu4hypers}
\mbox{\footnotesize\setlength{\arraycolsep}{1.4pt}$\begin{array}{l|l|l|l}
\text{state} & \text{in }P'_i & \text{in 6d variables} & \text{6d state}\\ \hline
e'_1+e'_2+M'_i & P'_2+M'_i & 2\hat\phi_0+\phi_1+\tfrac{2\tau}9-m_i & \text{string, }\lambda=1\\
e'_1+e'_3+M'_i & P'_1-P'_2+P'_3+M'_i & 2\hat\phi_0-\phi_1+\phi_2+\tfrac{2\tau}9-m_i & \text{string, }\lambda=1\\
e'_2+e'_3+M'_i & -P'_1+P'_3+M'_i & 2\hat\phi_0+\phi_2-\tfrac\tau9-m_i & \text{string, }\lambda=1\\
e'_1+e'_4+M'_i & P'_1-P'_3+M'_i & -2\hat\phi_0-\phi_2+\tfrac{4\tau}9-m_i & \text{string, }\lambda=-1\\
e'_2+e'_4+M'_i & -P'_1+P'_2-P'_3+M'_i & -2\hat\phi_0+\phi_1-\phi_2+\tfrac\tau9-m_i & \text{string, }\lambda=-1\\
e'_3+e'_4+M'_i & -P'_2+M'_i & -2\hat\phi_0-\phi_1+\tfrac\tau9-m_i & \text{string, }\lambda=-1
\end{array}$}
\end{equation}
Conversely, the eight non-string states of \eqref{eq:so8trihypers} have instanton number $1$ in this frame, e.g.\ $2\phi_1-\phi_2+m_i=(e'_2-e'_3)+m_0'-M'_i$, since $m_i=m_0'-M'_i$. The flopping states of \eqref{eq:so8triflops} are the negatives of three of them. No frame map \eqref{eq:framemapgen} to $\su(4)+2\Lambda^2$ with $u=0$ gives a smaller $r$, under the assumptions stated in Appendix~\ref{app:scans}.

\paragraph{The dictionary on the tensor-flux family.} In the $\su(4)_6$ frame, the shifted tensor flux is the flux $a'$ of $\hat\phi_0$, and $\nu'(a')=a'(1,2,3)$. With $B_{m_0'}=0$ and $B_{M'_i}=-\tfrac12$, the $\su(4)_6$ prepotential gives $V=6m_0'\,a'^2-a'\big[6\epsilon_+-3(M'_1+M'_2)\big]$, with $a'\in\BZ$ or $a'\in\BZ+\tfrac12$. The coset $a'\in\BZ+\tfrac12$ is the 't~Hooft sector, since $\tfrac12(1,2,3)\in\omega_2+\BZ^3$, with leading power $\fq^{1/4}$. So \eqref{eq:dictionary} holds with $k$ on the tensor-flux family, $q(1)=k(\nu')/r=\tfrac32/6=\tfrac14$ at $\nu'=\tfrac12(1,2,3)$. It fails for $k(\ngen)=\tfrac12$, at $\ngen=(\tfrac12,1,\tfrac12)$ off the tensor-flux family. So $r$ is a property of the frame, through its $\U(1)_I$.

\paragraph{Which sector is the 't~Hooft sector.} By the \hyperref[it:fluxmap]{flux rule} of \sref{sec:framemap}, \eqref{eq:so8trifmap5d} maps the fluxes as
\begin{equation}
\label{eq:so8trifluxmap}
\nu'=N\nu+\big(\tfrac14,\tfrac12,\tfrac34\big)\Bmz-\tfrac12\big(B_{M_1}+B_{M_2}\big)(1,2,3)\,,\qquad N(1,1,1)^T=(1,2,3)^T\,.
\end{equation}
With $\Bmz=0$, $B_{M_i}=\tfrac12$ and $\nu=(n_0,n_0,n_0)$ in the basis $e_i$, $\nu'=(n_0-\tfrac12)(1,2,3)$ in $P'_i$. So $n_0\in\BZ$ is the trivial sector of $\sp(3)_0$ and the 't~Hooft sector of $\su(4)_6$, and vice versa. The offset $-\tfrac12(1,2,3)\in\omega_2+\BZ^3$ exchanges the labels, as for $\su(4)_0$ against $\su(4)_2$. It comes from $B_{M_i}=\tfrac12$, the spin$^c$ offset of the blow-up \cite[\S3.7, (3.43)]{Harding:2026gzj}, forced by the zero weights of the $\mathbf{14}$. Since the $\BZ_2$ acts trivially on the matter, the offset can be placed in the flavour or in the gauge flux, and the labels ``trivial'' and ``'t~Hooft'' depend on that placement, which the frame map does not preserve. As in \sref{sec:pfanomalies}, the hypermultiplet fermions couple to a bundle with $w_2=w_2(\widehat{\BC}{}^2)$, which \eqref{eq:so8trifluxmap} moves partly from the $\sp(3)_0$ flavour bundle into the $\su(4)$ gauge bundle. The resulting conjecture, a shift of the $\BZ_2^{(1)}$ background by $w_2(\widehat{\BC}{}^2)$, is stated in \sref{sec:conclusions} below. Open checks are whether the geometry of \cite[(4.92)]{Bhardwaj:2020gyu} is in the phase $\{4,5,6\}$ and whether the one-instanton spectrum of $\su(4)_6+2\Lambda^2$ has the flopping states.

\subsection{\texorpdfstring{$\so(8)$}{so(8)} with the \texorpdfstring{$\BZ_2$}{Z2} twist}
\label{sec:exso8Z2}

The 6d minimal SCFT with gauge algebra $\so(8)$ on a $-4$ curve, compactified on a circle with the $\BZ_2$ outer-automorphism twist, gives rise to a 5d KK theory with twisted affine algebra $D_4^{(2)}$ and invariant subalgebra $\so(7)$. Although this theory lacks a 5d gauge-theory description, it can be engineered geometrically in terms of four compact surfaces $S_{0,3} = \mathbb F_6$, $S_{1,2} = \mathbb F_1$, glued along $(e \subset \mathbb F_6, 2 h \subset \mathbb F_1)$, $(e \subset \mathbb F_1, e \subset \mathbb F_1)$, $(2 h \subset \mathbb F_1, e \subset \mathbb F_6)$, see \cite[(5.152) with $k=4$]{Bhardwaj:2019fzv}. In \cite[\S6.2]{Kim:2020hhh}, the effective prepotential associated with the 6d frame is reported, from which the solution of the blow-up equations, in terms of the BPS spectrum, is provided. Here, we derive the effective prepotential directly from the 5d geometric setting, and we match its Coulomb branch moduli and mass parameter with the 6d variables. 

\paragraph{6d frame.} Here, we have $\mathcal K=4$, $D=\BZ_4$, $q(\lambda) = \frac{\lambda^2}{8}$, $(\mathfrak a,\mathfrak b,\mathfrak c)=(-\frac{1}{2},1,\frac{3}{2})$, $\hat{\mathfrak g}=D_4^{(2)}$, and invariant subalgebra $\so(7)$. As for $\SU(4)_8$, the twist does not enter \eqref{eq:Vfamily}, since only the Green--Schwarz couplings do. So $V\big|_a=2\tau a^2-4a\epsilon_+$, as in \eqref{eq:su48Lambda}.\footnote{The tensor fluxes differ from those of $\SU(4)_8$. There are no masses, so $a=n_0$ by \eqref{eq:adef}. The Freed--Witten condition of the paragraph \hyperref[par:fluxquantso8Z2]{Flux quantisation} below gives $a\in\tfrac18+B+\BZ$, with $B\in\{0,\tfrac14,\tfrac12,\tfrac34\}$ the 't~Hooft class, as stated below \eqref{eq:so8Z2cosets}. So $\mathcal Ka$ is half-integral, and not integral as in \sref{sec:pfblowup}. Up to a constant phase, $\Lambda_B=\vartheta\big[\genfrac{}{}{0pt}{1}{B+1/8}{0}\big](4\tau,-4\epsilon_+)$.}

The adjoint of $\so(8)$ decomposes as $\mathbf{28}\to\mathbf{21}_0\oplus\mathbf 7_{1/2}$ \cite[(2.42)]{Kim:2020hhh}. In Dynkin components, the simple roots are $\alpha_1 = (2, -1, 0)$ and $\alpha_2 = (-1,2,-2)$, both long, and $\alpha_3 = (0,-1,2)$, short, with $\tfrac12\tilde K\phi\phi=\mathsf q:=\phi_1^2-\phi_1\phi_2+\phi_2^2-2 \phi_2 \phi_3+2 \phi_3^2$. The roots of $\so(7)$ give $\sum_{\mathbf{21}}(\mathsf w\cdot\phi)^2=20\mathsf q$, and the $\mathbf{7}$ (six short roots and a zero weight) gives $4 \mathsf q$. With $B_2(0)=\frac{1}{6}$ and $B_2(\tfrac12)=-\tfrac1{12}$, the towers of \eqref{eq:F1loop6d} yield $\frac{10}{3} \mathsf q - \frac{1}{3} \mathsf q = 3 \mathsf q$, times the factor $-\frac{\tau}{4}$. Hence, $\Phi=4\phi_0-\tfrac34\tau$, and, in the fundamental chamber of $\so(7)$, the prepotential reads \cite[(6.12)]{Kim:2020hhh}
\begin{equation}
\label{eq:so8Z2E6d}
\begin{aligned}
\epsilon_1\epsilon_2\,\mathcal E_{\rm 6d}=2\tau\phi_0^2{}+{}& \Phi\,\mathsf q+\tfrac43\phi_1^3-\tfrac12\phi_1^2\phi_2-\tfrac12\phi_1\phi_2^2+\tfrac43\phi_2^3-3\phi_2^2\phi_3+2\phi_2\phi_3^2+\tfrac43\phi_3^3\\
{}+{}& (6\phi_0+\phi_1+\phi_2+\phi_3)\,\epsilon_+^2-\big(\tfrac12\phi_0+\tfrac{\phi_1+\phi_2+\phi_3}{12}\big)(\epsilon_1^2+\epsilon_2^2)\,.
\end{aligned}
\end{equation}

\paragraph{5d frame.} From $h = e + n f$ and $K = -2 e - \left(n+2\right) f$ for $\mathbb F_n$, together with $e \cdot e = - h \cdot h = -n$, $f \cdot f = e \cdot h = 0$, $e \cdot f = f \cdot h = 1$, the triple intersections for this geometric description read
\begin{equation}
\label{eq:so8Z2triple}
S_i^3=8\,,\quad S_0^2S_1 = S_2 S_3^2=4\,,\quad S_0S_1^2 = S_2^2 S_3=-6\,,\quad S_1^2 S_2 = S_1 S_2^2=-1\,,
\end{equation}
with $i = 0,\ldots,3$, where all non-adjacent triple intersections vanish. Here and below, $e_i$, $f_i$, $h_i$ denote the curves $e$, $f$, $h$ of $S_i$. We can take the K\"ahler form as $J = \sum_{i = 0}^3 P_i S_i - m_0 N$, with K\"ahler parameters $P_0, \ldots, P_3$, mass parameter $-m_0$,\footnote{\label{fn:volfso8Z2}The sign is fixed such that $\mathrm{vol}\left(\sum_{i=0}^3 f_i\right) = m_0 > 0$.} and $N$ non-compact, meeting $S_0$ in a curve $C_N \subset \mathbb F_6$ with $C_N\cdot f=1$ that avoids the gluing curve $e$.\footnote{Other choices are related by $N \rightarrow N + \sum_i \varsigma_i S_i$, i.e.\ $P_i \rightarrow P_i + \varsigma_i m_0$, a change of frame, as explored in the paragraph \hyperref[par:Z2symmframesso8Z2]{$\BZ_2$ symmetric frames} below.} The choice $C_N = h$ works, since $e \cdot h = 0$, $f \cdot h = 1$. This implies $S_0^2 N = -8$, $S_0 N^2 = 6$, and the other triple intersections involving $N$ vanish. The term $N^3$ is a background constant, which we drop. The volumes of the 2-cycles $\mathrm{vol}(C) = -J \cdot C$, with $S_i \cdot C = K \cdot C\rvert_{i}$ for $C \subset S_i$ and $S_j \cdot C = C_{ij} \cdot C$ for $j$ adjacent to $i$, with $C_{ij}$ the gluing curve $S_i\cap S_j$ in $S_i$, read
\begin{equation}
\label{eq:so8Z2vol}
\begin{aligned}
\mathrm{vol}(f_0) &= 2 P_0 - P_1 + m_0\,, \quad \mathrm{vol}(f_1) = -2 P_0 + 2 P_1 - P_2\,,\\
\mathrm{vol}(f_2) &= -P_1 + 2 P_2 - 2 P_3\,, \quad \mathrm{vol}(f_3) = -P_2 + 2 P_3\,,
\end{aligned}
\end{equation}
and, for a smooth Calabi--Yau threefold $X$, the mixed gauge--gravitational and gauge--$\SU(2)_R$ Chern--Simons levels of \eqref{eq:E5d} are $C_i^G=c_2(X)\cdot S_i=-4$ and $C_i^R=2\chi(\mathcal O_{S_i})=2$ \cite[(A.4)]{Harding:2026gzj}. So $c_\epsilon=\epsilon_+^2-\tfrac1{12}(\epsilon_1^2+\epsilon_2^2)$, and the effective prepotential is
\begin{equation}
\label{eq:effectiveFFSO8Z25d}
\epsilon_1 \epsilon_2 \mathcal{E}_{\mathrm{5d}} = \mathcal{F} + c_{\epsilon} \sum_{i=0}^3 P_i\,,
\end{equation}
where the prepotential can be directly derived as $\frac{1}{6} J^3$, yielding
\begin{equation}
\label{eq:FFSO8Z25d}
\begin{aligned}
\mathcal{F} ={}&  4 m_0 P_0^2 + 3 m_0^2 P_0 + \mathcal{F}_{\text{cubic}}(P_i)\,, \\ \mathcal{F}_{\text{cubic}}(P_i)={}& \tfrac43 \sum_{i=0}^3 P_i^3 + 2 \big(P_0^2 P_1 + P_2 P_3^2\big) - 3 \big(P_0 P_1^2 + P_2^2 P_3\big) - \tfrac12 \big(P_1^2 P_2 + P_1 P_2^2\big)\,.
\end{aligned}
\end{equation}

\paragraph{$\BZ_2$ symmetric frames.}\phantomsection\label{par:Z2symmframesso8Z2} The geometry has a $\BZ_2$ symmetry $\xi$, which exchanges $S_0 \leftrightarrow S_3$, $S_1 \leftrightarrow S_2$ and the corresponding gluing curves. It acts as $P_0 \leftrightarrow P_3$, $P_1 \leftrightarrow P_2$, and the cubic part $\mathcal{F}_{\text{cubic}}$ of \eqref{eq:FFSO8Z25d} is invariant. The divisor $N$ meets only $S_0$, so the terms with $m_0$ in \eqref{eq:so8Z2vol} and \eqref{eq:FFSO8Z25d} are not invariant. The symmetry is recovered by a shift of the K\"ahler parameters, $P_i = P'_i + c_i m_0$, with $c = \left(-\frac{3}{8},-\frac{1}{4},\frac{1}{4},\frac{3}{8}\right)$. In the new variables, the prepotential is $\mathcal{F}' =  2 m_0 \left({P'_0}^2 + {P'_3}^2\right) + \tfrac34 m_0^2 \left(P'_0+P'_3\right) + \mathcal{F}_{\text{cubic}}(P'_i)$, up to terms independent of $P'_i$. The volumes are $\mathrm{vol}(f_0) = 2 P'_0 - P'_1 + \frac{m_0}{2}$ and $\mathrm{vol}(f_3) = -P'_2 + 2 P'_3 + \frac{m_0}{2}$, while $\mathrm{vol}(f_1)$ and $\mathrm{vol}(f_2)$ keep their form.

By further shifting $P''_i = P'_i + \tilde{c}_i m_0$, with $\tilde{c}=\left(\frac{3}{16},\frac{3}{8},\frac{3}{8},\frac{3}{16}\right)$, which is equivalent to $P''_i = P_i + m_0 \left(\frac{9}{16},\tfrac58,\tfrac18,-\frac{3}{16}\right)$, we can get rid of the quadratic terms in $m_0$ in the prepotential, and land on a frame whose prepotential is linear in $m_0$, up to constant terms independent of $P''_i$, and manifestly symmetric under $P''_0 \leftrightarrow P''_3$, $P''_1 \leftrightarrow P''_2$:
\begin{equation}
\label{eq:FFSO8Z25dsymmlin}
\begin{aligned}
\mathcal{F}'' ={}& m_0 \Big[\tfrac12\big({P''_0}^2 + {P''_3}^2\big)+\tfrac32\big(P''_0 P''_1 + P''_2 P''_3\big)-\tfrac34\big({P''_1}^2 + {P''_2}^2\big)+\tfrac34 P''_1 P''_2\Big]\\
& + \mathcal{F}_{\text{cubic}}(P''_i)\,.
\end{aligned}
\end{equation}
This follows from a rational class $N''$, with ${N''}^2\rvert_{S_i} = 0$ on every surface, and $N''= N' + \sum_{i=0}^3 \tilde{c}_i S_i$,  where $N'$ is defined as $N'= N - \sum_{i=0}^3 c_i S_i$, and the vectors $c$, $\tilde{c}$ are defined as above. Hence, $N'' \cdot f_0 = N'' \cdot f_3 = \tfrac12$, $N'' \cdot f_1 = N'' \cdot f_2 = 0$, which is consistent with the $\frac{m_0}{2}$ term in $\mathrm{vol}(f_0)$ and $\mathrm{vol}(f_3)$, and no mass terms in $\mathrm{vol}(f_1)$ and $\mathrm{vol}(f_2)$, since every fibre volume is unchanged.\footnote{\label{fn:so8Z2invvec}Observe that the vector $\tilde{c}$ can be expressed as $\tfrac3{16}\mathsf a^\vee$, with $\mathsf a^\vee = (1,2,2,1)$, the comarks of $D_4^{(2)}$, being the left null vector of the matrix $S_i\cdot f_j$. Therefore, the linear combination $\sum_{i=0}^3 \mathsf a^\vee_i S_i$ pairs trivially with every fibre.} Explicitly, $N''$ restricts to $\tfrac12 e + \tfrac32 f$ on $S_0$ and $S_3$, and to $-\tfrac34 f$ on $S_1$ and $S_2$. Its non-zero intersection numbers are $S_0^2 N'' = S_3^2 N'' = -1$, $S_1^2 N'' = S_2^2 N'' = \tfrac32$, $S_0 S_1 N'' = S_2 S_3 N'' = -\tfrac32$, $S_1 S_2 N'' = -\tfrac34$, thus reproducing the prepotential \eqref{eq:FFSO8Z25dsymmlin}.

\paragraph{The map.} Step 1 of \sref{sec:recipe} is the affine-node rule of \eqref{eq:affinenode}, with the comarks $\mathsf a^\vee=(1,2,2,1)$ of $D_4^{(2)}$, so $P_0 = \phi_0$ and $P_{1,2,3} = \phi_{1,2,3} + \mathsf a^\vee_{1,2,3} \phi_0$, up to multiples of $\tau$. Step 2 fixes the shifts and $r = 2$, i.e.\ $m_0 = \frac{\tau}{2}$. In the frame \eqref{eq:FFSO8Z25d}, the 6d variables are mapped to the geometric ones as
\begin{equation}
\label{eq:so8Z2map6dto5d}
P_0 = \phi_0 -\frac{3 \tau}{16}\,, \quad P_1 = 2 \phi_0 +\phi_1 -\frac{3 \tau}{8}\,, \quad P_2 = 2 \phi_0 +\phi_2 -\frac{3 \tau}{8}\,, \quad P_3 = \phi_0 +\phi_3 -\frac{3 \tau}{16} \,.
\end{equation}
The inverse, from the geometry to 6d, reads $\phi_0 = P_0 +\frac{3 \tau}{16}$, $\phi_1 = P_1 - 2 P_0$, $\phi_2 = P_2 - 2 P_0$, $\phi_3 = P_3 - P_0$, in agreement with \cite[below (6.12)]{Kim:2020hhh}. With the map \eqref{eq:so8Z2map6dto5d}, the effective prepotential \eqref{eq:effectiveFFSO8Z25d} matches \eqref{eq:so8Z2E6d} up to $\phi_i$-independent terms.

\paragraph{Reading of the map.} Let $\Phi=4\phi_0-\tfrac34\tau$ be the coefficient of $\mathsf q$ in \eqref{eq:so8Z2E6d}. By \eqref{eq:so8Z2map6dto5d}, the compact curves, in 6d variables, are\footnote{Note that, since $S_1$ and $S_2$ are glued along the curve $e \subset \mathbb F_1$, $e_1 = e_2$ are the same curve, hence we list only $e_1$ as a representative in \eqref{eq:so8Z2curves}. The same holds for the other gluings: the curve $e \subset \mathbb F_6$ in $S_0$ is $2 h \subset \mathbb F_1$ in $S_1$, and $2h \subset \mathbb F_1$ in $S_2$ is the curve $e \subset \mathbb F_6$ in $S_3$. On the other hand, $h_1$ and $h_2$ are different curves.}
\begin{equation}
\label{eq:so8Z2curves}
\mbox{\footnotesize\setlength{\arraycolsep}{4pt}$\begin{array}{l|l|l|l}
\text{curve} & \text{in } P_i & \text{in 6d variables} & \text{6d state}\\ \hline
f_0 & 2 P_0 - P_1 + m_0 & \tfrac\tau2-\phi_1=\alpha_0 & \text{lowest state of $\mathbf 7_{1/2}$}\\
f_1 & -2 P_0 + 2 P_1 - P_2 & 2\phi_1-\phi_2=\alpha_1 & \text{zero mode, long root}\\
f_2 & -P_1 + 2 P_2 - 2 P_3 & -\phi_1 + 2 \phi_2 - 2 \phi_3=\alpha_2 & \text{zero mode, long root}\\
f_3 & -P_2 + 2 P_3 & -\phi_2 + 2 \phi_3=\alpha_3 & \text{zero mode, short root}\\
e_1 & P_1 + P_2 & \Phi + \phi_1 + \phi_2  & \text{string}
\end{array}$}
\end{equation}
The curve with $\Phi$ contains $\mathcal{K} \phi_0$, so it is a string. In \eqref{eq:so8Z2curves}, we only list the basis $f_0, \ldots, f_3, e_1$, from which the other curves follow.\footnote{Over the integers, $H_2(\mathcal S)$, with $\mathcal S = S_0\cup S_1\cup S_2\cup S_3$ the union of the four compact surfaces of our geometric frame, has rank 5. Each Hirzebruch surface $\mathbb F_n$ has $H_2 = \BZ e \oplus \BZ f$, so rank 2, and four surfaces give rank 8. Each of the three gluing curves identifies one class in one surface with one class in its neighbour: $e \subset \mathbb F_6$ with $2 h \subset \mathbb F_1$, then $e \subset \mathbb F_1$ with $e \subset \mathbb F_1$, and $2h \subset \mathbb F_1$ with $e \subset \mathbb F_6$, providing three independent relations. Hence, the rank is $8-3=5$, and a basis is $(f_0, f_1, f_2, f_3, e_1)$. The other curves follow, as $h_1 = e_1 + f_1$, $h_2 = e_1 + f_2$, $e_0 = 2 h_1$, $e_3 = 2 h_2$. In 6d variables, $h_1$ and $h_2$ yield $\Phi + 3 \phi_1$ and $\Phi + 3 \phi_2 -2 \phi_3$, which are states of one string. Consequently, $e_0$ and $e_3$ are states of two strings, as they appear with $ 2 \Phi$, containing $2 \mathcal{K} \phi_0$.} All four fibres $f_0, \ldots, f_3$ are the affine root and zero modes, and none is a string. With the marks $(1,1,1,1)$ of $D^{(2)}_4$, $\delta = \alpha_0+\alpha_1+\alpha_2+\alpha_3=\frac{\tau}{2}$, the $\phi_i$-dependence cancelling, and $\sum_{i=0}^3 \mathrm{vol}(f_i) = m_0$, in agreement with footnote~\ref{fn:volfso8Z2}. Since $N \cdot \sum_{i=0}^3 f_i=1$, $k(\delta)=1$, so \eqref{eq:m0delta} gives $m_0 = \delta$ and $r = r_{\rm tw}\,k(\delta) = 2$. 

The rule \eqref{eq:rrule} uses the frame of \eqref{eq:FFSO8Z25d}, in which only $\alpha_0$ contains $m_0$. In the \hyperref[par:Z2symmframesso8Z2]{$\BZ_2$ symmetric frames}, $m_0$ is split equally between $f_0$ and $f_3$, so $\alpha_0$ and $\alpha_3$ each have $k=\tfrac12$. This is possible since $N'$ and $N''$ are rational classes. Only $k(\delta)=1$ is the same in every frame, since $\sum_i\mathrm{vol}(f_i)=m_0$ contains no $P_i$.

\paragraph{The $\BZ_2$ symmetry in the 6d frame.} The symmetry $\xi$: $P'_0 \leftrightarrow P'_3$, $P'_1 \leftrightarrow P'_2$ of the symmetric frame, which is also a symmetry of the $P''$ frame, since the shift is along the invariant vector $(1,2,2,1)$, see footnote~\ref{fn:so8Z2invvec}, can be expressed in 6d variables as the affine map\footnote{This map can be derived as follows. We first start with the inverse map from the geometric frame to the 6d frame, reported below \eqref{eq:so8Z2map6dto5d}. Then, we express everything in terms of the $P'_i$ variables by applying the map $P_i=P'_i+c_i \frac{\tau}{2}$, and we perform the swaps $P'_i \rightarrow P'_{3-i}$. Finally, we sequentially apply the map from the $P'_i$ variables to the $P_i$ variables first, followed by the map \eqref{eq:so8Z2map6dto5d}, which gives the affine map \eqref{eq:Z2symmso8Z26d} of the 6d frame to itself.}
\begin{equation}
\label{eq:Z2symmso8Z26d}
\phi_0 \rightarrow \phi_0 + \phi_3 -\frac{3 \tau}{8}\,, \quad \phi_1 \rightarrow \phi_2 -2\phi_3 +\frac{\tau}{2}\,, \quad \phi_2 \rightarrow \phi_1 -2\phi_3 +\tau\,, \quad \phi_3 \rightarrow -\phi_3 +\frac{3 \tau}{4} \,.
\end{equation}
This map correctly leaves \eqref{eq:so8Z2E6d} invariant, squares to the identity,\footnote{For instance, $\phi_0 \overset{\text{\eqref{eq:Z2symmso8Z26d}}}{\rightarrow} \phi_0 + \phi_3 -\frac{3 \tau}{8} \overset{\text{\eqref{eq:Z2symmso8Z26d}}}{\rightarrow} \left(\phi_0 + \phi_3 -\frac{3 \tau}{8}\right) + \left(-\phi_3 +\frac{3 \tau}{4}\right) -\frac{3 \tau}{8} = \phi_0$.} and exchanges $\alpha_0 \leftrightarrow \alpha_3$, $\alpha_1 \leftrightarrow \alpha_2$, i.e.\ it is a diagram flip exchanging the two end nodes of the $D^{(2)}_4$ Dynkin diagram. The string $e_1$, realised as $\Phi+\phi_1+\phi_2$, is left invariant as well. It also preserves the linear combination $10 \phi_0 + 2 \phi_1 + 2 \phi_2 + \phi_3 - \frac{15 \tau}{8}$, which is $\sum_{i=0}^3 \mathsf a^\vee_i P'_i$ expressed in terms of the 6d variables $\phi_i$, since $\mathsf a^\vee$ is symmetric under the $\BZ_2$ symmetry $\xi$.

\paragraph{The blow-up function in the 5d frame.} With the fluxes $n_i$ of the $P''_i$ through $E$ and the flux $\Bmz$ of $m_0$, the finite difference \eqref{eq:Vdef} of the 5d effective prepotential in the $P''$ frame yields the blow-up function $q^{k(n)} (p_1 p_2)^{P(n)} \prod_{i=0}^3 z_i^{Z_i(n)}$, with $z_i = e^{P''_i}$. At $\Bmz=0$, which follows from $B_\tau=0$ by the \hyperref[it:fluxmap]{flux rule} of \sref{sec:framemap}, the exponents of the blow-up function read
\begin{equation}
\label{eq:so8Z2blowupfunction}
\begin{aligned}
k(n) &= \tfrac14 \left[2 n_0^2 + 2 n_3^2 + 6 \left(n_0 n_1 + n_2 n_3\right) - 3 \left(n_1^2 + n_2^2\right) + 3 n_1 n_2\right]\,,\\
P(n) &= \mathcal{F}_{\text{cubic}}(n) - \tfrac13 \left(n_0 + n_1 + n_2 + n_3\right) \,, \\
Z_0(n) &= -(2 n_0 + 3 n_1) (2 n_0 - n_1)\,, \\
Z_1(n) &= -\tfrac12 (2 n_0 - 2 n_1 + n_2) (2 n_0 - 4 n_1 - n_2)\,,\\
Z_2(n) &= -\tfrac12 (n_1 - 2 n_2 + 2 n_3) (-n_1 - 4 n_2 + 2 n_3)\,, \\
Z_3(n) &= -(3 n_2 + 2 n_3) (-n_2 + 2 n_3)\,.
\end{aligned}
\end{equation}
In the $P$ frame of \eqref{eq:FFSO8Z25d}, $P(n)$ and $Z_i(n)$ are the same, and $k(n)=4n_0^2$. The value $k(n)$ of \eqref{eq:so8Z2blowupfunction}, in the $P''$ frame, is $4n_0^2+\sum_i(\tilde c_i-c_i)Z_i(n)$, with $\tilde c-c=\left(\frac9{16},\frac58,\frac18,-\frac3{16}\right)$.

In the 6d frame, with $a=n_0$ and $b_i=n_i-\mathsf a^\vee_in_0$, $i=1,2,3$, the function $V$ of \eqref{eq:Vdef}, computed from \eqref{eq:so8Z2E6d}, equals $V(n)$ exactly, under the inverse of \eqref{eq:so8Z2map6dto5d} and the shift from the $P$ frame to the $P''$ frame. The exponents do not match one by one. The map adds $\mathsf a^\vee_i\phi_0$ to $P_i$ and shifts every $P''_i$ by a multiple of $\tau$, so
\begin{equation}
\label{eq:so8Z2Z6d}
\sum_{i=0}^3Z_iP''_i=\sum_{i=0}^3Z^{\rm 6d}_i\phi_i+\tfrac{\tau}{32}\left(3Z_0-2Z_1-10Z_2-9Z_3\right)\,,
\end{equation}
with $Z^{\rm 6d}_0=Z_0+2Z_1+2Z_2+Z_3$ and $Z^{\rm 6d}_{1,2,3}=Z_{1,2,3}$. With $q=\fq^{1/2}$ and $k(n)$ of \eqref{eq:so8Z2blowupfunction}, the power of $\fq=e^{-\tau}$ is $\tfrac12k(n)-\tfrac1{32}\left(3Z_0-2Z_1-10Z_2-9Z_3\right)=2a^2-\tfrac34\,\mathsf q(b_1,b_2,b_3)$.

\paragraph{Flux quantisation.}\phantomsection\label{par:fluxquantso8Z2} The quantisation of the 5d fluxes is fixed by the geometry. For every compact curve $C$, the fluxes must obey the Freed--Witten condition $\langle n,C\rangle\in\BZ+\tfrac12C^2$. Here, $\langle n,C\rangle=-\sum_in_i\,S_i\cdot C$ is the flux through $E$ seen by an M2-brane on $C$ at $\Bmz=0$, and $C^2$ is the self-intersection of $C$ in a compact surface containing it.\footnote{The parity of $C^2$ does not depend on the surface: a rational curve $C=S_i\cap S_j$ has $C^2_{S_i}+C^2_{S_j}=-2$.} The fibres have $f^2=0$, while $e_1$ and $h_{1,2}$ are odd, with $e^2=-1$ and $h^2=1$ in $\mathbb F_1$.
This condition is equivalent to the integrality of the exponents $Z_i$ of \eqref{eq:so8Z2blowupfunction} on the whole coset, i.e.\ $Z_i(n+\mu)\in\BZ$ for every $\mu\in\BZ^4$. Integrality at a single flux is not enough: $n=(0,0,0,0)$ and $n=\tfrac14\mathsf a^\vee$ have integer $Z_i$, but they violate the Freed--Witten condition, and some of their integer shifts have non-integer $Z_i$. Both conditions give exactly eight cosets mod $\BZ^4$:
\begin{equation}
\label{eq:solfluxesso8Z25d}
n = \left(2j + 1\right) \frac{\left(1,2,2,1\right)}{8} + \frac{l}{2} \left(0,0,0,1\right)\,, \quad j=0,\ldots,3\,, \quad l=0,1\,.
\end{equation}
The general solution of the integrality condition can be derived as follows. Write $Z_i(n)=-\tfrac12 \sum_{j,k} c_{ijk} n_j n_k$, where the coefficients $c_{ijk}$ are integers from \eqref{eq:so8Z2blowupfunction}. The condition $Z_i(n+\mu) \in \BZ$ for all $\mu \in \BZ^4$ splits into two conditions, 1) $Z_i(n) \in \BZ$ ($\mu =0$), and 2) $\sum_{j,k} c_{ijk} n_j \mu_k + \tfrac12 \sum_{j,k} c_{ijk} \mu_j \mu_k \in \BZ$ for all $\mu$. Observe that, for $\mu=\hat e_k$, the $k$-th unit vector, condition 2) reduces to 2$'$) $\sum_{j} c_{ijk} n_j+ \tfrac12 c_{ikk} \in \BZ$ for all $i,k$. Conversely, if 2$'$) holds, we can define $\sum_{j} c_{ijk} n_j = \mathsf z_{ik}-\tfrac12 c_{ikk}$ with $\mathsf z_{ik} \in \BZ$. Then, the left-hand side of 2) equals $\sum_{k} \mathsf z_{ik} \mu_k + \tfrac12 \sum_k c_{ikk} (\mu_k^2-\mu_k) + \sum_{j<k} c_{ijk} \mu_j \mu_k$, in which every term is an integer ($\mu_k^2 - \mu_k$ is even and $c_{ijk} \in \BZ$). So 2$'$) implies 2) for every $\mu \in \BZ^4$, and the two conditions are equivalent. Given that 2$'$) is linear in $n$ with 16 equations, whereas 2) is a condition over an infinite set of $\mu$, the integrality condition can be simplified into solving 1) + 2$'$). The linear system 2$'$) can be solved exactly with a Smith decomposition of the $16 \times 4$ matrix $A_{(ik),j}=c_{ijk}$. Its invariant factors are $(1,1,2,32)$, so 2$'$) has $1\cdot1\cdot2\cdot32=64$ solutions mod $\BZ^4$. Condition 1) keeps eight of them.\footnote{The flux $n=(0,0,0,0)$ solves 1) but not 2$'$).} The solution is found to be given by \eqref{eq:solfluxesso8Z25d}, where the vector $(0,0,0,1)$ associated with $S_3$ can be equivalently replaced by $(1,0,0,0)$ corresponding to $S_0$. A family such as $\left(2j + 1\right) \frac{\left(1,2,2,1\right)}{8} + \frac{l}{2} \left(1,0,0,1\right)$ is also a solution of conditions 1) + 2$'$), and it is symmetric under $\xi$: $S_0 \leftrightarrow S_3$, $S_1 \leftrightarrow S_2$. However, $\tfrac12(1,0,0,1)=\tfrac12\mathsf a^\vee-(0,1,1,0)\equiv\tfrac12\mathsf a^\vee$ mod $\BZ^4$, so this family equals the $l=0$ family of \eqref{eq:solfluxesso8Z25d} with $j \rightarrow j+2$, and gives only four of the eight cosets.

Let us list the eight cosets explicitly:
\begin{equation}
\vcenter{\hbox{%
\begin{tabular}{ccc}
\toprule
coset & $(n_0, n_1, n_2, n_3)$ & $(Z_0, Z_1, Z_2, Z_3)$ \\
\midrule
$j$, $l=0$ & $(2 j + 1) \bigl(\tfrac{1}{8}, \tfrac{1}{4}, \tfrac{1}{4}, \tfrac{1}{8}\bigr)$ & $(0,0,0,0)$ \\
$j=0$, $l=1$ & $\bigl(\tfrac{1}{8}, \tfrac{1}{4}, \tfrac{1}{4}, \tfrac{5}{8}\bigr)$ & $(0,0,0,-2)$ \\
$j=1$, $l=1$ & $\bigl(\tfrac{3}{8}, \tfrac{3}{4}, \tfrac{3}{4}, \tfrac{7}{8}\bigr)$ & $(0,0,1,-4)$ \\
$j=2$, $l=1$ & $\bigl(\tfrac{5}{8}, \tfrac{1}{4}, \tfrac{1}{4}, \tfrac{1}{8}\bigr)$ & $(-2,0,0,0)$ \\
$j=3$, $l=1$ & $\bigl(\tfrac{7}{8}, \tfrac{3}{4}, \tfrac{3}{4}, \tfrac{3}{8}\bigr)$ & $(-4,1,0,0)$ \\
\bottomrule
\end{tabular}}}
\label{eq:so8Z2cosets}
\end{equation}
For $l=0$, the cosets are $a=n_0\in\tfrac18+\tfrac14\BZ$ and $b_{1,2,3}\in\BZ$ in 6d variables. Each contains the member $\nu(a)=a\,(1,2,2,1)$ of the tensor-flux family, with $Z_i(\nu(a)) = 0$, $P(\nu(a)) = -2 a$ and $k(\nu(a)) = 4 a^2$. The classes $a\in\BZ+\{-\tfrac18,\tfrac18,\tfrac38\}$ are the flux sets of \cite[(6.13)]{Kim:2020hhh}, and $a\in\BZ+\tfrac58$ is a fourth one. For $l=1$, the flux has $b_3 \in \BZ + \tfrac12$, a half-integral coweight flux of $\Spin(7)$, the vector class, off the tensor-flux family. The offset $\tfrac18$ of $a$ comes from the spin$^c$ shift $n_\ast=\nu(\tfrac18)=\tfrac18(1,2,2,1)$, called $\delta$ in \cite[(3.43)]{Harding:2026gzj}. It is forced by the odd curves, and we obtain it from \eqref{eq:deltashift} below. The 't~Hooft class is the bare flux $a-\tfrac18\in\tfrac14\BZ$ mod $1$.

\paragraph{The $\BZ_2$ symmetry $\xi$ preserves the set of cosets, not each coset.} The invariance of $\mathcal{F}''$ in \eqref{eq:FFSO8Z25dsymmlin} in the $P''$ frame implies that the blow-up function $e^{-V(\xi n)}$ equals $e^{-V(n)}$ with $z_0 \leftrightarrow z_3$, $z_1 \leftrightarrow z_2$. Indeed, $Z_i (\xi n) = Z_{3-i} (n)$ for $i=0,\ldots,3$,\footnote{For instance, $n = \left(\tfrac18,\tfrac14,\tfrac14,\tfrac58\right)$ yields $Z_i=(0,0,0,-2)$, whereas $\xi n = \left(\tfrac58,\tfrac14,\tfrac14,\tfrac18\right)$ gives $Z_i=(-2,0,0,0)$, see \eqref{eq:so8Z2cosets}.} $P(\xi n) = P(n)$, and $k(\xi n) = k(n)$. Hence, the $\BZ_2$ symmetry $\xi$ requires the set of allowed fluxes to be invariant, not each coset. Indeed, $\xi$: $(j,l) \rightarrow (j + 2l, l)$. The $l=0$ cosets are individually invariant, while the $l=1$ cosets are exchanged in pairs, i.e.\ $j \leftrightarrow j+2$, so no $l=1$ coset is invariant on its own.

\paragraph{The 1-form symmetry from the pairing matrix.} The flux quantisation yielding eight different cosets labelled by $j=0,\ldots,3$ and $l=0,1$ signals that the 1-form symmetry is $\BZ_4 \times \BZ_2$. This also follows from the pairing matrix $M$ of the intersection numbers of the divisors $S_0,\ldots,S_3,N$ (rows) with the compact curves $f_0,\ldots,f_3,e_1,h_1$ (columns):
\begin{equation}
\label{eq:pairmatso8Z2}
M=\begin{pmatrix}-2&2&0&0&0&2\\1&-2&1&0&-1&-3\\0&1&-2&1&-1&0\\0&0&2&-2&0&0\\1&0&0&0&0&0\end{pmatrix}\,,
\end{equation}
whose Smith normal form $\diag(1,1,1,2,4)$ confirms the $\BZ_4 \times \BZ_2$ 1-form symmetry. If only the fibres are kept, the pairing matrix is the $5\times4$ matrix of the first four columns of \eqref{eq:pairmatso8Z2}, with Smith normal form $\diag(1,1,1,2)$. It sees only the $\BZ_2$ part of the 1-form symmetry, which is there with or without the row of $N$. Since the fibres carry no string charge, see \eqref{eq:so8Z2curves}, this matrix misses the $\BZ_4$ part, which descends from the 6d 2-form symmetry.

Given $x_i$ corresponding to the $S_i$ charges, 
the relations arising from the columns of the pairing matrix (the last row, corresponding to $N$, is not used here) are given explicitly by $f_0$: $-2 x_0 + x_1 = 0$, $f_1$: $2 x_0 -2 x_1 + x_2 = 0$, $f_2$: $x_1 -2 x_2 + 2 x_3 = 0$, $f_3$: $x_2 -2 x_3 = 0$, $e_1$: $- x_1 - x_2= 0$, $h_1$: $2 x_0 -3 x_1 = 0$. We can eliminate $x_1$ and $x_2$ using the first two relations, namely $x_1 = 2 x_0$ and $x_2 = 2 x_1 - 2 x_0 = 2 x_0$. The third and fourth relations both yield $2 (x_3 - x_0) = 0$, and the last two relations both give $4 x_0 = 0$. Thus, the 1-form symmetry is generated by $x_0$ and $x_3$, with the two independent relations $4 x_0=0$ and $2 (x_3 - x_0)=0$. However, in the basis $(x_0,x_3)$, the relation vectors are $(4,0)$ and $(-2,2)$, which is not diagonal. In the basis $(x_0,y)$, with $y=x_3-x_0$, the relations become $4 x_0 = 0$ and $2 y =0$, which is diagonal instead. The $\BZ_4$ part of the 1-form symmetry is thus generated by $x_0$, whereas the generator of the $\BZ_2$ part is $x_3-x_0$. Accordingly, $x_0$ has order four and $y$ has order two. Indeed, $2 x_0 = (2,0)$ would vanish only if $(2,0) = \zeta_1 (4,0) + \zeta_2 (-2,2)$ for integers $\zeta_1$, $\zeta_2$. The second component forces $\zeta_2=0$, and $2 = 4 \zeta_1$ has no integer solution. Moreover, $y$ is independent of $x_0$. For an integer $\zeta_0$, $y - \zeta_0\,x_0 = 0$ would require $(-1-\zeta_0,1) = \zeta_1 (4,0) + \zeta_2 (-2,2)$, i.e.\ $2 \zeta_2 = 1$, which is impossible for integer $\zeta_2$.\footnote{As a further check, observe that the relation matrix $\begin{pmatrix}4&0\\-2&2\end{pmatrix}$ has Smith normal form $\diag(2,4)$ and determinant 8.}

Finally, the central divisors $\sum_{i=0}^3 k^{(m)}_i S_i$, with $m=1,2$ labelling the $\BZ_4$ and $\BZ_2$ factors of the 1-form symmetry, respectively, and the spin$^c$ shift $n_\ast$ can also be recovered from the pairing matrix, as in \cite[(A.11), (A.13)]{Harding:2026gzj} for a single $\BZ_\fm$:
\begin{equation}
\label{eq:deltashift}
\begin{aligned}
{k^{(m)}}^{\mathsf T} M &= 0 \pmod{\fm^{(m)}}\,, \quad k^{(m)}\notin\fm^{(m)}\BZ^5\,, \quad \text{with $\fm^{(1)} = 4$, $\fm^{(2)} = 2$} \,, \\
M^{\mathsf T} n_\ast &= \mathsf s \pmod 1\,, \quad \mathsf s_C = \tfrac12 C^2\,.
\end{aligned}
\end{equation}
Here, $C$ runs over the curves labelling the columns of $M$. A solution is $k^{(1)}=(1,2,2,1,0)$, $k^{(2)}=(0,0,0,1,0)$,\footnote{Such central divisors $Z^{(m)} = \sum_{i=0}^3 k^{(m)}_i S_i$ satisfy $Z^{(m)} \cdot C = 0 \pmod{\fm^{(m)}}$ for every compact curve $C$.} and $n_\ast = \left(\tfrac18, \tfrac14, \tfrac14, \tfrac18, 0\right)$, as anticipated below \eqref{eq:so8Z2cosets}. Its last entry, for $N$, vanishes mod $1$ by the congruence itself, since $\sum_if_i$ pairs only with $N$.

\paragraph{Matching higher-form symmetries with the 6d frame.} From Appendix~\ref{app:twists} and Table~\ref{tab:twistcentre}, the twist acts trivially on $D = \BZ_4$ (twist of the first kind), so the tensor part of the 1-form symmetry is $\BZ_4$, the 2-form symmetry of the $-4$ curve, with anomaly \eqref{eq:KKterm}. The $\sigma$-fixed centre of $\Spin(8)$ is $\{1,v\}=\BZ_2$, with spinor and cospinor exchanged. No hypermultiplets and no KK modes of the $\so(7)$ representations screen it, since the $\mathbf{21}_0$ and the $\mathbf{7}_{1/2}$ are neutral under the centre of $\Spin(7)$. It follows that the 1-form symmetry is $\BZ_4 \times \BZ_2$ (for comparison, the untwisted case has $\BZ_2 \times \BZ_2$ instead of $\BZ_2$), in agreement with the pairing matrix \eqref{eq:pairmatso8Z2}. The gauge-descended part $\BZ_2$ is non-trivial, so the \hyperref[par:tensorpart]{criterion} in Appendix~\ref{app:twists} does not apply. Still, all $|D|=4$ sectors are admissible flux sets of this one frame. They are the cosets with $l=0$ in \eqref{eq:solfluxesso8Z25d}, one for each class of $D=\BZ_4$, and three of them, the flux sets of \cite[(6.13)]{Kim:2020hhh}, already fix the elliptic genus \cite[\S6.2]{Kim:2020hhh}. So the criterion is not necessary. The cosets with $l=1$, with $b_3\in\BZ+\tfrac12$, are further 't~Hooft sectors of the gauge-descended $\BZ_2$.

\paragraph{Anomalies.} We conclude with the cubic self-anomaly of the 1-form symmetry and its mixed anomalies with $\U(1)_I$ and $\SU(2)_R$, in the $P$ frame of \eqref{eq:FFSO8Z25d}.

The cubic self-anomaly $\exp\big[2\pi i\,\Omega\int\mathcal B\cup\mathcal B\cup\mathcal B\big]$, with $\mathcal B$ the background of the 1-form symmetry, follows from the geometry as $\Omega=-\frac{Z^3}{6\,\fm^3}-\frac{c_2(X)\cdot Z}{12\,\fm}\pmod 1$ \cite[(4.21)]{Apruzzi:2021nmk}, which we apply to the local threefold $X$ of this KK theory.\footnote{The $p_1(TX_6)$ of that reference equals $-2c_2(X)$ here, since $c_1(X)=0$. See also \cite[(A.2)]{Harding:2026gzj}.} Here, $Z$ is a central divisor of order $\fm$. For the $\BZ_4$ of tensor origin, it is $Z^{(1)}=S_0+2S_1+2S_2+S_3$, with $\fm=4$, since $\tfrac14Z^{(1)}=\nu(\tfrac14)$ lies on the tensor-flux family. For the gauge-descended $\BZ_2$, it is $Z^{(2)}=S_3$, with $\fm=2$. From ${Z^{(1)}}^3=0$, ${Z^{(2)}}^3=8$, $c_2(X)\cdot Z^{(1)}=-24$ and $c_2(X)\cdot Z^{(2)}=-4$, we find $\Omega=\tfrac12$ for the $\BZ_4$ and $\Omega=0$ for the $\BZ_2$.

The blow-up function gives $\Omega$ on every coset \cite[(4.11)--(4.13)]{Harding:2026gzj}. It is evaluated on the bare flux of the coset $(j,l)$ of \eqref{eq:solfluxesso8Z25d}, i.e.\ on $n$ without the spin$^c$ shift $n_\ast$ of \eqref{eq:deltashift} \cite[(3.44)]{Harding:2026gzj}:
\begin{equation}
\label{eq:so8Z2bareflux}
\fn=n-n_\ast=\tfrac j4\,(1,2,2,1)+\tfrac l2\,(0,0,0,1)\,.
\end{equation}
The linear term of $P(n)$ in \eqref{eq:so8Z2blowupfunction} is the sum of the gravitational piece $\sum_iC^G_in_i/48=-\tfrac1{12}\sum_in_i$ and the $\SU(2)_R$ piece $-\sum_iC^R_in_i/8=-\tfrac14\sum_in_i$. The value $\Omega(j,l)$ on $\fn$ is minus the cubic piece minus four times the gravitational piece:
\begin{equation}
\label{eq:cubicanomso8Z2}
\Omega(j,l) = -\mathcal{F}_{\text{cubic}}(\fn) +\frac{1}{3} \sum_{i=0}^3 \fn_i = \left(\frac{j}{2}+\frac{l}{6}\right)(1-l^2) = \frac{j}{2} (1-l^2)\,.
\end{equation}
It reduces to $\frac j2$ for $l=0$ and to $0$ for $l=1$. At the generators, $(j,l)=(1,0)$ and $(0,1)$, these are the geometric values above.\footnote{\label{foot:centraldivZ1Z2}This same result can be obtained by considering the linear combination of central divisors $Z(j,l) = j Z^{(1)} + 2 l Z^{(2)}$, and computing $\Omega$ as $-\frac{Z(j,l)^3}{6\,\fm^3} - \frac{c_2(X)\cdot Z(j,l)}{12\,\fm} \pmod 1$, with $\fm=4$. The factor $2$ in $2l$ is needed because $Z^{(2)}$ has order $2$: its flux is $\tfrac12Z^{(2)}=\tfrac14\big(2Z^{(2)}\big)$, so that $\tfrac14Z(j,l)$ is the bare flux $\fn$ of \eqref{eq:so8Z2bareflux}. Without it, $Z(j,l)$ would not pair with $f_2$ and $f_3$ to a multiple of $4$. Moreover, with $2l$, the cross term of $Z(j,l)^3$ is $3j(2l)^2\,Z^{(1)}\cdot Z^{(2)}\cdot Z^{(2)}=192\,jl^2$, from $Z^{(1)}\cdot Z^{(2)}\cdot Z^{(2)}=16$. Divided by $-6\cdot4^3$, it gives $-\tfrac12jl^2$, the mixed term of \eqref{eq:cubicanomso8Z2}. The other cross term vanishes, since $Z^{(1)}\cdot Z^{(1)}\cdot Z^{(2)}=0$.}

The mixed anomaly with $\SU(2)_R$, the phase $\exp\big[2\pi i\,\mathcal A_R\int c_2(R)\big]$, is minus four times the $\SU(2)_R$ piece of $P(\fn)$ \cite[(4.19)]{Harding:2026gzj}:
\begin{equation}
\label{eq:ARso8Z2}
\mathcal A_R(j,l)=\sum_{i=0}^3\fn_i=\frac{3j+l}2\equiv\frac{j+l}2\pmod 1\,.
\end{equation}

Finally, the mixed 1-form--$\U(1)_I$ anomaly, i.e.\ the fractional instanton number arising when a non-trivial background field for the 1-form symmetry is activated, is given by $\Omega_{ii,N}=-\frac{Z^2 \cdot N}{2 \fm^2} \pmod 1$, with $N$ the non-compact divisor dual to $\U(1)_I$,\footnote{The divisor is $N$ and not $-N$, although $J$ contains $-m_0N$. By footnote~\ref{fn:volfso8Z2}, the volume of a curve $C$ contains $+m_0\,N\cdot C$, so $N\cdot C$ is the instanton number of $C$. With this sign, $-\tfrac12\,N\cdot\big(\sum_in_iS_i\big)^2=4n_0^2$ is the $q$-exponent $k(n)$ in the $P$ frame.} see \cite[(4.21)]{Apruzzi:2021nmk} and \cite[(4.24)]{Harding:2026gzj}. The only non-zero $S_iS_jN$ is $S_0^2N=-8$. So, with $Z^{(1)}$ and $Z^{(2)}$ as above, $\Omega_{ii,N}$ is $\tfrac14$ for the $\BZ_4$ of tensor origin and $0$ for the gauge-descended $\BZ_2$. For the coset $(j,l)$, with $Z(j,l) = j Z^{(1)} + 2 l Z^{(2)}$ as in footnote~\ref{foot:centraldivZ1Z2},
\begin{equation}
\label{eq:mixed1forminstso8Z2}
\Omega_{ii,N}(j,l)=-\frac{Z(j,l)^2 \cdot N}{2 \times 4^2} = \frac{j^2}{4}\,.
\end{equation}
Equivalently, $\Omega_{ii,N}(j,l)$ is the $q$-exponent of the blow-up function on the bare flux in the $P$ frame, $k(\fn)=4\fn_0^2=\tfrac{j^2}4$ \cite[(4.3)]{Harding:2026gzj}, as in \sref{sec:setup5d}. The result differs if evaluated in the $P''$ frame, where, from \eqref{eq:so8Z2blowupfunction}, $k(\fn) = \frac{j^2}{4}+\frac{j l}{2}+\frac{l^2}{8}$. The reason is that the divisor dual to $\U(1)_I$ is fixed only up to $N\to N+\sum_i\varsigma_iS_i$, which mixes $\U(1)_I$ with the Cartan subgroup of the gauge symmetry. This change shifts $k(n)$ by $\sum_i\varsigma_iZ_i(n)$, with $Z_i$ of \eqref{eq:so8Z2blowupfunction}. For the $P''$ frame, $\varsigma=\tilde c-c$, and $N''$ is a rational class, as pointed out below \eqref{eq:FFSO8Z25dsymmlin}. For $l=0$, the value is the same in every frame, since $Z_i(\fn)=0$. For $l=1$, it is not, since $Z_i(\fn)=\left(0,0,j-\tfrac12,-2j-1\right)$. The $l=1$ value changes even if $N$ is replaced by another class that pairs integrally with every compact curve. For instance, $N+\tfrac14Z^{(1)}$ is such a class, and it shifts the $l=1$ values by $\tfrac12$ mod $1$. So the $l=1$ values depend on the choice of $\U(1)_I$, and the table \eqref{eq:so8Z2anomalies} below lists those of the $P$ frame.

To summarise, on the bare flux \eqref{eq:so8Z2bareflux} the anomalies are, mod $1$,
\begin{equation}
\label{eq:so8Z2anomalies}
\vcenter{\hbox{%
\begin{tabular}{ccccc}
\toprule
coset & bare flux $\fn$ & cubic & mixed with & mixed with $\U(1)_I$ \\ & & self-anomaly & $\SU(2)_R$ & ($P$ frame) \\
\midrule
$l=0$ & $j (\tfrac14,\tfrac12,\tfrac12,\tfrac14)$ & $\frac{j}{2}$ & $\frac{j}{2}$ & $\frac{j^2}{4}$ \\
$j=0$, $l=1$ & $(0,0,0,\tfrac12)$ & $0$ & $\frac{1}{2}$ & $0$ \\
$j \neq 0$, $l=1$ & $\left(\frac{j}{4},\frac{j}{2},\frac{j}{2},\frac{j+2}{4}\right)$ & $0$ & $\frac{j+1}{2}$ & $\frac{j^2}{4}$ \\
\bottomrule
\end{tabular}}}
\end{equation}

The 7d anomaly theory of the 6d parent reproduces the terms of \eqref{eq:cubicanomso8Z2}--\eqref{eq:mixed1forminstso8Z2} that depend on $j$, i.e.\ those that involve the $\BZ_4$ of tensor origin. They come from the 7d Chern--Simons term for the 2-form symmetry \eqref{eq:CS7d} and from the Green--Schwarz coupling \eqref{eq:GS7d}. On the circle, \eqref{eq:CS7d} gives \eqref{eq:KKterm} with $\mathcal K=4$, i.e.\ $2\,d\mathcal A\wedge\mathcal B\cup\mathcal B$, with $\mathcal A$ the background of $\U(1)_{\rm KK}$ and $\mathcal B$ that of the 1-form symmetry of tensor origin. The coupling \eqref{eq:GS7d} gives $\mathcal B\cup I$, with $I = \tfrac14 \left[4 \Tr F^2 + 2 p_1(T)\right] + 6 c_2(R) = \Tr F^2 + \tfrac12 p_1(T) + 6 c_2(R)$, given by \eqref{eq:GSdata} with $\mathcal{K}=4$ and $h^\vee = 6$. The twist acts trivially on $D=\BZ_4$ and enters only through $r=2$. On $E$, $\mathcal B$ has the bare flux $\tfrac j4$. The four terms give the following.
\begin{itemize}
\item The term $2\,d\mathcal A\wedge\mathcal B\cup\mathcal B$ gives the mixed anomaly with $\U(1)_{\rm KK}$. Its value is the KK momentum $\tfrac12\cdot4\cdot\big(\tfrac j4\big)^2=\tfrac{j^2}8=q(j)$. With $r=2$, this is $k(\fn)=\tfrac{j^2}4$, the dictionary \eqref{eq:dictionary} at $\lambda=j$.
\item The term $6\,\mathcal B\cup c_2(R)$ gives $\mathcal A_R=\tfrac{3j}2\equiv\tfrac j2$, which is \eqref{eq:ARso8Z2} at $l=0$.
\item The term $\tfrac12\,\mathcal B\cup p_1(T)$ is gravitational, and it gives the cubic self-anomaly. With $\mathcal B=\tfrac j4\,x$ and $x$ an integral class, it is $\tfrac j8\,x\cup p_1(T)$. On a spin manifold, the index theorem gives the congruence $x\cup p_1(T)\equiv4x^3\pmod{24}$ \cite[(4.14)]{Apruzzi:2021nmk}, see also below \cite[(A.2)]{Harding:2026gzj}. So $\tfrac j8\,x\cup p_1(T)\equiv\tfrac j2\,x^3\pmod 1$, which is the cubic self-anomaly $\Omega(j,0)=\tfrac j2$ of \eqref{eq:cubicanomso8Z2}. The coefficient $\tfrac j8$ is minus the gravitational piece of $P(\fn)$ at $l=0$, and the factor $4$ of the congruence is the factor four above \eqref{eq:cubicanomso8Z2}.
\item The term of $I$ with the gauge field, $4\,\mathcal B\cup\tfrac14\Tr F^2$, gives the mixed $\BZ_4$--$\BZ_2$ term. In 6d variables, the gauge part of $\fn$ is the $\so(7)$ flux $(b_1,b_2,b_3)=(0,0,\tfrac l2)$, which differs from $l$ times the vector coweight $(1,1,\tfrac12)$ by a coroot. Its instanton number is $\tfrac14\Tr F^2\to\mathsf q(0,0,\tfrac l2)=\tfrac{l^2}2$, with $\mathsf q$ defined above \eqref{eq:so8Z2E6d}. So this term gives $4\cdot\tfrac j4\cdot\tfrac{l^2}2=\tfrac{jl^2}2\equiv-\tfrac{jl^2}2\pmod 1$, the mixed term of \eqref{eq:cubicanomso8Z2}.
\end{itemize}
The first three items match the $l=0$ entries of \eqref{eq:cubicanomso8Z2}--\eqref{eq:mixed1forminstso8Z2}. Every term above contains $\mathcal B$, whose flux is $\tfrac j4$. So these 7d terms give nothing at $j=0$, and they do not give the term $\tfrac l2$ of \eqref{eq:ARso8Z2}. The terms without $j$ come from the one-loop terms of the $\so(7)$ zero modes in \eqref{eq:so8Z2E6d}, which do not contain $\phi_0$. On the flux $(b_1,b_2,b_3)=(0,0,\tfrac l2)$, the term $(\phi_1+\phi_2+\phi_3)\,\epsilon_+^2$ gives $\mathcal A_R=\tfrac l2$. The cubic terms and the term with $\epsilon_1^2+\epsilon_2^2$ give $\tfrac16(l-l^3)$ in $\Omega$, which vanishes for $l=0,1$.

% ===========================================================================
% Section 6: Faithful global symmetries and 2-groups
% ===========================================================================
\section{Faithful global symmetries and 2-groups}
\label{sec:faithful}

We treat the seven 5d frames of Table~\ref{tab:master} that have a 1-form symmetry $\BZ_2^{(1)}$ and a 6d parent with a gauge algebra. They are $\USp(4)_0+3\Lambda^2$, $\so(7)+5\Fund$, $\su(4)_{0,2,4}+4\Lambda^2$, $\su(4)_6+2\Lambda^2$ and $\sp(3)_0+2\Lambda^2$ (\sref{sec:exZ2twist} and \sref{sec:exsu40}--\sref{sec:exso8tri}). The M-string frame $\SU(2)_0+\adj$ also has $\BZ_2^{(1)}$, and we leave it aside. In each frame, $\BZ_2^{(1)}$ descends from the 2-form symmetry $D=\BZ_2$ of the 6d parent, and its charged lines are 6d surface defects wrapped on the circle. We determine the faithful 0-form symmetry group of each frame, and ask whether it forms a 2-group with $\BZ_2^{(1)}$ \cite{Cordova:2018cvg,Benini:2018reh}.

We adapt the criterion of \cite[\S6.1]{Harding:2026gzj} to KK theories, with two new inputs. The allowed centre twists of the 6d parent predict the faithful group \cite{Heckman:2022suy}. The frame map ties the instanton number of a state to its flavour charge and KK momentum, and so constrains how the flavour centre acts beyond perturbation theory. For 5d SCFTs, this input comes from the one-instanton spectrum or the geometry \cite{Apruzzi:2021vcu,Harding:2026gzj}. The faithful groups are continuous and act on the gauge-invariant bosonic operators \cite[(3.68)]{Heckman:2022suy}, \cite[(5.6)]{Harding:2026gzj}. Discrete 0-form symmetries, such as charge conjugation, are left out, although they could enter the Postnikov class. For global symmetries, we write $\Sp(N):=\USp(2N)$.

% ---------------------------------------------------------------------------
\subsection{Method}
\label{sec:faithfulmethod}

\paragraph{6d: centre twists.}
Let $G_{\rm 6d}$ and $F_{\rm 6d}$ be the simply connected gauge and flavour groups of the 6d parent. Write a central element of $G_{\rm 6d}\times F_{\rm 6d}\times\SU(2)_R$ as $g=(x,y,z)$, with $z\in\{0,1\}$ the exponent of $-1\in\SU(2)_R$. A \emph{centre twist} by $g$ is a background of the quotient of this group by $\langle g\rangle$. For $z=1$, it comes with the \emph{Lorentz twist}, the $\BZ_2$ twist of the Lorentz group that allows backgrounds without spin structure. The centre twist by $g$ is allowed if every field is invariant under $g$ combined with the Lorentz twist, i.e.\ if $g$ acts as $1$, with $z=0$, or as $(-1)^F$, with $z=1$, and if the Green--Schwarz four-form stays integral \cite[(3.65)--(3.68)]{Heckman:2022suy}.\footnote{\label{fn:GSfrac}The twist makes the characteristic classes in it fractional. If $\tilde G$ is simply connected, a $\tilde G/Z(\tilde G)$ bundle with class $w\in H^2(M_6,Z(\tilde G))$ has $c_2\equiv-\alpha_{\tilde G}\,\mathcal P(w)$ mod 1, with $\mathcal P$ the Pontryagin square, $\alpha_{\SU(n)}=\tfrac{n-1}{2n}$ and $\alpha_{\Sp(n)}=\tfrac n4$ \cite[(3.4), (3.5)]{Heckman:2022suy}. So $c_2(R)\equiv-\tfrac14\mathcal P(w_2)$, with $w_2:=w_2(R)$. The Lorentz twist sets $w_2(T)=w_2$, and then $\tfrac14p_1(T)\equiv\tfrac34\mathcal P(w_2)$ mod 1 \cite[(3.66)]{Heckman:2022suy}.} Gauge-invariant operators do not see $x$. So the image $T_F$ of the allowed centre twists under $(x,y,z)\mapsto(y,z)$ acts trivially on the gauge-invariant bosons, and the faithful group of the 6d parent is \cite[(3.68)]{Heckman:2022suy}
\begin{equation}
\label{eq:centretwist}
G_{\rm global}=\big(F_{\rm 6d}\times\SU(2)_R\big)/T_F\,.
\end{equation}
Let $F'$ be the quotient of $F_{\rm 6d}$ by the image of the allowed twists with $z=0$. All allowed twists with $z=1$ have the same image $c$ in $Z(F')$, since two of them differ by one with $z=0$, and $c^2=1$. So \eqref{eq:centretwist} is $[F'\times\SU(2)_R]/\BZ_2$, with $c$ identified with $-1\in\SU(2)_R$, if $c\neq1$, and $F'\times\SO(3)_R$ if $c=1$. A \emph{witness} of an identification of $-1\in F_{\rm 6d}$ with $-1\in\SU(2)_R$ is a gauge-invariant boson on which both act as $-1$.

\paragraph{The 6d parents.}\phantomsection\label{par:faithful6dparents}
All parents but one sit on a $-2$ curve.

\emph{$\su(N)+2N\Fund$}, $N=3,4,5$, is the parent of $\USp(4)_0+3\Lambda^2$, $\su(4)_0+4\Lambda^2$ and $\so(7)+5\Fund$ \cite[(3.69)--(3.76)]{Heckman:2022suy}. Let $g=(x,y,z)\in\BZ_N\times\BZ_{2N}\times\BZ_2$ act as $e^{2\pi ix/N}$, $e^{\pi iy/N}$ and $(-1)^z$ on the fundamentals of $\SU(N)_{\rm gauge}$, $\SU(2N)_{\rm flavour}$ and $\SU(2)_R$. It multiplies the hypermultiplet scalars, in $(\mathbf N,\mathbf{2N},\mathbf 2)$, by $e^{\pi i(2x+y+Nz)/N}$, the hypermultiplet fermions, which are $\SU(2)_R$ singlets, by $e^{\pi i(2x+y)/N}$, and the gaugini, tensorini and supercharges by $(-1)^z$. So every $g$ acts on the vector and tensor multiplets as $(-1)^{zF}$, and the only condition comes from the hypermultiplet: the twist is allowed if and only if the hypermultiplet scalars are invariant, $2x+y\equiv Nz$ mod $2N$. The hypermultiplet fermions then get $(-1)^z$ as well. The solutions are generated by $g_1=(1,-2,0)$ and $g_2=(0,N,1)$, and the Green--Schwarz condition adds nothing \cite[(3.74)--(3.75)]{Heckman:2022suy}. The image $(-2,0)$ of $g_1$ makes the flavour group $\SU(2N)/\BZ_N$. The image of $g_2$ is $(N\bmod2,\,1)$ in the centre $\BZ_2\times\BZ_2$ of $\SU(2N)/\BZ_N\times\SU(2)_R$. Hence \cite[(3.76)]{Heckman:2022suy}
\begin{equation}
\label{eq:faithfulsuNglobal}
G_{\rm global}=\begin{cases}\big[\SU(2N)/\BZ_N\times\SU(2)_R\big]/\BZ_2\,, & N\text{ odd}\,,\\[2pt] \SU(2N)/\BZ_N\times\SO(3)_R\,, & N\text{ even}\,.\end{cases}
\end{equation}
Here, $\BZ_N$ consists of the scalar matrices $\omega\,\ID_{2N}$ with $\omega^N=1$, i.e.\ the even $y$, so $-1\in\SU(2N)$, i.e.\ $y=N$, lies in it if and only if $N$ is even. For $N$ odd, both entries of $(N\bmod2,1)$ are non-trivial: $-1\in\SU(2N)/\BZ_N$ is identified with $-1\in\SU(2)_R$, and the two factors share a $\BZ_2$ quotient. For $N$ even, $-1\in\SU(2N)$ already lies in $\BZ_N$, so only $\SU(2)_R$ is divided, to $\SO(3)_R$. For $N$ odd, the baryon $\epsilon\,q^N$ of the hypermultiplet scalars $q$, in $\Lambda^N$ of $\SU(2N)$ with $R$-spin $\tfrac N2$, is a witness: $-1\in\SU(2N)$ and $-1\in\SU(2)_R$ each act on it as $(-1)^N=-1$.

\emph{The E-string glued to $\su(3)$ on a $-3$ curve} is the parent of $\su(4)_4+4\Lambda^2$. The E-string has $E_8\times\SO(3)_R$ \cite[(A.4)]{Heckman:2022suy}. Gauging $\su(3)$ leaves $E_6$, and $\mathbf{248}\supset(\mathbf 3,\mathbf{27})$ \cite[(3.20)]{Heckman:2022suy} correlates the centres, so gauge invariants see $E_6/\BZ_3$. The four-form stays integral under this diagonal $\BZ_3$ twist: by footnote~\ref{fn:GSfrac} with $\alpha_{E_6}=\tfrac23$ \cite[(3.5)]{Heckman:2022suy}, $-c_2(E_6)-c_2(\SU(3))$ on the $-1$ node shifts by $(\tfrac23+\tfrac13)\mathcal P(w)$ and $3c_2(\SU(3))$ on the $-3$ node by $-\mathcal P(w)$, unlike in \cite[(3.19)]{Heckman:2022suy}, where only $\su(2)\subset\su(3)$ is gauged. For the twist by $-1\in\SU(2)_R$ alone, only the terms $h^\vee c_2(R)-(2-\mathcal K_{\alpha\alpha})\tfrac14p_1(T)$ of the four-form can be fractional \cite[(3.65)]{Heckman:2022suy}, with $h^\vee=1$ for the E-string. By footnote~\ref{fn:GSfrac}, on the $-3$ node, $3c_2(R)+\tfrac14p_1(T)\equiv(-\tfrac34+\tfrac34)\mathcal P(w_2)=0$. On the $-1$ node, $c_2(R)-\tfrac14p_1(T)\equiv(-\tfrac14-\tfrac34)\mathcal P(w_2)=-\mathcal P(w_2)$ \cite[(A.3)]{Heckman:2022suy}. Both are integral. So $c=1$, and $G_{\rm global}=E_6/\BZ_3\times\SO(3)_R$.

\emph{$\mathfrak g_2+4\cdot\mathbf 7$} is the parent of $\su(4)_2+4\Lambda^2$. The matter is a half-hypermultiplet in $(\mathbf 7,\mathbf 8)$ of $G_2\times\Sp(4)$, and $G_2$ has no centre. The only non-trivial allowed twist is $-1$ in $\Sp(4)$ and in $\SU(2)_R$, which acts as $(-1)^F$. In the Green--Schwarz term $2c_2(G_2)-c_2(\Sp(4))+4c_2(R)$, its fractional parts $\pm\mathcal P(w_2)$ are integral \cite[(3.4), (3.5) and below (3.65)]{Heckman:2022suy}. So $G_{\rm global}=[\Sp(4)\times\SU(2)_R]/\BZ_2$. A witness is the $G_2$ cubic invariant of three hypermultiplet scalars, which contains $\Lambda^3\mathbf 8$ with $R$-spin $\tfrac32$.

\phantomsection\label{it:faithfulso86d}\emph{$\so(8)+2(\mathbf 8_v\oplus\mathbf 8_s\oplus\mathbf 8_c)$} is the parent of $\su(4)_6+2\Lambda^2$ and $\sp(3)_0+2\Lambda^2$. The matter is a half-hypermultiplet in $(\mathbf 8_x,\mathbf 4_x)$ of $\Spin(8)\times\Sp(2)_x$, $x=v,s,c$. The element of $Z(\Spin(8))$ trivial on $\mathbf 8_x$ compensates $-1$ in the two other $\Sp(2)$'s. So the even-parity $\BZ_2^2$ of the flavour centre $\BZ_2^3$ acts as $1$. An odd-parity element, such as $-1$ in every $\Sp(2)_x$, must be paired with $-1\in\SU(2)_R$. The Green--Schwarz term $2c_2(\Spin(8))-\sum_xc_2(\Sp(2)_x)+6c_2(R)$ has fractional part $3\cdot\tfrac12\mathcal P(w_2)-\tfrac32\mathcal P(w_2)=0$, since $2c_2(\Spin(8))$ is integral: the odd-parity twist needs no gauge centre, and, in any case, the fractional part of $c_2(\Spin(8))$ lies in $\tfrac12\BZ$ \cite[(3.6)]{Heckman:2022suy}. So $G_{\rm global}=[\Sp(2)^3/\BZ_2^2\times\SU(2)_R]/\BZ_2$, witnessed by the trilinear invariant in $(\mathbf 4,\mathbf 4,\mathbf 4)$ with $R$-spin $\tfrac32$.

\paragraph{The twist $\sigma$ and the 6d prediction.}\phantomsection\label{par:faithfultwistsigma}
The reduction keeps the subgroup $F_\sigma\subset F_{\rm 6d}$ fixed by the twist $\sigma$ (Appendix~\ref{app:twists}). The \emph{6d prediction} for a 5d frame is the image of $F_\sigma\times\SU(2)_R$ in $G_{\rm global}$. $F_\sigma$ depends on the flavour holonomy that $\sigma$ may contain. A holonomy $\exp(2\pi i\,\chi\cdot H)$, with $H$ the Cartan generators, shifts the KK momentum of a state of flavour weight $\mathsf f$ by $\mathsf f\cdot\chi$, i.e.\ it shifts the masses by $\chi\tau$. So a point in mass space fixes the twist with its holonomy, and the flavour symmetry there is generated by the currents massless at that point.\footnote{\label{fn:masslesscurrent}The 6d flavour current multiplet gives one 5d current for each flavour weight $\mathsf f$ and allowed KK momentum $p$. Such a current is massless at a point if its central charge $\mathsf f\cdot m+p\tau$ vanishes there. It then commutes with the background masses and holonomy, so it generates an unbroken symmetry. If the flavour symmetry were gauged, it would give a massless gauge boson, and the other currents W-bosons of mass $|\mathsf f\cdot m+p\tau|$.} In all seven frames, the frame map \eqref{eq:framemapgen} gives the 5d masses as
\begin{equation}
\label{eq:faithfulMv}
M_i=\pm m_i+v\,\tau\,,
\end{equation}
with $m_i$ the 6d masses. All groups below are stated at the point $M_i=0$ of the frame named. The instanton symmetry $\U(1)_I$ is not a factor of $F_{\rm 6d}$, since, by \eqref{eq:framemapgen}, it combines $\U(1)_{\rm KK}$ with flavour Cartan generators. It is left out of the 6d prediction and enters through the instanton parity below.

\paragraph{5d: the perturbative extension.}
Let $G$ be the simply connected 5d gauge group and $F=\Sp(N)$ the 5d flavour group. Let $\mathbb G\subset Z(G)\times Z(F)$ be the subgroup that acts trivially on all the matter \cite[(6.2)]{Harding:2026gzj}. Its intersection with $Z(G)$ is the 1-form symmetry, here $\BZ_2^{(1)}$. Its image $\mathcal Z_F\subset Z(F)$ acts trivially on every perturbative gauge invariant, so $F/\mathcal Z_F$ is the perturbative faithful flavour group. They form the extension
\begin{equation}
\label{eq:faithfulext}
0\to\BZ_2\to\mathbb G\to\mathcal Z_F\to0\,.
\end{equation}
A non-split \eqref{eq:faithfulext} is a candidate 2-group. Instanton operators can carry charges that perturbative operators do not. To include them, we take $F=\Sp(N)\times\U(1)_I$, and $\mathbb G\subset Z(G)\times Z(\Sp(N))\times\U(1)_I$ the subgroup that acts trivially on every state.

\paragraph{KK momentum and instanton parity.}
Let a 5d state have gauge weight $\mathsf w$ in Dynkin labels, instanton number $k$ and flavour weight $\mathsf f$ in the orthonormal basis of $\Sp(N)$. Its KK momentum $p$ is the coefficient of $\tau$ in its mass in 6d variables. By \eqref{eq:massin6d} and \eqref{eq:faithfulMv}, with $\mathsf f\cdot v=v\sum_i\mathsf f_i=:v\,|\mathsf f|$,
\begin{equation}
\label{eq:parity}
p=\mathsf w\cdot s+\frac kr+v\,|\mathsf f|\,.
\end{equation}
A state with $\mathsf w\neq0$ may carry string charge, and then $p$ depends on the origin of the tensor scalar (\hyperref[it:validity]{range of validity}, \sref{sec:framemap}). A gauge-invariant operator has $\mathsf w=0$ and $p=k/r+v|\mathsf f|$, a genuine KK momentum. Two facts constrain it. Let $r_{\rm tw}$ be the order of the twist, as in \eqref{eq:m0delta}, with $r_{\rm tw}=1$ without a twist. If $\sigma^{r_{\rm tw}}$ is trivial on gauge invariants, the eigenvalue $e^{2\pi ip}$ of $\sigma$ obeys $e^{2\pi ir_{\rm tw}p}=1$. Moreover, $e^{2\pi i\alpha}\in\U(1)_I$ and $e^{2\pi i\beta\sum_iH_i}\in\Sp(N)$ act on the operator as $e^{2\pi i\alpha k}$ and $e^{2\pi i\beta|\mathsf f|}$. We call $(-1)^k$, the action of $e^{i\pi}\in\U(1)_I$, the instanton parity, and $(-1)^{|\mathsf f|}$, the action of $-1\in\Sp(N)$, the flavour parity. Together they say that
\begin{equation}
\label{eq:hrule}
h=\Big(e^{2\pi ir_{\rm tw}/r}\in\U(1)_I,\ e^{2\pi ir_{\rm tw}v\sum_iH_i}\in\Sp(N)\Big)
\end{equation}
acts trivially on every local operator, so the faithful group quotients by $h$. Three cases occur.\footnote{\label{fn:faithfulsu40analogue}The 5d analogue is the parity rule \eqref{eq:E1parity} of pure $\SU(4)_0$, which we derive in \sref{sec:E1N2}. It says that, when $\SU(4)_0$ is written as an $\SU(2)$ gauging of two $E_1$ theories, the $E_1$ charges of every state have the parity of its $\SU(2)$ instanton number. Its derivation from the centre charge of $\SU(4)$ (Appendix~\ref{app:su40}) is the analogue of the argument here. There, the frame map writes the instanton number of one frame through the charges of the other, and the $N$-ality plays the role of the allowed KK momenta.}
\begin{itemize}
\item[(a)] \emph{Both parts of $h$ are non-trivial.} Then $\U(1)_I$ is tied to the flavour centre. In the frames below, $h=(e^{i\pi},-1)$,\footnote{\label{fn:hcasea}These are the frames $\su(4)_2+4\Lambda^2$ and $\su(4)_6+2\Lambda^2$ of \sref{sec:faithfullock}, with $(r,v,r_{\rm tw})=(2,-\tfrac12,1)$ and $(6,\tfrac16,3)$. So $r_{\rm tw}/r=\tfrac12$ and $r_{\rm tw}v=\mp\tfrac12$ in \eqref{eq:hrule}. The element $e^{\pm i\pi\sum_iH_i}$ acts on a state as $(-1)^{|\mathsf f|}$, i.e.\ as $-1$ on the fundamental, so it is $-1\in\Sp(N)$.} so every local operator obeys\footnote{\label{fn:lockparity}$h$ acts on an operator as $(-1)^{k+|\mathsf f|}$. Equivalently, by \eqref{eq:parity}, $k=r(p-v|\mathsf f|)$ is fixed by the KK momentum and the flavour charge, and the allowed $p$ fix its parity: $k=2p+|\mathsf f|$ with $p\in\BZ$ for $\su(4)_2+4\Lambda^2$, and $k=6p-|\mathsf f|$ with $p\in\tfrac13\BZ$ for $\su(4)_6+2\Lambda^2$.}
\begin{equation}
\label{eq:faithfullockrule}
(-1)^k=(-1)^{|\mathsf f|}\,.
\end{equation}
\item[(b)] \emph{Only the flavour part is non-trivial.} That flavour element acts trivially by itself.
\item[(c)] \emph{$h$ is trivial.} Then \eqref{eq:parity} gives no constraint, and 6d input decides.
\end{itemize}
If $\sigma^{r_{\rm tw}}$ acts on gauge invariants as a flavour element $g_F$ rather than trivially, $e^{2\pi ir_{\rm tw}p}$ equals the eigenvalue of $g_F$, and the element that acts trivially is $h\,g_F^{-1}$ instead of $h$ (\sref{sec:faithfuloddN}).

\paragraph{The 2-group criterion.}\phantomsection\label{par:faithfulcriterion}
In the 't~Hooft sector of the generator of $\BZ_2^{(1)}$, the blow-up fixes a fractional flux $\mu$ of $\U(1)_I$, of the flavour symmetry and of $\SU(2)_R$, the \emph{forced flux} \cite[\S6.1]{Harding:2026gzj}. In \eqref{eq:sectorexponents}, this is the sector $B=\tfrac12$, or $B=(\tfrac12,\tfrac12)$ for the two tensors of $\su(4)_4+4\Lambda^2$. We evaluate \eqref{eq:Vuniversal} at the representative $a=B$.\footnote{\label{fn:faithfulvariables}As in \sref{sec:pfblowup}, $a$ is the shifted tensor flux \eqref{eq:adef}, and the masses are the 5d masses $M_i$, held fixed. In the 6d masses of \eqref{eq:faithfulMv}, each $\tilde y_i^{1/2}$ gives an extra $\fq^{-v/2}$, so the instanton part $\fq^{1/4}$ of \eqref{eq:forcedflux} becomes $\fq^{1/4-Nv/2}$.} In all seven frames, we can check that $\tfrac12a^T\mathcal Ka=\tfrac14$, that every flavour Green--Schwarz level $\kappa_m$ is odd, and that $\tw\cdot a=0$. So, at $a=B$, the $\tilde y_m$-exponent $-\kappa_m a_{\alpha(m)}=-\kappa_m/2$ is $\tfrac12$ mod 1, and \eqref{eq:Vuniversal} gives
\begin{equation}
\label{eq:forcedflux}
\mu=\fq^{1/4}\prod_i\tilde y_i^{1/2}\,,\qquad \tilde y_i:=e^{M_i+\epsilon_+}\,,
\end{equation}
up to integer powers of $q$, $p_1p_2$ and the $\tilde y_i$, and Weyl reflections of the masses.\footnote{\label{fn:faithfulfluxes}These are the fluxes of the tensor-flux family (\sref{sec:pfblowup}), with $B_m=\tfrac12$ by \eqref{eq:familycond} and $\Bmz=0$ by the \hyperref[it:fluxmap]{flux rule} of \sref{sec:framemap} with $B_\tau=0$. On it, the gauge-fugacity exponents vanish, so their integrality, used in \cite[\S6.1]{Harding:2026gzj}, imposes nothing. Admissibility rests on the pairing rule of \cite[(3.9), (3.10)]{Kim:2020hhh}: the particle on each generator $C$ of the Mori cone must see an integral total flux through $E$, or a half-integral one if $C$ has odd self-intersection, i.e.\ integral on W-bosons and half-integral on hypermultiplets. Its integral case is \eqref{eq:pairingrule} below. The fluxes of $\USp(4)_0+3\Lambda^2$ obey it \cite[(5.12)]{Kim:2020hhh}, and we have checked the other six frames on all the states of \sref{sec:examples}. In general, a state of string charge $\lambda$ sees, in the sector $B$, the flux of the trivial sector plus $\lambda^T\mathcal KB\in\BZ$, so the 't~Hooft sector is admissible exactly when the trivial sector is.} The instanton part $\fq^{1/4}=q^{r/4}$ is a flux $\tfrac r4$ of $\U(1)_I$, with $r$ of \eqref{eq:rdef}. The flavour part is half a unit of every flavour flux, with $\epsilon_+$ attached.

$\mu$ is a fractional flux through the exceptional curve $E$. Whether it is a background of the faithful flavour group, $\U(1)_I$ included, and with which class in $\pi_1$, decides the outcome:
\begin{equation}
\label{eq:verdicts}
\text{\footnotesize$\begin{array}{l|l|l}
\text{class of }\mu & \text{extension \eqref{eq:faithfulext}} & \text{verdict}\\ \hline
w_2(F/\mathcal Z_F)\neq0\text{ in a non-Abelian factor} & \text{non-split} & \text{2-group}\\
\text{locked to }\U(1)_I\text{, as in }[\Sp(N)\times\U(1)_I]/\BZ_2 & \text{non-split, }{\rm Bock}(w_2)=0 & \text{mixed anomaly with }\U(1)_I\\
\mathcal Z_F\text{ trivial} & \text{split} & \text{mixed anomaly with flavour}
\end{array}$}
\end{equation}
In the first row, the Postnikov class is $\delta\mathcal B={\rm Bock}\big(w_2(F/\mathcal Z_F)\big)$ \cite[(6.1)]{Harding:2026gzj}, with $\mathcal B$ the background of $\BZ_2^{(1)}$, as in \eqref{eq:KKterm}, and ${\rm Bock}$ the Bockstein map of \eqref{eq:faithfulext}. In the second row, $w_2$ is the reduction of an integral class, so ${\rm Bock}(w_2)=0$. In the third row, we read the fractional flavour fluxes (\sref{sec:pfanomalies}) as mixed anomalies of $\BZ_2^{(1)}$ with the flavour Cartan directions.

\paragraph{The instanton part of $\mu$.}
$\U(1)_I$ is not enhanced at $M_i=0$ in any of the seven frames, so the instanton part of $\mu$ is a mixed anomaly of $\BZ_2^{(1)}$ with $\U(1)_I$. With $\U(1)_I$ normalised so that gauge-neutral states have integer charge, $\tfrac r4$ is $\tfrac12$ mod $1$ for $r=2,6$. This is the anomaly coefficient of \cite[(4.3), \S4.4]{Harding:2026gzj}. We write it as the term $\tfrac12\int c_1\cup\mathcal B\cup\mathcal B$ mod $1$, with $c_1$ the first Chern class of $\U(1)_I$. It is an integer on $S^2\times M_4$ with $M_4$ spin, but it equals $\tfrac12$ on the spin manifold $\BP^3$, with $c_1$ and $\mathcal B$ the hyperplane class. A properly quantised counterterm \cite[\S4.1, (4.3)]{Harding:2026gzj} is integral on closed 6-manifolds, so none removes it. For $r=3$, in $\sp(3)_0+2\Lambda^2$, the term is $\tfrac34\int c_1\cup\mathcal P(\mathcal B)$, with $\mathcal P$ the Pontryagin square. What remains to decide is the flavour part of $\mu$.

% ---------------------------------------------------------------------------
\subsection{\texorpdfstring{$\su(N)+2N\Fund$ with $N$ odd: no 2-group}{su(N)+2NF with N odd: no 2-group}}
\label{sec:faithfuloddN}

The frames are $\USp(4)_0+3\Lambda^2$ ($N=3$) and $\so(7)+5\Fund$ ($N=5$) of \sref{sec:exZ2twist}. Their frame maps have $m_0=\tfrac\tau2$ and $M_i=m_i+\tfrac\tau4$, i.e.\ $r=2$ and $v=\tfrac14$.

\paragraph{6d prediction.} $G_{\rm global}$ is \eqref{eq:faithfulsuNglobal} for $N$ odd. The points $M_i=0$ and $m_i=0$ differ by the holonomy $\chi=(\tfrac14,\dots,\tfrac14)$. The twist exchanges $\mathbf N$ and $\bar{\mathbf N}$ of $\SU(N)$, so it acts on $\SU(2N)$ as complex conjugation followed by conjugation with a unitary matrix $J$. Complex conjugation squares to $-1$ on the $\SU(2)_R$ doublet of scalars and on the 6d symplectic Majorana fermions. So, up to a gauge transformation, $\sigma^2=-J\bar J$ on the hypermultiplets.\footnote{\label{fn:sigmasquare}Write $\sigma=J\circ K$, with $K$ complex conjugation and $\bar J$ the complex conjugate of $J$. Then $\sigma^2=J\bar J\,K^2$, and $K^2=-1$ on the hypermultiplets for two reasons. The scalars are $\SU(2)_R$ doublets, and the doublet is pseudoreal. Its conjugation is $q^A\mapsto\epsilon^{AB}\bar q_B$, so $K^2=\epsilon\bar\epsilon=\epsilon^2=-1$. The fermions are $\SU(2)_R$ singlets, but they are 6d Weyl spinors, which are also pseudoreal. Their charge conjugation matrix obeys $C\bar C=-1$, which is why they are symplectic Majorana. Unitarity gives $\bar J=(J^T)^{-1}$, so $J\bar J=-1$ for antisymmetric $J$ and $J\bar J=1$ for symmetric $J$. The fixed subgroup, $\sigma(g)=J\bar gJ^{-1}=g$, is defined by $gJg^T=J$.} An antisymmetric $J$ gives a twist $\sigma_A$ with $\sigma_A^2=1$, whose fixed subgroup in $\SU(2N)$ is $\Sp(N)$. A symmetric $J$ gives a twist $\sigma_S$ with $\sigma_S^2=-1\in\SU(2N)$, whose fixed subgroup is $\SO(2N)$. At $M_i=0$, the massless currents span the perturbative $\sp(N)$ of the 5d frame, so the subgroup fixed by the twist is $\Sp(N)$ and $\sigma=\sigma_A$. At $m_i=0$, the $\u(N)$ currents ($k=0$, $\mathsf f=\mathsf e_i-\mathsf e_j$, with $\mathsf e_i$ the unit flavour weights) and the instanton modes with $k=\mp1$, $\mathsf f=\pm(\mathsf e_i+\mathsf e_j)$, of mass $\mp m_0\pm(M_i+M_j)=\pm(m_i+m_j)$ by \eqref{eq:usp4map} and \eqref{eq:so7map}, span $\so(2N)$,\footnote{\label{fn:so2Ncurrents}These are KK modes of the 6d $\su(2N)$ currents. Under $\sigma_A$, $\su(2N)=\sp(N)\oplus\Lambda^2_0(\mathbf{2N})$, with $\Lambda^l_0$ the symplectic-traceless part of $\Lambda^l$, and, in 5d variables, the two parts carry even and odd $k$. At $m_i=0$, the massless currents are those of 6d KK momentum $p=\tfrac k2+\tfrac{|\mathsf f|}4=0$: the $\u(N)$ part of $\sp(N)$ at $k=0$, and the weights $\pm(\mathsf e_i+\mathsf e_j)$ of $\Lambda^2_0(\mathbf{2N})$ at $k=\mp1$. The roots $\pm2\mathsf e_i$ of $\sp(N)$ have $p=\pm\tfrac12$, and $\Lambda^2_0(\mathbf{2N})$ has no weight $\pm2\mathsf e_i$. So the massless currents are the $N$ Cartan generators and the roots $\pm\mathsf e_i\pm\mathsf e_j$ of $\so(2N)$, $N(2N-1)$ in all.} so $\sigma=\sigma_S$. This action on the flavour, not specified in \cite[\S2.1]{Bhardwaj:2020gyu}, follows from the frame map. The element $-1\in\Sp(N)$ is $-1\in\SU(2N)$. Since $N$ is odd, it is non-trivial in $\SU(2N)/\BZ_N$, and $G_{\rm global}$ identifies it with $-1\in\SU(2)_R$ (below \eqref{eq:faithfulsuNglobal}). For $N$ even, it is trivial in $\SU(2N)/\BZ_N$ (\sref{sec:faithfulPSp4}). The prediction is $[\Sp(N)\times\SU(2)_R]/\BZ_2$.

\paragraph{5d, perturbative.} The matter is a half-hypermultiplet in $(\mathbf 5,\mathbf 6)$ of $\USp(4)\times\Sp(3)$, or $(\mathbf 7,\mathbf{10})$ of $\Spin(7)\times\Sp(5)$, on which the gauge centre is trivial. Let $y,z\in\{0,1\}$ be the exponents of $-1$ in $\Sp(N)$ and $\SU(2)_R$ (here $-1\in\Sp(N)$ is $y=1$, not $y=N$ as in 6d). The hypermultiplet scalars get $(-1)^{y+z}$, the hypermultiplet fermions $(-1)^y$ and the gaugini $(-1)^z$. Without $\SU(2)_R$, i.e.\ at $z=0$, the matter is invariant only for $y=0$. So $\mathbb G\cong\BZ_2$ is the gauge centre, and $\mathcal Z_F$ is trivial. With $\SU(2)_R$, $(y,z)=(1,1)$ leaves the scalars invariant and acts as $-1$ on all fermions, i.e.\ as $(-1)^F$. The perturbative faithful group is $[\Sp(N)\times\SU(2)_R]/\BZ_2$, as predicted.

\paragraph{Instanton parity.} At $\mathsf w=0$, \eqref{eq:parity} gives $p=\tfrac k2+\tfrac{|\mathsf f|}4$. Here, $p$ is measured at fixed 6d masses, so the relevant twist is the one at $m_i=0$, which is $\sigma_S$. A state of KK momentum $p$ has eigenvalue $e^{2\pi ip}$ under $\sigma_S$, hence $e^{4\pi ip}$ under $\sigma_S^2$. By footnote~\ref{fn:sigmasquare}, $\sigma_S^2=-1\in\SU(2N)$, and $-1=e^{i\pi\sum_iH_i}$ acts on a state of flavour weight $\mathsf f$ as $(-1)^{|\mathsf f|}$. So, on gauge-invariant bosons, the allowed momenta obey $e^{4\pi ip}=(-1)^{|\mathsf f|}$, instead of $e^{4\pi ip}=1$. The value $p=\tfrac k2+\tfrac{|\mathsf f|}4$ gives $e^{4\pi ip}=e^{2\pi ik}(-1)^{|\mathsf f|}=(-1)^{|\mathsf f|}$ for every $k$ and $\mathsf f$, so the condition does not relate $k$ to $|\mathsf f|$. In the notation of \eqref{eq:hrule}, $g_F=\sigma_S^2=-1\in\Sp(N)$ and $h=(1,-1\in\Sp(N))$, so $h\,g_F^{-1}$ is trivial. This is case~(c), and both flavour parities are allowed at every $k$. Odd gauge invariants exist already at $k=0$, e.g.\ the $\epsilon$-contraction of one hypermultiplet scalar with two vector-multiplet scalars of $\USp(4)\cong\Spin(5)$, or three of $\Spin(7)$, with $R$-spin $\tfrac12$. So instanton operators do not change the perturbative result: $-1\in\Sp(N)$ acts on local operators neither trivially nor as an element of $\U(1)_I$, and $\mathcal Z_F$ stays trivial.

\paragraph{Verdict.}\phantomsection\label{par:faithfuloddNverdict}
Without $\SU(2)_R$, $F/\mathcal Z_F=\Sp(N)\times\U(1)_I$ with $\Sp(N)$ simply connected. This is the third row of \eqref{eq:verdicts}, as for $\USp(4)_0+2\Lambda^2$ in \cite[Table~3]{Harding:2026gzj}. Including $\SU(2)_R$ does not change it. By \eqref{eq:forcedflux}, the flavour part of $\mu$ is the flux $\tfrac12$ along each Cartan direction of $\Sp(N)$ and $\tfrac N2$ along that of $\SU(2)_R$. For $N$ odd, it ends at the central element $(y,z)=(1,1)$, since $e^{i\pi\sum_iH_i}=-1\in\Sp(N)$ and $(-1)^N=-1\in\SU(2)_R$. So it is the non-trivial class in $\pi_1\big([\Sp(N)\times\SU(2)_R]/\BZ_2\big)=\BZ_2$. By the perturbative analysis above, $(1,1)$ acts as $(-1)^F$ without any gauge-centre element. So its lift to $\mathbb G$ has order $2$, and \eqref{eq:faithfulext} splits. In $\su(4)_{0,4}+4\Lambda^2$, by contrast, the lift $(i,-1)$ needs the gauge element $i$ and has order $4$ (\sref{sec:faithfulPSp4}). We therefore read the flavour part of $\mu$, as in \sref{sec:pfanomalies}, as a mixed anomaly of $\BZ_2^{(1)}$ with flavour and $\SU(2)_R$ jointly, and with the Lorentz group through $\epsilon_+$. As explained there, the blow-up cannot separate it from a term in $\mathcal B\cup w_2(\widehat{\BC}{}^2)$.

% ---------------------------------------------------------------------------
\subsection{\texorpdfstring{$\su(4)_{0}+4\Lambda^2$ and $\su(4)_{4}+4\Lambda^2$: a non-split $\BZ_4$ 2-group}{su(4)0+4AS and su(4)4+4AS: a non-split Z4 2-group}}
\label{sec:faithfulPSp4}

$\su(4)_0+4\Lambda^2$ descends from $\su(4)+8\Fund$ with the $\BZ_2$ twist (\sref{sec:exsu40}). $\su(4)_4+4\Lambda^2$ descends from the E-string glued to $\su(3)$ on a $-3$ curve, with the $\BZ_2$ twist on $\su(3)$ (\sref{sec:exsu44}). Both frame maps have $m_0=\tfrac\tau2$, so $r=2$. The masses are $M_i=m_i+\tfrac\tau4$ and $M_i=m_i$, respectively.

\paragraph{6d predictions.}
For $\su(4)+8\Fund$, \eqref{eq:faithfulsuNglobal} gives $G_{\rm global}=\SU(8)/\BZ_4\times\SO(3)_R$. At $M_i=0$, the twist leaves $\Sp(4)$. Its element $-1$, $y=4$, is trivial in $\SU(8)/\BZ_4$, since it equals $-1\in\SU(4)_{\rm gauge}$ on the hypermultiplets. On the circle, only the gauge transformations fixed by the twist remain. They form $G^\sigma$, here $\USp(4)$ by the first row of Table~\ref{tab:twistcentre}, which we write $\USp(4)^\sigma$. The same row gives $\ker_\sigma Z(\SU(4))=\BZ_2$: $-1\in\SU(4)_{\rm gauge}$ is fixed by $\sigma$ and generates the centre of $\USp(4)^\sigma$. So it is still a gauge transformation, and $-1\in\Sp(4)$ stays trivial on gauge invariants. The prediction is $\Sp(4)/\BZ_2\times\SO(3)_R$.\footnote{\label{fn:faithfuleighth}The $\sigma$-invariant part of $\SU(8)/\BZ_4$ also contains the discrete $\BZ_2$ of the class of $e^{2\pi i/8}\,\ID_8$, which is $-1$ on the baryons. We do not determine its 5d image.} For the E-string glued to $\su(3)$, $G_{\rm global}=E_6/\BZ_3\times\SO(3)_R$ (\hyperref[par:faithful6dparents]{6d parents}). The twist acts on $E_6$ by the outer involution with fixed subgroup of type $C_4$ (\sref{sec:exsu44}). Under $\usp(8)\subset\mathfrak e_6$, $\mathbf{27}\to\Lambda^2_0\mathbf 8$ and $\mathbf{78}\to\mathbf{36}\oplus\Lambda^4_0\mathbf 8$. All are even under $-1\in\Sp(4)$. So $-1\in\Sp(4)$ is the identity of $E_6$, and the prediction is again $\Sp(4)/\BZ_2\times\SO(3)_R$.

\paragraph{5d, perturbative.}\phantomsection\label{par:faithfulsu4pert}
Consider $\SU(4)_{\kCS}+N\Lambda^2$. The element $i\in Z(\SU(4))$ is $-1$ on the $\mathbf 6$. Write $g=(x,y,z)$ with $x\in\{\pm1,\pm i\}$, and $y,z\in\{0,1\}$ the exponents of $-1\in\Sp(N)$ and $-1\in\SU(2)_R$. The hypermultiplet scalars get $x^2(-1)^{y+z}$, the hypermultiplet fermions $x^2(-1)^y$ and the gaugini $(-1)^z$. So $g$ acts as $1$ for $(\pm1,0,0)$ and $(\pm i,1,0)$, and as $(-1)^F$ for $(\pm1,1,1)$ and $(\pm i,0,1)$. All four pairs $(y,z)$ occur, so the perturbative faithful group is $\Sp(N)/\BZ_2\times\SO(3)_R$. In multiplicative notation, $\mathbb G\cong\BZ_4$ is generated by $(i,-1)\in Z(\SU(4))\times Z(\Sp(N))$. Its square $(-1,1)$ generates $\BZ_2^{(1)}$, so \eqref{eq:faithfulext} does not split. This holds for every $\kCS$ and $N$, but only for perturbative states.

\paragraph{Instanton parity.} Both frames have $r=r_{\rm tw}=2$. In $\su(4)_0+4\Lambda^2$, $v=\tfrac14$ as for $N$ odd, but now $\sigma_S^2=-1\in\SU(8)$ equals $-1\in\SU(4)_{\rm gauge}$ on the hypermultiplets. It is trivial on gauge invariants, so $g_F$ is trivial and $h=(1,-1)$ itself acts trivially, case~(b). Since $h=\sigma_S^2$, this restates the 6d prediction, and \eqref{eq:parity} adds no constraint, as for $N$ odd. The independent 5d evidence is the one-instanton parity below. In $\su(4)_4+4\Lambda^2$, $v=0$ and $h$ is trivial, case~(c). There, a state with $\mathsf w=0$ has no string charge and no weight under $\su(2)^\sigma\subset\su(3)$, since $m_0$ and $M_i$ in \eqref{eq:su44map} contain no $\phi$. So it descends from a 6d local operator in a representation of $E_6/\BZ_3$, which is even under $-1\in\Sp(4)$ as shown above. In both frames, $\Sp(4)/\BZ_2$ holds beyond perturbation theory.

\paragraph{One-instanton states.}\phantomsection\label{par:faithfulzeromodes}
The fermion zero modes of one instanton give an independent check, for $\SU(4)_{\kCS}+N\Lambda^2$ in general \cite[\S\S2.2, 3.1, 4.1]{Tachikawa:2015mha}, \cite[\S4.1]{Yonekura:2015ksa}, \cite[app.~A]{Zafrir:2015rga}, \cite[\S5.2.2]{Harding:2026gzj}. We embed the one-instanton in the $\SU(2)$ of the upper-left $2\times2$ block of $\SU(4)$. Its commutant, $\SU(2)'\times\U(1)_T$, with $\SU(2)'$ the lower-right block and $T=\diag(1,1,-1,-1)$, is left unbroken \cite[(2.2)]{Yonekura:2015ksa}. Besides the broken supercharges, the gaugino gives four raising operators $\mathsf b$ in $(\mathbf 2',\mathbf 2_R)$ with $T=2$. The $\Lambda^2$ hypermultiplets give $4N$ real modes in $(\mathbf{2N},\mathbf 2')$ with $T=0$ (for $N=2$, \cite[Table~1]{Zafrir:2015rga}). A state with $n_{\mathsf b}$ raising operators has $T$-charge $\kCS-4+2n_{\mathsf b}$ \cite[below (2.2)]{Yonekura:2015ksa}. A gauge-invariant state is an $\SU(2)'$ singlet with $T$-charge $0$, so Gauss's law fixes
\begin{equation}
\label{eq:faithfulnb}
n_{\mathsf b}=2-\tfrac{\kCS}2\,.
\end{equation}
The $4N$ real modes form a Clifford algebra, whose states are $\bigoplus_{l=0}^N\Lambda^l_0(\mathbf{2N})\otimes{\rm Sym}^{N-l}(\mathbf 2')$ under $\Sp(N)\times\SU(2)'$. The $\mathsf b$-states $\Lambda^{n_{\mathsf b}}(\mathbf 2'\otimes\mathbf 2_R)$ have integer $\SU(2)'$ spin and integer $\SU(2)_R$ spin $j_R$ exactly when $n_{\mathsf b}$ is even. So an $\SU(2)'$ singlet needs
\begin{equation}
\label{eq:faithfullparity}
l\equiv N-n_{\mathsf b}\equiv N+\tfrac{\kCS}2\pmod 2\,.
\end{equation}
Since $-1\in\Sp(N)$ is $(-1)^l$ on $\Lambda^l_0$, and $(-1)^{2j_R}=(-1)^{n_{\mathsf b}}$ comes from the $\mathsf b$-states alone, the $\Lambda^2$ modes being $\SU(2)_R$ singlets, the gauge-invariant one-instanton bosons obey
\begin{equation}
\label{eq:faithfuloneinst}
(-1)^{|\mathsf f|}=(-1)^{N+\kCS/2}\,,\qquad(-1)^{2j_R}=(-1)^{\kCS/2}\,.
\end{equation}
For $N=4$, the flavour parity is even at $\kCS=0,4$ and odd at $\kCS=2$. Odd parity ties $-1\in\Sp(N)$ to $e^{i\pi}\in\U(1)_I$, and even parity gives $\Sp(N)/\BZ_2$, in agreement with \cite[(5.6)]{Harding:2026gzj} for $N=1,2$ and \cite[(4.87), (4.90), (4.93)]{Apruzzi:2021vcu} for $N=3$. The flavour parity holds also for $|\kCS|>4$, where gauge-invariant one-instanton states contain perturbative fields.\footnote{\label{fn:faithfulimone}Dressing does not change the flavour parity. On a gauge-invariant state, $i\in\SU(4)$ is trivial, so $(i,-1)$ acts as the flavour parity. It is trivial on perturbative fields (\hyperref[par:faithfulsu4pert]{5d, perturbative}). On zero-mode states, write $i\,\ID_4=e^{i\pi T/2}\cdot\diag(1,1,-1,-1)$, the second factor being $-1\in\SU(2)'$. A zero-mode state has $T=\kCS-4+2n_{\mathsf b}$ and $\SU(2)'$ parity $(-1)^{N-l+n_{\mathsf b}}$, so $(i,-1)$ acts on it as $i^{\kCS-4+2n_{\mathsf b}}(-1)^{N-l+n_{\mathsf b}}(-1)^l=(-1)^{N+\kCS/2}$, for every $n_{\mathsf b}$ and $l$. So every gauge-invariant one-instanton state has flavour parity $(-1)^{N+\kCS/2}$.}

\paragraph{Where the extension lives.}\phantomsection\label{par:faithfulextensionlives}
Every state has even $N$-ality, since $\BZ_2^{(1)}$ is unscreened. Dressing it with W-bosons and $\Lambda^2$ states, on which $(i,-1)$ is trivial, gives a gauge invariant. So $(i,-1)$ is trivial on every state. We write its gauge part $i$ in 6d variables. An element with Dynkin angles $\theta$ multiplies a state of Dynkin labels $\mathsf w$ by $e^{2\pi i\mathsf w\cdot\theta}$. Angles transform like the variables, as fluxes do by the \hyperref[it:fluxmap]{flux rule} of \sref{sec:framemap}. So, by \eqref{eq:framemapgen}, $\theta^{\rm 5d}=M\,\theta^{\rm 6d}$ mod $\BZ^3$, with $\theta^{\rm 5d}=(\tfrac14,\tfrac12,\tfrac34)$ for $i$.
\begin{itemize}
\item \emph{$\su(4)_0+4\Lambda^2$.} The frame matrix \eqref{eq:su40M} sends $(\phi_0,\phi_1,\phi_2)$ to $(\phi_0,\,2\phi_0+\phi_2,\,\phi_0+\phi_1)$, so $\theta^{\rm 6d}=(\tfrac14,\tfrac12,0)$. The angles $(\tfrac12,0)$ give $-1\in\USp(4)^\sigma$. The angle $\tfrac14$ on $\phi_0$ gives $i^{2\lambda}=(-1)^\lambda$ on a state of string charge $\lambda$, whose $\phi_0$-coefficient is $\mathcal K\lambda=2\lambda$. Since $(i,-1)$ is trivial,
\begin{equation}
\label{eq:faithfulstringodd}
(-1)_{\USp(4)^\sigma}\,(-1)_{\Sp(4)}=(-1)^{\lambda}\,,
\end{equation}
with $(-1)_{\Sp(4)}=\exp(i\pi\sum_iH_i)$ in 6d masses.\footnote{\label{fn:faithfullifts}Here, $(-1)_{\Sp(4)}$ is defined through the 6d masses. In 5d variables, it is the element $(-1,-1)\in Z(\SU(4))\times Z(\Sp(4))$: by \eqref{eq:su40map}, the angle $\tfrac12$ on every $m_i$ gives the 5d gauge angles $(\tfrac12,1,\tfrac12)$, i.e.\ $-1\in\SU(4)$, and the angle $\tfrac12$ on every $M_i$. The lift matters on strings: a single string has charge $\tfrac12$ under each $H_i$, the $\Sigma$ in the string rows of \eqref{eq:su40W}, so the other lifts $\exp(i\pi\sum_i\pm H_i)$ differ on it by the product of the signs.}
\item \emph{$\su(4)_4+4\Lambda^2$.} The frame map \eqref{eq:su44map} sends $(\phi_{0,1},\phi_{0,2},\phi_1)$ to $(\phi_{0,2},\,2\phi_{0,2}+\phi_1,\,\phi_{0,1})$, so $\theta^{\rm 6d}=(\tfrac34,\tfrac14,0)$. The angle on $\phi_1$ vanishes, so $i$ has no part in $\SU(2)^\sigma$, in line with $\ker_\sigma Z(\SU(3))=1$ (Table~\ref{tab:twistcentre}). A string of charge $\lambda=(\lambda_1,\lambda_2)$, with $\lambda_1$ E-strings, has tensor coefficients $\mathcal K\lambda=(\lambda_1-\lambda_2,\,3\lambda_2-\lambda_1)$. Since $\tfrac34(\lambda_1-\lambda_2)+\tfrac14(3\lambda_2-\lambda_1)=\tfrac{\lambda_1}2$, $i$ acts as $(-1)^{\lambda_1}$ and
\begin{equation}
\label{eq:faithfulEstringodd}
(-1)_{\Sp(4)}=(-1)^{\lambda_1}\,.
\end{equation}
The simplest odd state carries the $\mathbf 8$ of $\usp(8)$ on one E-string (\sref{sec:exsu44}). It comes from the $\sigma$-twisted sector of the level-one $E_8$ current algebra of the E-string. Since $E_8$ has no outer automorphisms, $\sigma$ acts on $E_8$ as an inner element, here the one of order four with centraliser $[\SU(8)\times\SU(2)]/\BZ_4$, where $\SU(2)$ is the invariant gauge group $\SU(2)^\sigma$ and $\SU(8)\supset\Sp(4)$. In the lattice description of the level-one $E_8$ theory, a state of momentum $\gamma\in\Gamma_8$ has conformal weight $L_0=\tfrac12|\gamma|^2$, and the twist $\sigma=e^{2\pi i\varrho\cdot H}$ shifts it to $L_0=\tfrac12|\gamma+\varrho|^2$, here with $|\varrho|^2=\tfrac78$. The minimum, $L_0=\tfrac7{16}$, is reached by eight states, which form the $\mathbf 8$ of $\SU(8)$, singlets of $\SU(2)^\sigma$. The one-E-string elliptic genus is $\Theta_{E_8}$ divided by $\eta^6\theta_1(\epsilon_1)\theta_1(\epsilon_2)$, of leading power $\fq^{1/2}$, with $\fq=e^{-\tau}$ \cite[(6.30)]{Haghighat:2014pva}, so a state of weight $L_0$ has KK momentum $L_0-\tfrac12$. This is $-\tfrac1{16}$ for the ground states, as for $e_2+e_3+M_i$ in \eqref{eq:su44states}, and $\tfrac3{16}$ for the $\SU(2)^\sigma$ doublets at $L_0=\tfrac{11}{16}$, as for $e_1+e_3+M_i$. The element $-1\in\SU(8)$ is the identity of $E_8$, but acts as $-1$ on the $\mathbf 8$. So the twisted sector carries a projective representation, which is \eqref{eq:faithfulEstringodd}.
\end{itemize}
The left-hand sides of \eqref{eq:faithfulstringodd} and \eqref{eq:faithfulEstringodd} are central elements of the 6d gauge and flavour groups, trivial on 6d local operators, but odd on single strings. In this sense, the extension lives on the strings.

\paragraph{Verdict.}\phantomsection\label{par:faithfulPSp4verdict}
By the instanton parity, $F/\mathcal Z_F=\Sp(4)/\BZ_2\times\U(1)_I$. With $r=2$, \eqref{eq:forcedflux} reads $\mu=q^{1/2}\prod_{i=1}^4\tilde y_i^{1/2}$, the forced flux of \cite[(6.7)]{Harding:2026gzj} up to integer powers and the $\epsilon_+$ in $\tilde y_i$. Its flavour flux $(\tfrac12,\tfrac12,\tfrac12,\tfrac12)$ is not in the coroot lattice $\BZ^4$ of $\Sp(4)$, so $w_2(\Sp(4)/\BZ_2)\neq0$. Its $\epsilon_+$-exponent is $2$, so $\SU(2)_R$ does not enter. $\mathbb G=\langle(i,-1)\rangle\cong\BZ_4$ does not split. Only its subgroup $\mathbb G\cap Z(\SU(4))\cong\BZ_2$ is a 1-form symmetry, and its quotient $Z(\Sp(4))$ is a 0-form symmetry. So $\su(4)_0+4\Lambda^2$ and $\su(4)_4+4\Lambda^2$ carry a non-split $\BZ_4$ 2-group, the first row of \eqref{eq:verdicts}, with
\begin{equation}
\label{eq:faithfulPostnikovPSp4}
\delta\mathcal B={\rm Bock}\big(w_2(\Sp(4)/\BZ_2)\big)\,.
\end{equation}
The closest 5d SCFT in \cite[Table~3]{Harding:2026gzj} is $\SU(4)_2+1\Lambda^2$, with $\mathbb G=\BZ_4$ and flavour symmetry $\SO(3)\times\U(1)_I$.

% ---------------------------------------------------------------------------
\subsection{\texorpdfstring{$\su(4)_2+4\Lambda^2$ and $\su(4)_6+2\Lambda^2$: the instanton symmetry locks to the flavour centre}{su(4)2+4AS and su(4)6+2AS: the instanton symmetry locks to the flavour centre}}
\label{sec:faithfullock}

$\su(4)_2+4\Lambda^2$ descends from $\mathfrak g_2+4\cdot\mathbf 7$ without a twist (\sref{sec:exsu42}). $\su(4)_6+2\Lambda^2$ and $\sp(3)_0+2\Lambda^2$ descend from $\so(8)+2(\mathbf 8_v\oplus\mathbf 8_s\oplus\mathbf 8_c)$ with the triality twist (\sref{sec:exso8tri}).

\paragraph{6d predictions.}\phantomsection\label{par:faithfullockpred}
The prediction is the image of $F_\sigma\times\SU(2)_R$ in $G_{\rm global}$, with $F_\sigma$ generated by the currents massless at the point $M_i=0$ of the frame (\hyperref[par:faithfultwistsigma]{\S\ref*{sec:faithfulmethod}}).

\emph{$\su(4)_2+4\Lambda^2$.} The reduction of $\mathfrak g_2+4\cdot\mathbf 7$ is untwisted, so only the flavour holonomy acts. By \eqref{eq:su42map}, $M_i=m_i-\tfrac\tau2$, so $M_i=0$ is $m_i=\tfrac\tau2$, i.e.\ the holonomy $\chi=(\tfrac12,\dots,\tfrac12)$, which is $e^{i\pi\sum_iH_i}=-1\in\Sp(4)$. It is central, so all of $\Sp(4)$ survives: a current of flavour weight $\mathsf f$ and KK momentum $p$ has mass $\mathsf f\cdot m+p\tau=(\tfrac12|\mathsf f|+p)\tau$ there, so the currents $\mathsf e_i-\mathsf e_j$ and the Cartan are massless at KK momentum $0$, and $\pm(\mathsf e_i+\mathsf e_j)$, $\pm2\mathsf e_i$ at $\mp1$. So $F_\sigma=\Sp(4)$, and the prediction is all of $G_{\rm global}=[\Sp(4)\times\SU(2)_R]/\BZ_2$.

\emph{$\sp(3)_0+2\Lambda^2$ and $\su(4)_6+2\Lambda^2$.} Triality permutes the factors $\Sp(2)_{v,s,c}$, so it fixes the diagonal $\Sp(2)$, with masses $m_{1,2}$ (\sref{sec:exso8tri}). In $\sp(3)_0$, $M_i=m_i$ by \eqref{eq:so8trimap}, so $M_i=0$ has no holonomy and $F_\sigma$ is the diagonal $\Sp(2)$. In $\su(4)_6$, $M'_i=\tfrac\tau6-m_i$ by \eqref{eq:so8trisu4map}, so $M'_i=0$ adds the holonomy $h_0=e^{2\pi i\sum_iH_i/6}$ in each factor. Since $h_0^3=-1$ is central, the twist then fixes the diagonal $\Sp(2)$ conjugated by $(1,h_0,h_0^2)$.\footnote{\label{fn:twisteddiag}With $\sigma(g_v,g_s,g_c)=(g_c,g_v,g_s)$ and $\hat h=(h_0,h_0,h_0)$, the fixed points of $g\mapsto\hat h\,\sigma(g)\,\hat h^{-1}$ obey $g_s=h_0g_vh_0^{-1}$, $g_c=h_0g_sh_0^{-1}$ and $g_v=h_0g_ch_0^{-1}$. The first two give $g=(g_0,\,h_0g_0h_0^{-1},\,h_0^2g_0h_0^{-2})$ with $g_0=g_v$. The third then reads $g_0=h_0^3g_0h_0^{-3}$, which holds for every $g_0\in\Sp(2)$ since $h_0^3=-1$ is central. So the fixed points are $(1,h_0,h_0^2)\,(g_0,g_0,g_0)\,(1,h_0,h_0^2)^{-1}$, $g_0\in\Sp(2)$.} In both frames, the only elements of $F_\sigma$ central in $\Sp(2)^3$ are $\pm(1,1,1)$. The even-parity elements, which $G_{\rm global}$ divides out, are not among them, since $-(1,1,1)$ has odd parity. So $F_\sigma$ stays $\Sp(2)$ in $G_{\rm global}$, and its $-1$ is identified with $-1\in\SU(2)_R$ (\hyperref[it:faithfulso86d]{6d parents}). The prediction is $[\Sp(2)\times\SU(2)_R]/\BZ_2$ in both frames.

The witness is the trilinear invariant in $(\mathbf 4,\mathbf 4,\mathbf 4)$ of the three $\Sp(2)_x$, with $R$-spin $\tfrac32$. Every weight of $(\mathbf 4,\mathbf 4,\mathbf 4)$ has odd $|\mathsf f|$, so $-1\in\Sp(2)$ and $-1\in\SU(2)_R$ both act on it as $-1$. Under the diagonal $\Sp(2)$ and the cyclic permutation, it splits into ${\rm Sym}^3\mathbf 4$ and $\Lambda^3\mathbf 4$, at KK momenta in $\BZ$, and two mixed parts, at $\BZ\pm\tfrac13$. The mode of ${\rm Sym}^3\mathbf 4$ at KK momentum $0$ is a 5d local operator; in $\sp(3)_0$, it is the cubic invariant of the $\mathbf{14}$ (\hyperref[par:faithfullockpert]{5d, perturbative}, below). So $-1\in\Sp(2)$ stays non-trivial, paired with $-1\in\SU(2)_R$.

\paragraph{5d, perturbative.}\phantomsection\label{par:faithfullockpert}
For $\su(4)_2+4\Lambda^2$ and $\su(4)_6+2\Lambda^2$, the \hyperref[par:faithfulsu4pert]{perturbative analysis} of \sref{sec:faithfulPSp4} gives $\Sp(N)/\BZ_2\times\SO(3)_R$, with $N=4,2$, and the non-split $\mathbb G=\langle(i,-1)\rangle\cong\BZ_4$. This quotients too much. Since $i\in Z(\SU(4))$ is $-1$ on the $\mathbf 6$, every perturbative gauge invariant contains an even number of hypermultiplet fields. So it is even under $-1\in\Sp(N)$ and, if bosonic, has integer $R$-spin. But the 6d prediction needs local operators odd under $-1\in\Sp(N)$, with half-integer $R$-spin, such as the witnesses of the \hyperref[par:faithful6dparents]{6d parents}. In the 5d frame, these cannot be perturbative, so they must carry instanton charge. The instanton parity below confirms this. In $\sp(3)_0+2\Lambda^2$, $-1\in\USp(6)$ is trivial on the $\mathbf{14}$, so $\mathbb G\cong\BZ_2$ and $\mathcal Z_F$ is trivial. The cubic invariant of the $\mathbf{14}$, in ${\rm Sym}^3\mathbf 4$ with $R$-spin $\tfrac32$, witnesses $[\Sp(2)\times\SU(2)_R]/\BZ_2$.

\paragraph{Instanton parity, $\su(4)_2+4\Lambda^2$.}
For a weight $\mathsf w$ of $\SU(4)$, let $\eta(\mathsf w):=\mathsf w_1+2\mathsf w_2+3\mathsf w_3$, whose residue mod $4$ is the $N$-ality. The frame map of \sref{sec:exsu42} has $s=(\tfrac14,\tfrac12,\tfrac34)$, $r=2$ and $v=-\tfrac12$, so \eqref{eq:parity} gives $p=\tfrac14\eta(\mathsf w)+\tfrac12(k-|\mathsf f|)$. Without a twist, $p\in\BZ$ on states with $\mathsf w=0$. In the origin of $\phi_0$ of that map, $p\in\BZ$ also on all perturbative states, strings included, by \eqref{eq:su42roots} and \eqref{eq:su42hypers}. Every state has even $N$-ality and becomes gauge-neutral after dressing with perturbative states, as in \sref{sec:faithfulPSp4}, so $p\in\BZ$ for every state. Hence
\begin{equation}
\label{eq:faithfulsu42parity}
\tfrac12\eta(\mathsf w)\equiv k+|\mathsf f|\pmod 2\,.
\end{equation}
On a state $(\mathsf w,k,\mathsf f)$, $i\in\SU(4)$ acts as $i^{\eta(\mathsf w)}$, $-1\in\Sp(4)$ as $(-1)^{|\mathsf f|}$ and $e^{i\pi}\in\U(1)_I$ as $(-1)^k$. By \eqref{eq:faithfulsu42parity}, $(i,-1,e^{i\pi})$ acts trivially on every state. At $\mathsf w=0$, this is \eqref{eq:faithfullockrule}, case~(a) with $r_{\rm tw}=1$. Two witnesses confirm it. Both have $k=1$, odd flavour parity and $R$-spin $\tfrac12$, as the 6d prediction requires.
\begin{itemize}
\item \emph{A perturbative 6d state.} The last row of \eqref{eq:su42hypers}, of mass $M_i+m_0$, has $\mathsf w=0$, $k=1$ and $\mathsf f=\mathsf e_i$. In 6d, it is the zero mode of the zero weight of the $\mathbf 7$, with flavour weight $\mathsf e_i$. The KK mode of momentum $1$ with flavour weight $-\mathsf e_i$ has mass $\tau-m_i=m_0-M_i$, so it also has $k=1$. Together they give the eight weights $\pm\mathsf e_i$ of the $\mathbf 8$ of $\Sp(4)$ at $k=1$. Their scalars, from the half-hypermultiplet in $(\mathbf 7,\mathbf 8)$ of $G_2\times\Sp(4)$, are $\SU(2)_R$ doublets, so they form an $(\mathbf 8,\mathbf 2)$ of $\Sp(4)\times\SU(2)_R$.
\item \emph{A one-instanton state.} By \eqref{eq:faithfuloneinst} at $N=4$ and $\kCS=2$, the gauge-invariant one-instanton states have odd flavour parity and $R$-spin $\tfrac12$. They form the $\Lambda^3_0\mathbf 8=\mathbf{48}$ of $\Sp(4)$ in an $R$-doublet.\footnote{\label{fn:su42oneinst}By \eqref{eq:faithfulnb}, $n_{\mathsf b}=1$, so the $\mathsf b$-states are $(\mathbf 2',\mathbf 2_R)$, and, by \eqref{eq:faithfullparity}, $l$ is odd. For $l=1$, $\mathbf 8\otimes{\rm Sym}^3\mathbf 2'\otimes(\mathbf 2',\mathbf 2_R)$ has no $\SU(2)'$ singlet. For $l=3$, $\Lambda^3_0\mathbf 8\otimes\mathbf 2'\otimes(\mathbf 2',\mathbf 2_R)$ contains one, $\Lambda^3_0\mathbf 8\otimes\mathbf 2_R$, with $\dim\Lambda^3_0\mathbf 8=56-8=48$. This is up to the multiplet of the broken supercharges.}
\end{itemize}

\paragraph{Instanton parity, $\su(4)_6+2\Lambda^2$.}
Here, the 5d-to-5d frame map \eqref{eq:so8trifmap5d} gives the rule directly. As in \sref{sec:exso8tri}, primes denote the $\su(4)_6$ variables, and $e$, $m_0$, $M_i$ are those of $\sp(3)_0$. In \eqref{eq:so8trifmap5d}, the coefficients of $m_0$ are $(\tfrac14,\tfrac12,\tfrac34)$ in $P'$, $\tfrac12$ in $m_0'$ and $\tfrac12$ in each $M_i'$. Let an $\su(4)_6$ state have weight $\mathsf w$, instanton number $k'$ and flavour weight $\mathsf f'$. Its instanton number in the $\sp(3)_0$ frame is its coefficient of $m_0$:
\begin{equation}
\label{eq:faithfulksp}
k=\tfrac14\eta(\mathsf w)+\tfrac12\big(k'+|\mathsf f'|\big)\,.
\end{equation}
It is an integer, since $\pi_3(\Sp(3))=\BZ$. Hence, $\eta(\mathsf w)\equiv2(k'+|\mathsf f'|)$ mod $4$ for every state, whatever the origin of $\phi_0$. In 6d variables, $m_0'=\tfrac\tau6$ and $M_i'=\tfrac\tau6-m_i$ \eqref{eq:so8trisu4map}, so $(r,v,r_{\rm tw})=(6,\tfrac16,3)$ and $h=(e^{i\pi},-1)$, case~(a). Two witnesses confirm it. The gauge-neutral 6d hypermultiplet state of mass $m_i=m_0'-M_i'$ has $\mathsf w=0$, $k'=1$ and charge $-1$ under $M_i'$ only, so $(-1)^{k'}=(-1)^{|\mathsf f'|}=-1$. It is perturbative in the $\sp(3)_0$ frame. The bare one-instanton state of \cite[\S2.2.3]{Apruzzi:2021vcu} has $N$-ality $2$, $T$-charge $\kCS-4=2$ and even flavour parity. A gauge-invariant dressing needs an odd number of $\Lambda^2$ fields, e.g.\ the $T=-2$ scalar of one $\Lambda^2$ hypermultiplet. So the dressed state is odd, as footnote~\ref{fn:faithfulimone} predicts. In the $\sp(3)_0$ frame, $m_0=\tfrac\tau3$ and $M_i=m_i$, so $(r,v,r_{\rm tw})=(3,0,3)$ and $h$ is trivial, case~(c).

\paragraph{Verdict.}\phantomsection\label{par:faithfullockverdict}
In $\su(4)_2+4\Lambda^2$ and $\su(4)_6+2\Lambda^2$, the element $(i,-1,e^{i\pi})$ of the group $Z(\SU(4))\times Z(\Sp(N))\times\U(1)_I$ acts trivially on every state, by \eqref{eq:faithfulsu42parity} and by the integrality of \eqref{eq:faithfulksp}, respectively. On local operators, where $\mathsf w=0$, this is the lock \eqref{eq:faithfullockrule}: $-1\in\Sp(N)$ acts as $e^{i\pi}\in\U(1)_I$.\footnote{\label{fn:lockfuture}The lock rests on the frame-map parity and the witnesses above. The one-instanton spectra of $\su(4)_2+4\Lambda^2$ and $\su(4)_6+2\Lambda^2$ would show the flavour-odd $R$-doublets directly. We leave their computation to future work.} So $F/\mathcal Z_F=[\Sp(N)\times\U(1)_I]/\BZ_2$, with $N=4,2$, and $\mathbb G\cong\BZ_4$ is generated by $(i,-1,e^{i\pi})$. Its square generates $\BZ_2^{(1)}$, so \eqref{eq:faithfulext} does not split. But the $w_2$ of $F/\mathcal Z_F$ is the reduction of the integral class $c_1\big(\U(1)_I/\BZ_2\big)$, so ${\rm Bock}(w_2)=0$. This is the second row of \eqref{eq:verdicts}, a mixed anomaly with $\U(1)_I$, as for $\SU(4)_2+2\Lambda^2$ in \cite[Table~3]{Harding:2026gzj}.

Perturbation theory sees a 2-group instead. The perturbative generator $(i,-1)$ differs from $(i,-1,e^{i\pi})$ by $e^{i\pi}\in\U(1)_I$, so it acts as $(-1)^k$, or $(-1)^{k'}$ in $\su(4)_6$. It is trivial on the $k=0$ states, which include all states of mass well below the instanton mass, $m_0=\tfrac\tau2$ or $m_0'=\tfrac\tau6$, in the gauge-theory regime where the Coulomb moduli and the masses $M_i$ or $M_i'$ are much smaller than it. In that regime, the gauge theory therefore has the $\BZ_4$ 2-group of the \hyperref[par:faithfulsu4pert]{perturbative analysis} in \sref{sec:faithfulPSp4}, with flavour group $\Sp(N)/\BZ_2$. The $k=1$ witnesses are odd under $(i,-1)$, so this 2-group is accidental in the sense of \cite[fn.~3]{Zafrir:2025xca}, broken by instanton states as in \cite[fn.~6]{Zafrir:2025xca}. This differs from the accidental 1-form $\BZ_2$ of pure $\SU(4)_0$ written as an $\SU(2)$ gauging of two $E_1$ theories (\sref{sec:E1N2}), where instantons break the 1-form symmetry itself. Here, the 1-form symmetry survives, and instantons break only the relation that gave the perturbative 2-group its Postnikov class.

In $\sp(3)_0+2\Lambda^2$, $\mathcal Z_F$ is trivial, the third row of \eqref{eq:verdicts}.

\paragraph{Two frames of one KK theory.}\phantomsection\label{par:faithfultwoframes}
$\sp(3)_0+2\Lambda^2$ and $\su(4)_6+2\Lambda^2$ are frames of the same KK theory, yet their faithful flavour groups, $\Sp(2)\times\U(1)_I$ and $[\Sp(2)\times\U(1)_I]/\BZ_2$, are not isomorphic, with centres $\BZ_2\times\U(1)$ and $\U(1)$. There is no contradiction. First, the charges are the same, written in different bases. By \eqref{eq:so8trifmap5d} and \eqref{eq:faithfulksp} at $\eta=0$, the $\sp(3)_0$ charges $(\mathsf f,k)$ of a local operator become $(\mathsf f',k')=(-\mathsf f,\,2k+|\mathsf f|)$. This maps the weight lattice $\BZ^2\oplus\BZ$ of $\Sp(2)\times\U(1)_I$ onto the sublattice with $k'+|\mathsf f'|$ even, that of $[\Sp(2)\times\U(1)_I]/\BZ_2$. Second, each group is the symmetry unbroken at the frame's point, $M_i=0$ or $M'_i=0$, i.e.\ at the 6d masses $m_i=0$ and $m_i=\tfrac\tau6$, by \eqref{eq:so8trimap} and \eqref{eq:so8trisu4map}, with the diagonal and the twisted-diagonal $\Sp(2)$ as $F_\sigma$ (\hyperref[par:faithfullockpred]{6d predictions} above and footnote~\ref{fn:twisteddiag}). No symmetry relates these points: the affine Weyl group of the flavour currents $\mathbf{10}_0\oplus\mathbf{10}_{1/3}\oplus\mathbf{10}_{2/3}$ moves $m_i=0$ only within $\tfrac\tau3\BZ^2$. The translation by $(\tfrac\tau6,\tfrac\tau6)$ is a coweight translation, the holonomy $h_0$ with $h_0^3=-1\in\Sp(2)$. It would be a symmetry only if $-1\in\Sp(2)$ acted trivially, which the flavour-odd witness excludes. So the $\U(1)$ that commutes with the massless $\Sp(2)$ differs, and with it the global form. What the frames share is the element $-1\in\Sp(2)$: the two $\sp(2)$'s have the same Cartan generators, so the same $-1$, identified with $-1\in\SU(2)_R$, and neither frame has a 2-group.

% ---------------------------------------------------------------------------
\subsection{Results and relation to the literature}
\label{sec:faithfulresults}

Table~\ref{tab:faithfulresults} collects the results. The non-split $\BZ_4$ 2-group appears only after a twist.

\begin{table}[t]
\centering
\footnotesize
\setlength{\tabcolsep}{2.5pt}
\renewcommand{\arraystretch}{1.2}
\begin{tabular}{l|l|l|l|c|l}
5d frame & 6d prediction & 5d pert. & $F/\mathcal Z_F$ with $\U(1)_I$ & $\mathbb G$ & verdict\\ \hline
$\USp(4)_0+3\Lambda^2$ & $[\Sp(3)\times\SU(2)_R]/\BZ_2$ & same & $\Sp(3)\times\U(1)_I$ & $\BZ_2$ & anomaly with flavour\\
$\so(7)+5\Fund$ & $[\Sp(5)\times\SU(2)_R]/\BZ_2$ & same & $\Sp(5)\times\U(1)_I$ & $\BZ_2$ & anomaly with flavour\\ \hline
$\su(4)_0+4\Lambda^2$ & $\Sp(4)/\BZ_2\times\SO(3)_R$ & same & $\Sp(4)/\BZ_2\times\U(1)_I$ & $\BZ_4$ & 2-group, \eqref{eq:faithfulPostnikovPSp4}\\
$\su(4)_4+4\Lambda^2$ & $\Sp(4)/\BZ_2\times\SO(3)_R$ & same & $\Sp(4)/\BZ_2\times\U(1)_I$ & $\BZ_4$ & 2-group, \eqref{eq:faithfulPostnikovPSp4}\\ \hline
$\su(4)_2+4\Lambda^2$ & $[\Sp(4)\times\SU(2)_R]/\BZ_2$ & $\Sp(4)/\BZ_2\times\SO(3)_R$ & $[\Sp(4)\times\U(1)_I]/\BZ_2$ & $\BZ_4$ & anomaly with $\U(1)_I$\\
$\su(4)_6+2\Lambda^2$ & $[\Sp(2)\times\SU(2)_R]/\BZ_2$ & $\Sp(2)/\BZ_2\times\SO(3)_R$ & $[\Sp(2)\times\U(1)_I]/\BZ_2$ & $\BZ_4$ & anomaly with $\U(1)_I$\\
$\sp(3)_0+2\Lambda^2$ & $[\Sp(2)\times\SU(2)_R]/\BZ_2$ & same & $\Sp(2)\times\U(1)_I$ & $\BZ_2$ & anomaly with flavour
\end{tabular}
\caption{Faithful groups and 2-groups of the seven frames with $\BZ_2^{(1)}$. The 6d prediction at $M_i=0$ and the perturbative 5d group include $\SU(2)_R$ and leave out $\U(1)_I$ (``same'': equal to the prediction). $F/\mathcal Z_F$ includes $\U(1)_I$ and leaves out $\SU(2)_R$. $\mathbb G$ of \eqref{eq:faithfulext}, with instanton states included, is generated by $(i,-1)$ in the $\su(4)_{0,4}$ rows and by $(i,-1,e^{i\pi})$ in the $\su(4)_{2,6}$ rows. The verdict is that of \eqref{eq:verdicts} for the flavour part of $\mu$. For $\sp(3)_0+2\Lambda^2$, the $\epsilon_+$-exponent of $\mu$ is an integer, so, with $\SU(2)_R$ included, $\mu$ ends at $(y,z)=(1,0)$, which is not an allowed twist, unlike $(1,1)$ for $N$ odd. We do not pursue this.}
\label{tab:faithfulresults}
\end{table}

\paragraph{6d origin.}\phantomsection\label{par:faithful6dorigin}
The verdicts concern the full KK theory, instanton states included, not only the 5d gauge theory. Their ingredients come from 6d as follows. $\BZ_2^{(1)}$ is the defect group $D$ of the 6d theory on the circle. The instanton part $\fq^{1/4}$ of $\mu$ is the KK momentum $\tfrac12B^T\mathcal KB$ of the sector $B$, which the 7d Chern--Simons term gives through \eqref{eq:KKterm} and the dictionary \eqref{eq:dictionary}. The flavour part of $\mu$ comes from the odd flavour Green--Schwarz levels $\kappa_m$ in \eqref{eq:GS7d}. In terms of $\fq$ and the $\tilde y_i$, both parts are the same in all seven frames, so the Green--Schwarz levels alone do not tell the verdicts apart. What does is the faithful group $F/\mathcal Z_F$ of Table~\ref{tab:faithfulresults}, i.e.\ the image of $-1\in\Sp(N)$ in it:
\begin{itemize}
\item non-trivial and independent of $\U(1)_I$ in $\USp(4)_0+3\Lambda^2$, $\so(7)+5\Fund$ and $\sp(3)_0+2\Lambda^2$: a mixed anomaly with flavour;
\item equal to the image of $e^{i\pi}\in\U(1)_I$ in $\su(4)_2+4\Lambda^2$ and $\su(4)_6+2\Lambda^2$: a mixed anomaly with $\U(1)_I$;
\item trivial in $\su(4)_{0,4}+4\Lambda^2$: a 2-group. In 6d, $-1\in\Sp(4)$ is $-1\in\SU(4)_{\rm gauge}$ for $\su(4)+8\Fund$, and the identity of $E_6$ for the E-string glued to $\su(3)$ (\sref{sec:faithfulPSp4}).
\end{itemize}
The 6d parent fixes this image through $G_{\rm global}$ and the instanton parity of the frame map (\sref{sec:faithfuloddN}--\sref{sec:faithfullock}). How the 7d symmetry theory may produce the Postnikov class \eqref{eq:faithfulPostnikovPSp4} is discussed in \sref{sec:conclusions}.

\paragraph{Relation to the literature.}
Ref.~\cite[\S7]{Heckman:2022suy} proposes the tensor-branch method as a cross-check of the geometric computation of 1-form and 2-group symmetries of circle reductions \cite{Cvetic:2022imb}. This section extends it to twisted reductions. In untwisted reductions, the 1-form symmetry of tensor origin does not form a 2-group with the continuous flavour symmetry \cite[\S4.2, (4.21)--(4.22)]{Cvetic:2022imb}, in agreement with $\su(4)_2+4\Lambda^2$. It can form one with a discrete 0-form symmetry, as in the untwisted $\so(8)$ KK theory, \eqref{eq:so8twogroup} below. The 2-groups of $\su(4)_{0,4}+4\Lambda^2$ arise after a twist, and a geometric computation of $\mathbb G$ for twisted reductions would test them independently. They have no 6d 2-group as parent, since a 6d 2-group needs a 6d 1-form symmetry \cite{Apruzzi:2021mlh}. None of the 6d parents has one. $G_2$ has no centre, E-string states screen the centre of $\su(3)$ on the $-3$ curve, and the matter screens it in the other parents. Twists also change 2-group structure elsewhere: for the continuous 2-groups of little string theories \cite{DelZotto:2020sop,Cordova:2020tij}, a twist that permutes tensor multiplets changes the structure constants \cite[\S6.2]{DelZotto:2020sop}. Our twists permute no tensors; the 2-groups above come instead from their action on the flavour and gauge centres.

% =====================================================================
% Section 7: SU(2) gauging copies of E_1   (key E1)
% =====================================================================
\section{\texorpdfstring{$\SU(2)$}{SU(2)} gauging copies of \texorpdfstring{$E_1$}{E1}}
\label{sec:E1}

Three theories of this paper have a low-energy description of the same kind: an $\SU(2)$ gauge theory that gauges the diagonal $\SO(3)_I$ of $N$ copies of the $E_1$ SCFT \cite{Seiberg:1996bd}. Here, $\SO(3)_I$ is the enhanced instanton symmetry of pure $\SU(2)_0$. We call this description the E-frame. The three cases are
\begin{equation}
\label{eq:E1cases}
\mbox{\small$\begin{array}{c|l|l}
N & \text{theory} & \text{E-frame}\\ \hline
2 & \text{5d SCFT of pure }\SU(4)_0 & \text{\cite[app.~A]{Zafrir:2025xca}}\\
4 & \text{6d }\so(8)\text{ SCFT on a }{-4}\text{ curve, untwisted circle} & \text{\cite[\S3.2]{DelZotto:2015rca}, \cite[(5.3)]{Hayashi:2017jze}, \cite[\S2.1]{Zafrir:2025xca}}\\
3 & \text{5d SCFT of pure }\Spin(8) & \text{\cite[(2.3)]{Hayashi:2017jze}, \cite[(2.52)]{Bhardwaj:2019ngx}}
\end{array}$}
\end{equation}
Each $E_1$ has a $\BZ_2$ 1-form symmetry that forms a 2-group with $\SO(3)_I$ \cite{Apruzzi:2021vcu}. Once $\SO(3)_I$ is gauged, the 2-groups become extensions of 1-form symmetries, and part of the resulting 1-form symmetry is accidental \cite[\S2.1, app.~A]{Zafrir:2025xca}. Ref.~\cite{Zafrir:2025xca} matches the symmetries of the two descriptions with a mock symmetry TFT. Here, we derive the matching from the blow-up equations. This gives the frame maps, the image of every flux sector, and the states that break each accidental symmetry and the background to which it is locked. For $N=4$, it also gives $r=2$, the mixed anomalies and the action of the 0-form symmetry in the 2-group to which the $S_3$-invariant 3-group reduces.

We set up the E-frame in \sref{sec:E1toolkit} and treat $N=2$ in full in \sref{sec:E1N2}, where both frames are 5d gauge theories. \sref{sec:E1N4} and \sref{sec:E1N3} apply the same mechanisms to $N=4$ and $N=3$, and \sref{sec:E1summary} collects the results.

% ---------------------------------------------------------------------
\subsection{The toolkit}
\label{sec:E1toolkit}

\paragraph{Geometry and prepotential.} The E-frame is engineered by a compact surface $S_0=\BF_0$ and $N$ compact surfaces $S_i=\BF_2$. Each $S_i$ is glued along its $(-2)$-curve $e_i$ to a fibre $e$ of one ruling of $S_0$, and the $N$ gluing curves are disjoint. Let $F_0$ be the fibre of the other ruling ($e^2=F_0^2=0$, $e\cdot F_0=1$, $K_{S_0}=-2e-2F_0$) and $F_i$ the fibre of $S_i$. The rules $S^3=K_S^2$ and $S^2S'=K_S\cdot C_{SS'}$ \cite[(2.29), (2.30)]{Bhardwaj:2019ngx}, with $C_{SS'}\subset S$ the gluing curve, give
\begin{equation}
\label{eq:E1triple}
S_0^3=S_i^3=8\,,\qquad S_0^2S_i=-2\,,\qquad S_0S_i^2=0\,,\qquad S_iS_j=0\ \ (i\neq j)\,.
\end{equation}
For $N=2,3$, this is the geometry of pure $\SU(4)_0$ and of pure $\Spin(8)$ \cite[(2.44), (2.52)]{Bhardwaj:2019ngx}. Their gauge-theory frames read $S_0$ with the ruling of fibre $F_0$, and the E-frame reads it with the other ruling.

Let $f_0$, $f_i$ be the moduli of $S_0$, $S_i$ in the divisor basis, and $J=f_0S_0+\sum_if_iS_i$. Each $E_1$ is pure $\SU(2)_0$, with gauge variable $\varphi_i$ (W-boson mass $2\varphi_i$) and instanton mass $m_E$. In the conventions of \eqref{eq:E5d}, its prepotential is
\begin{equation}
\label{eq:E1block}
\epsilon_1\epsilon_2\,\mathcal E_{E_1}=m_E\varphi_i^2+\tfrac43\varphi_i^3-\tfrac1{24}m_E^3+c_\epsilon\big(\varphi_i+\tfrac14m_E\big)=\tfrac43f_i^3-\tfrac14m_E^2f_i+c_\epsilon f_i\,,
\end{equation}
with $c_\epsilon:=-\tfrac1{12}(\epsilon_1^2+\epsilon_2^2)+\epsilon_+^2$ and $f_i:=\varphi_i+\tfrac14m_E$. The flavour terms, $-\tfrac1{24}m_E^3$ and the levels $C^G_{m_E}=-1$, $C^R_{m_E}=\tfrac12$ in the last bracket, are fixed by the $\SO(3)_I$ Weyl reflection $(\varphi_i,m_E)\to(\varphi_i+\tfrac12m_E,-m_E)$, which leaves $f_i$ invariant. The geometry fixes $\varphi_i$ and $m_E$. Since $S_i\cdot F_i=-2$, $S_0\cdot F_i=1$, $S_i\cdot e_i=0$ and $S_0\cdot e_i=e_i^2=-2$, the volumes are $-J\cdot F_i=2f_i-f_0$ and $-J\cdot e_i=2f_0$. So $\varphi_i=f_i-\tfrac12f_0$ and $m_E=2f_0$, and the root of each $\SO(3)_I$ is the W-boson of the central $\SU(2)$. Adding the central block $m_0'f_0^2+\tfrac43f_0^3+c_\epsilon f_0$, with $m_0'$ the central instanton mass, gives
\begin{equation}
\label{eq:E1prep}
\epsilon_1\epsilon_2\,\mathcal E_{\rm E}=m_0'f_0^2+\tfrac43\Big(f_0^3+\sum_if_i^3\Big)-f_0^2\sum_if_i+c_\epsilon\Big(f_0+\sum_if_i\Big)\,,
\end{equation}
whose cubic part is $\tfrac16J^3$ by \eqref{eq:E1triple}. The Mori cone is generated by $e$, $F_0$ and the $F_i$:
\begin{equation}
\label{eq:E1masses}
\begin{array}{l|l|l}
\text{curve} & \text{mass} & \text{E-frame state}\\ \hline
e & 2f_0 & \text{W-boson of the central }\SU(2)\\
F_0 & m_0'+2f_0-\sum_if_i & \text{one-instanton of the central }\SU(2)\\
F_i & 2f_i-f_0 & \text{W-boson of the }i\text{-th }E_1\\
e_i+F_i & 2f_i+f_0 & \text{instanton of the }i\text{-th }E_1
\end{array}
\end{equation}
The mass of $F_0$ is $2f_0$ plus the effective coupling $m_0'-\sum_if_i$, the coefficient of $f_0^2$ in \eqref{eq:E1prep}. The last two rows differ by $e$, so they form a doublet of the central $\SU(2)$, i.e.\ of $\SU(2)_I$ before gauging, and both have charge $Q_i=2$ under $\varphi_i$. This is the $E_1$ 2-group in the language of states: $-1\in\SU(2)_I$ acts on them as $-1$, and the gauge element of the $i$-th $E_1$ that compensates it is $e^{i\pi Q_i/2}$, a quarter turn of its Cartan $\U(1)$, whose square $e^{i\pi Q_i}=-1\in Z(\SU(2))$ is trivial on all states of the $E_1$, since every $Q_i$ is even, and generates the $\BZ_2$ 1-form symmetry of the $E_1$, acting as $-1$ on lines of odd $Q_i$. So the lift of $-1\in\SU(2)_I$ has order $4$, as $(i,-1)$ in \sref{sec:faithfulPSp4}.\footnote{\label{fn:E1wilson}The same 2-group is visible on line operators. Inserting the fundamental Wilson loop breaks the $\BZ_2$ 1-form symmetry explicitly, and, through the 2-group, this turns $\SO(3)_I$ into $\SU(2)_I$ on the line. For pure $\SU(2)$ at $\theta=0$, the loop has only odd powers of $e^{\varphi}$ in every instanton sector \cite[(4.8)]{Kim:2021gyj}, since no instanton state screens it. Its states have odd $Q$, so $e^{i\pi Q}=-1$, and the gauge quarter turn no longer compensates $-1\in\SU(2)_I$. At $\theta=\pi$, the $k=1$ terms have even powers \cite[(4.9)]{Kim:2021gyj}: the instantons screen the loop, and there is no 1-form symmetry.}

\paragraph{The pairing.} Let $n_\xi$ be the flux through $E$ of a modulus or parameter $\xi$, as in \eqref{eq:Vdef}. In the E-frame, these are $n'_0$, $n'_i$ and $B':=B_{m_0'}$, the fluxes of $f_0$, $f_i$ and $m_0'$. We write $n'=(n'_0,n'_i)$. For a compact curve $C$ of mass $\sum_\xi c_\xi(C)\,\xi$, we define
\begin{equation}
\label{eq:pairingdef}
\langle n',C\rangle:=\sum_\xi c_\xi(C)\,n_\xi\,.
\end{equation}
The coefficient $c_\xi(C)$ is the charge of the BPS particle of an M2-brane on $C$ under the $\U(1)$ of $\xi$, whose flux through $E$ is $n_\xi$. So $\langle n',C\rangle$ is the total flux through $E$ seen by the particle on $C$, the background flux $B'$ included. The particle couples to a well-defined line bundle on $E$ only if
\begin{equation}
\label{eq:pairingrule}
\langle n',C\rangle\in\BZ\qquad\text{for every generator }C\text{ of the Mori cone}\,.
\end{equation}
This is the rule of \cite[(3.9)]{Kim:2020hhh}. Its half-integral case concerns curves of odd self-intersection, which do not occur here.

\paragraph{Exponents.} The part of $V$ in \eqref{eq:Vdef} linear in a variable $\xi$ comes from the cubic part of \eqref{eq:E1prep} alone, since the $c_\epsilon$ terms drop out. It equals $\tfrac12\,\xi\,\partial_\xi(n\cdot\partial)^2(\epsilon_1\epsilon_2\mathcal E)$, with $n\cdot\partial:=\sum_{\xi'}n_{\xi'}\partial_{\xi'}$. So the blow-up function contains $q'^{\,k'(n')}\exp\big(Z'_0f_0+\sum_iZ'_if_i\big)$, with $q':=e^{-m_0'}$ and
\begin{equation}
\label{eq:E1Zprime}
k'(n')=n_0'^{\,2}\,,\qquad Z'_0=-\langle n',e\rangle\,\langle n',F_0\rangle\,,\qquad Z'_i=-\langle n',F_i\rangle\,\langle n',e_i+F_i\rangle\,.
\end{equation}
Here, $k'(n')$ is the instanton number \eqref{eq:kQ} of the central flux.

\paragraph{Labels.}\phantomsection\label{par:E1labels} Each $\SU(2)$ gauge flux is a coweight. So the flux $n'_i-\tfrac12n'_0$ of $\varphi_i$ and the central flux $n'_0$ lie in $\tfrac12\BZ$, and $n'_i\in\tfrac14\BZ$. The background flux $B'$ lies in $\tfrac1d\BZ$, with $d$ fixed by the frame map. We label a flux by
\begin{equation}
\label{eq:E1labels}
X_i:=4n'_i\ \mathrm{mod}\ 4\,,\qquad \hat B:=2n'_0\ \mathrm{mod}\ 2\,,\qquad B'\ \mathrm{mod}\ 1\,,
\end{equation}
with representatives $X_i\in\{0,1,2,3\}$ and $\hat B\in\{0,1\}$. They are the values on the exceptional curve $E$ of the backgrounds $\hat{\mathcal B}+2\mathcal B_i$ (the $\BZ_4$ class of the $i$-th $E_1$, see below) and $\hat{\mathcal B}$ (the 't~Hooft flux of the central $\SU(2)$). The instanton symmetry of the central $\SU(2)$ is $\U(1)_I'$, and $B'$ mod $1$ records whether its background is a $\U(1)_I'$ bundle. The labels are invariant under integer shifts of $n'$ and under the central Weyl reflection $f_0\to-f_0$. The Weyl reflection $f_i\to f_0-f_i$ of the $i$-th $E_1$ maps $X_i\to2\hat B-X_i$, so it preserves the labels on the classes with $X_i\equiv\hat B$ mod $2$. These are the classes that the $E_1$ 2-groups select in \eqref{eq:E1group} below.

\paragraph{The \texorpdfstring{$E_1$}{E1} 2-groups.} By \eqref{eq:E1masses}, $\langle n',F_i\rangle=2n'_i-n'_0\equiv\tfrac12(X_i-\hat B)$ mod $1$. The pairing $\langle n',e_i+F_i\rangle=2n'_i+n'_0$ differs from it by the integer $2n'_0=\langle n',e\rangle$. So the two are both integral or both in $\BZ+\tfrac12$, and $Z'_i$ of \eqref{eq:E1Zprime} is an integer exactly when $\langle n',F_i\rangle$ is. Both conditions are equivalent to
\begin{equation}
\label{eq:E1group}
X_i\equiv\hat B\ \ \mathrm{mod}\ 2\,.
\end{equation}
A $\BZ_4$ class of order $4$ for an $E_1$ thus requires a half-integral flux of the gauged $\SO(3)_I$. This is the 2-group $\delta\mathcal B_i={\rm Bock}(\hat{\mathcal B})$ of the $i$-th $E_1$ \cite{Apruzzi:2021vcu}, \cite[(A.3)--(A.5)]{Zafrir:2025xca}, as a statement about fluxes, with ${\rm Bock}$ the Bockstein map of $0\to\BZ_2\to\BZ_4\to\BZ_2\to0$. Since $X_i$ is the value of $\hat{\mathcal B}+2\mathcal B_i$ on $E$, the value of $\mathcal B_i$ on $E$ is
\begin{equation}
\label{eq:Bidef}
B_i:=\tfrac12\big(X_i-\hat B\big)\ \ \mathrm{mod}\ 2\,,
\end{equation}
with the same orientation for every $E_1$. The remaining tools hold for every $N$, and we introduce them on the example $N=2$.

% ---------------------------------------------------------------------
\subsection{Two copies: \texorpdfstring{$\SU(4)_0$}{SU(4)0}}
\label{sec:E1N2}

\paragraph{The \texorpdfstring{$\SU(4)_0$}{SU(4)0} frame.} In Dynkin variables $\phi_i$ and in the Weyl chamber,
\begin{equation}
\label{eq:su4zprep}
\epsilon_1\epsilon_2\,\mathcal E_{\SU(4)_0}=m_0\cdot\tfrac12\phi^TC\phi+\tfrac43\big(\phi_1^3+\phi_2^3+\phi_3^3\big)-\phi_2^2(\phi_1+\phi_3)+c_\epsilon(\phi_1+\phi_2+\phi_3)\,,
\end{equation}
with $C$ the Cartan matrix and $c_\epsilon$ of \eqref{eq:E1block}. Its blow-up function is \eqref{eq:expV} at $\kCS=0$, with $\Bmz\in\BZ$ \cite[(3.2), (3.3), (3.14)]{Harding:2026gzj}. The gauge flux $n$ lies in a coset $\omega_j$, $j\in\BZ_4$, of the coweight lattice modulo the coroot lattice, with representatives $n\equiv(\tfrac j4,\tfrac j2,\tfrac{3j}4)$ mod $1$ \cite[(3.10), (3.11)]{Harding:2026gzj}. Its instanton number \eqref{eq:kQ} obeys
\begin{equation}
\label{eq:su4zk}
k(\omega_j)\equiv\tfrac38j^2\ \ \mathrm{mod}\ 1\,.
\end{equation}
This is the fractional instanton number $-\tfrac38\int\mathcal P(\mathcal X)$ of a $\PSU(4)$ bundle, with $\mathcal X$ the background of the 1-form $\BZ_4$, of value $j$ on $E$, and $\int\mathcal P(\mathcal X)=j^2\,E\cdot E=-j^2$. Since $\U(1)_I$ couples to the instanton number, a shift of its holonomy by a full period multiplies the sector $\omega_j$ by $e^{2\pi ik(\omega_j)}$. This is the mixed anomaly $\tfrac{3\pi i}4\int\tilde{\mathcal C}_1\cup\mathcal P(\mathcal X)$ of \cite[(A.1)]{Zafrir:2025xca}, with $\tilde{\mathcal C}_1$ the mod-$8$ reduction of $c_1(\U(1)_I)$; only $c_1$ mod $8$ enters, since $\mathcal P(\mathcal X)$ is defined mod $8$. The forced flux $q^{3/8}$ at $\omega_1$ is a fractional flux of $\U(1)_I$, i.e.\ of a $\U(1)_I/\BZ_8$ bundle. Since $\U(1)_I$ is not enhanced, the class of this bundle is the reduction of the integral class $c_1$, so its Bockstein vanishes and it cannot be a Postnikov class. By the criterion of \cite[\S\S5.1, 6.1]{Harding:2026gzj}, it gives this mixed anomaly and no 2-group.

\paragraph{The frame map.} At $N=2$, the geometry \eqref{eq:E1triple} is $\BF_2\cup\BF_0\cup\BF_2$ of $\SU(4)_0$ \cite[(2.44)]{Bhardwaj:2019ngx}, with $S_1$, $S_0$, $S_2$ the surfaces of $\phi_1$, $\phi_2$, $\phi_3$. The $\SU(4)$ frame reads every surface with its fibre. So $F_1$, $F_0$, $F_2$ are the W-bosons of $\alpha_1$, $\alpha_2$, $\alpha_3$, and $e$ is a one-instanton state of weight $(0,2,0)$. The ansatz is $\phi=Mf+s\,m_0'$ and $m_0=u\,m_0'$, with $f=(f_1,f_0,f_2)$. Step~1 of \tref{tab:framemaprecipe} fixes $M$ to the identity, up to the quiver reflection $f_1\leftrightarrow f_2$ (Appendix~\ref{app:su40}). Step~2, the monomials $f_if_jm_0'$, then fixes $(s,u)$ uniquely:
\begin{equation}
\label{eq:su4zmap}
\phi_1=f_1+\tfrac12m_0'\,,\qquad \phi_2=f_0+m_0'\,,\qquad \phi_3=f_2+\tfrac12m_0'\,,\qquad m_0=-2m_0'\,.
\end{equation}
The prepotentials then differ by $\epsilon_1\epsilon_2(\mathcal E_{\SU(4)_0}-\mathcal E_{\rm E})=2c_\epsilon m_0'-\tfrac13m_0'^3$, which does not depend on the moduli. \hyperref[it:volmatch]{Volume matching} of $F_1$, $F_2$, $F_0$ and $e$ gives the same map. Since $m_0=-2m_0'$, the weak-coupling regions of the two frames lie at opposite signs of the single parameter \cite[app.~A.1]{Zafrir:2025xca}.

A state of $\SU(4)$ weight $\mathsf w$, in Dynkin labels, and instanton number $k$ has mass $\mathsf w\cdot\phi+km_0$. By \eqref{eq:su4zmap}, its E-frame mass is $\mathsf w_1f_1+\mathsf w_2f_0+\mathsf w_3f_2+k'm_0'$, with
\begin{equation}
\label{eq:su4zcharge}
k'=\tfrac12(\mathsf w_1+\mathsf w_3)+\mathsf w_2-2k\,.
\end{equation}
Here, $k'$ is the $\U(1)_I'$ charge, the instanton number of the central $\SU(2)$. A state of zero weight has $k'=-2k$. So $\U(1)_I=\U(1)_I'/\BZ_2$, and the fluxes obey $B'=-\tfrac12\Bmz$.

\paragraph{The dictionary.} Fluxes transform like the variables (the \hyperref[it:fluxmap]{flux rule} of \sref{sec:framemap}). So \eqref{eq:su4zmap} gives
\begin{equation}
\label{eq:su4zfluxmap}
n'_0=n_2+\tfrac12\Bmz\,,\qquad n'_1=n_1+\tfrac14\Bmz\,,\qquad n'_2=n_3+\tfrac14\Bmz\,,\qquad B'=-\tfrac12\Bmz\,,
\end{equation}
hence $d=2$. With the representatives of $\omega_j$, the labels \eqref{eq:E1labels} are
\begin{equation}
\label{eq:su4zdict}
(X_1,X_2,\hat B;B')=\big(j+\Bmz,\ -j+\Bmz,\ j+\Bmz;\ -\tfrac12\Bmz\big)\,,
\end{equation}
independent of the representatives. For example, at $\Bmz=0$, the flux $n=(\tfrac14,\tfrac12,\tfrac34)\in\omega_1$ maps to $n'=(n'_0;n'_1,n'_2)=(\tfrac12;\tfrac14,\tfrac34)$, with labels $(1,3,1)$, $k=\tfrac38$ by \eqref{eq:su4zk} and $k'(n')=n_0'^{\,2}=\tfrac14$ by \eqref{eq:E1Zprime}. The sector $\omega_2$, with $n=(\tfrac12,1,\tfrac32)$, maps to $n'=(1;\tfrac12,\tfrac32)$, with labels $(2,2,0)$, $k=\tfrac32$ and $k'(n')=1$, i.e.\ $k\equiv\tfrac12$ and $k'\equiv0$ mod $1$.

\paragraph{The \texorpdfstring{$\BZ_4$}{Z4} from the \texorpdfstring{$E_1$}{E1} 2-groups.} Perturbatively, the E-frame has three $\BZ_2$ 1-form symmetries, with backgrounds $\hat{\mathcal B}$, $\mathcal B_1$ and $\mathcal B_2$. The two 2-groups \eqref{eq:E1group} combine them into $\BZ_4\times\BZ_2$, the eight classes $(X_1,X_2,\hat B)$ with $X_i\equiv\hat B$ mod $2$. The generator $\omega_1$ maps to $(1,3,1)$. So the 1-form $\BZ_4$ of $\SU(4)_0$ has the value $X:=X_1=\hat B+2B_1$, by \eqref{eq:Bidef}. This is the value of $\mathcal X=\hat{\mathcal B}+2\mathcal B_1$ below \cite[(A.5)]{Zafrir:2025xca}. A class of order $4$ exists only because the 2-group of the first $E_1$ ties $\mathcal B_1$ to the 't~Hooft flux of the central $\SU(2)$. So the non-split extension $0\to\BZ_2\to\BZ_4\to\BZ_2\to0$ is the image of the $E_1$ 2-groups once $\SO(3)_I$ is gauged, the central point of \cite[app.~A]{Zafrir:2025xca}. At $\Bmz=0$, \eqref{eq:su4zdict} gives $X_2=-X_1$, so the $\BZ_4$ couples to the difference of the two $E_1$ charges in our orientation, as below \cite[(A.10)]{Zafrir:2025xca}. The quiver reflection $f_1\leftrightarrow f_2$ is charge conjugation, $\phi_1\leftrightarrow\phi_3$, which maps $j\to-j$. We now identify the remaining $\BZ_2$.

\paragraph{The pairing with the central instanton.} Take the generator $F_0$ of \eqref{eq:E1masses}, the central one-instanton. For any $N$, with $2n'_0\in\BZ$ and $n'_i\equiv\tfrac14X_i$ mod $1$, its pairing is $\langle n',F_0\rangle=B'+2n'_0-\sum_in'_i\equiv B'-\tfrac14\sum_iX_i$ mod $1$. So \eqref{eq:pairingrule} requires
\begin{equation}
\label{eq:E1paircond}
\sum_iX_i\equiv4B'\ \ \mathrm{mod}\ 4\,.
\end{equation}
By \eqref{eq:Bidef}, $\sum_iX_i\equiv N\hat B+2\sum_iB_i$ mod $4$, and \eqref{eq:E1paircond} becomes
\begin{equation}
\label{eq:E1quarters}
\tfrac N4\,\hat B+\tfrac12\sum_iB_i\equiv B'\ \ \mathrm{mod}\ 1\,.
\end{equation}
The labels allowed by \eqref{eq:E1group} form $2^{N+1}$ classes of $(X_i;\hat B)$, each with $d$ values of $B'$ mod $1$. In all three cases, $\tfrac14\sum_iX_i\in\tfrac1d\BZ$, so \eqref{eq:E1paircond} fixes $B'$ mod $1$ and keeps $2^{N+1}$ of the $2^{N+1}d$ classes. We find, in each case, that these are exactly the images of the sectors of the other frame. At $N=2$, $F_0$ is the W-boson of $\alpha_2$, and, by \eqref{eq:su4zmap}, $\langle n',F_0\rangle=\langle n,\alpha_2\rangle = (n_2-n_1)-(n_3-n_2)=2n_2-n_1-n_3$, an integer for every coweight, since $n_{i+1}-n_i\equiv\tfrac j4$ mod $1$ on $\omega_j$.

The integrality of $Z'$ is weaker. By \eqref{eq:E1group}, $Z'_i\in\BZ$ on all $16=2^{N+1}d$ classes allowed at $N=2$, where $d=2$ by \eqref{eq:su4zfluxmap}. In $Z'_0=-\langle n',e\rangle\langle n',F_0\rangle$, the first factor $2n'_0$ is even at $\hat B=0$, so all $8$ classes pass. At $\hat B=1$, it is odd, and $Z'_0\in\BZ$ is \eqref{eq:E1paircond}, which keeps $4$ of the $8$. So $Z'\in\BZ$ keeps $12$ classes, and \eqref{eq:E1paircond} keeps $8$:
\begin{equation}
\label{eq:su4zclasses}
\mbox{\small$\begin{array}{c|c|c|c|l}
\hat B & (X_1,X_2) & B' & \hat B+B_1+B_2 & \SU(4)_0\text{ sector}\\ \hline
0 & (0,0),\ (2,2) & \BZ & 0 & \omega_0,\ \omega_2;\ \Bmz=0\\
0 & (2,0),\ (0,2) & \BZ+\tfrac12 & 1 & \omega_1,\ \omega_3;\ \Bmz=1\\
1 & (1,3),\ (3,1) & \BZ & 0 & \omega_1,\ \omega_3;\ \Bmz=0\\
1 & (1,1),\ (3,3) & \BZ+\tfrac12 & 1 & \omega_0,\ \omega_2;\ \Bmz=1\\ \hline
0 & (2,0),\ (0,2) & \BZ & 1 & \text{none}\\
0 & (0,0),\ (2,2) & \BZ+\tfrac12 & 0 & \text{none}
\end{array}$}
\end{equation}
The last column lists, in order, the sectors that map to the classes by \eqref{eq:su4zdict}. The eight classes above the line are the images. Only \eqref{eq:E1paircond} excludes the four below it. In $\SU(4)$ variables, they are not coweights. For example, the class $(X_1,X_2,\hat B)=(2,0,0)$ with $B'\in\BZ$ contains the flux $n=(n_1,n_2,n_3)=(\tfrac12,0,0)$ at $\Bmz=0$, whose gauge exponents are the integers $(-1,0,0)$ \cite[(3.9b)--(3.9e)]{Harding:2026gzj}, while $\langle n,\alpha_2\rangle=-\tfrac12$. So \eqref{eq:E1paircond} is the image of the coweight condition of \cite[\S3.1]{Harding:2026gzj}, which is stronger than the integrality of the gauge exponents.

\paragraph{Charges.} Let a state have E-frame mass $k'm_0'+\mathsf c_0f_0+\sum_i\mathsf c_if_i$, with $k'$ its $\U(1)_I'$ charge and $\mathsf c_i$ its charge $Q_i$ under $\varphi_i$ (\sref{sec:E1toolkit}). By \eqref{eq:pairingdef}, it pairs with a flux of labels \eqref{eq:E1labels} as
\begin{equation}
\label{eq:E1charges}
\langle n',C\rangle\equiv k'B'+\tfrac12\mathsf c_0\hat B+\tfrac14\sum_i\mathsf c_iX_i\ \ \mathrm{mod}\ 1\,,
\end{equation}
and the coefficient of each label is the charge of the state under the corresponding symmetry. The generators \eqref{eq:E1masses} have $(k',\mathsf c_i)=(0,0)$ for $e$, $(1,-1)$ for $F_0$, and $k'=0$, $\mathsf c_i\in\{0,2\}$ for $F_j$. Since charges add, for any $N$,
\begin{equation}
\label{eq:E1parity}
\mathsf c_i\equiv k'\ \ \mathrm{mod}\ 2\qquad\text{for every }i\text{ and every BPS state}\,.
\end{equation}
At $N=2$, it also follows from the centre charge of $\SU(4)$ (Appendix~\ref{app:su40}).

\paragraph{The accidental \texorpdfstring{$\BZ_2$}{Z2}.} By \eqref{eq:E1parity}, perturbative states ($k'=0$) have even $\mathsf c_i$ and see $X_i$ only mod $2$, i.e.\ only $\hat B$. For them, all $16$ classes allowed by \eqref{eq:E1group} are admissible. By \eqref{eq:E1paircond}, the images are the classes with $\tfrac14(X_1+X_2)\equiv B'$ mod $1$. Each of the shifts $\delta n'_1=\tfrac12$, $\delta n'_2=\tfrac12$ and $\delta B'=\tfrac12$ preserves \eqref{eq:E1group} and changes $\tfrac14(X_1+X_2)-B'$ by $\tfrac12$, whereas a shift between two images changes it by $0$ mod $1$. So each of the three maps the images onto the excluded classes. A combination $\gamma_1\,\delta n'_1+\gamma_2\,\delta n'_2+\gamma_3\,\delta B'$, with $\gamma_1,\gamma_2,\gamma_3\in\{0,1\}$, changes the pairing of a state by $\tfrac12(\gamma_1\mathsf c_1+\gamma_2\mathsf c_2+\gamma_3k')\equiv\tfrac12(\gamma_1+\gamma_2+\gamma_3)k'$ mod $1$. By \eqref{eq:Bidef}, it changes $\hat B+B_1+B_2$ by $\gamma_1+\gamma_2$, and it changes $2B'$ by $\gamma_3$. So it pairs integrally with every state if and only if $\gamma_1+\gamma_2+\gamma_3$ is even, i.e.\ if and only if it preserves
\begin{equation}
\label{eq:su4zlock}
\hat B+B_1+B_2\equiv2B'\equiv\Bmz\ \ \mathrm{mod}\ 2\,.
\end{equation}
This is \eqref{eq:E1quarters} at $N=2$, and the fourth column of \eqref{eq:su4zclasses} shows it on every image. Three conclusions follow.
\begin{itemize}
\item \phantomsection\label{it:su4zaccidental}\emph{Accidental.} The $\BZ_2$ with background $\hat{\mathcal B}+\mathcal B_1+\mathcal B_2$ is a 1-form symmetry of the perturbative E-frame. Switching on its background alone, with $\gamma_1+\gamma_2=1$ and $\gamma_3=0$, changes the pairing of a state by $\tfrac12k'$ mod $1$. So the central instantons of odd $k'$ are charged under it. Hence, it holds only below the instanton mass, of order $m_0'$, and is not a 1-form symmetry of the SCFT. This is what accidental means in \cite[fn.~3]{Zafrir:2025xca}. One screening state is $F_0$, in $\SU(4)$ variables the W-boson of $\alpha_2$.
\item \phantomsection\label{it:su4zlocked}\emph{Locked.} Its background cannot be switched on independently of the $\U(1)_I$ flux, since its value is $\Bmz$ mod $2$. An odd $\Bmz$ gives $B'\in\BZ+\tfrac12$, a flux of $\U(1)_I=\U(1)_I'/\BZ_2$ that does not lift to $\U(1)_I'$. The states of odd $k'$ pair half-integrally with it, and $\hat B+B_1+B_2=1$ restores integrality. We say that the accidental $\BZ_2$ is locked to $\U(1)_I$.
\item \phantomsection\label{it:su4zcomparison}\emph{Comparison.} Ref.~\cite[(A.4)]{Zafrir:2025xca} defines $\mathcal B_2^+:=\mathcal B_1+\mathcal B_2$ and calls it accidental, without naming the states that break it. With our $\mathcal B_2$, this is not the accidental background: by \eqref{eq:Bidef} and \eqref{eq:su4zdict}, $B_1+B_2\equiv j$ mod $2$ on the image of $\omega_j$. Reversing the orientation of the second $E_1$, $X_2\to-X_2$, replaces $\mathcal B_2$ by $\mathcal B_2+\hat{\mathcal B}$.\footnote{\label{fn:orientflip}With the orientation reversed, \eqref{eq:Bidef} gives the background of the second $E_1$ as $\tfrac12(-X_2-\hat B)=\tfrac12(X_2-\hat B)-X_2=B_2-X_2$ mod $2$. By \eqref{eq:E1group}, $-X_2\equiv X_2\equiv\hat B$ mod $2$, so it is $B_2+\hat B$.} In that orientation, $\mathcal B_2^+=\mathcal B_1+(\mathcal B_2+\hat{\mathcal B})=\hat{\mathcal B}+\mathcal B_1+\mathcal B_2$ is the background of the accidental $\BZ_2$, locked to $\U(1)_I$ by \eqref{eq:su4zlock}.\footnote{\label{fn:zafrirorient}The $\BZ_2$ backgrounds of \cite[app.~A]{Zafrir:2025xca} do not fix this orientation. In \sref{sec:E1N4}, the analogous $\mathcal B_i^+:=\mathcal B_i+\mathcal B_1$ of \cite[(2.3)]{Zafrir:2025xca} need no reversal.} The states that break it are the central instantons of odd $k'$, by the \hyperref[it:su4zaccidental]{\emph{Accidental} item}, as in \cite[fn.~6]{Zafrir:2025xca} for $\SU(2)_\pi$.
\end{itemize}

\paragraph{The role of the 2-groups.} The argument uses the $E_1$ 2-groups in two places: the classes allowed by \eqref{eq:E1group}, and the backgrounds $B_i$ of \eqref{eq:Bidef}, which exist only because $X_i\equiv\hat B$ mod $2$. The charges \eqref{eq:E1parity} do not use them. Without the 2-groups, $\hat B$, $B_1$ and $B_2$ would be independent, and the pairing with $F_0$ would require $B_1+B_2\equiv2B'$ mod $2$. The SCFT would then keep $\BZ_2\times\BZ_2$, with backgrounds $\hat{\mathcal B}$ and $\mathcal B_1=\mathcal B_2$, and the accidental background would be $\mathcal B_1+\mathcal B_2$. The 2-groups put $\hat{\mathcal B}$ into the accidental background and combine the surviving symmetries into the $\BZ_4$ found above. So they fix which combination is accidental, and that the 1-form symmetry of the SCFT is $\BZ_4$ rather than $\BZ_2^2$.

\paragraph{\texorpdfstring{$\theta=0$}{theta=0}.}\phantomsection\label{par:theta0} On the trivial class, $X_i=\hat B=0$, \eqref{eq:E1paircond} requires $B'\in\BZ$. This is the rule for $\SU(2)_0$ \cite[(3.14)]{Harding:2026gzj}, while $\SU(2)_\pi$ would need $B'\in\BZ+\tfrac12$. The integrality of $Z'$ does not decide, since $Z'_0\in\BZ$ at $\hat B=0$. The rule \eqref{eq:pairingrule} assumed $F_0^2=0$, i.e.\ $S_0=\BF_0$, but the other frame confirms it: the trivial class with $B'\in\BZ$ is the image of its zero flux, and the trivial class with $B'\in\BZ+\tfrac12$ is not an image, \eqref{eq:su4zclasses}. So the central $\SU(2)$ has $\theta=0$, in agreement with $S_0=\BF_0$ and \cite[fn.~28]{Zafrir:2025xca}. The same holds for $N=4$ and $N=3$ below, where the images are again the classes that obey \eqref{eq:E1paircond}.

\paragraph{Anomalies.} By \eqref{eq:E1Zprime}, $k'(n')=n_0'^2\equiv\tfrac14\hat B^2$ mod $1$. This is the anomaly $\tfrac{\pi i}2\int\tilde{\mathcal C}_1'\cup\mathcal P(\hat{\mathcal B})$ of the central $\SU(2)$ \cite[(A.6)]{Zafrir:2025xca}, with $\tilde{\mathcal C}_1'$ the mod-$4$ reduction of $c_1(\U(1)_I')$. Set $\tilde{\mathcal C}_1=\zeta\,\tilde{\mathcal C}_1'$. Then \eqref{eq:su4zk} and $k'(n')$ agree on every class if and only if $\zeta\cdot\tfrac38X^2\equiv\tfrac14\hat B^2$ mod $1$. Since $X\equiv\hat B$ mod $2$, $X^2\equiv\hat B^2$ mod $4$, so this is $\tfrac18(3\zeta-2)X^2\in\BZ$. It holds for every $X$ if it holds at $X=1$, i.e.\ if $3\zeta\equiv2$ mod $8$. Multiplying by $3$, with $3\cdot3\equiv1$ mod $8$,
\begin{equation}
\label{eq:su4zmatching}
\zeta\equiv-2\ \ \mathrm{mod}\ 8\,.
\end{equation}
This is $m_0=-2m_0'$, sign included, and the identification $2\tilde{\mathcal C}_1'=-\tilde{\mathcal C}_1$ of \cite[(A.6)]{Zafrir:2025xca}. It is a consistency check, not an independent test, since exponents transform with the variables. The generator $(1,3,1)$ has $B'\in\BZ$ and forced flux $q'^{1/4}$ (see below \eqref{eq:su4zdict}), which is $q^{3/8}=q'^{-3/4}$ up to an integer power of $q'$. On local operators, $\U(1)_I'$ acts as the Abelian $\U(1)_I$, so the E-frame also gives a mixed anomaly and no 2-group. Appendix~\ref{app:su40} compares the blow-up with the individual terms of these anomalies in \cite[(A.2)]{Zafrir:2025xca}, and with the anomalies of the two $E_1$'s \cite[(A.7), (A.8)]{Zafrir:2025xca}.

% ---------------------------------------------------------------------
\subsection{Four copies: the untwisted \texorpdfstring{$\so(8)$}{so(8)} KK theory}
\label{sec:E1N4}

The untwisted $\so(8)$ KK theory is the one example of this paper in which the whole 6d 1-form symmetry $Z(\Spin(8))$, with its 3-group, survives on the circle. With the triality twist, none of it survives (\sref{sec:exSU48}), and, with the $\BZ_2$ twist, only $\{1,v\}$ (\sref{sec:exso8Z2}). Now, one frame is 6d, so the frame map also fixes $r$, and each E-frame label acquires a 6d origin.

\paragraph{6d data and the 3-group.}\phantomsection\label{sec:E1N4sixd} The tensor branch is that of $\SU(4)_8$ (\sref{sec:exSU48}), the reduction with the triality twist. The symmetries are
\begin{equation}
\label{eq:so8data}
\mbox{\small$\begin{array}{l|l}
\text{2-form symmetry} & D=\BZ_4,\ q(j)=\tfrac{j^2}8\ (\mathcal K=4);\ \text{background }\mathcal C_3\\
\text{1-form symmetry} & Z(\Spin(8))=\{1,z_v,z_s,z_c\}\cong\BZ_2\times\BZ_2;\ \text{backgrounds }\mathcal B_s,\ \mathcal B_c\\
\text{0-form symmetry} & S_3\ \text{(outer automorphisms, permuting }v,s,c);\ G_{\rm global}=\SO(3)_R
\end{array}$}
\end{equation}
Here, $q(j)$ is \eqref{eq:qdef}, $z_x$ acts trivially on $\mathbf 8_x$, and the gauge group is $\Spin(8)$ as in \cite[\S2]{Zafrir:2025xca}. No field is charged under $Z(\Spin(8))$, and $-1\in\SU(2)_R$ comes with the Lorentz twist, \eqref{eq:centretwist}. The Green--Schwarz four-form $4c_2(\Spin(8))+\tfrac12p_1(T)+6c_2(R)$ stays integral \cite[(3.6), (3.65), (3.66)]{Heckman:2022suy},\footnote{\label{fn:so8GS}With $w_2:=w_2(R)=w_2(T)$ the class of the twist, \cite[(3.65), (3.66)]{Heckman:2022suy} give $\tfrac12p_1(T)\equiv\tfrac32\mathcal P(w_2)$ and $6c_2(R)\equiv-\tfrac32\mathcal P(w_2)$ mod $1$, which cancel. With a background for $Z(\Spin(8))$, $c_2(\Spin(8))$ has fractional parts in $\tfrac12\BZ$ \cite[(3.6)]{Heckman:2022suy}, which the coefficient $4$ removes.} so the 1-form symmetry is all of $Z(\Spin(8))$.

With 't~Hooft flux $\mathcal B=(\mathcal B_s,\mathcal B_c)$, the instanton number is fractional, $c_2(\Spin(8))\equiv-\tfrac12\mathcal P(\mathcal B_s+\mathcal B_c)-\tfrac12\mathcal B_s\cup\mathcal B_c$ mod $1$ \cite[(3.6)]{Heckman:2022suy}. The tensor couples to $\mathcal Kc_2=4c_2$, so the 2-form background shifts:
\begin{equation}
\label{eq:so8threegroup}
\delta\mathcal C_3=\eta(\mathcal B)\,,\qquad \eta(\mathcal B):=2\mathcal P(\mathcal B_s+\mathcal B_c)+2\,\mathcal B_s\cup\mathcal B_c\ \ \mathrm{mod}\ 4\,.
\end{equation}
This is \cite[(2.10)]{Apruzzi:2022dlm} for $\so(8)$ on a $-4$ curve. As a function on $Z(\Spin(8))$, $\eta$ takes the value $\mathcal K\cdot\tfrac12=2$ on each of $v,s,c$, because the fractional instanton number $\tfrac12n^TCn$ is $\tfrac12$ on each class, with $C$ the Cartan matrix and $n$ the representative coweight of the class, listed in \eqref{eq:so8cosets} below. So \eqref{eq:so8threegroup} is $S_3$-invariant. Ref.~\cite[(2.5)]{Zafrir:2025xca} quotes the relation with coefficient $1$ in front of $\mathcal P(\mathcal B_s+\mathcal B_c)$. That form takes the values $1,1,2$ on $s,c,v$ and is not $S_3$-invariant,\footnote{\label{fn:so8values}As functions on $Z(\Spin(8))$, with $(\mathcal B_s,\mathcal B_c)=(1,0),(0,1),(1,1)$ for $s,c,v$. At $v$, the first term vanishes, since $\mathcal B_s+\mathcal B_c\equiv0$ mod $2$, and the cross term gives $2$.} so the coefficient should be $2$, as in \eqref{eq:so8threegroup} and \cite[(2.10)]{Apruzzi:2022dlm}.

On the circle, the 1-form $\BZ_2\times\BZ_2$ gives a 5d 1-form $\BZ_2\times\BZ_2$ (backgrounds $\mathcal B_x$) and a 5d 0-form $\BZ_2\times\BZ_2$ (holonomies, backgrounds $\mathcal A_x$), $x=s,c$. The 2-form $\BZ_4$ gives a 5d 1-form $\BZ_4$ (background $\mathcal X$) in the polarisation of \cite[fn.~5]{Zafrir:2025xca}. The $S_3$ and the 0-form $\BZ_2\times\BZ_2$ combine into $S_4=(\BZ_2\times\BZ_2)\rtimes S_3$, and the continuous symmetry is $\SO(3)_R\times\U(1)_{\rm KK}$ \cite[\S2.1]{Zafrir:2025xca}.

\paragraph{The frame map.}\phantomsection\label{sec:E1N4frame} Let $\phi_0$ be the tensor scalar and $\phi=(\phi_1,\dots,\phi_4)$ the $\so(8)$ moduli in the Dynkin basis, node $2$ central, so that $\tfrac12\phi^TC\phi=\phi_1^2+\phi_2^2+\phi_3^2+\phi_4^2-\phi_2(\phi_1+\phi_3+\phi_4)$. By \eqref{eq:Etree} and \eqref{eq:E6d} without twist, in the fundamental chamber,
\begin{equation}
\label{eq:so8E6d}
\begin{aligned}
\epsilon_1\epsilon_2\,\mathcal E_{\rm 6d}={}&2\tau\phi_0^2+4\phi_0\Big(\tfrac12\phi^TC\phi+\tfrac32\epsilon_+^2-\tfrac18(\epsilon_1^2+\epsilon_2^2)\Big)+\tfrac16\sum_{\alpha>0}(\alpha\cdot\phi)^3\\
&-\tau\cdot\tfrac12\phi^TC\phi+c_\epsilon(\phi_1+\phi_2+\phi_3+\phi_4)\,.
\end{aligned}
\end{equation}
The term $-\tau\cdot\tfrac12\phi^TC\phi$ is the KK tower, by \eqref{eq:F1loop6d} and \eqref{eq:Ktilde} with $h^\vee=6$ and $B_2(0)=\tfrac16$. The effective $\so(8)$ coupling is $\Phi:=4\phi_0-\tau$. In the ansatz \eqref{eq:framemapgen}, $f=M\phi+s\,\tau$ and $m_0'=\tau/r$, we take $M$ from the affine-node rule \eqref{eq:affinenode} with the comarks $(1;1,2,1,1)$ of $D_4^{(1)}$, but the order is fixed by the cubic terms. A scan with $|M_{ij}|\le5$ finds that $f_0$ must sit on the central node, the only one with comark $2$, while $f_1,\dots,f_4$ can be permuted. We put $f_1$ on the affine node. Explicitly, with rows $(f_0;f_1,\dots,f_4)$ and columns $(\phi_0;\phi_1,\dots,\phi_4)$,
\begin{equation}
\label{eq:so8M}
M=\begin{pmatrix}2&0&1&0&0\\1&0&0&0&0\\1&1&0&0&0\\1&0&0&1&0\\1&0&0&0&1\end{pmatrix}\,.
\end{equation}
The cubic terms of \eqref{eq:E1prep} and \eqref{eq:so8E6d} then agree identically. The $12$ terms linear in $\tau$ and quadratic in the moduli then give linear equations for $(s_0,\dots,s_4,1/r)$, with the unique solution
\begin{equation}
\label{eq:so8map}
\begin{gathered}
f_0=2\phi_0+\phi_2-\tfrac\tau2\,,\qquad f_1=\phi_0+\tfrac\tau4\,,\\
f_i=\phi_0+\phi_{\iota(i)}-\tfrac\tau4\ \ (i=2,3,4)\,,\qquad m_0'=\tfrac\tau2\,,
\end{gathered}
\end{equation}
with $(\iota(2),\iota(3),\iota(4))=(1,3,4)$. The $5$ terms quadratic in $\tau$ and linear in the moduli then agree identically. So $r=2$, \eqref{eq:rdef}, and $\epsilon_1\epsilon_2(\mathcal E_{\rm E}-\mathcal E_{\rm 6d})=\tfrac\tau{24}\big(2\epsilon_1^2+2\epsilon_2^2-24\epsilon_+^2+\tau^2\big)$ is independent of the moduli. In 6d variables, the generators \eqref{eq:E1masses} are
\begin{equation}
\label{eq:so8states}
\mbox{\small$\begin{array}{l|l|l}
\text{curve} & \text{mass in 6d variables} & \text{6d state}\\ \hline
F_1 & \tau-\phi_2=\alpha_0\cdot\phi & \text{affine root (KK mode of the W-boson of }-\theta)\\
F_i & 2\phi_{\iota(i)}-\phi_2=\alpha_{\iota(i)}\cdot\phi & \text{W-boson of an outer simple root}\\
F_0 & 2\phi_2-\phi_1-\phi_3-\phi_4=\alpha_2\cdot\phi & \text{W-boson of the central simple root}\\
e & 4\phi_0+2\phi_2-\tau & \text{string}
\end{array}$}
\end{equation}
with $i=2,3,4$ and $\theta\cdot\phi=\phi_2$ the highest root. The $E_1$ instantons, $4\phi_0+\phi_2$ and $4\phi_0+2\phi_{\iota(i)}+\phi_2-\tau$, are strings as well. So the four $E_1$ W-bosons are the outer simple roots of $D_4^{(1)}$. The central simple root is the central instanton $F_0$, of $k'=1$ and mark $2$. Hence, $\delta=\alpha_0+\alpha_1+2\alpha_2+\alpha_3+\alpha_4=\tau$ has $k'(\delta)=2$, and $r=2$ is the mark of the instantonic root, \eqref{eq:rrule}.

The $S_4$ of \cite[\S2.1]{Zafrir:2025xca} is the automorphism group of the affine diagram. The $S_3$ fixing $f_1$ permutes $\phi_1,\phi_3,\phi_4$, so it is triality. The normal $\BZ_2\times\BZ_2$ consists of the double transpositions. By \eqref{eq:so8map}, $(f_1f_2)(f_3f_4)$ maps $\phi_0\to\phi_0+\phi_1-\tfrac\tau2$ and $\phi\to(-\phi_1,\,\phi_2-2\phi_1,\,\phi_4-\phi_1,\,\phi_3-\phi_1)+\tau\,\omega_1^\vee$, with $\omega_1^\vee=(1,1,\tfrac12,\tfrac12)$.\footnote{\label{fn:so8coweights}The fundamental coweights are fixed by $\alpha_b\cdot\omega_a^\vee=\delta_{ab}$. Since $\alpha_b\cdot\phi=(C\phi)_b$, $\omega_a^\vee$ is the $a$-th column of $C^{-1}$. For an outer node $a=1,3,4$, it has $1$ at nodes $a$ and $2$, and $\tfrac12$ at the other two. In the orthonormal components of the next sentence, $\omega_1^\vee$, $\omega_3^\vee$ and $\omega_4^\vee$ are $(1,0,0,0)$, $\tfrac12(1,1,1,-1)$ and $\tfrac12(1,1,1,1)$, the highest weights of $\mathbf 8_v$, $\mathbf 8_s$ and $\mathbf 8_c$. The coweight $\omega_2^\vee=(1,2,1,1)$ of the central node is the highest coroot, so its class in $Z(\Spin(8))$ is trivial.} The linear part flips two signs of the orthonormal components $(\phi_1,\,\phi_2-\phi_1,\,\phi_3+\phi_4-\phi_2,\,\phi_4-\phi_3)$, so it lies in $W(D_4)$. The shift by $\tau$ times a coweight that is not a coroot is the action of the centre element $z_v$ (Appendix~\ref{app:twists}). Likewise, $(f_1f_3)(f_2f_4)$ and $(f_1f_4)(f_2f_3)$ shift by $\tau\omega_3^\vee$ and $\tau\omega_4^\vee$, the representatives of $s$ and $c$ in \eqref{eq:so8cosets} below, so they are $z_s$ and $z_c$. The $\BZ_2\times\BZ_2\subset S_4$ is therefore the circle image of the 6d 1-form symmetry, as \cite[\S2.1]{Zafrir:2025xca} proposes.

\paragraph{Flux sectors.}\phantomsection\label{sec:E1N4sectors} A 6d flux has three parts. The tensor flux is $n_0\in\BZ+\tfrac j4$, the sector $B=\tfrac j4$ of \sref{sec:prefactor}. Here, $j\in\BZ_4$ is the class of the string charge $\lambda=4n_0$, the value of $\mathcal C_3$ on $E$. The KK flux is $B_\tau\in\BZ$, set to zero in \sref{sec:prefactor}. The gauge flux $n$ is a coweight in one of the classes of $Z(\Spin(8))$, with the representatives of \eqref{eq:so8cosets} below. The part of $V$ linear in $\phi_0$ is $4\phi_0\big(\tfrac12n^TCn+n_0B_\tau\big)$. It vanishes for all $n_0$ only if $n=0$ and $B_\tau=0$. On this tensor-flux family, $V=2\tau n_0^2-4n_0\epsilon_+$, as for $\SU(4)_8$ in \eqref{eq:su48Lambda}, since the twist does not enter \eqref{eq:Vfamily}. We have not computed the exponents of the other sectors, which do not enter the prefactor $\Lambda_B$. The flux map below is linear, so it labels them too, but, beyond the tensor-flux family, the dictionary is kinematic.

By the \hyperref[it:fluxmap]{flux rule} of \sref{sec:framemap}, \eqref{eq:so8map} gives
\begin{equation}
\label{eq:so8fluxmap}
\begin{split}
n'_0 &=2n_0+n_2-\tfrac12B_\tau\,,\quad n'_1=n_0+\tfrac14B_\tau\,, \\ n'_i &=n_0+n_{\iota(i)}-\tfrac14B_\tau\ \ (i=2,3,4)\,,\quad B'=\tfrac12B_\tau\,,
\end{split}
\end{equation}
so $B'\in\tfrac12\BZ$ and $d=2$. Let $\beta_i:=2n_{\iota(i)}$ mod $2$, which is $1$ if the gauge flux is half-integral on the node of the $i$-th $E_1$. With $n_0=\tfrac j4$, the labels \eqref{eq:E1labels} are
\begin{equation}
\label{eq:so8dict}
X_1=j+B_\tau\,,\qquad X_i=j-B_\tau+2\beta_i\ \ (i=2,3,4)\,,\qquad \hat B=j+B_\tau\,,\qquad B'=\tfrac12B_\tau\,,
\end{equation}
independent of the representatives. For the four gauge classes, at $B_\tau=0$ (so $\hat B=j$ and $B'=0$):
\begin{equation}
\label{eq:so8cosets}
\begin{array}{c|c|c|c}
\text{class} & \text{representative }n & (\beta_2,\beta_3,\beta_4) & (X_1;X_2,X_3,X_4)\\ \hline
1 & (0,0,0,0) & (0,0,0) & (j;\,j,\,j,\,j)\\
v & (1,1,\tfrac12,\tfrac12) & (0,1,1) & (j;\,j,\,j+2,\,j+2)\\
s & (\tfrac12,1,1,\tfrac12) & (1,0,1) & (j;\,j+2,\,j,\,j+2)\\
c & (\tfrac12,1,\tfrac12,1) & (1,1,0) & (j;\,j+2,\,j+2,\,j)
\end{array}
\end{equation}
The $32$ sectors $(j,\text{class},B_\tau)$ with $B_\tau\in\{0,1\}$ have $32$ different images, and shifting $j$ and $B_\tau$ both by $2$ gives the same class. As for $N=2$, the images obey \eqref{eq:E1paircond}, since by \eqref{eq:so8states} $\langle n',F_0\rangle=2n_2-n_1-n_3-n_4=\langle n,\alpha_2\rangle\in\BZ$. Of the $64$ classes allowed by \eqref{eq:E1group}, \eqref{eq:E1paircond} keeps the $32$ images.

\paragraph{The symmetries.} Ref.~\cite[(2.3), (2.4)]{Zafrir:2025xca} recombines the five $\BZ_2$ backgrounds into $\mathcal X:=\hat{\mathcal B}+2\mathcal B_1$ and $\mathcal B_i^+:=\mathcal B_i+\mathcal B_1$, $i=2,3,4$. By \eqref{eq:Bidef} and \eqref{eq:so8dict}, their values on $E$ are
\begin{equation}
\label{eq:so8Zafrir}
X=j+B_\tau\ \ \mathrm{mod}\ 4\,,\qquad \big(B_2^+,B_3^+,B_4^+\big)=(\beta_2,\beta_3,\beta_4)+B_\tau(1,1,1)\ \ \mathrm{mod}\ 2\,.
\end{equation}
At $B_\tau=0$, $X=j$ is the tensor class, so $\mathcal X$ is the reduction of $\mathcal C_3$, as \cite[\S2.1]{Zafrir:2025xca} proposes. The triple $(B_2^+,B_3^+,B_4^+)$ then runs over the four triples of even sum, one per gauge class. So the backgrounds with $\sum_i\mathcal B_i^+=0$ are those of the 6d centre $\BZ_2\times\BZ_2$, and $\sum_i\mathcal B_i^+$ is the background of one more $\BZ_2$. At $N=4$, the term $\hat B$ in \eqref{eq:E1quarters} is an integer, so $\tfrac12\sum_{i=1}^4B_i\equiv B'$ mod $1$. Since $\sum_{i=2}^4B_i^+\equiv\sum_{i=1}^4B_i$ mod $2$, on every sector it gives
\begin{equation}
\label{eq:so8acc}
B_2^++B_3^++B_4^+\equiv2B'\equiv B_\tau\ \ \mathrm{mod}\ 2\,.
\end{equation}
This is the lock \eqref{eq:su4zlock}, with the KK flux in the place of the $\U(1)_I$ flux. Since $m_0'=\tfrac\tau2$, a neutral state has $k'$ equal to twice its KK momentum. So $\U(1)_{\rm KK}=\U(1)_I'/\BZ_2$, and a half-integral $B'$ is a KK flux that does not lift to $\U(1)_I'$. By the argument of \sref{sec:E1N2}, this extra $\BZ_2$ is accidental, screened by the central instantons of odd $k'$, and locked to the parity of the KK flux. One screening state is $F_0$, in 6d variables the W-boson of $\alpha_2$. The admissible classes are $\BZ_4\,(\mathcal X)\times\BZ_2\times\BZ_2\,(\mathcal B_i^+)\times\{B'\in\BZ,\BZ+\tfrac12\}$ with \eqref{eq:so8acc}, the content proposed in \cite[\S2.1]{Zafrir:2025xca}, and the central $\SU(2)$ has $\theta=0$.

\paragraph{Anomalies.}\phantomsection\label{sec:E1N4anom} On the tensor-flux family, \eqref{eq:Vuniversal} gives the fractional exponents $\fq^{j^2/8}$, from $\mathcal K$, and the $p_1p_2$-exponent of the $\SU(4)_8$ row of \eqref{eq:sectorexponents}, $\tfrac12$ at odd $j$, from the gravitational and $\SU(2)_R$ couplings. As in \sref{sec:exso8Z2}, these give the cubic self-anomaly and the mixed $\SU(2)_R$ anomaly of the $\BZ_4$, $\tfrac j2$ each. In the E-frame, the exponent $\fq^{j^2/8}$ is $k'(n')=n_0'^2\equiv\tfrac14X^2$, by \eqref{eq:E1Zprime} and \eqref{eq:so8dict}, i.e.\ $q'^{X^2/4}=\fq^{X^2/8}$ with $q'=e^{-\tau/2}$. On the generator, it is $\fq^{1/8}$, the spin $q(1)=\tfrac18$, as \eqref{eq:dictionary} requires with $k'=\tfrac14$ and $r=2$. As for $N=2$, it is the anomaly of the central $\SU(2)$, here $\tfrac{\pi i}2\int\tilde{\mathcal C}_1'\cup\mathcal P(\mathcal X)$ of \cite[(2.6), (2.7)]{Zafrir:2025xca}, with $\tilde{\mathcal C}_1'$ the mod-$4$ reduction of $c_1(\U(1)_I')$. The only continuous flavour symmetry is the Abelian $\U(1)_{\rm KK}$. So, by \eqref{eq:verdicts}, the forced flux $\fq^{1/8}$ gives a mixed anomaly and no 2-group with a continuous symmetry.

\paragraph{The discrete 2-group.} Write the 6d backgrounds on $M_5\times S^1$ as
\begin{equation}
\label{eq:so8reduction}
\mathcal B^{\rm 6d}_x=\mathcal B_x+\mathcal A_x\cup e_{S^1}\,,\qquad \mathcal C_3=\mathcal C_3^{\rm 5d}+\mathcal X\cup e_{S^1}\,,\qquad x=s,c\,,
\end{equation}
with $e_{S^1}$ the generator of $H^1(S^1,\BZ)$. Since $2\mathcal P(y)\equiv2\,y\cup y$ mod $4$, $\eta(b)=2\,(b_s\cup b_s+b_s\cup b_c+b_c\cup b_c)$ for $b=(b_s,b_c)$, with the bracket mod $2$. Inserting \eqref{eq:so8reduction} and using $e_{S^1}\cup e_{S^1}=0$, the terms quadratic in $\mathcal A$ drop, and $\eta(\mathcal B^{\rm 6d})=\eta(\mathcal B)+2(\mathcal A_s\cup\mathcal B_c+\mathcal A_c\cup\mathcal B_s)\cup e_{S^1}$. The component along $e_{S^1}$ is
\begin{equation}
\label{eq:so8twogroup}
\delta\mathcal X=2\,\mathcal A_s\cup\mathcal B_c+2\,\mathcal A_c\cup\mathcal B_s\ \ \mathrm{mod}\ 4\,.
\end{equation}
It is the polarisation of $\eta$, so it is $S_3$-invariant. It pairs different classes, and it has no term in $\mathcal A$ alone because $\eta$ is even. Ref.~\cite[(2.5)]{Zafrir:2025xca} has instead the diagonal form $2\mathcal A_s\cup\mathcal B_s+2\mathcal A_c\cup\mathcal B_c$. It follows from the coefficient $1$ of $\mathcal P(\mathcal B_s+\mathcal B_c)$ quoted below \eqref{eq:so8threegroup}. With coefficient $c_{\mathcal P}$, the same reduction gives $2c_{\mathcal P}\,(\mathcal A_s+\mathcal A_c)\cup(\mathcal B_s+\mathcal B_c)+2(\mathcal A_s\cup\mathcal B_c+\mathcal A_c\cup\mathcal B_s)$ mod $4$, which is the cross form for $c_{\mathcal P}=2$ and the diagonal form for $c_{\mathcal P}=1$. With the correct coefficient $2$, the 2-group is the cross form \eqref{eq:so8twogroup}, which also agrees with the 4d relation of \cite[app.~C]{Zafrir:2025xca}, since that relation pairs different classes.\footnote{Label each double transposition and each pair of sign flips in $W(D_4)$ by the partition of $\{1,2,3,4\}$ into two pairs that it preserves. A double transposition and a pair of sign flips commute if their partitions agree, and commute up to $-1$ if they differ, as for the pairs of \cite[(C.3)]{Zafrir:2025xca}.}

The 2-group \eqref{eq:so8twogroup} is preserved under $\mathcal A_{s,c}\to\mathcal A_{s,c}+\delta\chi_{s,c}$ if $\mathcal X\to\mathcal X+2\chi_s\mathcal B_c+2\chi_c\mathcal B_s$. So the 0-form $\BZ_2\times\BZ_2$ acts on the $\BZ_4$ class: $z_s$, $z_c$ and $z_v=z_sz_c$ shift $\mathcal X$ by $2\mathcal B_c$, $2\mathcal B_s$ and $2(\mathcal B_s+\mathcal B_c)$. The blow-up labels show this action, together with further terms at odd KK flux. The value $X=\hat B+2B_1$ is built from the first $E_1$. The double transposition $z_x$ that exchanges the first and the $i$-th $E_1$, with $x=v,s,c$ for $i=2,3,4$, fixes $\hat B$ and replaces $B_1$ by $B_i$:
\begin{equation}
\label{eq:so8Xshift}
X\to\hat B+2B_i=X+2(B_i-B_1)\equiv X+2B_i^+\ \ \mathrm{mod}\ 4\,.
\end{equation}
The action of $z_x$ on the gauge class $b\in Z(\Spin(8))$ of the flux $n$ is most direct in 6d variables. By \eqref{eq:so8map}, $z_x$ maps $\phi_0\to\phi_0+\phi_{\iota(i)}-\tfrac\tau2$, and $\phi$ to its image under an element of $W(D_4)$ plus $\tau\omega^\vee_{\iota(i)}$, as shown above for $z_v$ and by triality for $z_s$, $z_c$. By the \hyperref[it:fluxmap]{flux rule}, the fluxes transform in the same way, with $\tau$ replaced by the KK flux $B_\tau$, which $z_x$ leaves unchanged. The Weyl group preserves the gauge class, and $\omega^\vee_{\iota(i)}$ lies in the class $x$, so $b\to x^{B_\tau}b$. Since $j\equiv4n_0$ and $4n_{\iota(i)}\equiv2\beta_i$ mod $4$, also $j\to j+2\beta_i-2B_\tau$. With $X=j+B_\tau$, this gives $X\to X+2\beta_i+2B_\tau$ mod $4$, which is \eqref{eq:so8Xshift} with $B_i^+=\beta_i+B_\tau$ from \eqref{eq:so8Zafrir}.

At $B_\tau=0$, $z_x$ keeps $b$ and shifts $X$ by $2\beta_i$. Since $(\mathcal B_s,\mathcal B_c)$ is $(1,0),(0,1),(1,1)$ on $s,c,v$, \eqref{eq:so8cosets} shows that $\beta_3$, $\beta_4$ and $\beta_2$ are the values of $\mathcal B_c$, $\mathcal B_s$ and $\mathcal B_s+\mathcal B_c$ on $b$. So $z_s$, $z_c$ and $z_v$ shift $X$ by $2\mathcal B_c$, $2\mathcal B_s$ and $2(\mathcal B_s+\mathcal B_c)$ on $E$. This is the action of \eqref{eq:so8twogroup}, in the cross form rather than the diagonal form of \cite[(2.5)]{Zafrir:2025xca}. The labels cannot test a term in $\mathcal A$ alone, since the blow-up equations contain no backgrounds $\mathcal A_{s,c}$ (\sref{sec:conclusions}). The reduction above excludes such a term.

At odd $B_\tau$, i.e.\ odd KK flux, $z_x$ also maps $b$ to $xb$ and shifts $X$ by a further $2$. The first comes from the shift $\tau\omega^\vee_{\iota(i)}$ of $\phi$, the second from the $-\tfrac\tau2$ in $\phi_0$.\footnote{\label{fn:so8furthershift}The $-\tfrac\tau2$ is $-\tau$ times the fractional instanton number $\tfrac12n^TCn=\tfrac12$ of $n=\omega^\vee_{\iota(i)}$, as required for $z_x$ to preserve the terms of \eqref{eq:so8E6d} linear in $\phi_0$. So the further shift of $X$ is $\mathcal K\cdot\tfrac12B_\tau=\eta(x)B_\tau$ mod $4$, with $\eta$ of \eqref{eq:so8threegroup}, the same for $x=v,s,c$.} Both act on fluxes only through $B_\tau$, so these effects involve the KK background and lie outside \eqref{eq:so8twogroup} (\sref{sec:conclusions}).

% ---------------------------------------------------------------------
\subsection{Three copies: pure \texorpdfstring{$\Spin(8)$}{Spin(8)}}
\label{sec:E1N3}

Send $\tau\to\infty$ with $\phi$ and $\Phi=4\phi_0-\tau$ fixed. By \eqref{eq:so8E6d}, $\Phi$ multiplies $\tfrac12\phi^TC\phi$, so it becomes the instanton mass $m_0$ of pure $\Spin(8)$, normalised as in \eqref{eq:kQ}, i.e.\ twice that of \cite[\S3.6]{Harding:2026gzj}. By \eqref{eq:so8map}, $f_1=\tfrac14\Phi+\tfrac12\tau\to\infty$. So the first $E_1$ decouples \cite[\S5.1]{Hayashi:2017jze}, leaving the $\SU(2)$ gauging of the $E_1$'s $i=2,3,4$. The coupling $m_0'-\sum_{i=1}^4f_i$ stays finite, so the central instanton mass becomes $m_0'':=m_0'-f_1$, with flux $B''$ and charge $k''$ in the roles of $m_0'$, $B'$ and $k'$. The map \eqref{eq:so8map} becomes, after replacing $\Phi=4\phi_0-\tau$ by $m_0$,
\begin{equation}
\label{eq:so8puremap}
f_0=\phi_2+\tfrac12m_0\,,\qquad f_i=\phi_{\iota(i)}+\tfrac14m_0\ \ (i=2,3,4)\,,\qquad m_0''=-\tfrac14m_0\,.
\end{equation}
The last equality follows from \eqref{eq:so8map}: $m_0''=m_0'-f_1=\tfrac\tau2-\big(\tfrac14\Phi+\tfrac12\tau\big)=-\tfrac14\Phi$. The map agrees with \cite[(3.21), (3.23)]{Hayashi:2017jze} at their $N=2$. Directly, \eqref{eq:so8puremap} takes the pure $\Spin(8)$ prepotential, \eqref{eq:so8E6d} with $\Phi\to m_0$ and without the other terms in $\phi_0$ and $\tau$, to \eqref{eq:E1prep} with three $E_1$'s, up to $\epsilon_1\epsilon_2(\mathcal E_{\Spin(8)}-\mathcal E_{\rm E})=5c_\epsilon m_0''-\tfrac43m_0''^3$. As for $N=2$, the weak-coupling regions of the two frames lie at opposite signs of $m_0$. In the $\Spin(8)$ frame, both parities of $\Bmz$ are allowed \cite[\S3.6]{Harding:2026gzj}, and there is no 2-group, since $\U(1)_I$ is not enhanced \cite[\S\S5.1, 6.1]{Harding:2026gzj}.

\paragraph{Labels and dictionary.} By \eqref{eq:so8puremap},
\begin{equation}
\label{eq:so8pureflux}
n'_0=n_2+\tfrac12\Bmz\,,\qquad n'_i=n_{\iota(i)}+\tfrac14\Bmz\ \ (i=2,3,4)\,,\qquad B''=-\tfrac14\Bmz\,,
\end{equation}
so $d=4$. The labels \eqref{eq:E1labels} are now $X_2,X_3,X_4$ mod $4$, $\hat B$ mod $2$ and $B''$ mod $1$. By \eqref{eq:E1group}, $\hat B$ fixes the parity of all three $X_i$, so a class of $(X_2,X_3,X_4;\hat B)$ is given by $\hat B$ and the three $B_i$ of \eqref{eq:Bidef}. There are $2\cdot2^3=16$ such classes, and, with the $d=4$ values of $B''$ mod $1$, there are $64$ classes in all.

Classes add like fluxes. The class $(X_2,X_3,X_4)=(1,1,1)$, with $\hat B=1$ and all $B_i=0$, has order $4$, since twice it is $(2,2,2)$, with $\hat B=0$ and all $B_i=1$. It generates a $\BZ_4$. The four classes $(0,0,0)$, $(2,2,0)$, $(2,0,2)$, $(0,2,2)$ form a complementary $\BZ_2\times\BZ_2$. So, as for $N=2$, the $E_1$ 2-groups make the $16$ classes $\BZ_4\times\BZ_2\times\BZ_2$ rather than $\BZ_2^4$. The $B_i$ are not additive, since they vanish on the generator, but not on its double. Additive labels for the two factors are
\begin{equation}
\label{eq:so8pureXt}
\tilde X:=\hat B+2(B_2+B_3+B_4)\ \ \mathrm{mod}\ 4\,,\qquad B_{ij}:=B_i+B_j\ \ \mathrm{mod}\ 2\ \ (2\le i<j\le4)\,,
\end{equation}
with $\hat B\equiv\tilde X$ mod $2$ and $B_{23}+B_{24}+B_{34}\equiv0$ on every class.\footnote{\label{fn:so8pureadditive}By \eqref{eq:Bidef} and \eqref{eq:E1group}, $\tilde X\equiv X_2+X_3+X_4-2\hat B\equiv-(X_2+X_3+X_4)$ mod $4$ and $B_{ij}\equiv\tfrac12(X_i-X_j)$ mod $2$. These forms show that both labels are additive. The relation $\hat B\equiv\tilde X$ mod $2$ holds because $\tilde X-\hat B \equiv (3\hat B -2 \hat B) - \hat B \equiv 0$ mod $2$ by \eqref{eq:E1group}, and $B_{23}+B_{24}+B_{34}\equiv0$ because each $B_i$ appears in two of the $B_{ij}$.} The label $\tilde X$ is $1$ on the generator and $0$ on the $\BZ_2\times\BZ_2$, while the $B_{ij}$ vanish on the $\BZ_4$ and distinguish the elements of the $\BZ_2\times\BZ_2$. These labels are the values on $E$ of the backgrounds $\tilde{\mathcal X}=\hat{\mathcal B}+2(\mathcal B_2+\mathcal B_3+\mathcal B_4)$ and $\mathcal B_{ij}=\mathcal B_i+\mathcal B_j$.

In \eqref{eq:so8puremap}, all three $f_i$ have the same shift $\tfrac14m_0$. So the $S_3$ that permutes the three $E_1$'s, i.e.\ $f_2,f_3,f_4$, permutes $\phi_1,\phi_3,\phi_4$ and fixes $\phi_2$ and $m_0$. It is triality, which fixes $\tilde{\mathcal X}$ and permutes the $\mathcal B_{ij}$.

On the images of the $\Spin(8)$ sectors, \eqref{eq:so8pureflux} gives $X_i=4n_{\iota(i)}+\Bmz\equiv2\beta_i+\Bmz$ mod $4$, with $\beta_i=2n_{\iota(i)}$ mod $2$ as defined below \eqref{eq:so8fluxmap}. By the closed forms of footnote~\ref{fn:so8pureadditive}, $B_{ij}\equiv\tfrac12(X_i-X_j)\equiv\beta_i+\beta_j$ mod $2$ and $\tilde X\equiv-(X_2+X_3+X_4)\equiv-3\Bmz-2\sum_i\beta_i\equiv\Bmz$ mod $4$, since $\sum_i\beta_i$ is even by \eqref{eq:so8cosets}. Hence
\begin{equation}
\label{eq:so8puredict}
B_{ij}\equiv\beta_i+\beta_j\ \ \mathrm{mod}\ 2\,,\qquad \tilde X\equiv\Bmz\ \ \mathrm{mod}\ 4\,.
\end{equation}
By \eqref{eq:so8cosets}, with $(\mathcal B_s,\mathcal B_c)=(1,0),(0,1),(1,1)$ on $s,c,v$ as in footnote~\ref{fn:so8values}, $B_{23}$, $B_{24}$ and $B_{34}$ are then the values of $\mathcal B_s$, $\mathcal B_c$ and $\mathcal B_s+\mathcal B_c$ on the gauge class. So the $\mathcal B_{ij}$ are the backgrounds of the 1-form symmetry $Z(\Spin(8))$. The $\BZ_4$ label $\tilde X$, by contrast, is fixed by the $\U(1)_I$ flux $\Bmz$. We now show that this $\BZ_4$ is accidental.

\paragraph{The accidental \texorpdfstring{$\BZ_4$}{Z4}.} With three $E_1$'s, \eqref{eq:Bidef} gives $\sum_iX_i\equiv3\hat B+2\sum_iB_i\equiv-\tilde X$ mod $4$, and the pairing condition \eqref{eq:E1paircond} becomes
\begin{equation}
\label{eq:so8purepair}
\tilde X\equiv-4B''\ \ \mathrm{mod}\ 4\,.
\end{equation}
There are $d=4$ values of $B''$ mod $1$, since $B''\in\tfrac14\BZ$ by \eqref{eq:so8pureflux}. Written as $B''\equiv-\tfrac14\tilde X$ mod $1$, \eqref{eq:so8purepair} selects exactly one of them on each of the $16$ classes of $(X_2,X_3,X_4;\hat B)$. So $16$ of the $16\cdot4=64$ classes satisfy it, as in the count below \eqref{eq:E1quarters}. The images satisfy it, since $-4B''=\Bmz$ by \eqref{eq:so8pureflux} and $\tilde X\equiv\Bmz$ mod $4$ by \eqref{eq:so8puredict}. A sector is labelled by a class of $Z(\Spin(8))$ and $\Bmz$ mod $4$, which \eqref{eq:so8puredict} recovers from the $B_{ij}$ and $\tilde X$. The $16$ sectors thus map one to one onto these $16$ classes.

By \eqref{eq:so8puremap}, a state of E-frame mass $k''m_0''+\mathsf c_0f_0+\sum_i\mathsf c_if_i$ has $\Spin(8)$ instanton number $k=\tfrac12\mathsf c_0+\tfrac14\sum_i\mathsf c_i-\tfrac14k''$, which is an integer. We insert $X_i\equiv\hat B+2B_i$ mod $4$ from \eqref{eq:Bidef} into \eqref{eq:E1charges}. Since $\mathsf c_i\equiv k''$ mod $2$ by \eqref{eq:E1parity}, the state pairs as $k''B''+\big(\tfrac12\mathsf c_0+\tfrac14\sum_i\mathsf c_i\big)\hat B+\tfrac12k''\sum_iB_i$ mod $1$. The middle term is $\big(k+\tfrac14k''\big)\hat B\equiv\tfrac14k''\hat B$ mod $1$, since $k$ is an integer. The pairing is therefore $k''\big(B''+\tfrac14\tilde X\big)$ mod $1$, and the state has charge $k''$ mod $4$ under the $\BZ_4$ with background $\tilde{\mathcal X}$. Since $F_0$ has $k''=1$, every state pairs integrally exactly when \eqref{eq:so8purepair} holds. Three conclusions follow, the first two as in \sref{sec:E1N2}.
\begin{itemize}
\item \emph{Accidental.} Perturbative E-frame states ($k''=0$) are neutral. The central one-instanton $F_0$ has $k''=1$, and, by \eqref{eq:so8puremap}, it is the W-boson of $\alpha_2$ in the $\Spin(8)$ frame. So, as in the \hyperref[it:su4zaccidental]{\emph{Accidental} item} of \sref{sec:E1N2}, the $\BZ_4$ with background $\tilde{\mathcal X}$ is a 1-form symmetry of the perturbative E-frame. The central instantons with $k''\not\equiv0$ mod $4$ are charged under it, and $F_0$, of charge $1$, screens all of it in pure $\Spin(8)$.
\item \emph{Locked to $\U(1)_I$.} A state of zero weight, $\mathsf c_0=\mathsf c_i=0$, has $k''=-4k$. So $\U(1)_I=\U(1)_I''/\BZ_4$, and a flux $\Bmz\not\equiv0$ mod $4$ does not lift to $\U(1)_I''$. As in the \hyperref[it:su4zlocked]{\emph{Locked} item} of \sref{sec:E1N2}, the value $\tilde X\equiv\Bmz$ mod $4$ of \eqref{eq:so8puredict} restores integrality.
\item \emph{Genuine.} The pairing $k''\big(B''+\tfrac14\tilde X\big)$ does not involve the $B_{ij}$, so no state is charged under the $\mathcal B_{ij}$. The 1-form symmetry of pure $\Spin(8)$ is therefore $Z(\Spin(8))$, with backgrounds $\mathcal B_{ij}$. The central $\SU(2)$ has $\theta=0$, by the \hyperref[par:theta0]{argument on the trivial class} in \sref{sec:E1N2}, with $B''$ in place of $B'$.
\end{itemize}

\paragraph{Comparison with the KK theory.} The E-frame of pure $\Spin(8)$ is that of the KK theory without the first $E_1$. With $\Bmz=4n_0-B_\tau$, the flux of $\Phi$, \eqref{eq:so8pureflux} reproduces the fluxes $n'_0,n'_2,n'_3,n'_4$ of \eqref{eq:so8fluxmap}, and $B''=B'-n'_1$. So the background $\tilde{\mathcal X}=\hat{\mathcal B}+2(\mathcal B_2+\mathcal B_3+\mathcal B_4)$ of \eqref{eq:so8pureXt} is also defined in the KK theory. In the variables $\mathcal X=\hat{\mathcal B}+2\mathcal B_1$ and $\mathcal B_i^+=\mathcal B_i+\mathcal B_1$ of \sref{sec:E1N4}, it reads $\tilde{\mathcal X}=\mathcal X+2(\mathcal B_2^++\mathcal B_3^++\mathcal B_4^+)$ mod $4$, which combines the genuine $\BZ_4$ of the KK theory with its accidental $\BZ_2$. By \eqref{eq:so8Zafrir} and \eqref{eq:so8acc}, $\tilde X\equiv j+3B_\tau\equiv j-B_\tau$ mod $4$. Since $4n_0\equiv j$, this is $\Bmz$ mod $4$, in agreement with \eqref{eq:so8puredict}. So $\Bmz$ mod $4$ takes over the role of the tensor class, and the 3-group \eqref{eq:so8threegroup}, which shifts the tensor class by $\eta(\mathcal B)$, becomes a mixed anomaly of $\U(1)_I$ with $Z(\Spin(8))$. Since $\eta$ is valued in $\BZ_4$, its coefficient on a class $b$ is $\tfrac14\eta(b)$. In pure $\Spin(8)$, this anomaly is the fractional instanton number, $\tfrac12$ on each of $v,s,c$ \cite[(B.5)]{Apruzzi:2021vcu}. The form \eqref{eq:so8threegroup} gives $\tfrac12$ on all three, while the coefficient $1$ of \cite[(2.5)]{Zafrir:2025xca} gives $\tfrac14,\tfrac14,\tfrac12$ on $s,c,v$. So the limit supports the $S_3$-invariant form. The 0-form $\BZ_2\times\BZ_2\subset S_4$ does not survive, since each double transposition exchanges the diverging $f_1$ with a finite $f_i$. Only the triality $S_3$ remains, and the 2-group \eqref{eq:so8twogroup}, which involves the backgrounds $\mathcal A_{s,c}$ of this $\BZ_2\times\BZ_2$, has no counterpart in pure $\Spin(8)$.

\begin{table}[t]
\centering
\small
\renewcommand{\arraystretch}{1.2}
\setlength{\tabcolsep}{4pt}
\begin{tabular}{c|L{2.0cm}|l|c|L{3.3cm}|L{4.5cm}}
$N$ & theory & frame map & classes & genuine 1-form & accidental 1-form; locked value\\ \hline
$2$ & pure $\SU(4)_0$ & $m_0=-2m_0'$ & $16\to8$ & $\BZ_4$: $\mathcal X=\hat{\mathcal B}+2\mathcal B_1$ & $\BZ_2$: $\hat{\mathcal B}+\mathcal B_1+\mathcal B_2$; $\Bmz$ mod $2$\\
$4$ & untwisted $\so(8)$ on $-4$ & $\tau=2m_0'$ & $64\to32$ & $\BZ_4$: $\mathcal X$; $\BZ_2\times\BZ_2$: $\mathcal B_i^+$ & $\BZ_2$: $\sum_i\mathcal B_i^+$; $B_\tau$ mod $2$\\
$3$ & pure $\Spin(8)$ & $m_0=-4m_0''$ & $64\to16$ & $Z(\Spin(8))$: $\mathcal B_{ij}$ & $\BZ_4$: $\tilde{\mathcal X}$; $\Bmz$ mod $4$
\end{tabular}
\caption{$\SU(2)$ gauging $N$ copies of $E_1$. ``Classes'' counts the E-frame classes allowed by \eqref{eq:E1group} and those that also obey \eqref{eq:E1paircond}, which are exactly the images of the sectors of the other frame. The last column gives the value on $E$ of the accidental background, \eqref{eq:su4zlock}, \eqref{eq:so8acc}, \eqref{eq:so8puredict}. In every case, the central $\SU(2)$ has $\theta=0$, and its instantons of odd charge screen the accidental symmetry.}
\label{tab:Nsummary}
\end{table}

% ---------------------------------------------------------------------
\subsection{Summary}
\label{sec:E1summary}

Table~\ref{tab:Nsummary} collects the three cases. The pattern is set by the quarters of \eqref{eq:E1quarters}, which the central instanton $F_0$ sees. For even $N$, the term $\tfrac N4\hat B$ has no quarter: it is $\tfrac12\hat B$ for $N=2$ and the integer $\hat B$ for $N=4$. So $F_0$ locks only a $\BZ_2$, by \eqref{eq:su4zlock} for $N=2$ and \eqref{eq:so8acc} for $N=4$, and a $\BZ_4$ containing $\hat{\mathcal B}$ survives. For $N=3$, a quarter remains, so $F_0$ sees $\tfrac14\tilde X$ and locks the $\BZ_4$ itself, by \eqref{eq:so8purepair}.

% ===========================================================================
% Section 8: Conclusions and outlook
% ===========================================================================
\section{Conclusions and outlook}
\label{sec:conclusions}

We have studied 6d SCFTs on a circle through the blow-up equations of their 5d frames. Without a twist or with a twist of the first kind, the frame map fixes $r$ through the lightest neutral KK mode, \eqref{eq:m0delta}. On the tensor-flux family, the Green--Schwarz couplings fix the blow-up exponent, \eqref{eq:Vfamily}, and the prefactor $\Lambda_B$ is a lattice theta function, \eqref{eq:LambdaTheta}, whose characteristics are the 't~Hooft class of the 2-form symmetry and a Wu vector. The KK momentum of each sector is its Wu-refined spin, \eqref{eq:dictionary}. With the centre twists allowed in 6d, the frame map also gives the faithful continuous 0-form symmetry of the seven frames of \sref{sec:faithful} and decides whether their 1-form symmetry forms a 2-group. For the $\SU(2)$ gaugings of $E_1$'s, it separates the genuine 1-form symmetries from the accidental ones (\sref{sec:E1}).

Several questions remain open.
\begin{itemize}[leftmargin=*,itemsep=2pt]
\item \emph{A 6d anomaly term on the blow-up.} We read the fractional exponents \eqref{eq:sectorexponents} as mixed anomalies of the 2-form symmetry (\sref{sec:pfanomalies}). A derivation needs the term \eqref{eq:GS7d} on the non-spin $\widehat{\BC}{}^2$, with flavour and $\SU(2)_R$ backgrounds correlated with $w_2(\widehat{\BC}{}^2)$. The result should match the faithful groups of \cite{Heckman:2022suy}. The terms cubic in $\mathcal B$, $\tfrac23$ for $\SU(3)_9$ and $\tfrac12$ for $\SU(4)_8$ \cite[(4.7)]{Harding:2026gzj}, are not produced by the quadratic term \eqref{eq:CS7d}. For $\so(8)$ on a $-4$ curve, \sref{sec:exso8Z2} obtains the value $\tfrac12$ from the gravitational term of \eqref{eq:GS7d}, by a congruence that holds on spin manifolds. The twist plays no role there. The derivation on $\widehat{\BC}{}^2$ should confirm this value, and give that of $\SU(3)_9$ in the same way. We conjecture that the backgrounds of the $\BZ_2$ 1-form symmetries of $\sp(3)_0+2\Lambda^2$ and $\su(4)_6+2\Lambda^2$ are related by $\mathcal B_{\su(4)}=\mathcal B_{\sp(3)}+w_2(\widehat{\BC}{}^2)$. This would explain why their frame map exchanges the trivial and 't~Hooft sectors on $\widehat{\BC}{}^2$ (\sref{sec:exso8tri}). On a spin spacetime, it predicts no exchange.

\item \emph{Twists of the second kind.} For these twists, \eqref{eq:m0delta} conflicts with the folding rule \cite[(2.43)]{Kim:2020hhh} (\sref{sec:rrule}). If a string exchanged by the twist gives a W-boson only after winding the circle twice, the exchanged pair is the long root of the 5d algebra $\sp(2)$ of the folded $A_3$ theory \cite[\S4]{Tachikawa:2011ch}, and $m_0=\tfrac\tau2$, in agreement with \eqref{eq:m0delta} at $k(\delta)=1$. For $A_2$, the doubly wound string has the mass of the invariant bound state of the two exchanged strings, cf.\ \cite[\S6]{Tachikawa:2011ch}, and $m_0=\tau$ remains. A resolution would also cover $\sp(m)_{0,\pi}+\adj$, $m\ge3$, and $G_2+\adj$, the $\BZ_3$ twist of the $(2,0)$ $D_4$ theory \cite[\S5.1.10]{Kim:2020hhh}.

\item \emph{Blow-up equations beyond the prefactor $\Lambda_B$.} Solving the blow-up equations at one-string order would test the statements based on the prefactor $\Lambda_B$ and the admissible flux sets. Both sectors of $\su(4)_{0,2,4}+4\Lambda^2$, $\so(7)+5\Fund$, $\sp(3)_0+2\Lambda^2$ and $\su(4)_6+2\Lambda^2$ should give equations of one 5d theory, as for $\USp(4)_0+3\Lambda^2$ \cite[\S5.1.1]{Kim:2020hhh}. The sector $B=\tfrac12$ of $\SU(4)_8$, absent from \cite[(6.9)]{Kim:2020hhh}, is checked at one-string order (\sref{sec:exSU48}). Beyond that order, it is tested only as far as \cite[Table~30]{Kim:2020hhh} reaches. The sector $B=\tfrac12$ of $\so(8)$ with the $\BZ_2$ twist is not among the flux sets of \cite[(6.13)]{Kim:2020hhh}. The sectors of the untwisted $\so(8)$ KK theory off the tensor-flux family rest on the flux map alone. The criterion of \sref{sec:pfcramer} is not necessary, as \sref{sec:exso8Z2} shows. What decides closure when a gauge centre survives is open.

\item \emph{The 2-group of the untwisted $\so(8)$ KK theory.} The blow-up labels give the action of the 0-form $\BZ_2\times\BZ_2$ in \eqref{eq:so8twogroup}. Testing a term in $\mathcal A_{s,c}$ alone would need these backgrounds, the holonomies of the 6d 1-form symmetry around the circle, which the blow-up equations do not contain. The further label shifts at odd KK flux hint at a term that couples them to $\U(1)_{\rm KK}$. The anomalies of the individual $E_1$'s \cite[(2.8)]{Zafrir:2025xca} should follow in the same way.

\item \emph{Origin of the 2-groups.} The non-split $\BZ_4$ 2-groups of $\su(4)_{0,4}+4\Lambda^2$ have no 6d 2-group as parent. We expect them to come from the Green--Schwarz coupling \eqref{eq:GS7d}, evaluated on backgrounds twisted by $\sigma$ around the circle. On the circle, $\mathcal C_3=\mathcal C_3^{\rm 5d}+\mathcal B\cup e_{S^1}$, as in \eqref{eq:so8reduction}. In the gauge-theory frame, $\mathcal B$ is a background and $\mathcal C_3^{\rm 5d}$ is summed over \cite[(4.12)]{Apruzzi:2022dlm}, so the component of the coupling along $e_{S^1}$ shifts $\delta\mathcal B$, as it shifts $\delta\mathcal X$ in \eqref{eq:so8twogroup}. On an untwisted circle, with backgrounds pulled back from 5d, this component vanishes. After the twist, it can contain a single 2-torsion term, built from the background of $\sigma$ and $w_2(\Sp(4)/\BZ_2)$, which gives \eqref{eq:faithfulPostnikovPSp4}. The 5d results \eqref{eq:faithfulstringodd} and \eqref{eq:faithfulEstringodd}, which make a single wrapped string odd under a central element, require its coefficient to be $1$. For the E-string parent, the parity of a single wrapped E-string follows in 6d from the $\sigma$-twisted sector of the level-one $E_8$ current algebra (\sref{sec:faithfulPSp4}). For $\su(4)+8\Fund$, a 6d derivation is open. A geometric computation as in \cite{Cvetic:2022imb,Apruzzi:2021vcu}, extended to twisted reductions, would test these 2-groups. The one-instanton spectra of $\su(4)_2+4\Lambda^2$ and $\su(4)_6+2\Lambda^2$ would test the lock: they should contain the flavour-odd $\SU(2)_R$ doublets that lock $\U(1)_I$ to the flavour centre. The criterion of \cite{Harding:2026gzj} should also be extended to $\SU(2)_R$, $(-1)^F$ and discrete 0-form symmetries.

\item \emph{Other theories.} Theories with flavour and a larger $D$, such as $\so(N)$ on a $-4$ curve with $N-8$ vectors, with $D=\BZ_4$ and flavour $\Sp(N-8)$, would test the mixed-anomaly reading beyond the value $\tfrac12$ shared by all our flavour examples. The frames $\su(m)_0+\mathbf S^2+(m-2)\Fund$ are reductions of the quiver of two $-2$ curves with $\su(m-1)$ on each, with the exchange twist \cite[\S6]{Hayashi:2015vhy}, \cite[(2.9)]{Bhardwaj:2020gyu}, a twist of the second kind. There, $D=\BZ_3$ and the twist inverts it, so the tensor part $D/(1-\sigma)D$ of Appendix~\ref{app:twists} is trivial and no 1-form symmetry of tensor origin survives, in agreement with the absence of a centre symmetry in 5d.
\end{itemize}

\acknowledgments
We are grateful to Hee-Cheol Kim for useful discussions. Both W.H. and N.M. gratefully acknowledge support from the Simons Center for Geometry and Physics, Stony Brook University, during the 23rd Simons Physics Summer Workshop (2026). W.H. also acknowledges, for hospitality and support during the realisation and final stages of this project, the following institutions: the POSTECH String Theory Group, Pohang, especially Hee-Cheol Kim; the UNIST Mathematical Physics Group, Ulsan, in particular Rak-Kyeong Seong; the Asia Pacific Center for Theoretical Physics, Pohang.

\appendix
% ===========================================================================
% Appendix A: Twists and the gauge centre
% ===========================================================================
\section{Twists and the gauge centre}
\label{app:twists}

We decide when the centre of a 6d gauge group survives a twist of the first kind as a 1-form symmetry of the KK theory. This is the input of \sref{sec:pfcramer} and of \sref{sec:faithful}.

Let $\mathfrak g$ be the gauge algebra on a curve, $G$ its simply connected group, and $\sigma$ the outer automorphism used for the twist. The compactification uses a lift $\gamma$ of $\sigma$ to $G$, defined up to an inner automorphism, i.e.\ a constant holonomy along the circle. Changing $\gamma$ by a constant holonomy $e^{2\pi i\theta\cdot H}$ shifts the KK momenta of a weight $\mathsf w$ by $\theta\cdot\mathsf w$, i.e.\ the Coulomb moduli by $\theta\tau$, which the shifts $s$ of the frame map \eqref{eq:framemapgen} absorb. We write $\mathfrak g^\sigma:=\mathfrak g^\gamma$ and $G^\sigma:=G^\gamma$ for its fixed algebra and group. $\mathfrak g^\sigma$ is the gauge algebra of the 6d frame on the circle, not that of a 5d gauge-theory frame. By the \hyperref[par:twists]{twist rule} of \sref{sec:prepot6d}, every 6d multiplet decomposes into representations $\mathbf R_{\fr}$ of $\mathfrak g^\sigma$, graded by the KK momentum $\fr\in[0,1)$. The gauge-descended 1-form symmetry follows in two steps.

\paragraph{Step 1: action on the centre.} The centre is $Z(G)\cong\Lambda_\cw/\Lambda_\crt$. The class of the fundamental coweight $\omega_i^\vee$ is the element $e^{2\pi i\omega_i^\vee}$, which acts on a representation of highest weight $\mathsf w$ by the phase $e^{2\pi i\langle\mathsf w,\omega_i^\vee\rangle}$. The automorphism $\sigma$ permutes the Dynkin nodes, hence the fundamental coweights and their classes. For $A_{N-1}$, it maps $\omega_1^\vee$ to $\omega_{N-1}^\vee\equiv-\omega_1^\vee$ modulo the coroot lattice, so it acts on $Z(G)=\BZ_N$ as $z\mapsto z^{-1}$. A centre element that is not fixed by $\sigma$ cannot be wrapped on the twisted circle, so only the fixed subgroup $\ker_\sigma Z(G):=\{z\in Z(G):\sigma(z)=z\}$, the kernel of $1-\sigma$, survives \cite[(3.89)]{Bhardwaj:2020phs}. Table~\ref{tab:twistcentre} lists the cases.

\begin{table}[ht]
\centering\small
\setlength{\tabcolsep}{4pt}
\renewcommand{\arraystretch}{1.25}
\begin{tabular}{L{2.1cm} C{0.7cm} C{1.25cm} C{1.25cm} C{2.15cm} L{2.4cm} L{2.2cm}}
\toprule
$\mathfrak g$ & $\sigma$ & $\hat{\mathfrak g}$ & $\mathfrak g^\sigma$ & $Z(G)$ & action of $\sigma$ & $\ker_\sigma Z(G)$ \\
\midrule
$A_{N-1}$, $N\ge3$ & $\BZ_2$ & $A_{N-1}^{(2)}$ & $C_{\lfloor N/2\rfloor}$ & $\BZ_N$ & $z\mapsto z^{-1}$ & $\BZ_2$ ($N$ even), $1$ ($N$ odd) \\
$D_N$, $N$ odd & $\BZ_2$ & $D_N^{(2)}$ & $B_{N-1}$ & $\BZ_4=\{1,s,v,c\}$ & $s\leftrightarrow c$, i.e.\ $z\mapsto z^{-1}$ & $\{1,v\}$ \\
$D_N$, $N$ even & $\BZ_2$ & $D_N^{(2)}$ & $B_{N-1}$ & $\BZ_2\times\BZ_2$ & $s\leftrightarrow c$ & $\{1,v\}$ \\
$D_4$ & $\BZ_3$ & $D_4^{(3)}$ & $G_2$ & $\BZ_2\times\BZ_2$ & $v\to s\to c\to v$ & $1$ \\
$E_6$ & $\BZ_2$ & $E_6^{(2)}$ & $F_4$ & $\BZ_3$ & $z\mapsto z^{-1}$ & $1$ \\
\bottomrule
\end{tabular}
\caption{Step 1: twisted affine algebra $\hat{\mathfrak g}$, fixed algebra $\mathfrak g^\sigma$ \cite[Table~1]{Kim:2020hhh}, and the $\sigma$-fixed part of the centre. $z\mapsto z^{-1}$ is the action on $Z(G)$ induced by the diagram automorphism, i.e.\ charge conjugation. In $Z(\Spin(2N))$, $v$ is the vector class and $s$, $c$ the spinor classes. For $A_{2L}$, $\mathfrak g^\sigma=C_L$ needs the lift of order four below.}
\label{tab:twistcentre}
\end{table}

\paragraph{Step 2: screening by the KK modes.} Keep the elements of $Z(G^\sigma)$ that act trivially on every $\mathbf R_{\fr}$ of the adjoint and of the matter, twisted sectors included. On the adjoint alone, this reproduces Step 1. Indeed, such an element acts trivially on $\mathfrak g$, so it lies in $Z(G)$, and it is fixed by $\gamma$, which acts on $Z(G)$ as $\sigma$. The matter can screen further. For $\su(3)$ with the $\BZ_2$ twist, $\mathfrak g^\sigma=\su(2)$, and the doublets $\mathbf 2_{1/4}$ and $\mathbf 2_{3/4}$ of \eqref{eq:su39decomp} carry the centre charge of $G^\sigma=\SU(2)$. So the centre of $G^\sigma$ is screened by KK modes of quarter-integer momentum, in agreement with $\ker_\sigma\BZ_3=1$ in Table~\ref{tab:twistcentre}.

\paragraph{The tensor part.}\phantomsection\label{par:tensorpart} The 2-form symmetry $D$ of \eqref{eq:Ddef}, of tensor origin, gives 5d 1-form symmetries through its operators placed at a point of the circle. Moving such an operator, of class $\lambda\in D$, around the twisted circle maps $\lambda$ to $\sigma(\lambda)$, so the tensor part is the coinvariants $D/(1-\sigma)D$, i.e.\ $D$ with each $\lambda$ identified with $\sigma(\lambda)$. For a twist of the first kind, $\sigma$ acts trivially on $D$, and the tensor part is $D$ itself. The 5d 1-form symmetry is the product of the tensor part and the gauge-descended part of Steps 1 and 2 \cite[(3.88), (3.89)]{Bhardwaj:2020phs}. If the frame has a 6d gauge algebra, the gauge-descended part is trivial and the twist fixes $D$, then the 5d 1-form symmetry is $D$. All $|D|$ classes are then 't~Hooft sectors of a single 5d frame, and each is expected to give a blow-up equation of it, the situation of \sref{sec:pfcramer}. An example is $\SU(3)_9$ \cite[(3.105)]{Bhardwaj:2020phs}. The converse fails (\sref{sec:exso8Z2}).

\paragraph{The lift for $A_{2L}$.}\phantomsection\label{par:liftnorm} For $\mathfrak g=\su(2L+1)$, the lift matters. An involution in the outer class fixes $\so(2L+1)$. The lift whose fixed algebra is $\usp(2L)$, the algebra $C_L$ of $A_{2L}^{(2)}$, has order four. These are $\sigma_2(L)$ and $\sigma_2(0)$ of \cite[(3.23)]{Tachikawa:2011ch}. In the fundamental representation, the involution is complex conjugation, $\gamma'(g)=\bar g$.\footnote{\label{fn:bar}Here, $\bar g$ denotes the complex conjugate of the matrix $g$, entry by entry.} Its fixed points are the real matrices, and a real unitary matrix of unit determinant lies in $\SO(2L+1)$, so its fixed group is $\SO(2L+1)$. The order-four lift is $\gamma(g)=u\,\bar g\,u^{-1}$, with $u={\rm diag}(J,\dots,J,1)$ and $J=\Kmat{0}{1}{-1}{0}$. Its fixed group is $\USp(2L)$ in the upper-left block, since $h=J_L\bar hJ_L^{-1}$, with $J_L={\rm diag}(J,\dots,J)$, means $hJ_Lh^T=J_L$ for unitary $h$. Since $u$ is real, $\gamma^2(g)=u^2g\,u^{-2}$, with $u^2={\rm diag}(-1,\dots,-1,1)$. So $u^2$ is the image of $-1\in\USp(2L)$ under the embedding $h\mapsto{\rm diag}(h,1)$, and it is central in $\USp(2L)$. It is not a multiple of the identity, hence not central in $\SU(2L+1)$, so $\gamma$ has order four. For $L=1$, $\gamma$ acts on $\mathbf 8=\mathbf 3\oplus\mathbf 2\oplus\bar{\mathbf 2}\oplus\mathbf 1$ as $1$, $i$, $-i$, $-1$, which is the grading \eqref{eq:su39decomp}. To see this, take $X\in\su(3)=\su(2L+1)$, a traceless anti-Hermitian $3\times3$ matrix. The $\mathbf 8$ is the space of such $X$. Write $X$ in blocks adapted to the fixed $\USp(2)=\SU(2)$ in the upper-left corner:
\begin{equation}
\label{eq:Xblocks}
X=\begin{pmatrix}A&b\\b'^T&d\end{pmatrix}\,,\qquad \operatorname{tr}A+d=0\,,
\end{equation}
with $A$ a $2\times2$ block, $b,b'\in\BC^2$ and $d$ a number. Since $X$ is anti-Hermitian, $\bar X=-X^T$, so $\gamma(e^X)=e^{\gamma(X)}$ with $\gamma(X)=u\bar Xu^{-1}=-uX^Tu^{-1}$, extended linearly to complex $X$. With $u={\rm diag}(J,1)$ and $J^{-1}=-J$,
\begin{equation}
\label{eq:gammaX}
\gamma(X)=\begin{pmatrix}-JA^TJ^{-1}&-Jb'\\-(Jb)^T&-d\end{pmatrix}\,.
\end{equation}
Every $2\times2$ matrix obeys $JA^TJ^{-1}=(\operatorname{tr}A)\mathbf 1-A$. So the pieces of $X$ transform as follows.
\begin{itemize}[leftmargin=*,itemsep=2pt]
\item The traceless part of $A$, the $\mathbf 3$ of $\SU(2)$, is fixed, with eigenvalue $1$.
\item The trace part of $A$ is tied to $d$ by $\operatorname{tr}A+d=0$. This leaves one direction, $X\propto{\rm diag}(1,1,-2)$, the $\mathbf 1$, which changes sign, with eigenvalue $-1$.
\item The off-diagonal entries transform as $(b,b')\mapsto(-Jb',-Jb)$. Applying this twice gives $b\mapsto J^2b=-b$, so $\gamma^2=-1$ on these four entries, and $\gamma$ has eigenvalues $\pm i$. Each eigenspace is a doublet of $\SU(2)$, the $\mathbf 2$ and the $\bar{\mathbf 2}$.
\end{itemize}
The phases $1,i,-i,-1$ are $e^{2\pi i\fr}$ with $\fr=0,\frac14,\frac34,\frac12$. So the twist and the centre of $\USp(2L)$ generate the non-split extension
\begin{equation}
\label{eq:Z4ext}
1\to Z(\USp(2L))\to\BZ_4\to\BZ_2^{\rm twist}\to1\,.
\end{equation}
Going twice around the circle acts as the centre element, so a mode of centre charge $c\in\{0,1\}$ has KK momentum in $\BZ+\fr$ with $\fr\in\tfrac c4+\tfrac12\BZ$,\footnote{\label{fn:twoturns}One turn around the circle multiplies the mode by $e^{2\pi i\fr}$, so two turns give $e^{4\pi i\fr}$. Two turns act as $\gamma^2$, i.e.\ as the centre element $-1\in\USp(2L)$, which multiplies a mode of centre charge $c$ by $(-1)^c$. So $e^{4\pi i\fr}=(-1)^c$, i.e.\ $2\fr\in\tfrac c2+\BZ$.} i.e.\ $\fr\in\{0,\tfrac12\}$ for $c=0$ and $\fr\in\{\tfrac14,\tfrac34\}$ for $c=1$. The involution gives instead the centreless group $\SO(2L+1)$, e.g.\ $\mathbf 8\to\mathbf 3_0\oplus\mathbf 5_{1/2}$ of $\so(3)$ for $L=1$. The centre $\BZ_2$ of $\USp(2L)$ is screened by the KK modes of centre charge $1$, such as the doublets $\mathbf 2_{1/4}$ and $\mathbf 2_{3/4}$ for $L=1$, while $\SO(2L+1)$ has no centre. So in both cases no element of $Z(G^\sigma)$ survives Step 2, and the 1-form symmetry does not depend on the lift.

% ===========================================================================
% Appendix B: one-string blow-up equations of SU(3)_9
% ===========================================================================
\section{One-string blow-up equations of \texorpdfstring{$\SU(3)_9$}{SU(3)9}}
\label{app:su39}

This appendix writes the blow-up equations of $\SU(3)_9$ at one-string order, in the form used in \sref{sec:pfcramer}. We use the 6d variables $\phi_0'$ and $\phi_1'$ of \cite[(4.113)--(4.114)]{Kim:2021gyj}, the tensor scalar and the $\su(2)$ Cartan of the 6d frame. They are related to the 5d moduli by $\phi^{\rm 5d}_1=\phi_0'+\frac\tau6$, $\phi^{\rm 5d}_2=2\phi_0'+\phi_1'+\frac\tau{12}$ and $m_0=\frac\tau2$. This is the frame map \eqref{eq:SU39map} with another origin of the tensor scalar, in which the string fugacity is $e^{-3\phi_0'}$. The index part is $\hat Z=\hat Z_{\rm pert}\,(1+e^{-3\phi_0'}\hat Z_1+\dots)$, with $\hat Z_{\rm pert}$ the vector-multiplet part of \cite[(4.113)]{Kim:2021gyj}.

\paragraph{Fluxes and exponent.} Let $(a,b)$ be the fluxes of $(\phi_0',\phi_1')$. They map to the 5d Dynkin-basis fluxes $n^{\rm 5d}=(a,2a+b)$. The three sectors are \cite[(4.115)]{Kim:2021gyj}
\begin{equation}
\label{eq:su39fluxes}
a\in\BZ+B\,,\qquad b\in\BZ\,,\qquad B_\tau=0\,,\qquad B\in\{0,\tfrac13,\tfrac23\}\,.
\end{equation}
The exponent \eqref{eq:Vdef}, computed from \eqref{eq:E250} in these variables, is
\begin{equation}
\label{eq:Vab}
V(a,b)=3b^2\,\phi_0'+2b(3a+2b)\,\phi_1'+\tfrac32\,a^2\tau+\tfrac13\big(9ab^2-3a+4b^3-b\big)(\epsilon_1+\epsilon_2)\,.
\end{equation}
It is independent of the Coulomb moduli if and only if $b=0$. On this tensor-flux family, it gives \eqref{eq:SU39Lambda}. The sign $(-1)^{a+b}$ of the 6d basis and the sign $(-1)^{|n^{\rm 5d}|}=(-1)^{3a+b}$ of the 5d basis differ by $e^{2\pi iB}$, a constant on each sector.

\paragraph{The one-string equation.} At order $e^{-3\phi_0'}$, \eqref{eq:blowupeq} reads, for each $B$,
\begin{equation}
\label{eq:BU394}
\begin{aligned}
\Lambda_B\,\hat Z_1&=\sum_{a\in\BZ+B}(-1)^{a}\,e^{-V(a,0)}\Big(p_1^{3a}\,\hat Z_1^N+p_2^{3a}\,\hat Z_1^S\Big)\\
&\quad+\sum_{a\in\BZ+B}\ \sum_{b=\pm1}(-1)^{a+b}\,e^{3\phi_0'-V(a,b)}\,R_b\,,\\
R_b&=\frac{\hat Z_{\rm pert}(\phi_1'+b\epsilon_1;\epsilon_1,\epsilon_2-\epsilon_1)\,\hat Z_{\rm pert}(\phi_1'+b\epsilon_2;\epsilon_1-\epsilon_2,\epsilon_2)}{\hat Z_{\rm pert}(\phi_1';\epsilon_1,\epsilon_2)}\,.
\end{aligned}
\end{equation}
Here, $\hat Z_1=\hat Z_1(\phi_1';\epsilon_1,\epsilon_2)$, $\hat Z_1^N=\hat Z_1(\phi_1';\epsilon_1,\epsilon_2-\epsilon_1)$ and $\hat Z_1^S=\hat Z_1(\phi_1';\epsilon_1-\epsilon_2,\epsilon_2)$. The ratio $R_0$ equals $1$, which is why none appears in the first sum. By \eqref{eq:Vab}, $e^{-V(a,\pm1)}$ contains the one-string fugacity $e^{-3\phi_0'}$, so these fluxes contribute at one-string order through $R_{\pm1}$ alone. Since \eqref{eq:BU394} is the coefficient of $e^{-3\phi_0'}$, the factor $e^{3\phi_0'}$ divides it out. The remaining $a$-dependence of $e^{-V(a,\pm1)}$ is $\fq^{3a^2/2}e^{\mp6a\phi_1'-4a\epsilon_+}$, which resums to $\vartheta[\genfrac{}{}{0pt}{}{B}{1/2}](3\tau,\pm6\phi_1'+4\epsilon_+)$. $R_{\pm1}$ is a ratio of Jacobi theta functions, since $\hat Z_{\rm pert}$ includes the KK towers. Up to a factor $\eta^6$ and monomials, $R_1$ is the inverse of the product of $\theta_1(2\phi_1'+c)$, $c=0,\epsilon_1,\epsilon_2,2\epsilon_+$, from the triplet tower, and $\theta_1(\phi_1'\pm\frac\tau4)$, from the doublet towers. $R_{-1}$ follows by $\phi_1'\to-\phi_1'$. So the second sum of \eqref{eq:BU394} is the source $F_B$ of \cite[(4.117)]{Kim:2021gyj}.\footnote{In \cite[(3.94)--(3.95)]{Kim:2020hhh}, the sign $(-1)^{|n|}$ should be included, the flux $(n,2n)$ should be $\nu(a)=(a,2a)$ with fractional $a\in\BZ+B$, and, in (3.94), the exponent of $p_{1,2}$ should be $3a=3(n+B)$ instead of $3n+B$. The follow-up \cite[(4.114)--(4.117)]{Kim:2021gyj} has the correct map, fluxes and closed form.}

\paragraph{Solution.} The first sum of \eqref{eq:BU394} has the form \eqref{eq:sectoreq}, with $\mathcal K=3$ and $x=-\epsilon_1-\epsilon_2$, so that $x+3\epsilon_1=2\epsilon_1-\epsilon_2$ and $x+3\epsilon_2=-\epsilon_1+2\epsilon_2$. The second sum is the source $S_B=F_B$. The three sectors determine $\hat Z_1$, $\hat Z_1^N$ and $\hat Z_1^S$ by \eqref{eq:cramer}. The result is \cite[(4.116)]{Kim:2021gyj}, whose expansion reproduces the one-string BPS states of \cite[Table~9]{Kim:2020hhh}.

% =====================================================================
% Appendix C: SU(4)_0 as SU(2) gauging two E_1's, technical details   (key E1)
% =====================================================================
\section{Details of \texorpdfstring{$\SU(4)_0$}{SU(4)0} as \texorpdfstring{$\SU(2)$}{SU(2)} gauging two \texorpdfstring{$E_1$}{E1}'s}
\label{app:su40}

This appendix collects four technical points of \sref{sec:E1N2}.

\paragraph{Uniqueness of the frame map.} Among integer matrices with $|M_{ij}|\le3$, the cubic monomials and the $c_\epsilon$ terms of \eqref{eq:su4zprep} and \eqref{eq:E1prep} allow exactly two matrices $M$ in $\phi=Mf+s\,m_0'$, with $f=(f_1,f_0,f_2)$:
\begin{equation}
\label{eq:su4zM}
M=\begin{pmatrix}1&0&0\\0&1&0\\0&0&1\end{pmatrix}\,,\qquad M=\begin{pmatrix}0&0&1\\0&1&0\\1&0&0\end{pmatrix}\,.
\end{equation}
The second is the first with $f_1\leftrightarrow f_2$, a relabelling of the two $E_1$'s. We use the identity. The monomials $f_if_jm_0'$ then give \eqref{eq:su4zmap} with no further freedom.

\paragraph{Charge parity from the centre.} At $N=2$, the rule \eqref{eq:E1parity} also follows from the centre charge of $\SU(4)$. Every state of $\SU(4)_0$ is neutral under the centre $\BZ_4$, so its weight lies in the root lattice, $\mathsf w_1+2\mathsf w_2+3\mathsf w_3\equiv0$ mod $4$. By \eqref{eq:su4zmap}, its E-frame charges are $\mathsf c_1=\mathsf w_1$ and $\mathsf c_2=\mathsf w_3$, which are congruent mod $2$ by the root-lattice condition. Then, \eqref{eq:su4zcharge} gives $k'\equiv\tfrac12(\mathsf w_1+2\mathsf w_2+3\mathsf w_3)+\mathsf w_3\equiv\mathsf w_3$ mod $2$.

\paragraph{The terms of the mixed anomaly.} In the blow-up, $\tilde{\mathcal C}_1$ enters through the $\U(1)_I$ holonomy rather than through the flux $\Bmz$ on $E$: a full period multiplies the sector $\omega_X$ by $e^{2\pi ik}$, with $k(\omega_X)\equiv\tfrac38X^2$ by \eqref{eq:su4zk}. With $X=\hat B+2B_1$,
\begin{equation}
\label{eq:su4zsplit}
\tfrac38X^2\equiv\tfrac38\hat B^2+\tfrac12\hat BB_1+\tfrac32B_1^2\ \ \mathrm{mod}\ 1\,.
\end{equation}
The three terms are the coefficients of $2\pi i\,\tilde{\mathcal C}_1$ in $\tfrac{3\pi i}4\int\tilde{\mathcal C}_1\cup\mathcal P(\hat{\mathcal B})$, $\pi i\int\tilde{\mathcal C}_1\cup\hat{\mathcal B}\cup\mathcal B_1$ and $3\pi i\int\tilde{\mathcal C}_1\cup\mathcal P(\mathcal B_1)$, the terms of \cite[(A.2)]{Zafrir:2025xca} before the spin argument. The blow-up needs all three. At $(\hat B,B_1)=(1,1)$, it needs the second, without which $X=3$ would give $\tfrac78$. At $(0,1)$, it needs the third, which \cite[(A.2)]{Zafrir:2025xca} drops on spin 4-manifolds, while the blow-up is not spin.\footnote{\label{fn:su40spin}For an odd $\U(1)_I$ flux, the term survives in the 6d anomaly theory as $\tfrac12\int c_1\cup\mathcal B_1\cup\mathcal B_1$ mod $1$, with $c_1$ the first Chern class of $\U(1)_I$. On a spin 6-manifold, Wu's formula turns it into $\tfrac12\int c_1^2\cup\mathcal B_1$, which is $\tfrac12$ on $\BP^3$ with $c_1$ and $\mathcal B_1$ the hyperplane class. A properly quantised counterterm \cite[\S4.1, (4.3)]{Harding:2026gzj} changes the anomaly only by integers on closed 6-manifolds, so none removes it. The spin argument of \cite[(A.2)]{Zafrir:2025xca} applies to 4-manifold integrals only. The same holds for the $q^{1/2}$ of $\SU(4)_6$ in \cite[Table~3]{Harding:2026gzj}.}

In the E-frame, the exponent of a sector is its $\U(1)_I'$ charge $k'$, measured by a full period of $\U(1)_I'$. Since $m_0=-2m_0'$, this is two periods of $\U(1)_I$, which is $\zeta=-2$ of \eqref{eq:su4zmatching}. Then, \eqref{eq:su4zsplit} becomes $-\tfrac34\hat B^2-\hat BB_1-3B_1^2$. The $B_1$ terms are integers, and $-\tfrac34\hat B^2$ differs from $\tfrac14\hat B^2$ by the integer $-\hat B^2$. So only $\tfrac14\hat B^2$ remains, with no spin assumption: the anomaly \cite[(A.6)]{Zafrir:2025xca} of the central $\SU(2)$. Dropping the integers requires an integral $\tilde{\mathcal C}_1'$, i.e.\ a $\U(1)_I$ background that lifts to $\U(1)_I'$. On $E$, $\tilde{\mathcal C}_1'$ and $\tilde{\mathcal C}_1$ take the values $B'$ and $\Bmz$, with $B'=-\tfrac12\Bmz$ by \eqref{eq:su4zfluxmap}. So an odd $\Bmz$ does not lift. For it, $\hat B+B_1+B_2=1$ by \eqref{eq:su4zlock}: the accidental background is on.

\paragraph{No cubic anomaly.} At $\kCS=0$, \eqref{eq:QP} gives $P-\Bmz k=\tfrac16\sum_{\alpha>0}\big(x_\alpha^3-x_\alpha\big)$ with $x_\alpha=\langle n,\alpha\rangle$. On every coweight, $x_\alpha\in\BZ$, and $x^3-x=(x-1)x(x+1)$ is divisible by $6$. So $P\equiv\Bmz k$ mod $1$ on every coweight, not only on the coroot lattice as below \cite[(3.13)]{Harding:2026gzj}. The prepotential difference below \eqref{eq:su4zmap} contributes $\delta P(B')=-\tfrac13B'(2+B'^2)$ to the $p_1p_2$-exponent, so the E-frame exponent is $P'\equiv\Bmz\,k(\omega_j)-\delta P(B')$ mod $1$. A cubic anomaly of the $\BZ_4$ would be a $j$-dependent fractional part of $P'$ at $\Bmz=0$. There, $\delta P=0$ and $P'\equiv0$, so there is none. The $E_1$ anomalies \cite[(A.7)]{Zafrir:2025xca} are written in the variables of \cite[(A.4)]{Zafrir:2025xca}, with $\mathcal B_2^+=\hat{\mathcal B}+\mathcal B_1+\mathcal B_2$ by the \hyperref[it:su4zcomparison]{\emph{Comparison} item} of \sref{sec:E1N2}. By $\mathcal P(a+b)=\mathcal P(a)+\mathcal P(b)+2a\cup b$, they expand as
\begin{equation}
\label{eq:su4zA7}
\tfrac{\pi i}2\int\hat{\mathcal B}\cup\big(\mathcal P(\mathcal B_1)+\mathcal P(\mathcal B_2^++\mathcal B_1)\big)=\tfrac{\pi i}2\int\hat{\mathcal B}\cup\big(\mathcal P(\mathcal B_2^+)+2\,\mathcal B_2^+\cup\mathcal B_1\big)+\pi i\int\hat{\mathcal B}\cup\mathcal P(\mathcal B_1)\,.
\end{equation}
The first two terms are those of \cite[(A.8)]{Zafrir:2025xca}. They contain the accidental background and vanish on the $\BZ_4$. The last term, from $2\mathcal P(\mathcal B_1)$, is not shown in \cite[(A.8)]{Zafrir:2025xca}. It contains no $\mathcal B_2^+$, but it depends on the $\BZ_4$ lift of $\hat{\mathcal B}$ that defines $\mathcal B_1=\tfrac12(\mathcal X-\hat{\mathcal B})$, so it is not a function of $\mathcal X$ alone. The blow-up therefore does not test \eqref{eq:su4zA7} term by term. It finds that the complete cubic anomaly vanishes on the $\BZ_4$, as anticipated below \cite[(A.8)]{Zafrir:2025xca}.

\section{Technical details}
\label{app:technical}
% ===========================================================================
% Appendix D.1: the finite-difference lemma behind eq:Vfamily
% ===========================================================================
\subsection{The blow-up exponent on the tensor-flux family}
\label{app:lemma}

This appendix proves the lemma behind \eqref{eq:Vfamily}. Write $\mathcal E=\mathsf e/(\epsilon_1\epsilon_2)$, with $\mathsf e$ a polynomial. On the tensor-flux family, the patches of \eqref{eq:Vdef} shift only $\phi_{0,\alpha}\to\phi_{0,\alpha}+a_\alpha\epsilon$ and $m\to m+B_m\epsilon$, with $\epsilon=\epsilon_1$ on $N$ and $\epsilon=\epsilon_2$ on $S$. The gauge moduli and $\tau$ are not shifted. The three terms of \eqref{eq:Vdef} carry the denominators $\epsilon_1\epsilon_2$, $\epsilon_1(\epsilon_2-\epsilon_1)$ and $(\epsilon_1-\epsilon_2)\epsilon_2$, with signs $+,-,-$, and
\begin{equation}
\label{eq:threeden}
\frac1{\epsilon_1\epsilon_2}-\frac1{\epsilon_1(\epsilon_2-\epsilon_1)}+\frac1{\epsilon_2(\epsilon_2-\epsilon_1)}=0\,.
\end{equation}
So a term of $\mathsf e$ that depends neither on $\epsilon_{1,2}$ nor on the shifted variables drops out of $V$.

\paragraph{Lemma.} The monomials of $\mathsf e$ in the left column contribute to $V$ as in the right column:
\begin{equation}
\label{eq:lemma}
\mbox{\small\begin{tabular}{ll}
\toprule
monomial of $\mathsf e$ & contribution to $V$\\
\midrule
$F(\phi,\tau)\,\phi_{0,\alpha}$, $F$ independent of $\epsilon_{1,2}$ & $0$\\
$\frac\tau2\,\phi_0^T\mathcal K\phi_0$ & $\frac\tau2\,a^T\mathcal Ka$\\
$\epsilon_+^2\,\phi_{0,\alpha}$ & $\frac14\,\phi_{0,\alpha}-\frac12\,a_\alpha\epsilon_+$\\
$(\epsilon_1^2+\epsilon_2^2)\,\phi_{0,\alpha}$ & $3\,\phi_{0,\alpha}+2\,a_\alpha\epsilon_+$\\
$m^2\,\phi_{0,\alpha}$ & $B_m^2\,\phi_{0,\alpha}+2B_m\,a_\alpha\,(m+B_m\epsilon_+)$\\
\bottomrule
\end{tabular}}
\end{equation}
With the coefficients of the second form of \eqref{eq:Etree}, the sum of the rows is \eqref{eq:Vfamily}.

\paragraph{Proof.} Expand in powers of $\epsilon$. Terms of order $\epsilon^0$ vanish by \eqref{eq:threeden}. A term of order $\epsilon^1$ with an $\epsilon$-independent coefficient $c$ gives $-c\big[\frac{\epsilon_1}{\epsilon_1(\epsilon_2-\epsilon_1)}-\frac{\epsilon_2}{\epsilon_2(\epsilon_2-\epsilon_1)}\big]=0$, which proves the first row. A term $c\,\epsilon^2$ gives $-c\big[\frac{\epsilon_1^2}{\epsilon_1(\epsilon_2-\epsilon_1)}-\frac{\epsilon_2^2}{\epsilon_2(\epsilon_2-\epsilon_1)}\big]=c$. Applied to $\frac\tau2(a\epsilon)^T\mathcal K(a\epsilon)$, cross terms included, this gives the second row, and, applied to $B_m^2\epsilon^2\phi_{0,\alpha}$, the $\phi_0$-part of the fifth. For a term $L(\epsilon_1,\epsilon_2)\,\phi_{0,\alpha}$, with $L^N=L(\epsilon_1,\epsilon_2-\epsilon_1)$ and $L^S=L(\epsilon_1-\epsilon_2,\epsilon_2)$,
\begin{equation}
\label{eq:findiff}
V\supset\phi_{0,\alpha}\Big[\frac{L}{\epsilon_1\epsilon_2}-\frac{L^N}{\epsilon_1(\epsilon_2-\epsilon_1)}+\frac{L^S}{\epsilon_2(\epsilon_2-\epsilon_1)}\Big]-a_\alpha\,\frac{L^N-L^S}{\epsilon_2-\epsilon_1}\,.
\end{equation}
For $L=\epsilon_+^2$, $L^N=\frac14\epsilon_2^2$ and $L^S=\frac14\epsilon_1^2$, so the bracket is $\frac14$ and the second term $-\frac12a_\alpha\epsilon_+$. For $L=\epsilon_1^2+\epsilon_2^2$, $L^N-L^S=\epsilon_1^2-\epsilon_2^2$, the bracket is $3$ and the second term $2a_\alpha\epsilon_+$. In the $a_\alpha$-part of $m^2\phi_{0,\alpha}$, the explicit $\epsilon_{1,2}$ cancel the denominators, leaving $a_\alpha\big[(m+B_m\epsilon_2)^2-(m+B_m\epsilon_1)^2\big]/(\epsilon_2-\epsilon_1)=2B_ma_\alpha(m+B_m\epsilon_+)$. \hfill$\square$

\paragraph{The gauge moduli.} The patches do not shift the gauge moduli $\phi$. Three kinds of terms of \eqref{eq:E6d} therefore give no $\phi$-dependence to $V$:
\begin{itemize}[leftmargin=*,itemsep=2pt]
\item the gauge term $\mathcal K_{\alpha\gamma}\phi_{0,\alpha}\,\tfrac12\tilde K_\gamma(\phi_\gamma,\phi_\gamma)$ of \eqref{eq:Etree}, linear in $\phi_0$ with an $\epsilon$-independent coefficient, by the first row of \eqref{eq:lemma};
\item the cubic terms of the vector multiplet and its KK towers in \eqref{eq:F1loop6d}, which depend only on $\phi$ and $\tau$, by \eqref{eq:threeden};
\item the KK towers of the hypermultiplets, $\tau(\mathsf w\cdot\phi+m_f)^2$ up to constants. With $m_f\to m_f+B_{m_f}\epsilon$, their order-$\epsilon$ part drops out as in the proof, and their order-$\epsilon^2$ part is independent of $\phi$.
\end{itemize}
Two kinds of terms remain, the levels and the cubic terms of the hypermultiplets. A root $e$ enters the levels of \eqref{eq:E5d} as $-\partial|e\cdot\phi|$ in $C^G$ and $\frac12\partial|e\cdot\phi|$ in $C^R$. By \eqref{eq:findiff} with $a=0$, the levels contribute $\frac1{16}(C^G_i+2C^R_i)\,\phi_i$, which vanishes root by root. A weight $\mathsf w$ of the hypermultiplet $f$ with $x:=\mathsf w\cdot\phi+m_f>0$ contributes $-\frac1{12}x^3$ to $\mathsf e$ and $\mathsf w$ to $C^G$, by \eqref{eq:E5d}. On the patches, $x\to x+B_{m_f}\epsilon$, and the order-$\epsilon^2$ term of $-\frac1{12}(x+B_{m_f}\epsilon)^3$ is $-\frac14B_{m_f}^2x\,\epsilon^2$. By the rule $c\,\epsilon^2\to c$ of the proof, it gives $-\frac14B_{m_f}^2\,\mathsf w\cdot\phi$, up to $\phi$-independent terms. The lower orders drop out, and the order-$\epsilon^3$ term does not depend on $\phi$. The level enters $\mathsf e$ as $\frac1{48}(\epsilon_1^2+\epsilon_2^2)\,\mathsf w\cdot\phi$. By \eqref{eq:findiff} with $a=0$, whose bracket is $3$ for $L=\epsilon_1^2+\epsilon_2^2$, it gives $\frac1{16}\,\mathsf w\cdot\phi$. The sum $\big(\frac1{16}-\frac14B_{m_f}^2\big)\,\mathsf w\cdot\phi$ vanishes if and only if $B_{m_f}=\pm\frac12$. A weight with $x<0$ flips the sign of both terms. So, at vanishing gauge flux, and with these flavour fluxes, $V$ depends on the Coulomb moduli only through $\sum_\alpha\Delta_\alpha\phi_{0,\alpha}$. Conversely, in every example, a non-zero gauge flux makes $V$ depend on the gauge moduli, as \eqref{eq:Vab} shows for $\SU(3)_9$.

% ===========================================================================
% Appendix D.2: chamber scans behind the uniqueness statements of Section 5.
% ===========================================================================

\subsection{Chamber scans}
\label{app:scans}

Step~1 of \tref{tab:framemaprecipe} fixes $M$ from the cubic terms, and we solve it by a scan. For each pair of chambers, one in each frame, we run over the unimodular integer matrices $M$ with $|M_{ij}|$ below the bound given in \tref{tab:scans}, and keep those for which the cubic terms of the two prepotentials agree. We also require the 5d cubic to have no terms quadratic or cubic in the tensor scalars, as in 6d. Steps~2 and~3 then fix the shifts. The uniqueness and non-existence statements of \sref{sec:exSU39}, \sref{sec:exZ2twist}, \sref{sec:exSU48} and \sref{sec:exsu40}--\sref{sec:exso8tri} hold within these ranges.

\begin{table}[t]
\centering\footnotesize
\setlength{\tabcolsep}{4pt}
\begin{tabular}{L{2.0cm} L{3.5cm} L{1.1cm} L{3.5cm} L{3.4cm}}
\toprule
frame map & chambers scanned & $|M_{ij}|\le$ & Step~1: cubic terms & Steps~2, 3: other terms\\
\midrule
6d$\,\to\SU(3)_9$ & Weyl chambers & $6$ & unique, $\Kmat{1}{0}{2}{1}$ & unique, \eqref{eq:SU39map}\\
6d$\,\to\USp(4)_0$ & 6d: small $m_i$\newline 5d: \cite[(5.3)]{Kim:2020hhh} & $6$ & unique, $\Kmat{1}{0}{2}{1}$ & \eqref{eq:usp4map}\\
6d$\,\to\Spin(7)$ & 6d: $0$--$2$ pairs of the $\mathbf 4$\newline 5d: $0$--$3$ pairs $\pm e_i$ & $6$ & three chamber pairs$^\ddagger$, comarks $(1,2,2)$ & up to $m_i\to m_i-\tfrac25\sum_jm_j$\\
6d$\,\to\SU(4)_8$ & Weyl chambers & $6$ & unique, comarks $(1,2,3)$ & unique, \eqref{eq:su48map}\\
6d$\,\to\SU(4)_0$ & 6d: small $m_i$\newline 5d: all & $6$ & one 5d chamber, \eqref{eq:su40M} up to $P_1\leftrightarrow P_3$ & up to $m_i\to m_i-\tfrac12\sum_jm_j$\\
6d$\,\to\SU(4)_4$ & 5d: all & $3$ & only small $M_i$ with $e_1+e_4<0$, three matrices $M^{(n)}$, $n=0,\pm1$ & only $n=0$, \eqref{eq:su44Q}\\
6d$\,\to\SU(4)_2$ & 6d: $\{\}$, $\{\mathsf w_1\}$, $\{\mathsf w_1,\mathsf w_2\}$, all\newline 5d: all & $6$ & three chamber pairs$^\dagger$, \eqref{eq:su42M} & two solutions$^\dagger$\\
6d$\,\to\sp(3)_0$ & 6d: $4$\newline 5d: the $10$ of \eqref{eq:so8trichambers}\newline columns of $M$ sum to $(6,1,1)$ & $3$ & 5d: none dominated, or only $e_2-e_3$ & up to $\mathrm O(2)$\\
$\sp(3)_0\to\su(4)_{\kCS}$ & $\sp(3)$: the $10$\newline $\kCS$ free, columns of $N$ sum to $(3,2,1)$ & $4$ & $\kCS=6$, $\sp(3)$ in $\{4,5,6\}$ or a larger chamber, \eqref{eq:so8trichambers} & up to $P'_1\leftrightarrow P'_3$, $\mathrm O(2)$\\
\bottomrule
\end{tabular}
\caption{Scans behind the frame maps of \sref{sec:examples}, one map per row. A chamber is labelled by its dominated pairs, or by their number. ``Weyl chambers'' means that the frame has no masses. $N$ is the matrix of the map $P'=Ne+\dots$ of \sref{sec:exso8tri}. The steps are those of \tref{tab:framemaprecipe}. The row labels omit the matter, e.g.\ $\SU(4)_{0,2,4}$ stand for $\su(4)_{0,2,4}+4\Lambda^2$. The conditions on column sums are the matching of the $\epsilon_+^2$ terms. Without this condition, six further chamber pairs match in the row 6d$\,\to\sp(3)_0$. \eqref{eq:so8trichambers} and \eqref{eq:su44Q} are given in this subsection. $^\ddagger$The matching chamber pairs, by the numbers of dominated pairs in 6d and 5d, are $(0,2)$, $(1,1)$ and $(2,0)$. $^\dagger$The matching chamber pairs are (6d $\{\mathsf w_1\}$, 5d the two smaller pairs), (6d $\{\mathsf w_1,\mathsf w_2\}$, 5d the smallest pair) and (6d all, 5d none, $\sigma=-1$). The second solution, $M_i=m_i-\tfrac12\sum_jm_j+\tfrac\tau2$, holds in these pairs with $M_i>0$, but not when different masses lie in different chambers. The map \eqref{eq:su42map} holds in all three, with $M_i<0$ dominating in the first two.}
\label{tab:scans}
\end{table}

\paragraph{Mass chambers.} The cubic terms depend on the chambers even with the masses off. In a chamber, each term $|x|^3$ of \eqref{eq:E5d} is a polynomial, and setting the masses to zero afterwards keeps that polynomial. For a hypermultiplet of mass $m_f$, a pair of weights $\pm\mathsf w$ gives the masses $\pm\mathsf w\cdot\phi+m_f$. If $|m_f|<|\mathsf w\cdot\phi|$, they have opposite signs. If $|m_f|>|\mathsf w\cdot\phi|$, both have the sign of $m_f$, and we say that the mass \emph{dominates} the pair. A mass that dominates a pair also dominates every pair with smaller $|\mathsf w\cdot\phi|$. We label a chamber by its set of dominated pairs. If the pairs are totally ordered by $|\mathsf w\cdot\phi|$, this set consists of the smallest pairs, so it is fixed by their number. The frames of \sref{sec:examples} have the following chambers.
\begin{itemize}[leftmargin=*,itemsep=2pt]
\item \emph{$\su(4)_{\kCS}+4\Lambda^2$.} The weights $e_a+e_b$ of $\Lambda^2$ in \eqref{eq:su40E5d} form three pairs, ordered as $e_1+e_2>e_1+e_3>|e_1+e_4|$. Zero to three of them are dominated, and $e_1+e_4$ has either sign $\sigma$. These are the eight sign patterns of \eqref{eq:su40E5d}. The sign $\sigma$ matters only when no pair is dominated, so five of the eight sign patterns are distinct.
\item \emph{$\so(7)+5\Fund$.} The pairs $\pm e_i$ of the vector are ordered, $e_1>e_2>e_3$. Zero to three of them are dominated, the smallest first, which gives four chambers.
\item \emph{$\sp(3)_0+2\Lambda^2$.} The six pairs \eqref{eq:sp3pairs} are only partially ordered. For example, $e_2+e_3$ and $e_1-e_2$ are not comparable. Ten chambers are realised, listed in \eqref{eq:so8trichambers} below.
\item \emph{6d frames.} The 6d masses dominate pairs of weights of the 6d hypermultiplets, e.g.\ the short-root pairs $\pm\mathsf w_k$ of $\mathfrak g_2$ in \sref{sec:exsu42}.
\end{itemize}

\paragraph{Chambers of $\sp(3)_0$ and $\su(4)_6$.} In the labels \eqref{eq:sp3pairs} of $\sp(3)$, the pairs $1,2,3$ are the sums $e_1+e_2$, $e_1+e_3$, $e_2+e_3$, and $4,5,6$ the differences $e_1-e_2$, $e_1-e_3$, $e_2-e_3$. In the labels \eqref{eq:su4pairs} of $\su(4)$, the pairs $1',2',3'$ are $\pm(e'_1+e'_2)$, $\pm(e'_1+e'_3)$, $\pm(e'_1+e'_4)$. The 5d-to-5d map \eqref{eq:so8trifmap5d} relates the chambers as
\begin{equation}
\label{eq:so8trichambers}
\mbox{\footnotesize$\begin{array}{c|c|c|c|c|c}
\sp(3) & \{\},\{4\},\{6\},\{3,6\},\{4,6\},\{3,4,6\} & \{4,5,6\} & \{3,4,5,6\} & \{2,3,4,5,6\} & \{1,\dots,6\}\\ \hline
\su(4) & \text{no frame map} & \{\} & \{3'\} & \{2',3'\} & \{1',2',3'\}
\end{array}$}
\end{equation}
So a frame map exists only when the masses dominate all three differences. In the chamber $\{4,5,6\}$, the masses lie between the differences and the sums, and it maps to the small-mass chamber $\{\}$ of $\su(4)$. Dominating, in addition, the sums $3$, $2$ and $1$ in turn dominates the $\su(4)$ pairs $3'$, $2'$ and $1'$.

\paragraph{No smaller $r$ for $\su(4)+2\Lambda^2$.}\phantomsection\label{lem:rge6} The scans of \tref{tab:scans} bound $|M_{ij}|$. For the map to $\su(4)_6+2\Lambda^2$ of \sref{sec:exso8tri}, the following statement holds for every $M$. \emph{Let \eqref{eq:framemapgen} with $u=0$ map the 6d frame \eqref{eq:so8triE6d}, continued across flops of states without string charge, to $\su(4)_{\kCS}+2\Lambda^2$. Let the two 6d sectors map to the two 5d sectors. Then $r\ge6$, with equality only if the tensor column of $M$ is $(1,2,3)$ or $(3,2,1)$.}

\noindent\emph{Proof.} Let $c\in\BZ^3$ be the tensor column of $M$, i.e.\ the coefficients of $\phi_0$ in $P'$. Let $\mathcal C$ be the 5d cubic, as a function of $P'$ and the 5d masses $M'_j$.
\begin{enumerate}[leftmargin=*,itemsep=2pt]
\item \emph{$r=k(c)$.} The cubic terms of \eqref{eq:framemapgen} carry no $\tau$. So $\mathcal C$ is evaluated at $P'=c\,\phi_0+g_1\phi_1+g_2\phi_2+t\cdot m$ and $M'=m$, with $g_{1,2}$ the gauge columns of $M$. Expanding in $\phi_0$,
\begin{equation}
\label{eq:Cexpand}
\mathcal C(M\phi+t\cdot m,\,m)=\mathcal C\,\phi_0^3+\Big(\sum_i\phi_i\,g_i\cdot\nabla_{P'}\mathcal C+\sum_jm_j\big(t_j\cdot\nabla_{P'}\mathcal C+\partial_{M'_j}\mathcal C\big)\Big)\phi_0^2+\dots\,.
\end{equation}
On the right, $\mathcal C$ and its derivatives are taken at $P'=c$ and $M'=0$, and the dots have at most one power of $\phi_0$. The flops do not involve $\phi_0$, by \eqref{eq:flopjump}, so the 6d cubic has no $\phi_0^3$, $\phi_0^2\phi_i$ or $\phi_0^2m_j$ terms, and every coefficient in \eqref{eq:Cexpand} vanishes. At zero masses, $\mathcal C$ is homogeneous of degree three in $P'$, so also $c\cdot\nabla_{P'}\mathcal C=3\,\mathcal C=0$. As $c,g_1,g_2$ span $\BR^3$, $\nabla_{P'}\mathcal C=0$ at $P'=c$, and then $\partial_{M'_j}\mathcal C=0$. So the shifts $s\tau$ and $v_j\tau$ produce no $\tau\phi_0^2$ term. Such a term comes only from $m_0\,k(P')$ in \eqref{eq:su40E5d}, which gives $\tfrac\tau r\,k(c)\,\phi_0^2$. Matching the term $\tau\phi_0^2$ of \eqref{eq:so8triE6d} gives $r=k(c)$.
\item \emph{Two conditions on $c$.} The $\phi_0\epsilon_+^2$ terms, $\epsilon_+^2(P'_1+P'_2+P'_3)$ in 5d (\sref{sec:exso8tri}) and $\mathcal K\mathfrak c\,\phi_0\epsilon_+^2=6\phi_0\epsilon_+^2$ in 6d, give $c_1+c_2+c_3=6$. By the \hyperref[it:fluxmap]{flux rule} of \sref{sec:framemap}, the shift $n_0\to n_0+\tfrac12$ between the two 6d sectors shifts the 5d flux by $\tfrac12c$. The 5d 1-form symmetry is the $\BZ_2$ generated by $-1\in\SU(4)$, the only non-trivial central element that acts trivially on $\Lambda^2$ (\sref{sec:setup5d}). So the two 5d sectors are $\BZ^3$ and $\omega_2+\BZ^3$, with $\omega_2$ of \eqref{eq:su40sectors}, and $\tfrac12c\in\omega_2+\BZ^3$, i.e.\ $c\equiv(1,0,1)$ mod $2$.
\item \emph{$k(c)\ge6$.} In the orthonormal components $e$ of \eqref{eq:su40E5d}, $c$ becomes $\hat c=(c_1,c_2-c_1,c_3-c_2,-c_3)$, and $k(c)=\tfrac12\sum_a\hat c_a^2$. By step~2, the entries of $\hat c$ are odd, and $3\hat c_1+2\hat c_2+\hat c_3=c_1+c_2+c_3=6$. They also sum to zero. If all $|\hat c_a|=1$, two entries are $+1$ and two are $-1$, and then $3\hat c_1+2\hat c_2+\hat c_3\in\{0,\pm2,\pm4\}$. So some $|\hat c_a|\ge3$, and $k(c)\ge\tfrac12(9+1+1+1)=6$. Equality needs one entry $\pm3$ and three entries $\mp1$. The condition $3\hat c_1+2\hat c_2+\hat c_3=6$ then leaves $\hat c=(1,1,1,-3)$ or $(3,-1,-1,-1)$, i.e.\ $c=(1,2,3)$ or $(3,2,1)$. $\square$
\end{enumerate}

\paragraph{Exclusion of $n=\pm1$ for $\su(4)_4+4\Lambda^2$.} Take the three matrices $M^{(n)}$ of \eqref{eq:su44Mn}. With $m_0=\tau/r$, the coupling term $m_0\,k(P)$ of \eqref{eq:su40E5d} becomes $\tfrac\tau r\,k(M^{(n)}\phi)$, where
\begin{equation}
\label{eq:su44Q}
k(M^{(n)}\phi)-\phi_0^T\mathcal K\phi_0=\begin{cases}(3\phi_{0,2}-\phi_{0,1}+\phi_1)\,\phi_1\,, & n=0\,,\\ (3\phi_{0,2}-5\phi_{0,1}+7\phi_1)\,\phi_1\,, & n=1\,,\\ (3\phi_{0,2}+3\phi_{0,1}+7\phi_1)\,\phi_1\,, & n=-1\,.\end{cases}
\end{equation}
With the shifts $s$ of \eqref{eq:framemapgen}, $P=M^{(n)}\phi+s\tau=M^{(n)}(\phi+\tau\hat s)$ with $\hat s=(M^{(n)})^{-1}s$. So, inside the 5d cubic, the shifts move the 6d variables by $\tau\hat s$. By Step~1 of \tref{tab:framemaprecipe}, this cubic is $T_2\phi_1^2+\tfrac43\phi_1^3$ for all three $M^{(n)}$, with $T_2$ of \eqref{eq:su44K}. At first order in $\tau$, only the shift $\tau\hat s_3$ of $\phi_1$ gives terms $\phi_{0,\alpha}\phi_1$, namely $2\tau\hat s_3T_2\phi_1$. All three $M^{(n)}$ have $\phi_1=P_2-2P_1$, so $\hat s_3=s_2-2s_1$, and these terms are $\tau X(\phi_{0,1}-3\phi_{0,2})\phi_1$ with $X:=4s_1-2s_2$. The terms $\tau\phi_0^T\mathcal K\phi_0$ fix $1/r=\tfrac12$. The terms $\tau\phi_{0,\alpha}\phi_1$ then come from $\tfrac\tau2$ times \eqref{eq:su44Q} and from the shifts. Let $\gamma_n=-1,-5,3$ be the coefficient of $\phi_{0,1}\phi_1$ in \eqref{eq:su44Q}, for $n=0,1,-1$. The coefficients of $\tau\phi_{0,1}\phi_1$ and $\tau\phi_{0,2}\phi_1$ are then $\tfrac12\gamma_n+X$ and $\tfrac32-3X$. Both must vanish, since \eqref{eq:su44E6d} has no such terms. The second condition gives $X=\tfrac12$, and the first gives $X=-\tfrac12\gamma_n=\tfrac12,\tfrac52,-\tfrac32$. So only $n=0$ is consistent. The matrices of the form \eqref{eq:su44Mn} with $|n|\ge2$ lie outside the scan range. For them, the same steps give $\gamma_n=-1-4n$, with the same coefficient $3$ of $\phi_{0,2}\phi_1$, so they are excluded as well.

\bibliographystyle{JHEP}
\bibliography{bibli}

\end{document}